\documentstyle[12pt]{article}
\newcommand{\mysection}[1]{\setcounter{equation}{0}\section{#1}}

\def\be{\begin{equation}}\def\ee{\end{equation}}\def\l{\label}\def\C{{\cal C}}
\def\F{{\cal F}}\def\M{{\cal M}}\def\S{{\cal S}}
\def\K{{\cal K}}\def\T{{\cal T}}\def\V{{\cal V}} 
\def\Z{{{}}}\def\H{{\cal H}}
\def\O{{\cal O}}\def\P{{\cal P}}

\def\U{{\cal U}}\def\W{{\cal W}}
\def\Q{{\cal Q}}\def\s{{\tt s}}
\def\t{{\tt t}}

\makeatletter\ifcase\@ptsize\font\teneufm=eufm10
\font\seveneufm=eufm7\font\fiveeufm=eufm5
\font\teneusm=eusm10\font\seveneusm=eusm7
\font\fiveeusm=eusm5\or\font\teneufm=eufm10 scaled
\magstephalf\font\seveneufm=eufm7\font\fiveeufm=eufm5
\font\teneusm=eusm10 scaled\magstephalf
\font\seveneusm=eusm7\font\fiveeusm=eusm5\or
\font\teneufm=eufm10 scaled\magstep1\font\seveneufm=eufm7
\font\fiveeufm=eufm5\font\teneusm=eusm10 scaled\magstep1
\font\seveneusm=eusm7\font\fiveeusm=eusm5\fi

\newfam\eufmfam\newfam\eusmfam\textfont\eufmfam=\teneufm
\scriptfont\eufmfam=\seveneufm
\scriptscriptfont\eufmfam=\fiveeufm
\textfont\eusmfam=\teneusm\scriptfont\eusmfam=\seveneusm
\scriptscriptfont\eusmfam=\fiveeusm

\def\frak{\ifmmode\let\next\frak@\else
\def\next{\errmessage{Use\string\frak\space only in math
mode}}\fi\next}\def\frak@#1{{\frak@@{#1}}}
\def\frak@@#1{\fam\eufmfam#1}
\def\sh{\ifmmode\let\next\sh@\else
\def\next{\errmessage{Use\string\sh\space only in math
mode}}\fi\next}\def\sh@#1{{\sh@@{#1}}}
\def\sh@@#1{\fam\eusmfam#1}

\ifcase\@ptsize\font\tenmsa=msam10\font\sevenmsa=msam7
\font\fivemsa=msam5\font\tenmsb=msbm10
\font\sevenmsb=msbm7\font\fivemsb=msbm5\or
\font\tenmsa=msam10 scaled\magstephalf
\font\sevenmsa=msam7\font\fivemsa=msam5
\font\tenmsb=msbm10 scaled\magstephalf
\font\sevenmsb=msbm7\font\fivemsb=msbm5\or
\font\tenmsa=msam10 scaled\magstep1\font\sevenmsa=msam7
\font\fivemsa=msam5\font\tenmsb=msbm10 scaled\magstep1
\font\sevenmsb=msbm7\font\fivemsb=msbm5\fi

\newfam\msafam\newfam\msbfam\textfont\msafam=\tenmsa
\scriptfont\msafam=\sevenmsa
\scriptscriptfont\msafam=\fivemsa\textfont\msbfam=\tenmsb
\scriptfont\msbfam=\sevenmsb
\scriptscriptfont\msbfam=\fivemsb

\def\Bbb{\ifmmode\let\next\Bbb@\else
\def\next{\errmessage{Use\string\Bbb\space only in math
mode}}\fi\next}\def\Bbb@#1{{\Bbb@@{#1}}}
\def\Bbb@@#1{\fam\msbfam#1}\def\hexnumber@#1{\ifnum#1<10
\number#1\else\ifnum#1=10 A\else\ifnum#1=11
B\else\ifnum#1=12 C\else\ifnum#1=13 D\else\ifnum#1=14
E\else\ifnum#1=15 F\fi\fi\fi\fi\fi\fi\fi}
\def\msa@{\hexnumber@\msafam}\def\msb@{\hexnumber@\msbfam}
\mathchardef\square="0\msa@03

\makeatother
\newcommand{\beq}{\begin{equation}}
\newcommand{\eeq}{\end{equation}}
\newcommand{\ba}{\begin{array}}
\newcommand{\ea}{\end{array}}
\newcommand{\bea}{\begin{eqnarray}}
\newcommand{\eea}{\end{eqnarray}}
\newcommand{\bean}{\begin{eqnarray*}}
\newcommand{\eean}{\end{eqnarray*}}
\newtheorem{theorem}{Theorem}[section]

\newtheorem{remark}[theorem]{Remark}

\newtheorem{proof}{Proof.}

\makeatother
\newcommand{\HH}{{\Bbb H}}\newcommand\RR{{\Bbb R}}
\newcommand{\CC}{{\Bbb C}}\newcommand{\PP}{{\Bbb P}}
\newcommand{\ZZ}{{\Bbb Z}}

\newcommand{\II}{{\Bbb I}}

\begin{document}
\begin{titlepage}

\rightline{UMN-TH-1720-98-TPI-MINN-98/15}
\rightline{UFIFT-HEP-97-18, DFPD98/TH/41}
\rightline{September 1998, \tt hep-th/9809127}


\begin{center}

{\Large\bf The Equivalence Postulate of Quantum Mechanics}

\vspace{.133cm}

{\large Alon E. Faraggi$^1$ $\,$and$\,$ Marco Matone$^2$\\}
\vspace{.1in}
{\it $^1$ Department of Physics\\
University of Minnesota, Minneapolis MN 55455, USA\\
e-mail: faraggi@mnhepo.hep.umn.edu\\}
\vspace{.02in}
{\it $^2$ Department of Physics ``G. Galilei'' -- Istituto
Nazionale di Fisica Nucleare\\
University of Padova, Via Marzolo, 8 -- 35131 Padova, Italy\\
e-mail: matone@padova.infn.it\\}

\end{center}

\vspace{.133cm}

\centerline{\large\bf Abstract}


\noindent
The removal of the peculiar degeneration arising in the classical concepts of
{\it rest frame} and {\it time parameterization} is at the heart of the recently
formulated Equivalence Principle (EP). The latter, stating that all physical
systems can be connected by a coordinate transformation to the free one with
vanishing energy, univocally leads to the Quantum Stationary HJ Equation
(QSHJE). This is a third--order non--linear differential equation which
provides a trajectory representation of Quantum Mechanics (QM). The trajectories
depend on the Planck length through hidden variables which arise as initial
conditions. The formulation has manifest $p$--$q$ duality, a consequence of the
involutive nature of the Legendre transformation and of its recently observed
relation with second--order linear differential equations. This reflects in an
intrinsic $\psi^D$--$\psi$ duality between linearly independent solutions of the
Schr\"odinger equation. Unlike Bohm's theory, there is a non--trivial action
even for bound states and no pilot--wave guide is present. A basic property of
the formulation is that no use of any axiomatic interpretation of the
wave--function is made. For example, tunnelling is a direct consequence of the
quantum potential which differs from the Bohmian one and plays the role of
particle's self--energy. Furthermore, the QSHJE is defined only if the ratio
$\psi^D/\psi$ is a local homeomorphism of the extended real line into itself.
This is an important feature as the $L^2(\RR)$ condition, which in the
Copenhagen formulation is a consequence of the axiomatic interpretation of the
wave--function, directly follows as a basic theorem which only uses the
geometrical gluing conditions of $\psi^D/\psi$ at $q=\pm\infty$ as implied by
the EP. As a result, the EP itself implies a dynamical equation that does not
require any further assumption and reproduces both tunnelling and energy
quantization. Several features of the formulation show how the Copenhagen
interpretation hides the underlying nature of QM. Finally, the non--stationary
higher dimensional quantum HJ equation and the relativistic extension are
derived.

\end{titlepage}
\setcounter{footnote}{0}
\renewcommand{\thefootnote}{\arabic{footnote}}

\pagenumbering{roman}

\tableofcontents

\newpage

\mysection{Introduction}\l{intro}

\pagenumbering{arabic}

Twentieth century physics has revolutionized our view and understanding of the
fundamental principles underlying all physical phenomena. At its core, General
Relativity (GR) and Quantum Mechanics (QM) reign unchallenged in their
respective domains. Yet, the mutual coexistence of these two fundamental
theories remains elusive after many decades since their inception. While GR is
based on underlying geometrical concepts, general covariance and the Equivalence
Principle, the traditional formulation of QM does not follow from such
principles. Rather, the conventional formulation of QM is based on concepts
concerning the measurement process and the related axiomatic interpretation of
the wave--function. This indicates that difficulties in understanding the
synthesis of these two fundamental theories trace back to the apparently
different nature of such principles. A deeper comprehension of such principles
constitutes the basic theoretical challenge of the new millennium.

Recently we showed in \cite{1}\cite{1l2} that QM can in fact be derived from an
Equivalence Principle (EP) which is reminiscent of Einstein's EP
\cite{Einstein}. In this paper we will expand and develop the approach
initiated in Refs.\cite{1}\cite{1l2}. We start in sect.\ref{stibd} with a
critical investigation which will lead to formulate the EP. This includes a
basic analysis of the role of time in Classical Mechanics (CM) and of the
distinguished nature of the rest frame. In this process we will reexamine the
basic relations among the canonical variables which are at the heart of
Hamilton--Jacobi (HJ) theory. In particular, we will uncover new algebraic
symmetries which shed surprising new view on the fundamental role of the
canonical variables.

A basic view point of our approach may be regarded as addressing an important
question. In order to formulate it, let us first recall that in HJ theory the
generalized coordinate and momentum $q$ and $p$ are initially regarded as
independent variables. The Classical HJ Equation (CHJE) follows from looking
for canonical transformations leading to the free system with vanishing energy.
The functional dependence among the canonical variables is only extracted
after the HJ equation is solved. Let us now consider a variation of this
approach and suppose that we start, for a stationary system, from $p=\partial_q
\S_0$, with $\S_0$ the Hamiltonian characteristic function, that we will also
call reduced action. Let $\S_0(q)$ and $\S_0^v(q^v)$ be the reduced actions of
two arbitrary physical systems. The question we are considering is to find the
coordinate transformation $q\longrightarrow q^v$ connecting them. This is
reminiscent of HJ theory as it would also imply the existence of the coordinate
transformation reducing an arbitrary physical system to the free one with
vanishing energy. The basic difference with respect to classical HJ theory is
that we are considering coordinate transformations only, with the transformation
of $p$ being induced by the relation $p=\partial_q\S_0$. Of course, to find the
transformation rule of $p$ we need to know how $\S_0$ transforms. On general
grounds we can just choose
\be
\S_0^v(q^v)=\S_0(q),
\l{DIRIFERIMENTO}\ee
which defines the ``$v$--transformation'' (VT) $q\longrightarrow q^v=v(q)=\S_0^{
v^{\;-1}}\circ\S_0(q)$. However, although (\ref{DIRIFERIMENTO}) seems innocuous,
requiring the existence of these transformations for any pair of states, is the
same that imposing equivalence of states under coordinate transformations. We
will show that this leads to QM. For the time being we note that the requirement
that (\ref{DIRIFERIMENTO}) be well--defined for any system rules out CM. This
follows because of the peculiar nature of the particle at rest, as can be
seen from the fact that for the free particle of vanishing energy we have
$\S_0^{cl}=cnst$ and no coordinate transformations can connect a constant to a
non--constant function. As we will see, this feature is related to the fact that
in our formulation, unlike in Bohm's theory \cite{Bohm}, the quantum potential,
as much as $\S_0$, is never trivial. This is a property at the heart of basic
phenomena such as tunnelling and energy quantization.

The impossibility of connecting different states in CM by a coordinate
transformation, a consequence of the peculiar nature of the rest frame,
disappears if one considers time--dependent coordinate transformations.
Observe that in order to introduce time parameterization in HJ theory,
one has to identify the conjugate momentum $p$ with the mechanical one
$m\dot q$. In CM this is equivalent to use Jacobi's theorem $t-t_0=
\partial_E\S_0^{cl}$. Before doing this, time--dependent coordinate
transformations cannot be defined. Describing
dynamics in terms of the functional relation between $p$ and $q$ is a deep
feature of HJ theory. As we said, once time parameterization is
introduced, the peculiar nature of the rest frame disappears. For example,
the effect of the time--dependent coordinate transformation
\be
q'=q-{1\over2}gt^2,
\l{IntroEinsWein}\ee
is that of reducing the motion of a particle in an external gravitational
field, $m\ddot q=mg$, to $m\ddot q'=0$, corresponding to a free particle
of energy $E$. This includes the particle at rest for which $E=0$ and
$\S_0^{cl}=cnst$, showing that there are no distinguished frames if one
uses time--dependent coordinate transformations. Therefore, while with the
dynamical description based on the reduced action it is not always
possible to connect two systems by a coordinate transformation, this is not
the case if one describes the dynamics using Newton's equation. Thus,
formulating the EP can be seen as the natural way to remove this asymmetry
between the role of space and time, a basic topic discussed in sect.\ref{stibd}.

Note that in (\ref{DIRIFERIMENTO}), both the old and new coordinates can be seen
as independent variables in their own systems. Thus, while $q\longrightarrow
q^v$ defines a functional relation between $q^v$ and $q$, the physics of the
old and new systems is described by the functional dependence of $\S_0$ and
$\S_0^v$ on $q$ and $q^v$ respectively. While a canonical transformation maps
an independent pair $(p,q)$ to a new one $(P,Q)$, we will see that requiring
that all systems be related by a coordinate transformation, fixes $\S_0$ to be
solution of the Quantum Stationary HJ Equation (QSHJE), which in turn implies
the Schr\"odinger Equation (SE).

A feature of CM concerns the symmetry between $q$ and $p$, a topic considered in
sect.\ref{pqdatlt}. However, this symmetry, which is reflected in the form of
the Hamilton equations of motion, is in general lost by the choice of a
potential for a given physical system. A central theme of our approach is to
seek a formulation in which duality between the canonical variables is manifest.
We will show that this is provided by the Legendre transformation. In
particular, we will introduce a new generating function
\be
\S_0=p{\partial\T_0\over\partial p}-\T_0,\qquad\qquad
\T_0=q{\partial\S_0\over\partial q}-\S_0,
\l{introlegendre}\ee
{\it i.e.} $q=\partial_p\T_0$. In the stationary case the Hamilton's principal
function can be expressed as $\S=\S_0-Et$. We also have $\T=\T_0+Et$, where
$\T=q\partial_q\S-\S$. In sect.\ref{pqdatlt} we will show that
\be
{\partial\S\over\partial t}=-{\partial\T\over\partial t},
\l{partialSandpartialT}\ee
which holds also in the non--stationary case. This equation guarantees stability
of the $\S$--$\T$ Legendre duality under time evolution. In this context there
appears an imaginary factor which arises by selecting the distinguished, or
self--dual, states defined as those states which are invariant under the
interchange of $q$ and $p$. Thus, loosely speaking, since the transformation is
of order two, one has to consider the ``square root'' of the minus sign in
(\ref{partialSandpartialT}). This results in an imaginary factor appearing in
the self--dual states, which in turn reflects in its appearance in the
expression of $\S_0$ in terms of solutions of the SE. So, Legendre duality,
and, in particular, the minus sign in (\ref{partialSandpartialT}), is at the
heart of the imaginary factor characterizing quantum time evolution.

A sequential important step in our construction concerns the relationship
between second--order linear differential equations and the Legendre
transformation. Its involutive nature reflects in the existence of dual
differential equations. In particular, taking the second derivative with respect
to $\s=\S_0(q)$ and $\t=\T_0(p)$ of the first and second equation in
(\ref{introlegendre}) respectively, we obtain
\be
\left({\partial^2\over\partial\s^2}+\U(\s)\right){{q\sqrt p}\choose\sqrt p}=0,
\qquad\qquad\left({\partial^2\over\partial\t^2}+\V(\t)\right){p\sqrt q\choose
\sqrt q}=0.
\l{introcanoneq}\ee
Apparently, there is no new information in these equations. In particular, since
they are equivalent to $\U(\s)=\{q,\s\}/2$ and $\V(\t)=\{p,\t\}/2$, where $\{f,x
\}=f'''/f'-3(f''/f')^2/2$ denotes the Schwarzian derivative, one may consider
(\ref{introcanoneq}) as definition of $\U$ and $\V$. However, a first signal
that these equations may be relevant, follows from their manifest $p$--$q$
duality, which is a direct consequence of the dual structure of the Legendre
transformations (\ref{introlegendre}). Furthermore, suppose that either $\U$ or
$\V$ is given. In this case one can consider (\ref{introcanoneq}) as equations
of motion to be solved with respect to the ``potential" $\U$ or $\V$. Thus, we
would have a dynamical description with manifest $p$--$q$ duality. Investigation
of these equations will shed light on our formulation of QM. In particular,
using the transformation properties of the Schwarzian derivative, we see that
the ``canonical potential" $\U(\s)$ and the quantum potential $Q=\hbar^2\{\S_0,q
\}/4m$, which, as we will see, appears in the QSHJE, satisfy the simple relation
\be
2mQ+\hbar^2p^2\U=0.
\l{bella}\ee
The issue of $p$--$q$ duality has an old origin (see {\it e.g.} Born's paper
\cite{Bornpq}) and is a matter of research also in recent literature. Actually,
the above relation between the Legendre transformation and second--order linear
differential equations, which can in fact be seen to arise from the basic
definition of generating function, was introduced in \cite{M1} in the framework
of Seiberg--Witten theory \cite{SW}. In particular, this relation was introduced
as a way of inverting the solution for the moduli parameter in Seiberg--Witten
theory \cite{SW} from $a(u)$ to $u(a)$, where $a$ is the VEV of the scalar
component of the $N=2$ adjoint superfield and $u=\langle{\rm tr}\,\phi^2\rangle$
is the gauge invariant parameter. In Ref.\cite{FM}, inspired by the use of the
inversion formula in Seiberg--Witten theory \cite{M1}, we applied the same
procedure to the SE. We introduced in QM a prepotential function $\F$, defined
by $\psi^D=\partial_\psi\F$, where $\psi^D$ and $\psi$ are two linearly
independent solutions of the SE. Thereby, we showed that in QM the spatial
coordinate can be regarded as the Legendre transform of $\F$ with respect to
the probability density, a proposal further investigated in
\cite{Carroll}\cite{Vancea}. This relationship between the Legendre
transformation and second--order linear differential equations, which shows
a $\psi^D$--$\psi$ duality just as the $p$--$q$ duality implied
(\ref{introcanoneq}), can be better understood by
considering the following basic question posed in \cite{M1}

\vspace{.233cm}

\noindent
{\it Given a second--order linear differential equation, with linearly
independent solutions $\psi^D$ and $\psi$, find the function $\F(\psi)$
$(\F_D(\psi^D))$ such that $\psi^D=\partial\F/\partial\psi$ $(\psi=
\partial\F_D/\partial\psi^D)$.}

\vspace{.233cm}

\noindent
The answer is that the variable $q$ of the differential equation is
proportional to the Legendre transform of $\F$ with respect to $\psi^2$.
Posing the same question with $\psi^D$ and
$\psi$ replaced by $q\sqrt p$ and $\sqrt p$, one sees that the corresponding
$\F$ is proportional to the dual reduced action $\T_0$ and
\be
2q\sqrt p={\partial\T_0\over\partial\sqrt p}.
\l{insommadunquemalodicevoiocheeraoutstandingsolochehalescarpebrutte}\ee
The association between the Legendre transformation and second--order
linear differential equations is seen to provide complementary ways to
access information on a physical theory, with the symmetry properties of the
Legendre transformation and its dual reflecting in symmetry properties of the
associated differential equations. It was shown in \cite{BOMA} that there are
natural ``Legendre brackets'' underlying this association. Let us consider a
``generating function'' $U(z)$ and its Legendre transform $V(w)=z\partial_zU(z)
-U(z)$, where $w=\partial_zU$ (note that $z=\partial_wV$). Set $u=U(z)$ and
\be
\tau={\partial z\sqrt w\over\partial\sqrt w}=
z+2{w\over\partial_zw}={1\over2}{\partial^2V\over\partial{\sqrt w}^2}.
\l{taurus}\ee
The Legendre brackets are \cite{BOMA}
\be
\{X,Y\}_{(u)}=(\partial_u\tau)^{-1}\left({\partial X\over\partial\sqrt w}
{\partial\over\partial u}{\partial Y\over\partial\sqrt w}-{\partial Y\over
\partial\sqrt w}{\partial\over\partial u}{\partial X\over\partial\sqrt
w}\right).
\l{bomas}\ee
If the generating function is $\S_0(q)$, then
the dual solutions of the first equation in (\ref{introcanoneq}) satisfy
\be
\{\sqrt p,\sqrt p\}_{(\s)}=0=\{q\sqrt p,q\sqrt p\}_{(\s)},\qquad
\{\sqrt p,q\sqrt p\}_{(\s)}=1.
\l{bracketsss1}\ee
Similarly, considering the generating function $\T_0(p)$, we have
\be
\{\sqrt q,\sqrt q\}_{(\t)}=0=\{p\sqrt q,p\sqrt q\}_{(\t)},\qquad
\{\sqrt q,p\sqrt q\}_{(\t)}=1.
\l{bracketsss2}\ee
Repeating the construction with $\F(\psi^2)$, we obtain
\be
\{\psi,\psi\}_{(q)}=0=\{\psi^D,\psi^D\}_{(q)},\qquad\quad\;
\{\psi,\psi^D\}_{(q)}=1.
\l{bracketsss3}\ee

Observe that (\ref{DIRIFERIMENTO}), under which $p$ transforms as
$\partial_q$, can be seen as the invariance of the Legendre transform of
$\T_0(p)$ under VTs.\footnote{In sect.\ref{ld} we will see that this
structure admits a dual formulation in which one considers the
$u$--transformations $p\longrightarrow p^u=u(p)$ defined by $\T_0^u(p^u)=\T_0
(p)$ which is the dual version of (\ref{DIRIFERIMENTO}).} This is reflected
in the covariance of the associated differential equation. In general, these
equations have a M\"obius symmetry which is exhibited by the representation
of the potential as a Schwarzian derivative. However, the potential does not
remain invariant if the transformations are extended to general coordinate
transformations. This indicates that different physical systems,
characterized by different potentials, can be connected by coordinate
transformations. This fact, based on $p$--$q$ duality, and the necessity of
removing the asymmetry between space and time, which appears in considering
the specific role of the rest frame in CM, constitute the basic motivations
to formulate the EP. The relationship between duality and time
parameterization is investigated in sect.\ref{pqdatlt}. These two sections,
together with sections \ref{gl2csatce}--\ref{sds}, devoted to the algebraic
structures underlying $p$--$q$ duality, constitute the investigation
culminating with the EP formulated in sect.\ref{tep}. In this respect we
stress that sections \ref{pqdatlt}--\ref{tep} considerably extend the
structures we introduced in \cite{1}\cite{1l2}\cite{3prima}. The content of
the equivalence postulate is \cite{1}\cite{1l2}

\vspace{.133cm}

\noindent
{\it For each pair $\W^a\equiv V^a(q^a)-E^a$ and $\W^b\equiv V^b(q^b)-E^b$,
there is a VT $q^a\longrightarrow q^b=v(q^a)$ such that}
\be
\W^a(q^a)\longrightarrow{\W^a}^v(q^b)=\W^b(q^b).
\l{equivalenceanticipo}\ee

\vspace{.133cm}

\noindent
Note that according to this postulate, all physical systems can be mapped to
the one corresponding to $\W^0\equiv 0$. In this respect we stress that while
we do not define a priori to which spaces the quantities $\W$ and $\S_0$
belong, these will be determined by the request that the QSHJE be defined.

A key point in \cite{1} concerns the transformation properties of the
Classical Stationary HJ Equation (CSHJE). These properties, discussed in
sect.\ref{tep}, can be summarized as follows

\vspace{.233cm}

\noindent
Given two physical systems with reduced actions $\S_0^{cl\,v}$
and $\S_0^{cl}$, consider the {\it classical} $v$--map
\be
q\longrightarrow q^v=v(q)=\S_0^{{cl\,v}^{\;-1}}\circ\S_0^{cl}(q),
\l{insomminino}\ee
that is $\S_0^{cl\,v}(q^v)=\S_0^{cl}(q)$. Then, from the CSHJEs
\be
{1\over2m}\left({\partial\S_0^{cl\,v}(q^v)\over\partial q^v}\right)^2+
\W^v(q^v)=0,
\qquad\qquad
{1\over2m}\left({\partial\S_0^{cl}(q)\over\partial q}\right)^2+\W(q)=0,
\l{aggiuntinatantopergradirla1}\ee
it follows that under {\it classical} $v$--maps $\W(q)\longrightarrow\W^v(q^v)=
\left(\partial_{q^v}q\right)^2\W(q)$. In particular
\be
\W^0(q^0)\longrightarrow\left(\partial_{q^v}q^0\right)^2\W^0(q^0)=0.
\l{aggiuntinatantopergradirla3}\ee

\vspace{.233cm}

\noindent
Hence, in CM it is not possible to connect by a coordinate transformation the
state $\W^0$ to other states with $\W\ne\W^0$. Of course, this is the same fact
we observed in considering the degeneration which arises for $\S_0^{cl}=cnst$.
Nevertheless, the above observation clearly indicates that the EP requires a
modification of the CSHJE. We therefore add to the CSHJE a function $Q(q)$ to
be determined. Existence of CM imposes that in some limit $Q\longrightarrow0$.
Repeating the above analysis with the additional term $Q$, we see that it is
now the term $\W+Q$ that should transform as a quadratic differential under
$v$--maps. That is, as discussed in sect.\ref{tep}, we have
\be
\left(\W^v(q^v)+Q^v(q^v)\right)(dq^v)^2=\left(\W(q)+Q(q)\right)(dq)^2.
\l{adfst2Hre}\ee
This property, together with the request that all possible states be
connected by a coordinate transformation to the trivial one corresponding to
$\W^0\equiv 0$, is the starting point of the investigation in sect.\ref{itep}.
In particular, this property indicates that $\W$ and $Q$ pick an inhomogeneous
term under coordinate transformations. Therefore, we have
\be
\W^v(q^v)=\left(\partial_{q^v}q^a\right)^2\W^a(q^a)+\Z(q^a;q^v),
\l{wvqv}\ee
and
\be
Q^v(q^v)=\left(\partial_{q^v}q^a\right)^2Q^a(q^a)-\Z(q^a;q^v).
\l{qvqv}\ee
Considering the consistency condition for which the transformation
connecting $\W^a$ and $\W^c$ should be equivalent to the composition of the two
transformations mapping $\W^a$ into $\W^b$, and then $\W^b$ into $\W^c$, we
arrive to the basic cocycle condition \cite{1}
\be
\Z(q^a;q^c)=\left(\partial_{q^c}q^b\right)^2\left[\Z(q^a;q^b)-\Z(q^c;q^b)
\right].
\l{inhomtrans}\ee
In sect.\ref{itep} we will show that (\ref{inhomtrans}) implies a basic M\"obius
invariance of $\Z(q^a;q^b)$. Next, in sect.\ref{tqhje} we will first show that
this symmetry fixes $\Z(q^a;q^b)\propto\{q^a,q^b\}$. More precisely, we will
prove that

\vspace{.333cm}

\noindent
{\it The cocycle condition (\ref{inhomtrans}) uniquely defines the Schwarzian
derivative up to a global constant and a coboundary term.}

\vspace{.333cm}

\noindent
We will see that this univocally implies that $Q=\beta^2\{\S_0,q\}/4m$ and
therefore the QSHJE. In sect.\ref{theschr} we will show that the QSHJE implies
the SE once $\beta$ is identified with Planck's reduced constant $\hbar$.
Therefore, we will arrive to the remarkable conclusion that one necessitates the
appearance of QM if the EP is to be implemented consistently.

In sect.\ref{tqshjeatcl} we will show that the formulation has a well--defined
classical limit. We will also examine the relation with other versions of the
QSHJE, in particular with Bohm's representation. In sect.\ref{tweocm} we will
derive the representation of the CSHJE in terms of a modified SE with the
resulting classical wave--function interpreted in terms of probability amplitude.

In sect.\ref{tseattm} the trivializing transformations, mapping the
SE with non--trivial potentials to the SE with the trivial one, will be derived
and discussed in analogy with the trivializing canonical transformations in the
classical HJ formalism. In this context we consider the basic analogy with
uniformization theory of Riemann surfaces.

The fact that QM may in fact arise from our EP suggests a reconsideration of the
role of the potential. Just as in GR the Equivalence Principle leads to a
deformation of the geometry induced by the stress tensor, also in QM it should
be possible to relate the EP to a deformation of the geometry induced by the
potential. In sect.\ref{tseattm} we will follow \cite{3prima} to show that both
the classical and quantum potentials are curvature terms
arising in projective geometry. Furthermore, we will see that the QSHJE takes
the classical form with the spatial derivative $\partial_q$
replaced by $\partial_{\hat q}$, where ($\beta^2=\hbar^2\U$)
\be
d\hat q={dq\over\sqrt{1-\beta^2(q)}}.
\l{odEPoinc}\ee
We will conclude sect.\ref{tseattm} by showing that the framework
of projective geometry considered in \cite{3prima}, allows us to express the
Heisenberg commutation relations by means of the area function.

The outcome of the EP is that the CSHJE is modified by a uniquely determined
function
\be
{1\over2m}\left({\partial\S_0\over\partial q}\right)^2+V(q)-E+{\hbar^2\over
4m}\{\S_0,q\}=0,
\l{intqhje}\ee
where, as we will see in sect.\ref{tqshjeatcl}, the term $Q=\hbar^2\{\S_0,q\}/
4m$ can be identified as the genuine quantum potential which plays the role of
particle's self--energy. Eq.(\ref{intqhje}) is related to the one considered in
Bohm's approach to QM \cite{Bohm}. However, $\S_0$ differs from Bohm's phase in
important ways. In particular, as a result of the EP, $\S_0$ is never a
constant. This reflects in the fact that a general solution of the SE, and
therefore also the
wave--function, will have the form\footnote{In the following, unless otherwise
specified, by wave--function we will mean the Hamiltonian eigenfunction.}
\be
\psi={1\over\sqrt{\S_0'}}\left(A e^{-{i\over\hbar}\S_0}+Be^{{i\over\hbar}
\S_0}\right).
\l{opiYh9}\ee
The basic difference with Bohm's theory \cite{Bohm} arises for bound states. In
this case the wave--function is proportional to a real function. This implies
that with Bohm's identification $\psi=R\exp(i\S_0/\hbar)$,
which is also usually considered in standard textbooks, one would have $\S_0
=cnst$.\footnote{To be precise, bound states would correspond to $\S_0=cnst$
outside the nodes of the wave--function.} This seems an unsatisfactory feature
of Bohm's theory as there are difficulties in getting a non--trivial classical
limit for $\S_0=cnst$, and therefore for $p=0$. Whereas this would be
consistent in the case of classically forbidden regions as there
$\S_0^{cl}=cnst$, one has troubles when considers regions
which are no classically forbidden. In this case the reduced classical
action is not trivial whereas with the standard identification
$\psi=R\exp(i\S_0/\hbar)$, one would have $\S_0=cnst$.
So, for example, while quantum mechanically the
particle would be at rest, after taking the $\hbar\to 0$ limit,
the particle should start moving. This problem
disappears if one uses Eq.(\ref{opiYh9}) which directly follows from the QSHJE:
the fact that $\psi\propto\bar\psi$ simply implies that $|A|=|B|$ and $\S_0\ne
cnst$. Furthermore, we would like to remark that in Bohm's approach some
interpretational aspects are related to the concept of pilot--wave guide.
There is no need for this in the present formulation. Nevertheless, there are
some similarities between our formulation and Bohm's interpretation of QM. In
particular, by solving the QSHJE for $\S_0$, we can evaluate $p=\partial_q\S_0$
as a function of the initial conditions, so that we have a determined orbit in
phase space. The solutions of the QSHJE, which is a third--order non--linear
differential equation, are obtained by utilizing two real linearly
independent solutions of the corresponding SE. In particular, denoting by
$\psi^D$ and $\psi$ these solutions, we will see in sect.\ref{cvamt} that
\be
e^{{2i\over\hbar}\S_0\{\delta\}}=e^{i\alpha}{w+i\bar\ell\over w-i\ell},
\l{ints0}\ee
where $\delta=\{\alpha,\ell\}$, with $\alpha$ and $\ell$, ${\rm Re}\,\ell\ne0$,
integration constants, and
\be
w={\psi^D\over\psi}\in\RR.
\l{wintroduzione}\ee
Denoting by $W=\psi'\psi^D-{\psi^D}'\psi=cnst$ the Wronskian, we have
\be
p={\hbar W (\ell+\bar\ell)\over2\left|\psi^D-i\ell\psi\right|^2}.
\l{intp}\ee

In sect.\ref{cvamt} we will consider the effect of the transformation of the
initial conditions $\alpha$ and $\ell$ on the wave--function. In particular,
following the second reference in \cite{1}, we will see that there are
transformations of $\alpha$ and $\ell$ that leave the wave--function invariant.
This is another important signal of the general fact that the SE contains less
information than the QSHJE.

A basic geometrical quantity of the formalism concerns the {\it trivializing
map} reducing the system with a given $\W$ to the one corresponding to $\W^0$.
In sect.\ref{tatep} we will see that this map is expressed in terms
of a M\"obius transformation of the ratio $w=\psi^D/\psi$, whose coefficients
depend on the intial conditions associated to the states $\W^0$ and $\W$. The
role of the phase $\alpha$ reflects in the possibility of reducing this
transformation of $w$ to an affine one. More generally, we will see that the
trivializing map has a M\"obius symmetry.

A property of the formulation is that $p=\partial_q\S_0\ne m{\dot q}$. In
particular, in sect.\ref{tatep} we will follow Floyd \cite{Floyd82b} who
defined time parameterization by means of Jacobi's theorem, according to which
$t-t_0=\partial_E\S_0$. Using the expression for $\dot q$, we set $p=m_Q{\dot
q}$, where $m_Q=m(1-\partial_E Q)$ can be seen as an effective quantum mass.
Furthermore, we will rewrite the QSHJE (\ref{intqhje}) in terms of $m_Q$ and
its derivatives with respect to $q$, and in terms of $q$ and its time
derivatives. This is the dynamical quantum
equation for $q$. An interesting property of this new equation concerns the
appearance of the third--order time derivative of $q$. This is of course related
to the appearance of the integration constant $\ell$. For this reason, we can
consider ${\rm Re}\,\ell$ and ${\rm Im}\,\ell$ as a sort of hidden variables.

We stress that the consistent implementation of the EP is related to the
existence of the Legendre transformation for all physical states. This brings to
another basic observation concerning the existence of the self--dual states
mentioned above and discussed in sect.\ref{sds}. These are the states that
remain invariant under the M\"obius transformations of $q$ and $p$ corresponding
to the common symmetry of the Legendre transformation and of its dual.
Alternatively, the self--dual states are precisely those states which are
simultaneous solutions of the two dual differential equations
(\ref{introcanoneq}) associated to the two dual Legendre transformations
(\ref{introlegendre}). From the view point of the QSHJE, we will see that among
the self--dual states there is a reference state which precisely corresponds to
the trivial one $\W^0\equiv 0$. While in the classical case this corresponds to
$\S_0^{cl}=cnst$, we will see that in the quantum case we always have $\S_0\ne
cnst$. On the other hand, since the free particle with $E=0$ can be obtained
{}from the one with $E\ne 0$ in the $E\longrightarrow0$ limit, we have
that even the more general case $\S_0=Aq+B$, for which Legendre duality breaks
down, is ruled out. Since the QSHJE contains the Schwarzian derivative of $\S_0$,
we have that existence of the QSHJE and the exclusion of the solution $\S_0=Aq
+B$, imply that the Legendre transformation is defined for any state. Following
\cite{3}, we will see in sect.\ref{epafc} that considering the classical limit
and the $E\longrightarrow0$ limit of the free particle, leads to introduce a
dependence of the initial conditions, or hidden variables, ${\rm Re}\,\ell$ and
${\rm Im}\,\ell$, of the QSHJE on the Planck length. This
suggests a connection between GR and our formulation of QM.

Besides the tunnel effect, another crucial phenomenon of QM concerns energy
quantization. In our formulation this arises as a consequence of the existence
of the QSHJE and therefore is a direct consequence of the EP itself! In
particular, since the identity
\be
(\partial_q\S_0)^2={\hbar^2\over2}(\{e^{{2i\over\hbar}\S_0},q\}-\{\S_0,q\}),
\l{oiqn}\ee
implies that the QSHJE is equivalent to the Schwarzian equation
\be
\{w,q\}=-{4m\over\hbar^2}(V-E),
\l{introwq}\ee
one has that the QSHJE is well--defined if and only if $w$ is a local
self--homeomorphism of the extended real line $\hat\RR=\RR\cup\{\infty\}$.
To show this we first note that existence of Eq.(\ref{introwq}) implies
\be
w\ne cnst,\;w\in C^2(\RR),\;and\;\partial_q^2w\;differentiable\;on\;\RR.
\l{senzahat}\ee
Observe that by (\ref{wintroduzione}) these conditions, which arise from the EP,
imply the continuity of both $\psi^D$, $\psi$ and their first derivative, with
$\partial_q\psi^D$ and $\partial_q\psi$ differentiable. Therefore, we have
\be
Equivalence\;Principle\;\longrightarrow\;continuity\;of\;(\psi^D,\psi),
\;and\;(\partial_q\psi^D,\partial_q\psi)\;differentiable.
\l{introequivalenzaederivata}\ee
Note that the QSHJE, and therefore (\ref{introwq}),
is a consequence of the cocycle condition (\ref{inhomtrans}) that should be
always satisfied. In particular, Eq.(\ref{introwq}) must be equivalent to
$\{w,q^{-1}\}=-4mq^4(V-E)/\hbar^2$. On the other hand, since under the map
$q\longrightarrow q^{-1}$, the point $0^-$ ($0^+$) maps to $-\infty$
($+\infty$), we have that gluing $0^-$ to $0^+$ corresponds gluing $-\infty$ to
$+\infty$. This means that (\ref{senzahat}) extends to $\hat\RR$, {\it i.e.}
\be
w\ne cnst,\;w\in C^2(\hat\RR),\;and\;\partial_q^2w\;differentiable\;on\;\hat\RR.
\l{introccnn}\ee
We now show that there is another relevant condition. Namely, equivalence under
coordinate transformations implies that these transformations must be locally
invertible. A property which can be also seen from the cocycle condition
(\ref{inhomtrans}). This amount to require that transformations such as
$\{w,q\}=-(\partial_qw)^2\{q,w\}$ be well--defined, so that $w(q)$ should be
locally invertible, that is $\partial_qw\ne 0$, $\forall q\in\RR$. On the
other hand, considering the ``$S$--transformation'' $q\longrightarrow q^{-1}$,
one sees that also this condition of local univalence should be extended to
$\hat\RR$. It is easy to see that this reflects in the following
joining conditions for $w$ at spatial infinities
\be
w(-\infty)=\left\{\begin{array}{ll} w(+\infty),& for\,w(-\infty)\ne\pm
\infty,\\ -w(+\infty),& for\,w(-\infty)=\pm\infty.\end{array}\right.
\l{introspecificandoccnn}\ee
As we will see, this condition is at the heart of energy quantization. Observe
that, as illustrated by the non--univalent function $w=q^2$, making the choice
$w(-\infty)=w(+\infty)$ in the case in which $w(-\infty)=\pm\infty$, would
break local univalence. The above analysis is the essence of the proof that,
according to the EP, $w$ is a local self--homeomorphism of $\hat\RR$.

The physical interest underlying the above results is due to the following
basic theorem \cite{1l2} that we will review in sect.\ref{epqsat}. Let us
denote by $q_-$ ($q_+$) the lowest (highest) $q$ for which $V(q)-E$ changes
sign. We have that

\vspace{.233cm}

\noindent
{\it If}
\be
V(q)-E\geq\left\{\begin{array}{ll}P_-^2 >0,&q<q_-,\\
P_+^2 >0,&q> q_+,\end{array}\right.
\l{perintroasintoticopiumeno}\ee
{\it then $w$ is a local self--homeomorphism of $\hat\RR$
if and only if the corresponding SE has an $L^2(\RR)$ solution.}

\vspace{.233cm}

\noindent
Thus, since the QSHJE is defined if and only if $w$ is a local
self--homeomorphism of $\hat\RR$, this theorem implies that energy
quantization {\it directly} follows from the QSHJE itself. Therefore, while in
the usual approach to QM one needs the SE with the {\it additional} $L^2(\RR)$
condition, in our formulation the EP itself implies a dynamical equation that
does not require any further assumption and reproduces both the tunnel effect
and energy quantization. Another feature of our approach is that, as we
will see in sect.\ref{epqsat}, the geometrical properties of $w$ define the
space of admissible potentials.

It is useful to provide an explicit example illustrating some features of our
approach. Other relevant examples are considered in sect.\ref{pwho}. We
consider the potential well
\be
V(q)=\left\{\begin{array}{ll}0,&|q|\leq L,\\ V_0,&|q|> L.\end{array}\right.
\l{introVu1}\ee
According to (\ref{ints0}), in order to solve Eq.(\ref{intqhje}) we need two
real linearly independent solutions of the SE. Let us set $k=\sqrt{2mE}/\hbar$,
$\kappa=\sqrt{2m(V_0-E)}/\hbar$ and $Q_\pm=\kappa(q\pm L)$. A solution of the
SE is
\be
\psi=k^{-1}\cdot\left\{\begin{array}{ll}-a\exp Q_+-b\exp-Q_+,& q<-L,\\
\sin(kq),&|q|\leq L,\\ a\exp-Q_-+b\exp Q_-,&q>L,\end{array}\right.
\l{introcasogenerale1}\ee
where for any $E\geq0$ the continuity conditions
(\ref{introequivalenzaederivata}), implied by the EP, give
\be
a={1\over2}\sin(kL)-{k\over2\kappa}\cos(kL),\qquad
b={1\over2}\sin(kL)+{k\over2\kappa}\cos(kL).
\l{introcasogeneraledue}\ee
It is easy to see that any solution of the SE can be expressed as a linear
combination of $\psi$ and
\be
\psi^D=\left\{\begin{array}{ll}c\exp Q_++d\exp-Q_+,&q<-L,\\ \cos(kq),&|q|
\leq L,\\ c\exp-Q_-+d\exp Q_-,&q>L,\end{array}\right.
\l{introcasogenerale3}\ee
where
\be
c={1\over2}\cos(kL)+{k\over2\kappa}\sin(kL),\qquad
d={1\over2}\cos(kL)-{k\over2\kappa}\sin(kL).
\l{introcasogeneralequattro}\ee
The ratio of the solutions is given by
\be
w=k\cdot\left\{\begin{array}{ll}-(c\exp Q_++d\exp-Q_+)/(a\exp Q_++b\exp-Q_+),
&q<-L,\\ \cot(kq),&|q|\leq L,\\ (c\exp-Q_-+d\exp Q_-)/(a\exp-Q_-+b\exp Q_-),&
q>L.\end{array}\right.
\l{introcasogenerale5}\ee
Since the asymptotic behavior of $w$ is
\be
\lim_{q\longrightarrow\pm\infty}w=\pm{d\over b}k,
\l{introlimitidoversisplendido2generale}\ee
we immediately see that the gluing conditions at $\pm\infty$,
at $q=\pm\infty$, Eq.(\ref{introspecificandoccnn}), imply that either
\be
b=0,
\l{introariprovace}\ee
so that $w(-\infty)=-{\rm sgn}\,d\cdot\infty=-w(+\infty)$, or
\be
d=0,
\l{introtoprocnc}\ee
so that $w(-\infty)=0=w(+\infty)$. Eq.(\ref{introariprovace}) and
(\ref{introtoprocnc}) precisely correspond to the spectrum derived in the
usual approach. In particular, according to the above general theorem, the
gluing conditions (\ref{introariprovace}) and (\ref{introtoprocnc}) correspond
to the unique cases in which the SE has an $L^2(\RR)$ solution. Furthermore,
note that if $b=0$, then $\psi\in L^2(\RR)$ (and, of course, $\psi^D\notin L^2
(\RR)$), while if $d=0$, then $\psi^D\in L^2(\RR)$ (and $\psi\notin L^2(\RR)
$).\footnote{Manifest duality of our formulation, reflects in the appearance in
many formulas of both $\psi^D$ and $\psi$. As illustrated in sect.\ref{theschr},
this is the effect of the underlying Legendre duality which reflects in the
$\psi^D$--$\psi$ duality. In this context we note that the above example
explicitly shows a general fact that, even if well--known, should be stressed.
Namely, since the SE is a second--order linear differential equation, it always
has a pair of linearly independent solutions. In particular, by Wronskian
arguments, illustrated in sect.\ref{epqsat}, it can be seen that if the SE has an
$L^2(\RR)$ solution, then any other linearly independent solution is divergent
at $q=\pm\infty$. Therefore, the fact that for some system the wave--function
belongs to the $L^2(\RR)$ space, does not mean that for the same $E$ the SE has
no divergent solutions. In other words, one should not confuse uniqueness of the
wave--function for bound states, which is the well--known non--degeneration
theorem of the spectrum for bound states, with the wrong one: ``the SE for bound
states has only one solution". Also note that the fact that the divergent
solution cannot be the wave--function does not imply that it cannot appear in
relevant expressions which in fact characterize the present formulation.} Let us
find the explicit form of the trivializing map in the case of the potential well.
A pair of linearly independent solutions of the SE with $V-E=0$, is given by
$\psi^{D^0}=q^0$, $\psi^0=1$. Therefore, $\exp(2i\S_0^0/\hbar)=\exp(i\alpha_0)
(q^0+i\bar\ell_0)/(q^0-i\ell_0)$, and by (\ref{DIRIFERIMENTO})
\be
e^{i\alpha_0}{q^0+i\bar\ell_0\over q^0-i\ell_0}=
e^{i\alpha}{w+i\bar\ell\over w-i\ell}.
\l{introKdT301}\ee
Since $\alpha_0$ and $\alpha$ have no effect on the conjugate
momenta, we set $\alpha=\alpha_0+2\pi k$ to have (see sect.\ref{tatep})
\be
q^0={(\ell_0+\bar\ell_0)w+i\ell_0\bar\ell-i\bar\ell_0\ell\over\ell+\bar\ell}.
\l{introqzeroo}\ee
Let us consider the potential well with $b=0$. We have
\be
q^0={k\over\ell+\bar\ell}\cdot\left\{\begin{array}{ll}-a^{-1}(\ell_0+\bar\ell_0
)(c+de^{-2Q_+})+i\ell_0\bar\ell-i\bar\ell_0\ell,&q<-L,\\(\ell_0+\bar\ell_0)\cot
(kq)+i\ell_0\bar\ell-i\bar\ell_0\ell,&|q|\leq L,\\a^{-1}(\ell_0+\bar\ell_0)(c+d
e^{2Q_-})+i\ell_0\bar\ell-i\bar\ell_0\ell,&q>L,\end{array}\right.
\l{introesempione}\ee
where
\be
a=-{k\over\kappa}\cos(kL),\qquad c={\kappa^2-k^2\over2\kappa^2}\cos(kL),\qquad
d={\kappa^2+k^2\over2\kappa^2}\cos(kL).
\l{acdquandob0}\ee
One sees that $q^0$ satisfies all the
properties of the trivializing map. In particular, we have
\be
\{q^0,q\}=-4m(V-E)/\hbar^2.
\l{leregoleehchedianime}\ee
As follows by (\ref{wvqv}) with $\W^a(q^a)=0$, this is in accordance with the
general fact that any $\W$ can be expressed in terms of the inhomogeneous part
in the transformation properties of $\W^0$.

We now consider a further simple example which clearly shows the relevance of
the continuity at spatial infinity. In particular, we provide an example of the
general fact, proved in sect.\ref{pwho}, according to which if $V(q)>E$,
$\forall q\in\RR$, then there are no admissible solutions of the QSHJE. In
other words, if there are no classically accessible regions in $\RR$, then this
also happens quantum mechanically. In the usual approach the absence of
solutions when $V-E>0$, $\forall q\in\RR$ is, once again, a consequence of the
axiomatic interpretation of the wave--function. Thus we have to understand how
this fact emerges (note that we have no turning points, so the previous theorem
does not concern this case). The example we consider is the QSHJE with
$V-E=a^2$. The associated SE
\be
{\hbar^2\over2m}\partial_q^2\psi=a^2\psi,
\l{introesempioill2}\ee
has the following linearly independent solutions $\psi^D=Ae^{aq}+Be^{-aq}$,
$\psi=Ce^{aq}+De^{-aq}$, $AD-BC\ne 0$. Their ratio $w=(Ae^{2aq}+B)/(Ce^{2aq}+
D)$, has the asymptotics (we consider $a>0$)
\be
\lim_{q\longrightarrow-\infty}w={B\over D},\qquad
\lim_{q\longrightarrow+\infty}w={A\over C},
\l{VxFTg}\ee
so that neither the $w(-\infty)=finite=w(+\infty)$ case, nor $w(-\infty)=-w(+
\infty)=\pm\infty$ can occur. Hence, for any $a\in\RR\backslash\{0\}$, $w$ is
not a local homeomorphism of $\hat\RR$ into itself and therefore the QSHJE does
not admit solutions. This means that the EP cannot be implemented in this case.

A peculiarity of QM is that many of its characterizing properties, such as
tunnel effect, energy quantization, Hilbert space structure and Heisenberg
uncertainty relations, already appear in one spatial dimension. Thus it is not
surprising that our formulation extends to the higher dimensional case. This
generalization, including the time--dependent case and the relativistic
extension, will be investigated in \cite{BFM} and is shortly considered in
sect.\ref{tdc}.

In sect.\ref{conclusions} we will make some concluding comments concerning
the results obtained and possible developments of our formulation.

We conclude this Introduction by making some bibliographic remarks. First, we
note that geometrical concepts have been used in relevant approaches to QM such
as geometric quantization, developed by Kirillov, Konstant, Guillemin, Soriau,
Sternberg et al. (see {\it e.g.} \cite{geometricquantization}), and coherent
states quantization (see {\it e.g.} \cite{KlauderPerelomov}). Furthermore,
besides Bohm's theory, another interesting formulation based on the quantum HJ
equation, seen as describing the Madelung fluid, concerns the stochastic
quantization initiated by Nelson and further developed by Guerra et al. (see
{\it e.g.} \cite{NelsonGuerra}). As shown by Gozzi \cite{Gozzi}, HJ theory
appears in QM also in deriving the SE from the anomalous conservation law
associated to the overall rescaling of the Lagrangian. We also mention Periwal's
paper \cite{Periwal}, where a version of the quantum HJ equation has been
obtained in the path--integral framework in a way related to our approach.
Furthermore, we note that an EP in the framework of QM has
been recently considered by Anandan \cite{Anandan}, while an interesting
investigation of Einstein's EP in QM has been done in \cite{OnofrioViola}, where
a free fall in a constant gravitational field is considered in the case of
quantum states with and without classical analogue. Another interesting approach
to QM has been recently considered by Cini \cite{Cini} who formulated QM as a
generalization of Classical Statistical Mechanics, and in which it is used
Feynman's proposal of dropping the assumption that the probability for an event
must always be a positive number. Last but not least, we mention some of the
relevant approaches to QM proposed in the last decade, such as the
decoherent histories \cite{GellMannHartle}, quantum state diffusion
\cite{GisinPercival}, quaternionic QM \cite{Adler} and the GRW spontaneous
localization model of wave--function's collapse \cite{GRWP}.

\mysection{Canonical and coordinate transformations}\l{stibd}

In this section we first consider basic properties of canonical transformations
in CM. This analysis will bring us to consider the problem of finding the
coordinate transformations connecting different systems in the case in which
$p$ is considered as {\it dependent} on $q$. In this context we will show how in
CM the rest frame is a privileged one. Existence of this frame can be seen as a
possible motivation for introducing time parameterization of trajectories: this
makes it possible to define non singular, although time--dependent, coordinate
transformations connecting different systems. This investigation, based on the
peculiar nature of the rest frame of CM, is the first step towards the
modification of classical HJ theory that will lead to a formalism with manifest
$p$--$q$ duality which, as we will see, underlies the EP. We conclude this
section by considering a geometrical picture in which transformations connecting
different states are seen as patch transformations on a suitable manifold.

\subsection{The classical HJ equation}

Let us consider some properties of classical canonical transformations. Given
Hamilton's equations
\be
\dot q={\partial H\over\partial p},\qquad\dot p=-{\partial H\over\partial q},
\l{equazione0}\ee
the canonical transformations are the transformations
\be
q\longrightarrow Q=Q(q,p,t),\qquad p\longrightarrow P=P(q,p,t),
\l{02}\ee
which leave Hamilton's equations form invariant, that is
\be
\dot Q={\partial\tilde H\over\partial P},\qquad\dot P=-{\partial\tilde H\over
\partial Q},
\l{unpoinino}\ee
where $\tilde H(Q,P,t)$ is the new Hamiltonian. Furthermore, there is the
condition that the dynamics must be equivalent to (\ref{equazione0}), that is
\be
p\dot q-H=P\dot Q-\tilde H+{dF\over dt},
\l{03}\ee
where the generating function $F$ depends on $t$ and on any of the pairs
$(q,Q)$, $(q,P)$, $(p,Q)$ and $(p,P)$ considered as independent
variables. Choosing $(q,Q)$ yields
\be
p={\partial F\over\partial q},\qquad P=-{\partial F\over\partial Q},
\l{W7U1}\ee
so that
\be
H(q,p,t)+{\partial F\over\partial t}=\tilde H(Q,P,t).
\l{ohdiissei}\ee
The HJ equation arises by considering the transformation that leads to a
vanishing Hamiltonian
\be
\tilde H=0,
\l{oimmeia}\ee
which is equivalent to $\dot Q=0$, $\dot P=0$. Once the canonical
transformation $Q=cnst=Q(q,p,t)$, $P=cnst=P(q,p,t)$ has been found, the
dynamical problem reduces to the inversion problem
\be
\left\{\begin{array}{ll}Q=cnst=Q(q,p,t),\\ P=cnst=P(q,p,t),\end{array}\right.
\qquad\Longrightarrow\qquad
\left\{\begin{array}{ll}p=p(Q,P,t),\\ q=q(Q,P,t),\end{array}\right.
\l{laiesa}\ee
so that $Q$ and $P$ play the role of initial conditions. By (\ref{W7U1}) the
associated generating function $\S^{cl}(q,Q,t)$, called Hamilton's principal
function, satisfies
\be
p={\partial\S^{cl}\over\partial q},\qquad P=cnst=-{\partial\S^{cl}\over
\partial Q}{|_{Q=cnst}}.
\l{W7U1de}\ee
Eqs.(\ref{ohdiissei})(\ref{oimmeia}) and (\ref{W7U1de}) imply the classical HJ
equation
\be
H\left(q,p={\partial\S^{cl}\over\partial q},t\right)+
{\partial\S^{cl}\over\partial t}=0,
\l{09}\ee
that for Hamiltonians of the form
\be
H={p^2\over2m}+V(q,t),
\l{W7U2}\ee
becomes
\be
{1\over2m}\left({\partial\S^{cl}\over\partial q}\right)^2+
V(q,t)+{\partial\S^{cl}\over\partial t}=0.
\l{09BV7U3}\ee
For a time--independent potential there is the decomposition
\be
\S^{cl}(q,Q,t)=\S_0^{cl}(q,Q)-Et,
\l{01}\ee
with $E$ the energy of the stationary state. The function $\S_0^{cl}$ is
called Hamilton's characteristic function or reduced action that by
(\ref{09}) satisfies the Classical Stationary HJ Equation (CSHJE)
\be
H\left(q,p={\partial\S_0^{cl}\over\partial q}\right)-E=0,
\l{010}\ee
that is ($\W(q)\equiv V(q)-E$)
\be
{1\over2m}\left({\partial\S_0^{cl}\over\partial q}\right)^2+\W=0.
\l{012}\ee

\subsection{Coordinate transformations and the distinguished frame}

At this stage it is worth making some remarks concerning the derivation of the
classical HJ equation. First of all, observe that while for an arbitrary
canonical transformation one passes from a couple of independent variables
$(q,p)$ to another one $(Q,P)$, in the case in which one uses the Hamilton's
principal function as generating function, one has that the right hand side of
(\ref{ohdiissei}) vanishes with the effect that the independent variables
become dependent, that is
\be
\tilde H(Q,P,t)=0\qquad\longrightarrow\qquad p=\partial_q\S^{cl}(q,Q,t).
\l{dipendentiii}\ee
Hence, the effect of having $\tilde H(Q,P,t)=0$ is just that of collapsing $p$
and $q$ to be dependent variables. In the case of stationary systems we have
\be
H(q,p)\qquad\stackrel{Canon.\;Transf.}{\longrightarrow}\qquad\tilde H(Q,P)
\qquad\stackrel{\tilde H=0}{\longrightarrow}\qquad CSHJE.
\l{staccorello}\ee
Let us now consider a similar question to that leading to the CSHJE but starting
with
\be
p={\partial\S_0^{cl}\over\partial q},
\l{W7U5}\ee
rather than with $p$ and $q$ considered as independent variables. In other
words, we are looking for the coordinate transformation defined by the following
analogue of the transformation (\ref{staccorello})
\be
\S_0^{cl}(q)\qquad\stackrel{Coord.\;Transf.}{\longrightarrow}\qquad\tilde
\S_0^{cl}(\tilde q),
\l{staccorello2}\ee
with $\tilde\S_0^{cl}(\tilde q)$ denoting the reduced action of the system
with vanishing Hamiltonian. Note that since we consider $p$ and $q$ dependent,
the transformation we are considering is not a canonical one.
Even if we still have to define the specific structure of the coordinate
transformation, since $\tilde\S_0^{cl}(\tilde q)=cnst$ it is clear that
(\ref{staccorello2}) is a degenerate transformation. The existence of this
degenerate case is a rather peculiar one. Let us consider two free particles
of mass $m_A$ and $m_B$ moving with relative velocity $v$. For an observer at
rest with respect to the particle $A$ the two reduced actions are
\be
\S_0^{cl\,A}(q_A)=cnst,\qquad\S_0^{cl\,B}(q_B)=m_Bvq_B.
\l{AB1s}\ee
It is clear that there is no way to have an equivalence under coordinate
transformations by setting $\S_0^{cl\,B}(q_B)=S_0^{cl\,A}(q_A)$. This means
that at the level of the reduced action there is no coordinate transformation
making the two systems equivalent. However, note that this coordinate
transformation exists if we consider the same problem described by an
observer in a frame in which both particles have a non--vanishing velocity so
that the two particles are described by non--constant reduced actions.
Therefore, in CM, it is possible to connect different systems
by a coordinate transformation except in the case in which one of the
systems is described by a constant reduced action. This means that in CM
equivalence under coordinate transformations is frame dependent.

Let us consider the case in which the two particles are described by an observer
in a frame which is accelerated with respect to them. The reduced actions of the
$A$ and $B$ particles as seen by the observer with constant acceleration $a$ are
\be
\tilde\S_0^{cl\,A}(Q_A)={m_A\over3a}(2aQ_A)^{3\over2},\qquad
\tilde\S_0^{cl\,B}(Q_B)={m_B\over3a}(v^2+2aQ_B)^{3\over2},
\l{AB2s}\ee
where $Q_A$ ($Q_B$) is the coordinate of the particle $A$ ($B$) in the
accelerated frame. If in describing the particle $B$ in the accelerated frame
one uses the coordinate $Q_A$ defined by $\tilde\S_0^{cl\,A}(Q_A)=\tilde
\S_0^{cl\,B}(Q_B)$, one has that the resulting dynamics coincides with the one
of the particle $A$, that is
\be
\tilde\S_0^{cl\,B}(Q_B(Q_A))=\tilde\S_0^{cl\,A}(Q_A).
\l{AB3s}\ee
This simply means that the system $B$, described in terms of the coordinate
$Q_A$, coincides with the system $A$. Hence, in CM the equivalence under
coordinate transformations requires choosing a frame in which no particle is at
rest. In other words, while in any frame two systems are equivalent under
coordinate transformations, this is not anymore true once the frame coincides
with the one in which the particle is at rest, so that in the CSHJE description
there is a distinguished frame. This seems peculiar as on general grounds what
is equivalent under coordinate transformations in all frames should remain so
even in the one at rest.

\subsection{Time parameterization and space in Classical Mechanics}

A quite remarkable property of the CSHJE is that it provides a functional
relation between $p$ and $q$. In particular, note that in HJ theory,
time--dependent coordinate transformations cannot be defined. Thus, loosely
speaking, dynamics is described without using time parameterization. The
relevance of this is that we can consider possible relations among different
systems without introducing time parameterization, a concept that, as we will
see, is related to the privileged nature of the rest frame. Experience in
Special and General Relativity indicates that privileged situations may in
fact be a consequence of underlying unjustified and somehow hidden assumptions.
In Special Relativity the concept of absolute time was shown to be inconsistent.
Here we are in similar situation, but our approach is more drastic as we need
to start in the framework of HJ theory in which the concept of time does not
appear directly. As a matter of fact, this property of HJ theory is in fact
at the heart of our formulation of Quantum Mechanics. HJ theory provides the
equation for the reduced action which in turns fixes
the relationship between $p$ and $q$. While in the Hamilton and Lagrange
equations time derivatives appear also in the stationary case, in HJ theory
the time parameterization is introduced only after one uses
Jacobi's theorem $t-t_0=\partial_E\S_0^{cl}$. In CM this is equivalent to
identify the conjugate momentum with the mechanical one. Namely, setting
\be
p=\partial_q\S_0^{cl}=m\dot q,
\l{pmdotq}\ee
yields
\be
t-t_0=m\int^q_{q_0}{dx\over\partial_x\S_0^{cl}(x)},
\l{tempo}\ee
which also provides the solution of the equation of motion $q=f(t)$. It is just
the implicit role of time in the CSHJE at the basis of the peculiar nature of
the frame at rest we considered before. Let us consider the equation of motion
of a particle in an external gravitational field
\be
m\ddot q=mg.
\l{Newton}\ee
Performing the time--dependent coordinate transformation
\be
q'=q-{1\over2}gt^2,
\l{EinsWein}\ee
we have
\be
m\ddot q'=0,
\l{newtoneinsteingalileo}\ee
for any value of the energy $E$ of the particle, including the free particle at
rest for which $E=0$. So that, depending on the initial conditions of
(\ref{newtoneinsteingalileo}), we may have $q'$ to be constant, say $q'=0$.
Therefore, there are no selected frames if one uses time--dependent coordinate
transformations. Hence, while with the CSHJE description it is not always
possible to connect two systems by a coordinate transformation, this is not the
case if one describes the dynamics using Newton's equation. In particular, in
finding the coordinate transformation reducing the CSHJE description of
Eq.(\ref{Newton}) one has
\be
{1\over2m}\left({\partial\S_0^{cl}(q)\over\partial q}\right)^2-mgq+E=0,
\l{inboccaallupoChiara}\ee
for which there is no coordinate transformation $q\longrightarrow\tilde q(q)$
such that $\S_0^{cl}(q)=\tilde\S_0^{cl}(\tilde q)$ with $\tilde\S_0^{cl}(\tilde
q)$ the reduced action of the free particle with $E=0$. In this context we
observe that if for some reason $V-E$ were never vanishing, then the above
peculiarity of the CSHJE would not exist.

Time parameterization can be seen as a way to express a constant,
say $0$, by means of the solution of the equation of motions, $q=f(t)$. For
example, for a particle with constant velocity, we have
\be
0=q-vt,
\l{0qvt}\ee
so that particle's position can be denoted by either $q$ itself or $vt$. In
this way one can always reduce to the particle at rest by simply setting $q'=q
-f(t)$. While in the case of the CSHJE description there is the degenerate case
$cnst=mvq$, corresponding to $\tilde\S_0^{cl}(\tilde q)=\S_0^{cl}(q)$, time
parameterization provides a well--defined and invertible transformation {\it
i.e.} $q'=q-f(t)\longrightarrow q=q'+f(t)$. The reason underlying the
differences in considering the role of space and time is that fixed values of
$q$ and $t$ correspond to quite different situations. Even if the particle is
at rest, say at $q=0$, {\it time continues to flow}. It is just the use of time
that allows to connect different systems by a coordinate transformation.

The basic difference in the nature of space and time is rather peculiar of CM.
On the other hand, in special relativity space and time are intrinsically
related. In particular, this property is not unrelated to the fact that in the
{}framework of special relativity the energy of the particle at rest is
non--zero. Recalling that in the relativistic case the CSHJE has the form
\be
{1\over2m}\left({\partial\S_0^{cl}\over\partial q}\right)^2+\W_{rel}=0,
\l{012relativistico}\ee
where
\be
\W_{rel}\equiv {1\over2mc^2}[m^2c^4-(V-E)^2],
\l{wrelativisticopr}\ee
we see that the critical $\W_{rel}=0$ case corresponds to
\be
V-E=\pm mc^2.
\l{emijxop}\ee
Our aim is to find a dynamical description in which there always exists
coordinate transformations connecting arbitrary systems. We can see from the
above discussion that such an investigation should imply the existence of a sort
of ``energy function'' which is never vanishing, whatever structure the
potential $V$ may have. This would avoid the degenerate situation in which the
reduced action is a constant and will remove the consequence of the asymmetric
role of space and time discussed above. We will see that imposing the EP will
univocally lead to the deformation of the CSHJE by a term which is identified
with the quantum potential.

We anticipated above some features underlying the EP, which we will formulate
later. As we will see, looking for coordinate transformations connecting
different physical systems, and in particular for those reducing any system to
that corresponding to $\W^0\equiv 0$, will bring us to the conclusion that the
CSHJE must be modified. This will define a new function $\S_0$ that by abuse of
language we will continue to call reduced action. Thus, we will have
\be
\S_0\ne\S_0^{cl}.
\l{conclusione}\ee

Information about a physical system is encoded in the functional dependence of
$\S_0$ on its argument. Therefore, a given transformation would reflect in a
change of the functional structure of $\S_0$. In this context the specific
choice of the coordinate reduces to a matter of notation with the new coordinate
playing the role of independent variable for the new system. The coordinate
transformation defined in (\ref{AB3s}) is a natural way to define an induced
transformation of $\S_0$. Let us consider the following simple example. Given
two functions, say
\be
f_1(x_1)=x_1^m,\qquad f_2(x_2)=x_2^n,
\l{1}\ee
there is the associated coordinate transformation $x_1\longrightarrow x_2=v(x_1)
=x_1^{m/n}$, defined by
\be
f_2(x_2)=f_1(x_1).
\l{2}\ee
This is equivalent to saying that starting from the function $f_1(x_1)=x_1^m$,
the transformation $x_1\longrightarrow x_2=v(x_1)=x_1^{m/n}$, induces the
functional transformation $f_1\longrightarrow f_2=f_1\circ v^{-1}$.

\subsection{$v$--transformations}

Let us now consider the case of the reduced action $\S_0$ and let
\be
q\longrightarrow q^v=v(q),
\l{6}\ee
be a locally invertible coordinate transformation. Similarly to Eq.(\ref{2}),
setting
\be
\S_0^v(q^v)=\S_0(q(q^v)),
\l{VIP}\ee
naturally defines a new reduced action $\S_0^v$ associated to such
``$v$--transformations'' (VTs). Note that we may also consider alternative
transformations of $\S_0$. In general, we can also choose
\be
\S_0\longrightarrow\tilde\S_0=\S_0\circ v^{-1}+f,
\l{7azz}\ee
for some function $f$. The construction in this case would be equivalent. In
particular, since for any transformation (\ref{7azz}) we can always find another
map $q\longrightarrow q^w=w(q)$, such that $\S_0\longrightarrow\tilde\S_0=\S_0^w
=\S_0\circ w^{-1}$, it is clear that we can directly consider (\ref{VIP}). This
is a convenient choice as our formulation will be particularly simplified and
will exhibit duality properties which are generally more difficult to recognize
in working with the form (\ref{7azz}). We note that this is not a restriction as
we are considering the problem of connecting different systems, with $q^v$ seen
as the new independent coordinate. Then, what is of interest in this context is
the functional change $\S_0\longrightarrow\S_0^v$. Therefore, considering
\be
\S_0\longrightarrow\S_0^v=\S_0\circ v^{-1},
\l{aaa7a7T2}\ee
which is equivalent to (\ref{VIP}), does not imply loss of generality. Note that
this is equivalent to saying that associated to $v=\S_0^{v^{\;-1}}\circ\S_0$
there is the induced map ${v^{-1}}^*:\S_0\mapsto{v^{-1}}^*(\S_0)$. In other
words, $\S_0^v$ is the pullback of $\S_0$ by ${v^{-1}}^*$.

Let us anticipate a motivation which illustrates why the formalism will simplify
in considering the induced transformation in the form (\ref{aaa7a7T2}). The
point is that as $p_v=\partial_{q^v}\S_0^v (q^v)$, we have
\be
p\longrightarrow p_v=(\partial_qq^v)^{-1}p,
\l{12xc}\ee
that is, from the point of view of the conjugate momentum, one has that the
choice (\ref{VIP}) implies that $p$ transforms as $\partial_q$. As a
consequence, the resulting formalism will be manifestly covariant.

\subsection{A geometrical picture}

Let us conclude this section by observing that Eq.(\ref{VIP}) admits the
following geometrical interpretation. Suppose that the system has an underlying
geometry induced by $\W\equiv V-E$ and that the use of $q$ corresponds to a
choice of a local coordinate representing a point $x$ of an underlying manifold.
Then the map $q\longrightarrow q^v=v(q)$ can be seen as a patch transformation
with $\S_0$ seen as a scalar function, that is a zero--differential. Since
$p=\partial_q\S_0$, we have that the conjugate momentum transforms as a
one--differential. Let us recall that a $\lambda$--differential $f^{(\lambda)}$
is a set of functions $f^{(\lambda)}=\{f_\alpha^{(\lambda)}(q_\alpha)|\alpha\in
I\}$, where each $f_\alpha^{(\lambda)}$ is defined on the patch $U_\alpha$ of
the atlas $\{(U_\alpha,q_\alpha)|\alpha\in I\}$, and satisfying
\be
f_\alpha^{(\lambda)}(q_\alpha)(dq_\alpha)^\lambda=
f_\beta^{(\lambda)}(q_\beta)(dq_\beta)^\lambda,
\l{oq2ij}\ee
in $U_\alpha\cap U_\beta$. In other words, $f^{(\lambda)}(q)$ transforms as $(d
q)^{-\lambda}$, that is $f_\alpha^{(\lambda)}(q_\alpha)=\left(\partial_{q_\alpha
}q_\beta\right)^\lambda f_\beta^{(\lambda)}(q_\beta)$. In this context the
structure of the transition functions between the patches $U_\alpha$ and
$U_\beta$ is determined by the $\alpha$-- and $\beta$--physical systems, so that
there is the correspondence
\be
(\W^\alpha,\W^\beta)\longleftrightarrow(U_\alpha,U_\beta).
\l{W7U6}\ee

\mysection{$p$--$q$ duality and the Legendre transformation}\l{pqdatlt}

A feature of CM is that while in considering canonical
transformations and the phase space there is a formal $p$--$q$ duality, this is
broken in the explicit solution of the equations of motion. For example, for the
Hamiltonian one usually considers the structure $H=p^2/2m+V(q)$, so that,
despite the symmetric structure of the equations of motion $\dot q=\partial_p
H$, $\dot p=-\partial_q H$, the effective dynamical solution breaks $p$--$q$
duality due to the difference between the kinetic term and $V(q)$ structures.
There is however a particular case in which $H$ has explicit $p$--$q$
duality. This happens for the harmonic oscillator for which $V(q)=m\omega^2q^2/
2$. This is an interesting property if one thinks that quantum field theories
are described by infinitely many interacting oscillators. However,
rather than explicit $p$--$q$ duality, what is lacking in CM is a formulation
in which the $p$ and $q$ descriptions have the same structure.

To understand $p$--$q$ duality of Hamilton's equations,
it is useful to first investigate some properties of CM's formalism.
This will be useful in setting the basis for a formulation with manifest duality
between the $p$ and $q$ descriptions. In doing this we will see that there is a
remarkable way to characterize the Hamilton equations of motion. This shows, in
a way illustrated in the next subsection, how the issue of dependence --
independence of the canonical variables is connected to dynamical features.

\subsection{Energy conservation and classical $p$--$q$ duality}

We have seen that there are interesting questions which arise in considering $p$
and $q$ as dependent quantities through the still unknown $\S_0$. We now show
that, considering the Hamiltonian equations of motion as a way to reduce
independent variables to dependent ones, will give a connection between energy
conservation and classical $p$--$q$ duality. A step in this subsection concerns
the analysis of the relation among the structures of relevant equations which
arise by considering the canonical variables as dependent on one side, and those
with the variables considered as independent on the other. We will start with an
arbitrary function and then we will derive the Hamilton equations of motion from
a perspective which is somewhat different from the standard one. This analysis
will connect energy conservation and $p$--$q$ duality.

Let $F(x_1,x_2,x_3)$ be an arbitrary function of the independent variables
$x_1,x_2,x_3$, so that
\be
{dF\over dx_k}={\partial F\over\partial x_k},
\l{Uz1}\ee
$k=1,2,3$, and suppose that $F$ satisfies a differential equation
\be
\P(F)=0,
\l{Uz1b}\ee
with the property that its solution defines
\be
x_1=x_1(x_3),\qquad x_2=x_2(x_3).
\l{Uz2}\ee
The pair $(F,\P)$ naturally defines the function in one variable
\be
G(x_3)=F(x_1(x_3),x_2(x_3),x_3),
\l{Uz3}\ee
whose derivative is
\be
{\partial G\over\partial x_3}={dG\over dx_3}={dF\over dx_3}={\partial F\over
\partial x_1}{\partial x_1\over\partial x_3}+{\partial F\over\partial x_2}
{\partial x_2\over\partial x_3}+{\partial F\over\partial x_3}.
\l{Uz4}\ee
Note that, while the total and partial derivatives of $F$ with respect to $x_3$
coincide for $F$ seen as function of the independent variables $x_1,x_2,x_3$,
in the case in which $F$ is considered as solution of (\ref{Uz1b}) one has
\be
{d F\over dx_3}\ne {\partial F\over\partial x_3}.
\l{Uz5}\ee
There is however an exception to this. This happens if (\ref{Uz1b}) implies
\be
{\partial F\over\partial x_1}{\partial x_1\over\partial x_3}=-
{\partial F\over\partial x_2}{\partial x_2\over\partial x_3},
\l{Uz6}\ee
so that
\be
{d F\over dx_3}={\partial F\over\partial x_3}.
\l{f3toteparziale}\ee
Therefore, Eq.(\ref{Uz1}) is preserved for $k=3$ even after the ``equations of
motion'' (\ref{Uz6}) are imposed. It follows that from the equation
\be
\left({d\over d x_3}-{\partial\over\partial x_3}\right)F=0,
\l{Uz6c}\ee
it is not possible to distinguish whether $F$ is to be considered as a function
of three independent variables $x_1,x_2,x_3$ or rather if these satisfy the
equations of motion (\ref{Uz6}) so that $x_1=x_1(x_3)$ and $x_2=x_2(x_3)$.
A property of Eq.(\ref{Uz6}) is that it is invariant under the interchange
\be
x_1\longleftrightarrow x_2.
\l{Uz21}\ee
However, if one considers
\be
{\partial x_1\over\partial x_3}={\partial F\over\partial x_2},\qquad
{\partial x_2\over\partial x_3}=-{\partial F\over\partial x_1},
\l{Uz7}\ee
as a particular solution of (\ref{Uz6}), then there is a slightly different
version of the duality (\ref{Uz21}), since the symmetry of Eq.(\ref{Uz7}) is now
\be
x_1\longrightarrow -x_2,\qquad x_2\longrightarrow x_1.
\l{6Tx}\ee
Making the identification
\be
F=H,
\l{Uz8}\ee
where $H$ is the Hamiltonian and $x_1=q,x_2=p,x_3=t$, the equation (\ref{Uz7})
corresponds to
\be
\dot q={\partial H\over\partial p},\qquad\dot p=-{\partial H\over\partial q}.
\l{Uz9}\ee
Hence, the Hamiltonian equations can be seen as a particular solution to
the problem of finding the structure of $\P(F)=0$ such that (\ref{Uz6c})
is satisfied irrespectively of considering $x_1$ and $x_2$ as dependent of
$x_3$ or not. In other words, both $H$ as function of the independent
variables $q,p,t,$ and $H$ evaluated on the classical trajectory, satisfy
\be
\left({d\over dt}-{\partial\over\partial t}\right)H(q,p,t)=0=\left({d\over dt}
-{\partial\over\partial t}\right)H(q^{cl}(t),p^{cl}(t),t).
\l{Uz10}\ee
This aspect is related to duality as interchanging $x_1\longleftrightarrow x_2$
leaves (\ref{Uz6}) invariant. As we have seen, since Hamilton's equations
(\ref{Uz9}) correspond to a particular solution of Eq.(\ref{Uz6}), it follows
that (\ref{Uz9}) is invariant under
\be
q\longrightarrow -p,\qquad p\longrightarrow q.
\l{ppqquad}\ee
The above discussion shows that classical $p$--$q$ duality is related to the way
in which Eq.(\ref{Uz6}), which makes (\ref{Uz6c}) insufficient to distinguish
between dependent and independent variables, is specifically satisfied. In
particular, we have seen that the particular solution (\ref{Uz7}) of (\ref{Uz6})
breaks the $p\longleftrightarrow q$ duality to the duality (\ref{ppqquad}).
Therefore, classical $p$--$q$ duality is related to the dependence that the
structure of relevant equations may have on the canonical variables in the
dependent and independent cases. In particular, since (\ref{Uz10}) implies
energy conservation if the Hamiltonian does not explicitly depend on time, we
see that classical $p$--$q$ duality is a manifestation of the role of time as
parameter for classical trajectories and to energy conservation. The fact that
in QM position and momentum cannot be simultaneously measured, indicates that
the issue of dependence and independence of the canonical variables is a basic
feature.\footnote{In this respect we note that according to the Kochen--Specker
theorem \cite{KochenSpecker} it is not possible to preserve functional relations
between physical quantities with assigned values. We note that this topic,
related to hidden variables \cite{Bell1}\cite{Brown}, has been recently
considered in \cite{IshamButterfield}.} In this context we will see that the
$p$--$q$ duality of Hamilton mechanics is a restrictive one. This lack of the
Hamiltonian formalism indicates that a manifest full $p$--$q$ duality deserves
formulating an alternative description of the dynamical problem. In fact, there
is a natural structure, with explicit $p$--$q$ duality, which is a consequence
of the involutive nature of the Legendre transformation, and that cannot be a
feature of CM. This duality, which arises by considering the canonical variables
on equal footing, will make it quite natural to formulate the EP. However, we
stress that the QSHJE will be derived from this principle, formulated in
sect.\ref{tep}, without assumptions about the existence of dualities.

\subsection{The dual reduced action}

As a first step towards a formulation with manifest $p$--$q$ duality, we note
that among the possible transformations of $\S_0$ we can consider that induced
by
\be
q\longrightarrow q^v=v(q)=p,
\l{ideuzza}\ee
with $p$ the momentum. According to (\ref{VIP}) we have
\be
\S_0(q)\longrightarrow\S_0^v(p)=\S_0(q(p)).
\l{FrakZappathereisanidentity}\ee
In order to understand the structure of the function $\S_0^v(p)$, we first
observe that there is a natural way to introduce the tools we need in going
towards a formulation with manifest $p$--$q$ duality. Let us introduce the dual
reduced action $\T_0(p)$ defined as the Legendre transform of $\S_0(q)$
\be
\T_0=q{{\partial\S_0\over\partial q}}-\S_0,\qquad\qquad
\S_0=p{{\partial\T_0\over\partial p}}-\T_0,
\l{lt00}\ee
\be
p={\partial\S_0\over\partial q},\qquad\qquad q={\partial\T_0\over\partial p}.
\l{pq}\ee
It follows that the transformation (\ref{FrakZappathereisanidentity})
is nothing but (\ref{lt00}) seen as function of $p$
\be
\S_0^v(p)=p{\partial\T_0(p)\over\partial p}-\T_0(p).
\l{W7U8}\ee

Above we discussed the dependence of the structure of some
relevant equations on the canonical variables considered either as dependent or
independent and noticed that this issue may be connected to the fact that in QM
position and momentum cannot be simultaneously measured. In this respect it is
interesting to note that for the pair $\S_0$--$\T_0$, in the case in
which $p$ and $q$ are considered as independent variables, we have $F(q,p)=qp-
\T_0(p)\leq F(q,p(q))=\S_0(q)$, $\forall q,p$, with $p=p(q)$ given by
$\partial_p\T_0(p)=q$. Therefore, considering $p$ and $q$ as independent
variables we have the Young inequality (see {\it e.g.} \cite{Arnold})
\be
\T_0(p)+\S_0(q)\geq qp,\qquad\forall q,p.
\l{Young}\ee

\subsection{The dual HJ equation}

Despite $\T_0$ is not usually considered in the literature, the involutive
character of the Legendre transformation makes it clear that $\T_0$ is a natural
function to consider in looking for a formulation with manifest $p$--$q$
duality. Let us introduce the Legendre transform of the Hamilton principal
function $\S$
\be
\T=q{\partial\S\over\partial q}-\S,\qquad\S=p{\partial\T\over\partial p}-\T,
\l{lt00g7U92}\ee
\be
p={\partial\S\over\partial q},\qquad\qquad q={\partial\T\over\partial p}.
\l{pqxvcfd07U9}\ee
Observe that in the stationary case
\be
\S(q,t)=\S_0(q)-Et,\qquad\T(p,t)=\T_0(p)+Et.
\l{timeind}\ee
A first question to investigate concerns of the classical HJ equation in
the $\T^{cl}$--representation. Let us consider the differentials
\be
d\S={\partial\S\over\partial q}dq+{\partial\S\over\partial t}dt=
pdq+{\partial\S\over\partial t}dt,
\l{diffs}\ee
\be
d\T={\partial\T\over\partial p}dp+{\partial\T\over\partial t}dt=
qdp+{\partial\T\over\partial t}dt,
\l{difft}\ee
which imply
\be
d\S=d(pq-\T)=pdq+qdp-qdp-{\partial\T\over\partial t}dt,
\l{implicanoa}\ee
that is
\be
{\partial\S\over\partial t}=-{\partial\T\over\partial t}.
\l{BtR}\ee
Relevance of this equation resides in the fact it connects the $\S$ and $\T$
pictures through time evolution. For example, we will see that Eq.(\ref{BtR})
will fix a sign ambiguity in determining the ``self--dual states''. Furthermore,
observe that in the classical case it is precisely the partial derivative of
$\S^{cl}$ with respect to time that appears in the HJ equation. In particular,
it follows from (\ref{09}) and (\ref{BtR}) that $\T^{cl}$ satisfies the
following dual version of the classical HJ equation
\be
H\left(q={\partial\T^{cl}\over\partial p},p,t\right)-
{\partial\T^{cl}\over\partial t}=0,
\l{09perT}\ee
which for general Hamiltonians
\be
H={p^2\over2m}+V(q,p,t),
\l{aa7U2}\ee
becomes
\be
{1\over2m}p^2+V\left(q={\partial\T^{cl}\over\partial p},p,t\right)-
{\partial\T^{cl}\over\partial t}=0.
\l{09BV7U3bbb}\ee
We can derive an equivalent version of (\ref{09BV7U3bbb}) with manifest
dual structure. In particular, by analogy with the HJ equation, we may
impose that its dual formulation contains the ``{\it dual kinetic term}"
\be
{m\over2}\alpha^2q^2.
\l{kinetic}\ee
The request of equivalence between the dual equations
defines a potential $U^{(\alpha)}$ dual to $V$, namely
\be
{m\alpha^2\over2}\left({\partial\T^{cl}\over\partial p}\right)^2+U^{(\alpha)}
\left(p,q={\partial\T^{cl}\over\partial p},t\right)+{\partial\T^{cl}\over
\partial t}=0,
\l{09BV7U3bvg}\ee
where by (\ref{09BV7U3bbb})
\be
U^{(\alpha)}(p,q,t)=-{1\over2m}(p^2+m^2\alpha^2q^2)-V(q,p,t),
\l{SPQR1}\ee
and $\alpha$ is a constant with the dimension of a frequency. Associated with
$\T^{cl}$ there is the ``Hamiltonian'' $K(p,q,t)$, dual to $H(q,p,t)$
\be
K(p,q,t)={m\alpha^2\over2}q^2+U^{(\alpha)}(p,q,t),
\l{cappa}\ee
which has the simple relation with $H$
\be
K(p,q,t)=-H(q,p,t).
\l{kugualemenoh}\ee
It follows that the classical equations of motion are equivalent to
\be
\dot q=-{\partial K\over\partial p},\qquad\dot p={\partial K\over\partial q},
\l{Uz9K}\ee
and the dual classical HJ equation can be written as
\be
K\left(p,q={\partial\T^{cl}\over\partial p},t\right)+{\partial\T^{cl}\over
\partial t}=0.
\l{xc09perT}\ee
There is a particular system in which the $\S^{cl}$ and $\T^{cl}$ pictures have
basically the same HJ equation. This happens for the harmonic oscillator
\be
H_\omega(q,p)={1\over2m}p^2+V_\omega(q),
\l{afsdt}\ee
where $V_\omega(q)=m\omega^2q^2/2$. By (\ref{SPQR1}) we have
\be
U_{\omega}^{(\alpha)}=-{1\over2m}p^2-{m(\omega^2+\alpha^2)\over2}q^2,
\l{SPQR2}\ee
so that $U_{\omega}^{(\alpha)}=-H_{\omega_D}$, where the frequency of the new
oscillator is $\omega_D^2={\omega^2}+\alpha^2$. It follows that for $\alpha=\pm
i\omega$, the dual potential corresponds to minus the free particle Hamiltonian
\be
U_{\omega}^{(\pm i\omega)}=-H_0=-{1\over2m}p^2.
\l{omegadb}\ee
We also observe that as $H_\omega(q,p)=H_\omega(p/m\omega,m\omega q)$, we have
\be
K_\omega(p,q)=-H_\omega(p/m\omega,m\omega q).
\l{ioj}\ee

\mysection{M\"obius symmetry and the canonical equation}\l{gl2csatce}

In this section we will first consider the effect of the VT on $\T_0$. Next, we
will observe that by (\ref{VIP}), it follows by construction that the Legendre
transform of $\T_0$ is invariant under VTs. We will see that a
$GL(2,\CC)$--transformation corresponds to a rotation of the two--dimensional
kernel of a second--order linear operator which we will determine by considering
the second derivative of the Legendre transform of $\T_0$ with respect to $\s=
\S_0(q)$. As a consequence, the potential $\U(\s)$ appearing in this operator
turns out to be invariant under the VT corresponding to the M\"obius group. In
this context covariance of the relevant second--order linear differential
equation follows by consistency.

\subsection{$\T_0$--transformation induced by $v$}

Since $\S_0$ and $\T_0$ are a Legendre pair, specifying the transformation
properties of $\S_0$ will fix the transformation of $\T_0$ and vice versa. In
particular, consistency implies that if $\S_0(q)\longrightarrow\tilde\S_0(\tilde
q)$ is an arbitrary transformation, not necessarily of $v$--type, that is in
general $\tilde\S_0(\tilde q)\ne\S_0(q)$, then the new dual reduced action will
be the Legendre transform of $\tilde\S_0(\tilde q)$. In other words, the
transformation $\S_0(q)\longrightarrow\tilde\S_0(\tilde q)$ induces the
following transformation of $\T_0(p)$
\be
\T_0(p)\longrightarrow\tilde\T_0(\tilde p)=\tilde q\tilde p-\tilde\S_0(\tilde
q).
\l{v4}\ee
In the case in which the transformation of the functional structure of $\S_0$ is
performed by a $v$--map we have $p_v=(\partial_qq^v)^{-1}p$, and
replacing $\S_0^v(q^v)$ with $pq-\T_0$ in (\ref{v4}) yields
\be
\delta_v\T_0=\delta_v(qp),
\l{v6}\ee
where $\delta_v\T_0\equiv\T_{0\,v}(p_v)-\T_0(p)$ and $\delta_v(qp)\equiv q^vp_v
-qp$. Observe that by Eq.(\ref{VIP}), the VTs can be equivalently seen as a
symmetry of the Legendre transform of $\T_0$.

\subsection{The canonical equation}

Let us now consider the VT corresponding to the M\"obius transformation
\be
q\longrightarrow q^v={Aq+B\over Cq+D},
\l{w4}\ee
that due to the dimensionality of $p$ and $q$, we consider Eq.(\ref{w4}) as a
$GL(2,\CC)$--transformation rather than a $PGL(2,\CC)\cong
PSL(2,\CC)$--transformation. Eqs.(\ref{VIP}) and (\ref{w4}) imply
\be
p\longrightarrow p_v=\rho^{-1}(Cq+D)^2p,
\l{w5}\ee
where $\rho\equiv AD-BC$. Observe that $\rho$ has the dimension
of a length times the dimension of $A^2$ so that $\rho$ would be dimensionless
if $Dim [A]=length^{-1/2}$. By (\ref{v6})(\ref{w4}) and (\ref{w5}) it follows
that the action on $\T_0$ induced by the $v$--maps is
\be
\delta_v\T_0=\rho^{-1}(ACq^2+2BCq+BD)p.
\l{7z}\ee
Taking the square root of (\ref{w5}), we see that ($\epsilon=\pm 1$)
\be
q^v\sqrt{p_v}=\epsilon\sqrt{\rho^{-1}}(Aq\sqrt p+B\sqrt p),
\quad\sqrt{p_v}=\epsilon\sqrt{\rho^{-1}}(Cq\sqrt p+D\sqrt p).
\l{ssss}\ee
This version of Eqs.(\ref{w4}) and (\ref{w5}) indicates
that $q^v\sqrt{p^v}$ and $\sqrt{p^v}$ are linear combinations of the solutions
of a second--order linear differential equation. In particular, Eq.(\ref{ssss})
can be seen as a rotation of the elements in the kernel of a second--order
linear operator which can be simply identified by observing that the second
derivative of the second equation in (\ref{lt00}) with respect to $\s=\S_0(q)$
\be
{\partial q\over\partial\s}{\partial p\over\partial\s}+p{\partial^2q\over
\partial\s^2}=0,
\l{ccchj}\ee
is equivalent to
\be
{1\over q\sqrt p}{\partial^2(q\sqrt p)\over\partial\s^2}=
{1\over\sqrt p}{\partial^2\sqrt p\over\partial\s^2}.
\l{sec}\ee
On the other hand, this is equivalent to the ``canonical equation''
\be
\left({\partial^2\over\partial\s^2}+\U(\s)\right)q\sqrt p
=0=\left({\partial^2\over\partial\s^2}+\U(\s)\right)\sqrt p,
\l{w10}\ee
where $\U$ is the ``canonical potential''
\be
\U(\s)={1\over2}\{q\sqrt p/\sqrt p,\s\}={1\over2}\{q,\s\},
\l{scharz}\ee
with
\be
\{h(x),x\}={h'''\over h'}-{3\over2}\left({h''\over h'}\right)^2=
(\ln h')''-{1\over2}{(\ln h')'}^2,
\l{Schwarzian}\ee
denoting the Schwarzian derivative whose chain rule is
\be
\{h(x),x(y)\}=\left({\partial y\over\partial x}\right)^2\{h(x),y\}
-\left({\partial y\over\partial x}\right)^2\{x,y\}.
\l{cometrasforma}\ee

\subsection{Legendre transformation as M\"obius transformation}

Another interesting property due to (\ref{VIP}) is that $\S_0$ can be expressed
in terms of a M\"obius transformation of $\T_0$. Let us first consider the
effect of an $S$--transformation
\be
S=\left(\begin{array}{c}0\\-1\end{array}\begin{array}{cc}1\\0\end{array}\right).
\l{Str}\ee
We have by (\ref{7z})
\be
S\circ\T_0=\T_0 -2pq,
\l{7dfx}\ee
so that the Legendre transform of $\T_0$ can be written in the form
\be
\S_0(q)=-{1\over2}(\T_0(p)+S\circ\T_0(p)).
\l{7dfxbb}\ee
Observe that from (\ref{7z}) it follows that for any $B$ and $C$, $BC\ne 0$, the
action of $R=\left(\begin{array}{c}0\\C\end{array}\begin{array}{cc}B\\0
\end{array}\right)$ on $\T_0$ does not depend on $B$ and $C$. However, by
(\ref{w4}) the action of $R$ on $q$ depends on $\delta=B/C$
\be
R:q\longrightarrow q^v=v(q)={\delta\over q},
\l{unosuq}\ee
whereas by (\ref{w5}) the action on $p$ is
\be
R:p\longrightarrow p_v=-{1\over\delta}q^2p.
\l{pinqduep}\ee
In particular, we have
\be
R:pq\longrightarrow p_vq^v=-pq.
\l{unosuqb}\ee

\subsection{M\"obius invariance of $\U$}

Eqs.(\ref{w10}) and (\ref{scharz}) follow from (\ref{sec}) or,
equivalently, from the observation that the identities
\be
{\partial\over\partial x}{h'}^{1/2}{h'}^{-1/2}=0={\partial\over\partial x}
{1\over h'}{\partial\over\partial x}{h'}^{1/2}{h'}^{-1/2}h,
\l{icuhyA}\ee
imply that the kernel of the second--order linear operator
\be
{h'}^{1/2}{\partial\over\partial x}{1\over h'}{\partial\over\partial x}
{h'}^{1/2}={\partial^2\over\partial x^2}+{1\over2}\{h,x\},
\l{easytosee}\ee
is given by the linear span of ${h'}^{-1/2}$ and ${h'}^{-1/2}h$, that is
\be
\left({\partial^2\over\partial x^2}+{1\over2}\{h,x\}\right){h'}^{-1/2}(Ah+B)=0.
\l{icuhy}\ee
Eqs.(\ref{icuhyA})--(\ref{icuhy}) show that the Schwarzian derivative of the
ratio of two linearly independent elements in the kernel of a second--order
linear operator $(\partial_x^2+V(x))$ is twice $V(x)$. Noticing that for
any $A$ and $B$, not simultaneously vanishing, $(\partial_x^2+V(x))f_k(x)=0$,
$k=1,2$, is equivalent to $V=-(Af_1''+Bf_2'')/(Af_1+Bf_2)$, we have the
well--known fact
\be
\{\gamma(h),x\}=\{h,x\},
\l{dj3e}\ee
where $\gamma(h)$ is the M\"obius transformation of $h$
\be
\gamma(h)={Ah+B\over Ch+D}.
\l{ppolo}\ee
Eq.(\ref{dj3e}) implies $\{\gamma(x),x\}=\{x,x\}=0$.
Conversely, if $\{h,x\}=0$, then, solving
\be
(\ln h')''-{1\over2}{(\ln h')'}^2=0,
\l{solving0}\ee
yields $h=\gamma(x)$.
By (\ref{cometrasforma}) these properties of the Schwarzian derivative
are equivalent to the fact that
\be
\{f,x\}=\{h,x\},
\l{dj3eccb}\ee
if and only if $f=\gamma(h)$. This implies that $\U(\s)$, associated to the
Legendre transform of $\T_0$, is invariant under $v$--maps corresponding to
the M\"obius transformations $q^v=\gamma(q)$, that is
\be
\U^v (\s^v)={1\over2}\{q^v,\s^v\}={1\over2}\{\gamma(q),\s\}=\U(\s),
\l{INVARIANTE}\ee
where $\s^v={\S_0}^v(q^v)=\S_0(q)=\s$. However, if $q^v$ is not a M\"obius
transformation of $q$, then
\be
\U^v(\s^v)={1\over2}\{q^v,\s^v\}\ne\U(\s).
\l{ieur}\ee
Therefore, although by definition any $v$--map leaves the Legendre
transform of $\T_0$ invariant, we have that $\U$ is invariant only under
$GL(2,{\CC})$ M\"obius transformations of $q$.

\subsection{Canonical equation, dynamics and initial conditions}

We have seen that the canonical equation associated with the Legendre
transform of $\T_0$ is $GL(2,{\CC})$--invariant. Another property of the
canonical equation (\ref{w10}) is that it can be seen as an equation of motion.
In fact, suppose that the canonical potential $\U(\s)$ is given. Then, if
$y_1(\s)$ and $y_2(\s)$ are two linearly independent solutions of the canonical
equation, we have
\be
q\sqrt p=Ay_1(\s)+By_2(\s),\qquad\sqrt p=Cy_1(\s)+Dy_2(\s).
\l{cow}\ee
In order to derive $\S_0(q)$, that is to solve the dynamical problem, we first
use (\ref{cow}) to obtain $q$ as function of $\s$ and then consider its
inverse. Taking the ratio of (\ref{cow}) yields
\be
q={Ah(\s)+B\over Ch(\s)+D},
\l{cow23}\ee
where $h(\s)=y_1(\s)/y_2(\s)$. Inverting (\ref{cow23}) we have $\S_0=h^{-1}
\circ{\gamma}^{-1}$ with $\gamma$ given in (\ref{ppolo}) so that the dynamical
problem is solved by
\be
\S_0(q)=h^{-1}(\gamma^{-1}(q)).
\l{doveat}\ee
Although the canonical equation is a second--order linear differential equation,
so that a particular solution is obtained by giving two conditions, we have that
as $q(\s)$ is the ratio of two linearly independent solutions, one needs three
initial conditions to fix it. This follows from the observation that
(\ref{cow23}) can be written in terms of three constants; {\it e.g.} if $A\ne
0$ and $C\ne 0$, then
\be
q={A\over C}\left({h(\s)+B/A\over h(\s)+D/C}\right).
\l{cow23esempio}\ee
Therefore, the dynamics is obtained by first considering the ratio of two
linearly independent solutions of the canonical equation (\ref{w10}) and then
evaluating its inverse. In particular, all the reduced actions $\S^v_0$ in the
$GL(2,\CC)$--orbit of $\S_0$ are the inverse of the ratio of two linearly
independent solutions of the same canonical equation.

\subsection{Covariance}

A property of the canonical equation is that it is manifestly covariant under
arbitrary transformations. In particular, we have seen that the transformation
of $\T_0$, induced by an arbitrary transformation of $\S_0$, not necessarily of
$v$--type, is determined by the fact that $\tilde\T_0$ is the Legendre transform
of $\tilde\S_0$. Then, by Eq.(\ref{v4}) the canonical equation (\ref{w10}) is
covariant under arbitrary transformations. In particular, the second derivative
of (\ref{v4}) with respect to $\tilde\s=\tilde\S_0(\tilde q)$ yields
\be
\left({\partial^2\over\partial{\tilde\s}^2}+\tilde\U(\tilde\s)\right)\tilde q
\sqrt{\tilde p}=0=\left({\partial^2\over\partial{\tilde\s}^2}+\tilde{\U}(\tilde
\s)\right)\sqrt{\tilde p}.
\l{w10tildato}\ee
Thus, the procedure to associate a second--order linear differential equation
with a Legendre transformation is manifestly covariant.
There is another property of $\U(\s)$. Namely, by the chain rule of the
Schwarzian derivative (\ref{cometrasforma}), we have
\be
\{q,(A\s+B)/(C\s+D)\}=\tau^{-2}(C\s+D)^4\{q,\s\}=2\tau^{-2}(C\s+D)^4\U(\s),
\l{msym2}\ee
where $\tau\equiv AD-BC\ne 0$. Hence, a M\"obius transformation of $\S_0$
corresponds to a rescaling of $\U(\s)$. Under such a transformation we have
\be
\tilde q=q,\qquad\qquad\tilde p={\partial\tilde\s\over\partial\tilde q}
={\partial\tilde\s\over\partial\s}p,
\l{s1}\ee
so that, setting $\phi(\s)=aq\sqrt p+b\sqrt p$, with $a$ and $b$ two non
simultaneously vanishing constants, we have $\tilde\phi(\tilde\s)=(\partial_\s
\tilde\s)^{1/2}\phi(\s)$. Therefore, $\phi$ transforms as a
$-1/2$--differential. Observe that
\be
{\partial^2\over\partial{\tilde\s}^2}=\left({\partial\tilde\s\over\partial\s}
\right)^{-2}{\partial^2\over\partial\s^2}-\left({\partial\tilde\s\over\partial\s
}\right)^{-3}\left({\partial^2\tilde\s\over\partial\s^2}\right){\partial\over
\partial\s},
\l{1234}\ee
so that under the transformation $\s\longrightarrow\tilde\s=(A\s+B)/(C\s+D)$,
we have
\be
{\partial^2\over\partial {\tilde\s}^2}=\tau^{-2}(C\s+D)^4\left({\partial^2\over
\partial\s^2}+{2C\over C\s+D}{\partial\over\partial\s}\right).
\l{eeert}\ee
Hence, by (\ref{msym2})
$$
\left({\partial^2\over\partial{\tilde\s}^2}+\tilde\U(\tilde\s)\right)\tilde\phi
(\tilde\s)=\tau^{-2}(C\s+D)^4\left({\partial^2\over\partial\s^2}+{2C\over C\s+D}
{\partial\over\partial\s}+\U(\s)\right)\tilde\phi(\tilde\s)=
$$
\be
\tau^{-3/2}(C\s+D)^3\left({\partial^2\over\partial\s^2}+\U(\s)\right)\phi(\s)=0.
\l{odjko}\ee
Comparing the VTs corresponding to M\"obius transformations
\be
q\longrightarrow {Aq+B\over Cq+D},\qquad\S_0\longrightarrow\S_0,
\qquad p\longrightarrow\rho^{-1}(Cq+D)^2 p,
\l{aafghy}\ee
with the above transformations of $\s$
\be
\S_0\longrightarrow{A\S_0+B\over C\S_0+D},\qquad
q\longrightarrow q,\qquad p\longrightarrow\tau(C\S_0+D)^{-2}p.
\l{BtDluci}\ee
we see that there is a sort of duality between $q$ and $\s$. In particular,
Eq.(\ref{BtDluci}) is nothing but the analogue of the VTs corresponding to the
M\"obius transformations, with the role of $\s$ and $q$ interchanged. This
duality between $\s$ and $q$ suggests considering the analogue of the canonical
equation but with the role of $\s$ and $q$ interchanged. In this case one has
\be
\left({\partial^2\over\partial q^2}+{1\over2}\{\S_0,q\}\right)\S_0p^{-1/2}=0=
\left({\partial^2\over\partial q^2}+{1\over2}\{\S_0,q\}\right)p^{-1/2}.
\l{ESQ1}\ee
We will see that $\{\S_0,q\}$ is proportional to the quantum potential.
Generalizing the above M\"obius transformations to arbitrary ones
\be
\s\longrightarrow\tilde\s,
\l{wtt}\ee
we obtain
\be
{\partial^2\over\partial\tilde\s^2}\left({\partial\tilde\s\over\partial\s}\right
)^{1/2}=\left({\partial\tilde\s\over\partial\s}\right)^{-3/2}\left({\partial^2
\over\partial\s^2}+{1\over2}\{\tilde\s,\s\}\right),
\l{derhalfdif}\ee
where now the $\{\tilde\s,\s\}$ does not vanish. In the case in which
the functional change of $q$ is given by
\be
q\longrightarrow\tilde q(\tilde\s)=q (\s),
\l{iduqd}\ee
we have by $\tilde\U(\tilde\s)=\{\tilde q,\tilde\s\}/2=\{q,\tilde\s\}/2$,
that is
\be
0=\left({\partial\tilde\s\over\partial\s}\right)^{3/2}\left({\partial^2\over
\partial{\tilde\s}^2}+{1\over2}\{q,\tilde\s\}\right)\tilde\phi(\tilde\s)=
\left({\partial^2\over\partial\s^2}+{1\over2}\{\tilde\s,\s\}+
{1\over2}(\{q,\s\}-\{\tilde\s,\s\})\right)\phi(\s),
\l{wehavesfd}\ee
that is the equation $\left(\partial_{\tilde\s}^2+{1\over2}\{\tilde q,\tilde
\s\}\right)\tilde\phi(\tilde\s)=0$ is equivalent to
\be
\left({\partial^2\over\partial\s^2}+{1\over2}\{q,\s\}\right)\phi(\s)=0.
\l{iseedf}\ee
The above analysis shows that though $\U$ transforms as the Schwarzian
derivative, its additive term is canceled by that coming from
$\partial^2_{\tilde\s}$ acting on $(\partial\tilde\s/\partial\s)^{1/2}$.

\mysection{Legendre duality}\l{ld}

The involutive nature of the Legendre transformation implies that the
construction considered in the previous section admits a dual formulation
obtained by the exchange
\be
\S_0\longleftrightarrow\T_0,\qquad\qquad q\longleftrightarrow p.
\l{stpq}\ee
In particular, we will see how both the VTs and the canonical equation have a
dual version. The correspondence manifests the equivalence of the $q$
and $p$ descriptions and shows how this duality relies on the properties of the
Legendre transformation.

\subsection{$u$--transformations}

Let us start by considering a locally invertible momentum transformation
\be
p\longrightarrow p^u=u(p).
\l{6w}\ee
Similarly to the transformation properties of the reduced action $\S_0$, we can
associate in a natural way a new $\T_0$ with the $u$--map $\T_0\longrightarrow
\T_0^u=\T_0\circ u^{-1}$, which is equivalent to
\be
\T_0^u(p^u)=\T_0(p(p^u)).
\l{8w}\ee
Since any pair $\T_0$ and $\T_0^u$ induces the $p$--transformation defined by
(\ref{8w}), we can consider the previous construction in the reversed order,
{\it i.e.}
\be
p\longrightarrow p^u=\T_0^{u^{\;-1}}\circ\T_0(p).
\l{9w}\ee
Since $q\longrightarrow q_u=\partial_{p^u}\T_0^u(p^u)$, under (\ref{6w}) $q$
transforms as $\partial_p$, that is $q_u=(\partial_pp^u)^{-1}q$.
In summary, under VTs we have
\be
q^v=v(q),\qquad\S_0^v(q^v)=\S_0(q),\qquad\delta_v\T_0=\delta_v(qp),
\l{9sa}\ee
\be
p\sim {\partial\over\partial q},
\l{pcomederq}\ee
where $\delta_v\T_0\equiv\T_{0\,v}(p_v)-\T_0(p)$
and $\delta_v(qp)\equiv q^vp_v-qp$. Similarly, for the $u$--transformations
\be
p^u=u(p),\qquad\T_0^u(p^u)=\T_0(p),\qquad\delta_u\S_0=\delta_u(pq),
\l{29sessaaurunca}\ee
\be
q\sim {\partial\over\partial p},
\l{qcomederp}\ee
where $\delta_u\S_0\equiv\S_{0\,u}(q_u)-\S_0(q)$
and $\delta_u (pq)\equiv p^uq_u-pq$.

\subsection{The dual canonical equation}

Let us now consider the $GL(2,\CC)$--transformations
\be
p\longrightarrow p^u={Ap+B\over Cp+D},
\l{w4w}\ee
$\sigma\equiv AD-BC$. Since $q_u=\partial_{p^u}\T_0^u(p^u)=\partial_{p^u}
\T_0(p)$, we have that the effect on $q$ of (\ref{w4w}) is
\be
q\longrightarrow q_u=\sigma^{-1}(Cp+D)^2q.
\l{w5w}\ee
Eqs.(\ref{w4w})(\ref{w5w}) are equivalent to
\be
p^u\sqrt{q_u}=\epsilon\sqrt{\sigma^{-1}}(Ap\sqrt q+B\sqrt q),
\quad\sqrt{q_u}=\epsilon\sqrt{\sigma^{-1}}(Cp\sqrt q+D\sqrt q),
\l{ssssdd}\ee
where $\epsilon=\pm 1$. As $\delta_u\S_0=\delta_u(pq)$ we have
\be
\delta_u\S_0=\sigma^{-1}(ACp^2+2BCp+BD)q.
\l{sessuona}\ee

Let us recall that the canonical equation (\ref{w10}) arises from the identity
(\ref{sec}). This identity can be also obtained by simply taking the second
derivative of $p=(\partial_\s q)^{-1}$ with respect to $\s$. Thus we see that
the canonical equation stems from the very basic definition of dual variable
$p=\partial_q\S_0$, thereby capturing the germ of $p$--$q$ duality. In this
context the fact that $\S_0$ and $\T_0$ are the Legendre transform of
each other, leads to the dual version of the canonical equation (\ref{w10})
\be
\left({\partial^2\over\partial\t^2}+\V(\t)\right)p\sqrt q=0=\left({\partial^2
\over\partial\t^2}+\V(\t)\right)\sqrt q,
\l{w10w}\ee
where $\t=\T_0(p)$, and
\be
{\V}(\t)={1\over2}\{p\sqrt q/\sqrt q,\t\}={1\over2}\{p,\t\}.
\l{scharzw}\ee

\subsection{Legendre brackets}

We now show, following \cite{BOMA}\cite{1}, that there are natural brackets
associated to the relationship between the Legendre transformation and
second--order linear differential equations observed in \cite{M1}. Let us start
by considering a function $U(z)$ and its Legendre transform
\be
V(w)=z\partial_zU(z)-U(z),\qquad U(z)=w\partial_wV(w)-V(w)
\l{dkqp99}\ee
where $w=\partial_zU$ (note that $z=\partial_wV$). Taking the second derivative
of the second equation in (\ref{dkqp99}) with respect to $u=U(z)$, we see that
$\sqrt{\partial_zU}$ and $z\sqrt{\partial_zU}$ correspond to linearly
independent solutions of a second--order linear differential equation. Let us
now set
\be
\tau={\partial z\sqrt w\over\partial\sqrt w}=
z+2{w\over\partial_zw}={1\over2}{\partial^2V\over\partial{\sqrt w}^2}.
\l{2taurus}\ee
The ``Legendre brackets'' introduced in \cite{BOMA} are
\be
\{X,Y\}_{(u)}=(\partial_u\tau)^{-1}\left({\partial X\over\partial\sqrt w}
{\partial\over\partial u}{\partial Y\over\partial\sqrt w}-{\partial Y\over
\partial\sqrt w}{\partial\over\partial u}{\partial X\over\partial\sqrt
w}\right).
\l{2bomas}\ee
In the case in which the generating function is $\S_0(q)$, we have that the
linearly independent solutions of the canonical equation satisfy
\be
\{\sqrt p,\sqrt p\}_{(\s)}=0=\{q\sqrt p,q\sqrt p\}_{(\s)},\qquad
\{\sqrt p,q\sqrt p\}_{(\s)}=1,
\l{2bracketsss1}\ee
where $\s=S_0(q)$. Similarly, starting from the generating function $\T_0(p)$
we have
\be
\{\sqrt q,\sqrt q\}_{(\t)}=0=\{p\sqrt q,p\sqrt q\}_{(\t)},\qquad
\{\sqrt q,p\sqrt q\}_{(\t)}=1.
\l{2bracketsss2}\ee

\subsection{$p$ and $q$ as Legendre pair}

We have seen that the Legendre transformation is a natural framework to
investigate the structure of dualities. In particular, with the ``Legendre
pair'' $\S_0$--$\T_0$ there is a naturally associated pair of second--order
linear differential equations. We now consider the Legendre pair $\S_0$--$\T_0$
{}from a slightly different point of view. First of all observe that $\S_0$ and
$\T_0$ define the function $\t=f(\s)$, where $\s=\S_0(q)$ and $\t=\T_0(p)$. Then
suppose that instead of deriving $\T_0$ from $\S_0$ we started from a given
$\t=f(\s)$. In this case, we can define the dual pair $p$--$q$ by requiring that
\be
\T_0=q{\partial\S_0\over\partial q}-\S_0,\qquad\S_0=p{\partial\T_0\over
\partial p}-\T_0.
\l{91}\ee
Similarly, we can introduce a new pair $x^D$--$x$ by considering $p$ as the
Legendre dual of $q$, that is
\be
\nu p=x{\partial q\over\partial x}-q,\qquad\nu^{-1} q=x^D{\partial p\over
\partial x^D}-p,
\l{deltaa}\ee
where $\nu$ is a constant which has been introduced for dimensional reasons.
Therefore, considering $p$ and $q$ as Legendre duals induces the definition of
the new pair $x^D$--$x$. Integrating (\ref{deltaa}) we obtain
\be
x=x^0\exp [{\int^q_{q^0}dz(z+\nu\partial_z\S_0(z))^{-1}}],
\l{93}\ee
and
\be
x^D={x^D}^0(q+\nu\partial_q\S_0(q))\exp[{-\int^q_{q^0}dz(z+\nu
\partial_z\S_0(z))^{-1}}],
\l{94}\ee
where $x^0\equiv x(q^0)$, ${x^D}^0\equiv {x^D}(q^0)$. Regarding $p$--$q$ duality
as Legendre duality suggests considering a sequence of dual variables. In
particular, we can consider also the $x^D$--$x$ pair as Legendre duals. This
defines the pair $y^D$--$y$ by
\be
\mu x^D=y{\partial x\over\partial y}-x,\qquad\mu^{-1}x=y^D{\partial x^D\over
\partial y^D}-x^D,
\l{adeltaa}\ee
where $\mu$ is a dimensional constant. Integrating (\ref{adeltaa}) yields
\be
y=y^0\exp [{\int^x_{x^0}dz(z+\mu\partial_zq(z))^{-1}}],
\l{a93}\ee
and
\be
y^D={y^D}^0(x+\mu\partial_xq(x))\exp[{-\int^x_{x^0}dz(z+\mu\partial_zq(z))^{-1}
}],
\l{a94}\ee
where $y^0\equiv y(x^0)$, ${y^D}^0\equiv {y^D}(x^0)$.
Iteration of this construction leads to the ``$\W$--sequence''
\be
\ldots\longrightarrow ({x^D}_{i-1},x_{i-1})\longrightarrow ({x^D}_i,x_i)
\longrightarrow ({x^D}_{i+1},x_{i+1})\longrightarrow\ldots ,
\l{1Qa}\ee
and if we convene to set $(\T_0,\S_0)=(x^D_i,x_i)$ for some $i$, then
$(p,q)=(x^D_{i+1},x_{i+1})$.

\mysection{Self--dual states}\l{sds}

We have seen that, as a consequence of $\S_0$--$\T_0$ duality, there is a full
correspondence among the $u$-- and $v$--transformations. This duality indicates
that any physical system with given $\W$ has two equivalent descriptions,
corresponding to the $\S_0$ and $\T_0$ pictures. This suggests to investigate
whether this dual structure by itself may select some distinguished states. We
will see that this question will lead to the determination of states with
rather peculiar properties. In particular, based on Legendre duality, we
will see that the structure of time evolution will imply the appearance of
imaginary numbers in considering the $p\longleftrightarrow q$ interchange
for the highest symmetric states.

\subsection{Where do the $\S_0$--$\T_0$ pictures overlap?}

Let us consider the naturally selected $\W$ states corresponding to the special
case in which the $\S_0$ and $\T_0$ pictures overlap. On general grounds we
should expect that the distinguished states, for which $\S_0$ and $\T_0$ have
the same functional structure, have some peculiar properties. In order to find
this common subspace we consider the interchange of the $\S_0$ and $\T_0$
pictures given by
\be
q\longrightarrow\tilde q=\alpha p,\qquad p\longrightarrow\tilde p=\beta q.
\l{sd1}\ee
This implies that
\be
{\partial\tilde\T_0\over\partial\tilde p}=\alpha{\partial\S_0\over\partial q},
\qquad{\partial\tilde\S_0\over\partial\tilde q}=\beta{\partial\T_0\over
\partial p},
\l{oiqhx}\ee
which is equivalent to
\be
{\partial\tilde\T_0\over\partial q}=\alpha\beta{\partial\S_0\over\partial q},
\qquad{\partial\tilde\S_0\over\partial p}=\alpha\beta{\partial\T_0
\over\partial p},
\l{zzoiqhx}\ee
that is
\be
\tilde\S_0(\tilde q)=\alpha\beta\T_0(p)+cnst,\qquad\tilde\T_0(\tilde p)=
\alpha\beta\S_0(q)+cnst ,
\l{sd111}\ee
so that $\tilde\S_0(\tilde q)$ and $\tilde\T_0(\tilde p)$ are essentially the
Legendre transform of $\S_0(q)$ and $\T_0(p)$ respectively. Since
interchanging the role of $q$ and $p$ twice has no effect on the functional
dependence relating $q$ and $p$, we have that up to an additive constant
\be
\tilde{\tilde\S_0}=\S_0,\qquad\tilde{\tilde\T_0}=\T_0,
\l{sd112}\ee
so that
\be
(\alpha\beta)^2=1.
\l{113}\ee
The distinguished $\W$ states are precisely those states which are left
invariant by (\ref{sd1}) and (\ref{sd111}). Hence, up to an additive constant
\be
\tilde\S_0(\tilde q)=\S_0(q),\qquad\tilde\T_0(\tilde p)=\T_0(p).
\l{sd114}\ee
Since $\S=\S_0-Et$ and $\T=\T_0+Et$, it follows by
(\ref{sd111})(\ref{113})(\ref{sd114}) that these states correspond to
\be
\S=\pm\T+cnst.
\l{sd6}\ee
As (\ref{sd6}) should be stable under time evolution, the relation
(\ref{BtR}) fixes the sign ambiguity and sets
\be
\alpha\beta=-1.
\l{cBtProla}\ee
Therefore, the distinguished $\W$ states correspond to
\be
\S=-\T+cnst.
\l{sd7}\ee
On the other hand, since $\S=pq-\T$, we have
\be
pq=\gamma,
\l{sd11}\ee
where $\gamma$ is a constant. By (\ref{sd11}) it follows that in the case of
distinguished states the transformation (\ref{sd1}) is equivalent to
\be
q\longrightarrow\tilde q={\alpha\gamma\over q},\qquad p\longrightarrow
\tilde p={\beta\gamma\over p}.
\l{xfsd1}\ee
Let us consider the $R$--transformation (\ref{unosuq})(\ref{pinqduep})
in the case of the distinguished states (\ref{sd11})
\be
R:q\longrightarrow q^v={\delta\over q},\qquad
R:p\longrightarrow p_v=-{\gamma^2\over\delta p}.
\l{RXC}\ee
By (\ref{xfsd1}) and (\ref{RXC}) it follows that for the distinguished states
the interchanging transformation (\ref{sd1}) corresponds to an
$R$--transformation with
\be
\delta=\alpha\gamma=-{\gamma\over\beta}.
\l{dddf}\ee
Among the possible transformations (\ref{sd1}), labeled by $\alpha$ and $\beta$,
with $\alpha\beta=-1$, the one with the highest symmetric structure corresponds
to the case in which (\ref{sd1}) remains invariant under the interchange
\be
p\longleftrightarrow q,
\l{pleftrightq}\ee
that is when
\be
\alpha=\beta.
\l{HgV}\ee
Observe that, without loss of generality, we are setting to $1$ a dimensional
constant one should consider in (\ref{HgV}). It follows by (\ref{cBtProla}) and
(\ref{HgV}) that the highest symmetric states are those with
\be
\alpha=\beta=\pm i.
\l{sameway}\ee
This shows a relation between stability of the Legendre transformation under
time evolution, expressed by (\ref{BtR}), which in turn implies the minus sign
in (\ref{cBtProla}), and the appearance of the imaginary factor in
(\ref{sameway}). The interesting outcome is that, based on Legendre duality,
we have seen that the structure of time evolution implies the appearance of
imaginary numbers for the highest symmetric states.

Let us now consider another way to select the distinguished states which will
shed new light on their basic nature. First of all note that if the $\S_0$ and
$\T_0$ pictures coincide, then, in particular, the $v$--action on $\T_0$ should
be equivalent to the $u$--action on $\S_0$, that is
\be
\delta_v\T_0=\delta_u\S_0,
\l{1Sa}\ee
which is satisfied by (\ref{sd114}). Then, to find the distinguished states we
may first look for the M\"obius transformations
\be
q^v={A_vq+B_v\over C_vq+D_v},\qquad p^u={A_up+B_u\over C_up+D_u},
\l{w4wqdqd}\ee
satisfying (\ref{1Sa}), that by (\ref{7z}) and (\ref{sessuona}) is equivalent to
\be
\rho^{-1}({A_vC_v}q^2+2B_vC_vq+{B_vD_v})p=
\sigma^{-1}({A_uC_u}p^2+2B_uC_up+{B_uD_u})q.
\l{sessualona}\ee

\subsection{Dilatations}

Eq.(\ref{sessualona}) shows that the answer to the above question has two
solutions which do not fix any functional relation between $p$ and $q$. One
solution is given by $B_v=0$, $C_v=0$ and $B_u=0$, $C_u=0$. This corresponds
to the dilatations
\be
q^v={A_v\over D_v}q,\qquad p_v={D_v\over A_v}p,
\l{ba1}\ee
and
\be
p^u={A_u\over D_u}p,\qquad q_u={D_u\over A_u}q,
\l{ba2}\ee
and
\be
\delta_v\T_0=0=\delta_u\S_0.
\l{1Saddd}\ee

\subsection{Changing sign: $pq\longrightarrow p^uq_u=-pq=p_vq^v$}

The other solution of (\ref{sessualona}), which does not fix the functional
dependence between $p$ and $q$, is given by $A_v=0$, $D_v=0$ and $A_u=0$,
$D_u=0$, corresponding to
\be
q^v={B_v\over C_v}{1\over q},\qquad p_v=-{C_v\over B_v}q^2p,
\l{ba1c}\ee
and
\be
p^u={B_u\over C_u}{1\over p},\qquad q_u=-{C_u\over B_u}p^2q.
\l{ba2c}\ee
Observe that the effect of this transformation
is that of changing the sign of $pq$
\be
pq\longrightarrow p^uq_u=-pq=p_vq^v.
\l{puqupvqv}\ee
We also have
\be
\delta_v\T_0=-2pq=\delta_u\S_0,
\l{1Saee}\ee
and as $\rho=-B_vC_v$ and $\sigma=-B_uC_u$, there are no constraints
on the coefficients $B_u,C_u,B_v,C_v$.

\subsection{Self--dual states: $q^v=q_u$, $p^u=p_v$}

The solutions of (\ref{sessualona}) we found did not imply any constraint on
the state. In other words, these solutions do not fix the dynamics. However,
this investigation naturally leads to extend (\ref{1Sa}) by imposing the
conditions with the highest symmetry, that is
\be
q^v=q_u,
\l{1nU1}\ee
and
\be
p^u=p_v.
\l{1nU2}\ee
In the case of dilatations, both (\ref{1nU1}) and (\ref{1nU2}) give
\be
{A_v\over D_v}={D_u\over A_u}.
\l{give}\ee
In the case of Eqs.(\ref{ba1c}) and (\ref{ba2c}), both (\ref{1nU1})
and (\ref{1nU2}) yield
\be
{B_vB_u\over C_vC_u}=-p^2q^2,
\l{giveabbastanza}\ee
implying that
\be
pq=\gamma,
\l{gamma}\ee
with $\gamma$ a constant. By (\ref{1Saee}) and (\ref{gamma}) we have
\be
\delta_v\T_0=-2\gamma=\delta_u\S_0.
\l{1Sabb}\ee
Therefore, the transformations (\ref{w4wqdqd}) meeting the requirements
(\ref{1Sa})(\ref{1nU1}) and (\ref{1nU2}) should satisfy
\be
{B_vB_u\over C_vC_u}=-\gamma^2.
\l{1Ta}\ee
Observe that this transformation corresponds to interchanging $p$ and $q$,
namely
\be
q\longrightarrow q^v={1\over\gamma}{B_v\over C_v}p=-\gamma{C_u\over B_u}p=q_u,
\l{1Tb}\ee
which is equivalent to
\be
p\longrightarrow p^u={1\over\gamma}{B_u\over C_u}q=-\gamma{C_v\over B_v}q=p_v.
\l{1Tbb}\ee
For these states the transformations (\ref{1Tb}) and
(\ref{1Tbb}) correspond to (\ref{sd1}) with
\be
\alpha={1\over\gamma}{B_v\over C_v},\qquad\beta=-\gamma{C_v\over B_v},
\l{1Tc}\ee
and $\alpha\beta=-1$. It follows by (\ref{cBtProla}) that these states
correspond to the distinguished states. Observe that by (\ref{1Ta}) it follows
that if $B_vB_u/C_vC_u>0$, then $\gamma$ is purely imaginary. Since
\be
\delta_v(pq)=\delta_u(qp)=-pq=-\gamma,
\l{oi4w}\ee
it follows that if $B_vB_u/C_vC_u>0$, then $\delta_v(pq)=\delta_u(qp)$
corresponds to the complex conjugation
\be
\delta_v(\gamma)=\delta_u(\gamma)=\bar{\gamma}.
\l{oi4wb}\ee
Based on purely symmetry arguments, we arrived to (\ref{sameway}) in which the
imaginary factor appears. This equation is essentially a consequence of
(\ref{cBtProla}) whose minus sign origins from (\ref{BtR}). Hence, the origin
of the imaginary factor in (\ref{sameway}) traces back to (\ref{BtR}) which
relates the time evolution of the $\S$ and $\T$ pictures. As we will see, this
relation between time evolution and imaginary number is at the heart of the
imaginary factor appearing in the relation between $\S_0$ and the solutions of
the SE. On the other hand, this factor is the signal that time evolution in QM
is related to imaginary numbers. Hence, time evolution in QM is related to time
evolution of the $\S$ and $\T$ dual pictures

Due to the highest symmetry (\ref{1nU1})(\ref{1nU2}), we will call self--dual
states the distinguished states. As for these states we have $p=\gamma/q$, the
self--dual states correspond to
\be
\S_0(q)=\gamma\ln\gamma_qq,\qquad\T_0(p)=\gamma\ln\gamma_p p,
\l{selfduals}\ee
where, due to
\be
\S_0+\T_0=pq=\gamma,
\l{s0t0}\ee
the dimensional constants $\gamma_p$, $\gamma_q$ and $\gamma$ satisfy
\be
\gamma_p\gamma_q\gamma=e.
\l{k}\ee
We also note that when $p=\gamma/q$, the solutions of (\ref{w10}) and
(\ref{w10w}) coincide and
\be
\U(\s)=-{1\over4\gamma^2}=\V(\t).
\l{s0t0bbb}\ee
It follows that for the self--dual states the canonical equation
\be
\left({\partial^2\over\partial\s^2}-{1\over4\gamma^2}\right)q\sqrt p=0=
\left({\partial^2\over\partial\s^2}-{1\over4\gamma^2}\right)\sqrt p,
\l{w10congamma}\ee
coincides with its dual version
\be
\left({\partial^2\over\partial\t^2}-{1\over4\gamma^2}\right)p\sqrt q
=0=\left({\partial^2\over\partial\t^2}-{1\over4\gamma^2}\right)\sqrt q.
\l{w10wcongamma}\ee
Considering the inverse of $\S_0$ and $\T_0$ in (\ref{selfduals}), we have $q=
\gamma_q^{-1}\exp(\s/\gamma)$, $p=\gamma_p^{-1}\exp(\t/\gamma)$, so that the
solutions of Eqs.(\ref{w10}) and (\ref{w10w}), with $\U(\s)=-1/4\gamma^2=\V(\t)
$, are
\be
q\sqrt p=\sqrt{\gamma\gamma_q^{-1}}e^{{1\over2\gamma}\s}=\sqrt\gamma\sqrt q=
\gamma\sqrt{\gamma_p}e^{-{1\over2\gamma}\t},
\l{primasol}\ee
and
\be
\sqrt p=\sqrt{\gamma\gamma_q}e^{-{1\over2\gamma}\s}=\sqrt{\gamma^{-1}}p
\sqrt q=\sqrt{{\gamma_p}^{-1}}e^{{1\over2\gamma}\t}.
\l{secondasol}\ee

\mysection{The Equivalence Principle}\l{tep}

In sect.\ref{stibd} we proposed that as in the search for canonical
transformations leading to a system with vanishing Hamiltonian one obtains the
HJ equation, we may similarly look for transformations $q$ and $p$, seen as
dependent variables, reducing to the free system with vanishing energy.
This is at the heart of the EP we will formulate in this section.

\subsection{Equivalence Principle and $v$--transformations}

The discussion concerning the way of inducing the transformations $\S_0
\longrightarrow\S_0^v$ made in sect.\ref{stibd}, allows us to formulate the
above question more precisely

\vspace{.333cm}

\noindent
{\it Given an arbitrary system with reduced action $\S_0(q)$,
find the coordinate transformation $q\longrightarrow q^{v_0}=v_0(q)$,
such that the new reduced action $\S_0^{v_0}$, defined by
\be
\S_0^{v_0}(q^{v_0})=\S_0(q),
\l{universalstate}\ee
corresponds to the system with $V-E=0$.}

\vspace{.333cm}

\noindent
In the following we will use the notation $q^0\equiv q^{v_0}$, $\S_0^0\equiv
\S_0^{v_0}$, and denote by $\W^0(q^0)$ the state corresponding to $\W=0$.
Observe that the structure of the states described by $\S_0^0$ and $\S_0$
determines the ``trivializing coordinate" $q^0$ to be
\be
q\longrightarrow q^0=\S_0^{0^{\;-1}}\circ\S_0(q),
\l{9thebasicidea}\ee
The problem (\ref{universalstate}) suggested the following Equivalence Principle
(EP) formulated in \cite{1}

\vspace{.333cm}

\noindent
{\it For each pair $\W^a,\W^b$, there is a $v$--transformation
$q^a\longrightarrow q^b=v(q^a)$ such that}
\be
\W^a(q^a)\longrightarrow\W^b(q^b).
\l{equivalence}\ee

\subsection{Classical Mechanics and the Equivalence Principle}

We now show the basic fact that the EP cannot be consistently implemented in CM.
In order to implement the EP in CM we should first understand the nature of
the transformation
\be
\W(q)\longrightarrow\W^v(q^v),
\l{natura}\ee
induced by
\be
\S_0^{cl}(q)\longrightarrow\S_0^{cl\,v}(q^v)=\S_0^{cl}(q(q^v)).
\l{8perSzeroclassico}\ee
To answer this question we use the fact that the CSHJE (\ref{012}) provides a
correspondence between $\W$ and $\S_0^{cl}$. In particular,
$\S_0^{cl\,v}(q^v)$ should satisfy the CSHJE
\be
{1\over2m}\left({\partial\S_0^{cl\,v}(q^v)\over\partial q^v}\right)^2
+\W^v(q^v)=0.
\l{eepf}\ee
Using $\S_0^{cl\,v}(q^v)=\S_0^{cl}(q)$, which defines the classical
$v$--transformations $q\longrightarrow q^v=v(q)$, and comparing (\ref{012}) with
(\ref{eepf}) we obtain
\be
\W(q)\longrightarrow\W^v(q^v)=(\partial_{q^v}q)^2\W(q),
\l{natura2}\ee
that is
\be
\W^v(q^v)(dq^v)^2=\W(q)(dq)^2.
\l{sty}\ee
Therefore, in CM consistency requires that $\W(q)$ belongs
to $\Q^{cl}$, the space of functions transforming as quadratic differentials
under the classical $v$--transformations, that is
\be
\W\in\Q^{cl}=\left\{g\left|g^v(q^v)=\left({\partial_{q^v}q}\right)^2g(q)\right.
\right\}.
\l{quadratici}\ee
Let us now consider the case of the state $\W^0$. By (\ref{natura2}) it follows
that
\be
\W^0(q^0)\longrightarrow\W^v(q^v)=\left(\partial_{q^v}q^0\right)^2\W^0(q^0)=0.
\l{qddd}\ee
Then we have \cite{1}

\vspace{0.333cm}

\noindent
{\it $\W$ states transform as quadratic differentials under classical $v$--maps.
It follows that $\W^0$ is a fixed point in $\H$. Equivalently, in CM the space
$\H$ cannot be reduced to a point upon factorization by the classical
$v$--transformations. Hence, the EP (\ref{equivalence}) cannot be consistently
implemented in CM.}

\subsection{Modifying the classical HJ equation}

It is clear that in order to implement the EP we have to modify the CSHJE. Let
us consider the properties that the equation satisfied by $\S_0$ should have.
First of all, observe that by Eqs.(\ref{lt00})--(\ref{pq}) it follows that
adding a constant to either $\S_0$ or $\T_0$ does not change the dynamics.
Therefore, the most general equation that $\S_0$ should satisfy has the form
\be
F(\S_0',\S_0'',\ldots)=0,
\l{6Yc}\ee
where $'\equiv\partial_q$. Since in the classical limit we have
\be
F(\S_0',\S_0'',\ldots)=0\longrightarrow
{1\over2m}\left({\partial\S_0^{cl}\over\partial q}\right)^2+\W(q)=0,
\l{6Ycclassic}\ee
it is useful to write Eq.(\ref{6Yc}) in the form
\be
{1\over2m}\left({\partial\S_0(q)\over\partial q}\right)^2+\W(q)+Q(q)=0,
\l{1Xv}\ee
so that the classical limit corresponds to
\be
Q\longrightarrow0.
\l{classicoqezero}\ee

\subsection{Covariance as consistency condition}

Let us derive the transformation properties of $\W+Q$ by consistency. Namely,
we first consider the transformed version of Eq.(\ref{1Xv})
\be
{1\over2m}\left({\partial\S_0^v(q^v)\over\partial q^v}\right)^2+\W^v(q^v)+
Q^v(q^v)=0,
\l{3H}\ee
then note that by (\ref{VIP})
$(\partial_{q^v}\S_0^v)^2=(\partial_{q^v}q)^2(\partial_q\S_0)^2$,
so that (\ref{1Xv}) and (\ref{3H}) yield
\be
\W^v(q^v)+Q^v(q^v)=\left({\partial_{q^v}q}\right)^2\left(\W(q)+Q(q)\right).
\l{1Td}\ee
Hence $\W+Q$ belongs to the space $\Q$ of functions
transforming as quadratic differentials under VTs
\be
(\W+Q)\in\Q.
\l{fiof}\ee
Therefore, choosing (\ref{VIP}) to represent the functional change
$\S_0\longrightarrow\S_0^v$ provides simple transformation properties for
$\W+Q$. In particular, Eq.(\ref{1Xv}) transforms in a manifestly covariant way
with the choice (\ref{VIP}). Had we used another representation,
{\it e.g.} (\ref{7azz}), then the resulting transformation rule for $\W+Q$ would
involve also $f$ and $\S_0$. We stress that there is nothing wrong in
considering this alternative representation, nevertheless it is clear that the
formalism would be far more cumbersome.

\mysection{Implementing the Equivalence Principle}\l{itep}

The relevance of the EP is manifest already in considering the transformation
properties of $\W$ and $Q$. Actually, according to the EP, all the $\W$'s are
connected with each other by a coordinate transformation. On the other hand, we
have seen that if $\W$ transforms as a quadratic differential, then the state
$\W^0$ would be a fixed point in the space $\H$. It follows that
\be
\W\notin\Q.
\l{WnotinH}\ee
On the other hand, we have seen that by consistency $(\W+Q)\in\Q$, so that by
(\ref{WnotinH})
\be
Q\notin\Q.
\l{iuwdf}\ee
In this section, following \cite{1}, we will see that the transformation
properties of $\W$ and $Q$ lead to a cocycle condition which will be crucial to
fix $Q$. In particular, we will first find the transformation properties
of an inhomogeneous term one is forced to introduce in the transformation
properties of $\W$. In the next section we will see that these properties fix
this term and therefore $Q$. In the present section we will also consider the
problem of selecting a self--dual state as ``reference state''.

\subsection{$\W$ states as inhomogeneous terms}

Note that the only possible way to reach an arbitrary state $\W^v\ne 0$
starting from the state $\W^0$, is that it transforms with an inhomogeneous
term. This is the physical content of Eq.(\ref{WnotinH}). This implies that
under a VT, in general we have
\be
\W^0\longrightarrow\W^v\ne 0.
\l{dazeroadiversodazero}\ee
In particular, any two arbitrary states $\W$ and $\W^v$ are related by
\be
\W^v(q^v)=\left({\partial_{q^v}q}\right)^2\W(q)+\Z(q;q^v),
\l{BtD1Td}\ee
where $\Z(q;q^v)$ is the inhomogeneous term to be determined.
By (\ref{1Td}) we have
\be
\W^v(q^v)+Q^v(q^v)=\left({\partial_{q^v}q}\right)^2\left(\W(q)+Q(q)\right),
\l{1Tda}\ee
so that, replacing $\W^v(q^v)$ in (\ref{1Tda}) with
the right hand side of (\ref{BtD1Td}), we obtain
\be
\left({\partial_{q^v}q}\right)^2\W(q)+\Z(q;q^v)+Q^v(q^v)=\left(
{\partial_{q^v}q}\right)^2\left(\W(q)+Q(q)\right),
\l{Te2}\ee
that is
\be
Q^v(q^v)=\left({\partial_{q^v}q}\right)^2Q(q)-\Z(q;q^v).
\l{BtD21Td}\ee
Replacing $\W(q)$ in (\ref{BtD1Td}) with $\W^0(q^0)$,
we have $\W^v(q^v)=\Z(q^0;q^v)$, so that
\be
\W(q)=\Z(q^0;q).
\l{1H}\ee
Therefore, all the states correspond to the inhomogeneous, and unique, part
in the transformation of $\W^0$ induced by $v$--maps. In other words, all
the states originate from the map $q^0\longrightarrow q=v_0^{-1}(q^0)$. We
also note that, in view of (\ref{1H}), $\W^0$ plays a distinguished role, as
any state can be written in the universal form (\ref{1H}) in which the
distinguished coordinate $q^0$, representing the state $\W^0$, appears.

Observe that by construction, Eq(\ref{1Xv}) has the simple transformation
property
\be
{1\over2m}\left({\partial\S_0^v(q^v)\over\partial q^v}\right)^2+\W^v(q^v)+
Q^v(q^v)=\left({\partial q\over\partial q^v}\right)^2\left[{1\over2m}\left(
{\partial\S_0(q)\over\partial q}\right)^2+\W(q)+Q(q)\right]=0.
\l{1Xvesempiod}\ee

We saw in sect.\ref{sds} that there is a set of distinguished, or self--dual,
states which are naturally selected by purely symmetry arguments. In doing this
we found an explicit expression for $\S_0$ and $\T_0$ corresponding to the
self--dual states. Nevertheless, we did not consider the problem of finding
the corresponding expression of $\W$. However, even if the form of $\W$ has not
been derived,\footnote{Note that finding the explicit form of $\W$ associated to
the self--dual states, would essentially be equivalent to finding the dynamical
equation for $\S_0$.} one expects that its structure should reflect the highest
symmetry we used to fix the explicit form (\ref{selfduals}) of $\S_0$ and
$\T_0$. We will see that the state $\W^0$ precisely corresponds to a self--dual
state. As such the state $\W^0$ can be seen as the state in which the $\S_0$ and
$\T_0$ pictures overlap.\footnote{We stress that in deriving the quantum version
of the HJ equation, we will only use the EP and the existence of CM. In doing
this we frequently discuss the role of the self--dual states. One aim of these
discussions is to show that the quantum HJ equation fully reflects the natural
dual structure which emerges in considering the Legendre pair $\S_0$--$\T_0$.
Therefore, in deriving the quantum HJ equation, we do not assume any kind of
duality to be satisfied, rather we will see that the solution following from the
EP satisfies $\S_0$--$\T_0$ duality.}

\subsection{The cocycle condition}

We now consider the transformation properties of $\Z(q^a;q^b)$. Comparing
\be
\W^b(q^b)=\left({\partial_{q^b}q^a}\right)^2\W^a(q^a)+\Z(q^a;q^b)=\Z(q^0;q^b),
\l{ganzate}\ee
with the same formula with $q^a$ and $q^b$ interchanged we have
\be
\Z(q^b;q^a)=-(\partial_{q^a}q^b)^2\Z(q^a;q^b),
\l{inparticolare}\ee
in particular
\be
\Z(q;q)=0.
\l{inparticolareb}\ee
More generally, comparing
$$
\W^b(q^b)=\left({\partial_{q^b}q^c}\right)^2\W^c(q^c)+\Z(q^c;q^b)=\left({
\partial_{q^b}q^c}\right)^2\left[\left({\partial_{q^c}q^a}\right)^2\W^a(q^a)+
\Z(q^a;q^c)\right]+\Z(q^c;q^b)=
$$
\be
\left({\partial_{q^b}q^a}\right)^2\W^a(q^a)+\left({\partial_{q^b}q^c}
\right)^2\Z(q^a;q^c)+\Z(q^c;q^b),
\l{DemetrioStratosZ}\ee
with Eq.(\ref{ganzate}), we obtain the basic cocycle condition
\be
\Z(q^a;q^c)=\left({\partial_{q^c}q^b}\right)^2\left[\Z(q^a;q^b)-\Z(q^c;q^b)
\right],
\l{cociclo3}\ee
which can be seen as the essence of the EP.

Before going further some remarks are in order. We started by noticing that
since the coordinate of a physical system can be seen as an independent
variable, its choice essentially reduces to a notational one. This suggested
considering the transformation among different physical systems to be performed
through VTs which are equivalent to consider $\S_0$ as a scalar function. The
power of this approach is evident in implementing the EP. In particular, we
obtained a manifestly covariant description where the scalar nature of $\S_0$
implied that $(\W+Q)\in\Q$. The whole investigation culminated in the basic
equation (\ref{cociclo3}), which makes clear the basic nature of the EP. As
already observed in sect.\ref{stibd}, even if considering another kind of
representation for the transformation of $\S_0$ would yield the same results,
these would not be manifestly covariant and the elegance of the construction
would be lost. In this sense representing the state transformation by the
pullback of $\S_0$ by ${v^{-1}}^*$ is the natural choice.

By eqs.(\ref{BtD1Td})(\ref{BtD21Td}) and (\ref{cociclo3}) we have
\be
\W\in\P,\qquad Q\in\P,\qquad\Z(\,\cdot\,;q)\in\P,
\l{projective}\ee
where $\P$ is the space of ``inhomogeneous" quadratic differentials
\be
\P=\left\{f\left|f^v(q^v)=\left({\partial_{q^v}q}\right)^2f(q)+
\alpha_f\Z(q;q^v)\right.\right\},
\l{pprroojjeeccttiivvee}\ee
with $\alpha_f$ a constant. Note that the difference between any two elements
$f_i$ and $f_j$ in $\P$ with $\alpha_{f_i}=\alpha_{f_j}$, belongs to the space
$\Q$, that is $(f_i-f_j)\in\Q$, $f_i,f_j\in\P$, $\alpha_{f_i}=\alpha_{f_j}$.

As we will see, Eq.(\ref{cociclo3}), which follows as a consistency condition
{}from the EP, univocally fixes the differential equation for $\S_0$. In
particular, in the following we will first show that Eq.(\ref{cociclo3}), and
(\ref{6Yc}) implies $\Z(q;\gamma(q))=0=\Z(\gamma(q);q)$, with $\gamma$ a
M\"obius transformation. In the next section,
we will prove that $\Z(q^a;q^b)$ is proportional to the Schwarzian
derivative $\{q^a,q^b\}$. It follows that $\P$ is the space of functions
transforming as projective connections under $v$--maps. The fact that
$\Z(q^a;q^b)\propto\{q^a,q^b\}$, is related to the so--called pseudogroup
property \cite{Gunning} associated to (\ref{cociclo3}). This is the basic
defining property of complex analytic structures as it implies that the
composition of two complex analytic local homeomorphisms is again a local
homeomorphism. In turns out that any element $g$ in the family of
one--dimensional complex analytic local homeomorphisms, solutions of a system
of differential equations involving only the first and higher derivatives, and
satisfying the pseudogroup property, satisfies either $g''(x)/g'(x)=0$ or
$\{g(x),x\}=0$ \cite{Gunning}\cite{Cartan}\cite{GuillStern}. It is worth
noticing that this issue is related to the cohomology of Lie algebras
\cite{Fuks}\cite{Segal}.

\subsection{Cocycle condition and M\"obius symmetry}

Let us start observing that if the cocycle condition (\ref{cociclo3}) is
satisfied by $\Z(f(q);q)$, then this is still satisfied by adding a coboundary
term
\be
\Z(f(q);q)\longrightarrow\Z(f(q);q)+(\partial_qf)^2G(f(q))-G(q).
\l{coboundary}\ee
Since $\Z(Aq;q)$ evaluated at $q=0$ is independent of $A$, we have
\be
0=\Z(q;q)=\Z(q;q)_{|q=0}=\Z(Aq;q)_{|q=0}.
\l{valezero}\ee
Therefore, if both $\Z(f(q);q)$ and (\ref{coboundary}) satisfy (\ref{cociclo3}),
then $G(0)=0$, which is the unique condition that $G$ should satisfy. We now use
(\ref{6Yc}) to fix the ambiguity (\ref{coboundary}). First of all observe that
by (\ref{1H}) the differential equation we are looking for is
\be
\Z(q^0;q)=\W(q).
\l{equazionewelook}\ee
Then, recalling that $q^0=\S_0^{0^{\;-1}}\circ\S_0(q)$ (see
Eq.(\ref{9thebasicidea})), we have that a necessary condition to satisfy
(\ref{6Yc}) is that $\Z(q^0;q)$ depends only on the first and higher derivatives
of $q^0$. This in turn implies that for any constant $B$ we have $\Z(q^a+B;q^b)
=\Z(q^a;q^b)$, that together with (\ref{inparticolare}) gives
\be
\Z(q^a+B;q^b)=\Z(q^a;q^b)=\Z(q^a;q^b+B).
\l{1XuBtR}\ee
Let $A$ be a non--vanishing constant and set $h(A,q)=\Z(Aq;q)$. By
(\ref{1XuBtR}) we have $h(A,q+B)=h(A,q)$, that is $h(A,q)$ is
independent of $q$. On the other hand by (\ref{valezero}) we have
$h(A,0)=0$. This fact and (\ref{inparticolare}), imply
\be
\Z(Aq;q)=0=\Z(q;Aq).
\l{pt11}\ee
Eq.(\ref{cociclo3}) implies $\Z(q^a;
Aq^b)=A^{-2}(\Z(q^a;q^b)-\Z(Aq^b;q^b))$, so that by (\ref{pt11})
\be
\Z(q^a;Aq^b)=A^{-2}\Z(q^a;q^b).
\l{1Xw}\ee
By (\ref{inparticolare}) and (\ref{1Xw}) we have $\Z(Aq^a;q^b)=-A^{-2}(
\partial_{q^b}q^a)^2\Z(q^b;Aq^a)=-(\partial_{q^b}q^a)^2\Z(q^b;q^a)=
\Z(q^a;q^b)$, that is
\be
\Z(Aq^a;q^b)=\Z(q^a;q^b).
\l{1Xu}\ee
Setting $f(q)=q^{-2}\Z(q;q^{-1})$ and noticing that by
(\ref{inparticolare}) and (\ref{1Xu}) $f(Aq)=-f(q^{-1})$, we obtain
\be
\Z(q;q^{-1})=0=\Z(q^{-1};q).
\l{kenzoexenasarco}\ee
Furthermore, since by (\ref{cociclo3}) and (\ref{kenzoexenasarco}) one has
$\Z(q^a;q^{b^{-1}})=q^{b^4}\Z(q^a;q^b)$, it follows that
\be
\Z(q^{a^{-1}};q^b)=-\left(\partial_{q^b}q^{a^{-1}}\right)^2\Z(q^b;
q^{a^{-1}})=-\left(\partial_{q^b}{q^a}\right)^2\Z(q^b;q^a)=\Z(q^a;q^b),
\l{ririaggiunto}\ee
so that
\be
\Z(q^{a^{-1}};q^b)=\Z(q^a;q^b)=q^{b^{-4}}\Z(q^a;q^{b^{-1}}).
\l{pt22}\ee
Since translations, dilatations and inversion are the generators of the M\"obius
group, it follows by (\ref{1XuBtR})(\ref{1Xw})(\ref{1Xu}) and (\ref{pt22}) that
\be
\Z(\gamma(q^a);q^b)=\Z(q^a;q^b),
\l{am21}\ee
and
\be
\Z(q^a;\gamma(q^b))=\left(\partial_{q^b}\gamma(q^b)\right)^{-2}\Z(q^a;q^b),
\l{am22}\ee
where
\be
\gamma(q)={Aq+B\over Cq+D},
\l{am23}\ee
with $\left(\begin{array}{c}A\\C\end{array}\begin{array}{cc}B\\D\end{array}
\right)\in GL(2,{\CC})$. In particular
\be
\Z(\gamma(q);q)=0=\Z(q;\gamma(q)).
\l{am24}\ee

For sake of completeness we recall that the infinitesimal versions of
translations, dilatations and special conformal transformations of $q$ are
$q\longrightarrow q-\epsilon$, $q\longrightarrow q-\epsilon q$ and $q
\longrightarrow q-\epsilon q^2$ respectively. Recall that a special
conformal transformation has the form $q\longrightarrow q/(1-Bq)$. In the
case of infinitesimal M\"obius transformations of $q+\epsilon f(q)$, we
have
\be
q+\epsilon f(q)\longrightarrow q+\epsilon f(q)-\epsilon(
q+\epsilon f(q))^n,\qquad n=0,1,2,
\l{trasformazioneperqeps}\ee
so that the first--order contribution to the variation of $q+\epsilon
f(q)$ coincides with that of $q$. 

\subsection{Selecting a self--dual state}

Let us now go back to the question posed at the beginning of sect.\ref{tep},
Eq.(\ref{universalstate}), and to the EP. We saw that, if $\S_0=cnst$,
then the EP cannot be consistently implemented. Similarly, the Legendre
transformation cannot be defined for $\S_0=Aq+B$ and $\T_0=Ap+B$. In CM, the
state with $\S_0^{cl}=cnst$ corresponds to the free particle with vanishing
energy, whereas in the case in which $\S_0^{cl}\propto q$ one has the free
particle with $E\ne 0$. Therefore, problems arise just for the free particle.
The $\S_0$--$\T_0$ duality investigated above, and the fact that the problems
which arise in applying the EP reflect in the impossibility of defining the
Legendre transformation, indicate that the solution to this question is
related to the observed dualities. Let us recall that the self--dual (or
distinguished) states
\be
\S_0(q)=\gamma\ln\gamma_qq,\qquad\T_0(p)=\gamma\ln\gamma_p p,
\l{selfdualsdfwf}\ee
parameterized by $\gamma$, are those with the highest symmetric properties
as they are characterized by
\be
\delta_v\T_0=\delta_u\S_0,\qquad q^v=q_u,\qquad p^u=p_v.
\l{univvv}\ee
We have seen that the states (\ref{selfdualsdfwf}) play a special role. On the
other hand, the fact that all the states satisfy the relation $\S_0^v(q^v)=\S_0
(q)$, suggests selecting a particular self--dual state (\ref{selfdualsdfwf}) as
reference state. This means choosing a given value $\gamma_{sd}$ for the
constant $\gamma$ in (\ref{selfdualsdfwf}). We denote by $q^{sd}$ and $p^{sd}$
the corresponding coordinate and the momentum respectively. Therefore, we have
the reference state
\be
\S_0^{sd}(q^{sd})=\gamma_{sd}\ln\gamma_qq^{sd},\qquad\T_0(p^{sd})=\gamma_{sd}
\ln\gamma_p p^{sd}.
\l{xxxzselfdualsdfwf}\ee
We will denote by $\W^{sd}$ the distinguished self--dual state $\W$
corresponding to $\S_0^{sd}(q^{sd})$. Therefore, all possible states have a
reduced action of the form
\be
\S_0(q)=\S_0^{sd}(q^{sd})=\gamma_{sd}\ln\gamma_qq^{sd}(q),
\l{sxc1}\ee
which implies that the coordinate transformation reducing an arbitrary system
with reduced action $\S_0(q)$ to the distinguished self--dual state
$\S_0^{sd}(q^{sd})$ is
\be
q\longrightarrow q^{sd}=v_{sd}(q)=\gamma_q^{-1}e^{{1\over\gamma_{sd}}\S_0(q)}.
\l{mapppa}\ee
Note that, according to the EP, the transformation should
be locally invertible. Therefore, there is also the map transforming
$\S_0^{sd}$ into $\S_0$, that is $q^{sd}\longrightarrow q=v^{-1}_{sd}(q^{sd})$.

\mysection{The quantum stationary HJ equation}\l{tqhje}

We can now determine the explicit expression of $Q$ in Eq.(\ref{1Xv}). We first
consider some facts about the Schwarzian derivative. Next, we will prove the
basic identification $\Z(q^a;q^b)=-\beta^2\{q^a,q^b\}/4m$. This result will
lead to the explicit determination of $Q$. After proving that $Q$ is univocally
fixed, we will show in the next section that the modified stationary HJ equation
implies the SE.

\subsection{Complex entries and real values of the Schwarzian derivative}

Let us recall a few facts about the Schwarzian derivative. Let us first
consider $\{e^h,x\}$. Since by (\ref{cometrasforma}) $\{y,x\}={y'}^2
(\{y,y\}-\{x,y\})=-{y'}^2\{x,y\}$, we have
\be
\{e^h,x\}=-\left({\partial e^h\over\partial x}\right)^2\{x,e^h\}.
\l{191166}\ee
On the other hand, (\ref{cometrasforma}) implies
\be
\{x,e^h\}=e^{-2h}\{x,h\}-\{e^h,h\}=e^{-2h}\{x,h\}+{1\over2},
\l{schawww}\ee
and by (\ref{191166})
\be
\{e^h,x\}=\{h,x\}-{1\over2}\left({\partial h\over\partial x}\right)^2.
\l{oier}\ee
A property of $\{h,x\}$ is that it may take real values even if $h$ and $x$ take
complex values. The basic reason is already evident from the invariance under
the complex M\"obius transformations $\{\gamma(h),x\}=\{h,x\}$. Another example
is provided by $\{e^{ih},x\}=\{h,x\}+(\partial_x h)^2/2$. This implies that if
$h$ and/or $x$ take real or purely imaginary values, then $\{e^{ih},x\}\in\RR$.
Similarly, if $\alpha$ is independent of $x$, we have
\be
\{e^{i\alpha h},x\}=
\{h,x\}+{\alpha^2\over2}\left({\partial h\over\partial x}\right)^2.
\l{xxisacons}\ee
Observe that if $h$ is independent of $\alpha$, then
\be
(\partial_x h)^2=\alpha^{-1}\partial_\alpha\{e^{i\alpha h},x\}.
\l{dametteresenzaltro}\ee
The fact that $\{e^{i\alpha h},x\}$ splits into two parts is a key point in our
construction. This is a consequence of the following property of the Schwarzian
derivative
\be
\{h^{\alpha},x\}=\{h,x\}+{1\over2}(1-\alpha^2)\left({h'\over h}\right)^2.
\l{alphah}\ee
The identity (\ref{xxisacons}) implies that $(\partial_q\S_0)^2$, which
transforms as a quadratic differential under $v$--maps, can be expressed
as the difference of two Schwarzian derivatives
\be
\left({\partial\S_0\over\partial q}\right)^2=
{\beta^2\over2}\left(\{e^{{2i\over\beta}\S_0},q\}-\{\S_0,q\}\right).
\l{expoid}\ee
Observe that this basic identity forces us to introduce the dimensional
constant $\beta$. We also note the appearance of the imaginary factor.

\subsection{$\Z(q^a;q^b)=-{\beta^2\over 4m}\{q^a,q^b\}$}

We now start showing that $\Z(q^a;q^b)\propto\{q^a,q^b\}$. Since we are
interested to what corresponds $\Z(\,\cdot\,;q)$, we can consider,
without loss of generality, the case in which $q^a$ differs
infinitesimally from $q\equiv q^b$, that is $q^a=q+\epsilon f(q)$. Since
$\Z(q^a;q^b)$ depends only on the first and higher derivatives of $q^a$,
we have
\be
\Z(q+\epsilon f(q);q)=c_1\epsilon f^{(k)}(q)+\O(\epsilon^2),
\l{preliminare1}\ee
where $f^{(k)}\equiv\partial_q^kf$. We will see that
\be
c_1\ne0.
\l{enneecuno}\ee
Let us first fix the value of $k$ in (\ref{preliminare1}). To this end we
note that
\be
\Z(Aq+\epsilon Af(q);Aq)=\Z(q+\epsilon f(q);Aq)=A^{-2}\Z(q+\epsilon
f(q);q),
\l{preliminare2}\ee
on the other hand, setting $F(Aq)=Af(q)$, by Eq.(\ref{preliminare1}) we have
\be
\Z(Aq+\epsilon Af(q);Aq)=\Z(Aq+\epsilon F(Aq);Aq)=
c_1\epsilon\partial_{Aq}^kF(Aq)=A^{1-k}c_1\epsilon f^{(k)}(q),
\l{preliminare3}\ee
so that $k=3$, that is
\be
\Z(q+\epsilon f(q);q)=c_1\epsilon f^{(3)}(q)+\O(\epsilon^2).
\l{allordineespilon}\ee
The above scaling property generalizes to higher--order contributions in
$\epsilon$. In particular, the contribution at order $\epsilon^n$ to
$\Z(Aq+\epsilon Af(q);Aq)$ is given by a sum of terms of the form
\be
c_{i_1\ldots i_n}\partial_{Aq}^{i_1}\epsilon F(Aq)\cdot\cdot\cdot
\partial_{Aq}^{i_n}\epsilon F(Aq)=c_{i_1\ldots i_n}\epsilon^nA^{n-\sum i_k}
f^{(i_1)}(q)\cdot\cdot\cdot f^{(i_n)}(q),
\l{tisognocalifornia}\ee
and by (\ref{preliminare2})
\be
\sum_{k=1}^ni_k=n+2.
\l{lordineepsilon}\ee
On the other hand, since $\Z(q^a;q^b)$ depends only on the first and
higher derivatives of $q^a$, we have
\be
i_k\geq 1,\qquad k\in[1,n].
\l{pietruccio}\ee
Eqs.(\ref{lordineepsilon})(\ref{pietruccio}) imply that either
\be
i_k=3,\qquad i_j=1,\qquad j\in[1,n],\;\; j\ne k,
\l{eitheruno}\ee
or
\be
i_k=i_j=2,\qquad i_l=1,\qquad l\in[1,n],\;\; l\ne k,\,l\ne j,
\l{orunodue}\ee
so that
\be
\Z(q+\epsilon f(q);q)=\sum_{n=1}^\infty\epsilon^n\left(c_nf^{(3)}
f^{(1)^{n-1}}+d_nf^{(2)^2}f^{(1)^{n-2}}\right),\qquad d_1=0.
\l{preliminare4}\ee

\vspace{.333cm}

We now show that either $c_1\ne0$ or $\Z(q+\epsilon f(q);q)=0$.
Let us consider the transformations
\be
q^b=v^{ba}(q^a),\qquad q^c=v^{cb}(q^b)=v^{cb}\circ v^{ba}(q^a),
\qquad q^c=v^{ca}(q^a).
\l{abcevx}\ee
Note that $v^{ab}=v^{ba^{-1}}$, and
\be
v^{ca}=v^{cb}\circ v^{ba}.
\l{vcacbba}\ee
We can write these transformations in the form
\be
q^b=q^a+\epsilon^{ba}(q^a),\qquad q^c=q^b+\epsilon^{cb}(q^b)=q^b+
\epsilon^{cb}(q^a+\epsilon^{ba}(q^a)),\qquad q^c=q^a+\epsilon^{ca}(q^a).
\l{epsilonabcvx}\ee
Since $q^b=q^a-\epsilon^{ab}(q^b)$, we have $q^b=q^a-\epsilon^{ab}(q^a+
\epsilon^{ba}(q^a))$ that compared with $q^b=q^a+\epsilon^{ba}(q^a)$
yields
\be
\epsilon^{ba}+\epsilon^{ab}\circ({\bf 1}+\epsilon^{ba})=0,
\l{abbabccbxft}\ee
where ${\bf 1}$ denotes the identity map. More generally,
Eq.(\ref{epsilonabcvx}) gives
\be
\epsilon^{ca}(q^a)= \epsilon^{cb}(q^b)+\epsilon^{ba}(q^a)=
\epsilon^{cb}(q^b)-\epsilon^{ab}(q^b),
\l{unpoininoFBT1}\ee
so that we obtain (\ref{vcacbba}) with $v^{yx}={\bf 1}+\epsilon^{yx}$
\be
\epsilon^{ca}=\epsilon^{cb}\circ({\bf 1}+\epsilon^{ba})+\epsilon^{ba}=
({\bf 1}+\epsilon^{cb})\circ({\bf 1}+\epsilon^{ba})-{\bf 1}.
\l{unpoininoFBT2}\ee
Let us consider the case in which $\epsilon^{yx}(q^x)=\epsilon f_{yx}
(q^x)$, with $\epsilon$ infinitesimal. At first--order in $\epsilon$
Eq.(\ref{unpoininoFBT2}) reads
\be
\epsilon^{ca}=\epsilon^{cb}+\epsilon^{ba},
\l{unpoininoFBT3}\ee
in particular, $\epsilon^{ab}=-\epsilon^{ba}$. Since $(q^a;q^b)=c_1
{\epsilon^{ab}}'''(q^b)+\O^{ab}(\epsilon^2)$, where $'$ denotes the
derivative with respect to the argument, we can use the cocycle
condition (\ref{cociclo3}) to get
\be
c_1{\epsilon^{ac}}'''(q^c)+\O^{ac}(\epsilon^2)=(1+{\epsilon^{bc}}'(q^c))^2
\left(c_1{\epsilon^{ab}}'''(q^b)+\O^{ab}(\epsilon^2)
-c_1{\epsilon^{cb}}'''(q^b)-\O^{cb}(\epsilon^2)\right),
\l{unpoininoFBT4}\ee
that at first--order in $\epsilon$ corresponds to (\ref{unpoininoFBT3}).
We see that $c_1\ne0$. For, if $c_1=0$, then by (\ref{unpoininoFBT4}), at
second--order in $\epsilon$ one would have
\be
\O^{ac}(\epsilon^2)=\O^{ab}(\epsilon^2)-\O^{cb}(\epsilon^2),
\l{unpoininoFBT5}\ee
which contradicts (\ref{unpoininoFBT3}). In fact, by (\ref{preliminare4}) we
have
\be
\O^{ab}(\epsilon^2)=c_2{\epsilon^{ab}}'''(q^b){\epsilon^{ab}}'(q^b)+
d_2{{\epsilon^{ab}}''}^2(q^b)+\O^{ab}(\epsilon^3),
\l{unpoininoFBT5bisse}\ee
that together with (\ref{unpoininoFBT5}) provides a relation which cannot
be consistent with $\epsilon^{ac}(q^c)=\epsilon^{ab}(q^b)-\epsilon^{cb}
(q^b)$. A possibility is that $(q^a;q^b)=0$. However, this is ruled out by
the EP, so that $c_1\ne0$. Higher--order terms due to a non--vanishing
$c_1$ are obtained by using $q^c=q^b+\epsilon^{cb}(q^b)$, $\epsilon^{ac}
(q^c)=\epsilon^{ab}(q^b)-\epsilon^{cb}(q^b)$ and $\epsilon^{bc}(q^c)=-
\epsilon^{cb}(q^b)$ in $c_1\partial_{q^c}^3\epsilon^{ac}(q^c)$ and in
$c_1\left(2\partial_{q^c}\epsilon^{bc}(q^c)+{\partial_{q^c}\epsilon^{bc}
(q^c)}^2\right)\partial^3_{q^b}\left({\epsilon^{ab}}(q^b)-{\epsilon^{cb}}
(q^b)\right)$. Note that one can also consider the case in which both the
first-- and second--order contributions to $(q^a;q^b)$ are vanishing.
However, this possibility is ruled out by a similar analysis. In general,
one has that if the first non--vanishing contribution to $(q^a;q^b)$ is
of order $\epsilon^n$, $n\geq2$, then, unless $(q^a;q^b)=0$, the cocycle
condition (\ref{cociclo3}) cannot be consistent with linearity of
(\ref{unpoininoFBT3}). Observe that we proved that $c_1\ne0$ is a necessary
condition for the existence of solutions $(q^a;q^b)$ of the cocycle
condition (\ref{cociclo3}), depending only on the first and higher
derivatives of $q^a$. Existence of solutions follows from the fact that
the Schwarzian derivative $\{q^a,q^b\}$ solves (\ref{cociclo3}) and
depends only on the first and higher derivatives of $q^a$.

The fact that $c_1=0$ implies $(q^a;q^b)=0$, can be also seen by explicitly
evaluating the coefficients $c_n$ and $d_n$. These can be obtained using the
same procedure considered above to prove that $c_1\ne0$. Namely, inserting
the expansion (\ref{preliminare4}) in (\ref{cociclo3}) and using $q^c=q^b+
\epsilon^{cb}(q^b)$, $\epsilon^{ac}(q^c)=\epsilon^{ab}(q^b)-\epsilon^{cb}
(q^b)$ and $\epsilon^{bc}(q^c)=-\epsilon^{cb}(q^b)$, we obtain
\be
c_n=(-1)^{n-1}c_1,\qquad\qquad d_n={3\over2}(-1)^{n-1}(n-1)c_1,
\l{cenneedennexf}\ee
which in fact are the coefficients one obtains expanding
$c_1\{q+\epsilon f(q),q\}$. However, we now use only the fact that
$c_1\ne0$, as the relation $\Z(q+\epsilon f(q);q)=c_1\{q+\epsilon f(q),q\}$
can be proved without making the calculations leading to
(\ref{cenneedennexf}). Summarizing, we have

\vspace{.333cm}

\noindent
\underline{Lemma}

\vspace{.133cm}

\noindent
{\it The cocycle condition (\ref{cociclo3}) admits solutions $(q^a;q^b)$
depending only on the first and higher derivatives of $q^a$, if and only if
in the case $q^a=q^b+\epsilon^{ab}(q^b)$, one has $(q^a;q^b)=
c_1{\epsilon^{ab}}'''(q^b)+\O^{ab}(\epsilon^2)$, with $c_1\ne0$.}

\vspace{.333cm}

We are now ready to prove that, up to a multiplicative constant and
a coboundary term, the Schwarzian derivative is the unique solution
of the cocycle condition (\ref{cociclo3}). Let us first note that
\be
[q^a;q^b]=\Z(q^a;q^b)-c_1\{q^a;q^b\},
\l{diordinesuperiore2}\ee
satisfies the cocycle condition
\be
[q^a;q^c]=\left({\partial_{q^c}q^b}\right)^2\left([q^a;q^b]-[q^c;q^b]
\right).
\l{nuovacociclo3}\ee
In particular, since both $\Z(q^a;q^b)$ and $\{q^a;q^b\}$ depend only
on the first and higher derivatives of $q^a$, we have, as in the case of
$\Z(q+\epsilon f(q);q)$, that
\be
[q+\epsilon f(q);q]=\tilde c_1\epsilon f^{(3)}(q)+\O(\epsilon^2),
\l{allordineespilon2}\ee
where either $\tilde c_1\ne0$ or $[q+\epsilon f(q);q]=0$. However, since
$\{q+\epsilon f(q);q\}=\epsilon f^{(3)}(q)+\O(\epsilon^2)$, by
Eq.(\ref{allordineespilon}) we have $\tilde c_1=0$ and the Lemma yields
$[q+\epsilon f(q);q]=0$. Therefore, we have that the EP univocally implies
the following central result
\be
\Z(q^a;q^b)=-{\beta^2\over4m}\{q^a,q^b\},
\l{equazione3}\ee
where for convenience we replaced $c_1$ by $-\beta^2/4m$. Observe that from
a mathematical point of view we proved the following

\vspace{.333cm}

\noindent
\underline{Theorem}

\vspace{.133cm}

\noindent
{\it The cocycle condition (\ref{cociclo3}) uniquely defines the Schwarzian
derivative up to a global constant and a coboundary term.}

\vspace{.333cm}

We observe that
despite some claims \cite{Ovsienko1}, we have not be able to find in the
literature a complete and close proof of the above theorem.\footnote{We
thank D.B. Fuchs for a bibliographic comment concerning the above theorem.}

Let us now consider the issue of the classical limit.
Comparing $\W^v(q^v)=\left(\partial_{q^v}q\right)^2\W(q)+\Z(q;q^v)$ with the
classical case $\W^v(q^v)=\left(\partial_{q^v}q\right)^2\W(q)$, we have that in
the classical limit $\Z(q^a;q^b)\longrightarrow0$. Hence, by (\ref{equazione3})
\be
{\beta^2\over4m}\{q,q^v\}\longrightarrow0.
\l{qqex2}\ee
Thus $\beta$ is precisely the parameter we are looking for.
Since in the classical limit $Q\longrightarrow0$, we have
\be
\lim_{\beta\longrightarrow0}Q=0.
\l{JiW}\ee
Note that in this limit $\W$ is kept fixed. Since for $Q\longrightarrow0$ one
obtains the CSHJE (\ref{012}), we have
\be
\lim_{\beta\longrightarrow0}\S_0=\S_0^{cl}.
\l{asd3}\ee
{}From (\ref{1H}) and (\ref{equazione3}) it follows that $\W$ itself is a
Schwarzian derivative
\be
\W(q)=-{\beta^2\over4m}\{q^0,q\}.
\l{bastanza}\ee
On the other hand, by (\ref{1Xv}) and (\ref{expoid}) we have
\be
\W(q)={\beta^2\over4m}\left(\{\S_0,q\}-\{e^{{2i\over\beta}\S_0}
,q\}\right)-Q(q),
\l{g1}\ee
so that by (\ref{bastanza})
\be
Q(q)={\beta^2\over4m}\left(\{\S_0,q\}-\{e^{{2i\over\beta}\S_0},q\}+
\{q^0,q\}\right).
\l{XwE}\ee
In particular, the correction to the CSHJE in the case of the state $\W^0$ has
the form
\be
Q^0(q^0)={\beta^2\over4m}\left(\{\S_0^0,q^0\}-\{e^{{2i\over\beta}
\S_0^0},q^0\}\right),
\l{XwE2}\ee
where $\S_0\equiv\S_0(q)$, $\S_0^0\equiv\S_0^0(q^0)$. Observe that we still
have to determine what $q^0$ is. Namely, we have to find the expression of the
reduced action $\S_0^0(q^0)$. Since $\S_0^0(q^0)=\S_0(q)$ this will also give
$q^0=v_0(q)$. We will see that the unique possible solution is
\be
q^0={Ae^{{2i\over\beta}\S_0(q)}+B\over Ce^{{2i\over\beta}\S_0(q)}+D},
\l{egia}\ee
so that, Eq.(\ref{bastanza}) and the M\"obius symmetry
$\left\{(Ah+B)/(Ch+D),x\right\}=\left\{h,x\right\}$, yield
\be
\W(q)=-{\beta^2\over4m}\{e^{{2i\over\beta}\S_0},q\},
\l{sothat}\ee
and by (\ref{XwE})
\be
Q(q)={\beta^2\over4m}\{\S_0,q\}.
\l{sothat2}\ee
It follows from (\ref{1Xv}) and (\ref{sothat2}) that the equation for
$\S_0$ we were looking for is
\be
{1\over2m}\left({\partial\S_0(q)\over\partial q}\right)^2+V(q)-E
+{\beta^2\over4m}\{\S_0,q\}=0.
\l{1Q}\ee

\subsection{Uniqueness of the solution}

To show the uniqueness of the solution (\ref{sothat}), that by (\ref{bastanza})
is equivalent to (\ref{egia}), we first set
\be
Q={\beta^2\over4m}\{\S_0,q\}-g(q),
\l{hf123s}\ee
so that by (\ref{g1})
\be
\W=-{\beta^2\over4m}\{e^{{2i\over\beta}\S_0},q\}+g(q).
\l{f123srf}\ee
We now consider, for any fixed $\W$, the $\beta\longrightarrow0$ limit of $Q$
in (\ref{hf123s}). Suppose that $\lim_{\beta\to 0}\{\S_0,q\}=
\{\lim_{\beta\to0}\S_0,q\}$, so that
\be
\lim_{\beta\longrightarrow0}\{\S_0,q\}=\{\S_0^{cl},q\}.
\l{limite}\ee
The exception arises when $\{\S_0^{cl},q\}$ is not defined so that one has to
consider $\lim_{\beta\to 0}\{\S_0,q\}$. This happens, for example, in the case
of the state $\W^0$ corresponding to $\S_0^{0\,cl}=cnst$. Let us then first
consider an arbitrary state for which $\{\S_0^{cl},q\}$ exists. By
(\ref{limite}) we have
\be
\lim_{\beta\longrightarrow0}\left({\beta^2\over4m}\{\S_0,q\}-g(q)\right)=
\lim_{\beta\longrightarrow0}{\beta^2\over4m}\{\S_0^{cl},q\}-g^{cl}(q)=
-g^{cl}(q),
\l{f123sxx}\ee
and by (\ref{JiW})
\be
g^{cl}=0.
\l{daiiije4}\ee
Next, note that since $\{\S_0,q\}\in\P$, it follows from (\ref{projective}) that
$g(q)$ must transform as a quadratic differential under $v$--maps, that is
consistency requires
\be
g\in\Q.
\l{ginQ}\ee
On the other hand, the only quadratic differential that can be built from
$\S_0$ has the form
\be
g(q)={1\over4m}\left({\partial_q\S_0}\right)^2G(\S_0),
\l{da4}\ee
with $G$ a function of $\S_0$. In other words, there is no way to construct a
quadratic differential by means of higher derivatives of $\S_0$, as these
terms would break the consistency condition (\ref{ginQ}). Furthermore,
(\ref{6Yc}) implies $G(\S_0)=c$, where $c$ is a constant. On the other hand,
since by (\ref{da4}) $c$ is dimensionless whereas $\beta$ has the dimension of
an action, Eq.(\ref{daiiije4}) gives $c=0$. Hence
\be
g=0.
\l{finale}\ee
The extension of (\ref{finale}) to arbitrary states simply follows from the
observation that by (\ref{ginQ}) we have $g(q)\longrightarrow g^v(q^v)=
(\partial_qq^v)^{-2}g(q)$, so that by (\ref{finale}) $g^v(q^v)=0$. In other
words, (\ref{finale}) holds for all $\W\in\H$.

\mysection{The Schr\"odinger equation}\l{theschr}

We have seen that the EP univocally leads to Eq.(\ref{1Q}). In this section we
will first see that this equation implies the Schr\"odinger equation, so that
we will call Eq.(\ref{1Q}) the Quantum Stationary HJ Equation (QSHJE). Next,
we will consider the role of the wave--function in comparison with the
appearance of a pair of linearly independent solutions of the SE arising in
the present formulation. Furthermore, we will derive a dual SE in which $\W$
is replaced by (minus) the quantum potential $Q$. Furthermore, we will see that
the self--dual state $\W^{sd}$, we selected by symmetry arguments, coincides
with $\W^0$. Subsequently, we will consider the explicit form of the
trivializing map, an investigation that will be further considered later on.

\subsection{QSHJE and Schr\"odinger equation}

It is easy to see that Eq.(\ref{1Q}) implies the SE. It simply arises by
considering Eqs.(\ref{icuhyA})--(\ref{icuhy}) with $h=e^{{2i\over\beta}\S_0}$.
In particular, we immediately see that (\ref{1Q}) yields
\be
e^{{2i\over\beta}\S_0}={\psi^D\over\psi},
\l{dfgtp}\ee
where $\psi^D$ and $\psi$ are linearly independent solutions of the stationary
SE
\be
\left(-{\beta^2\over2m}{\partial^2\over\partial q^2}+V(q)\right)\psi=E\psi.
\l{ES}\ee
Thus, for the ``covariantizing parameter'' we have
\be
\beta=\hbar,
\l{Planck}\ee
where $\hbar=h/2\pi$ is Planck's reduced constant. Therefore, in our
formulation the SE arises as the equation which linearizes the QSHJE.

Although the QSHJE (\ref{1Q}) is not familiar, it can be found in
\cite{Floyd82b}\cite{Messiah}\cite{Holland}\cite{Floyd8X}\cite{Floyd86}.
In this respect we note that (\ref{1Q}) differs from other more familiar
versions of the quantum HJ equations. A well--known version arises by
identifying the wave--function with $Re^{{i\over\hbar}\hat\S_0}$, $R,\hat\S_0
\in\RR$, in the SE. As we will see, the basic difference between this
stationary quantum HJ equation and (\ref{1Q}) is that while in the standard
approach one considers $Re^{{i\over\hbar}\hat\S_0}$ to be the wave--function,
(\ref{1Q}) is obtained
by considering the wave--function as a linear combination of $Re^{{i\over\hbar}
\hat\S_0}$ and $Re^{-{i\over\hbar}\hat\S_0}$ and $\S_0$ is never a constant. In
particular, in the cases in which the wave--function is proportional to a real
function, the standard formulation leads to $\hat\S_0=cnst$ even for
non--trivial systems such as the harmonic oscillator. As a result, while the
QSHJE has a well--defined classical limit, in the standard version one has
situations in which quantum mechanically the reduced action is trivial even if
its classical analogue is not. Another version of the quantum analogue of the
HJ equation arises by setting $\psi=e^{{i\over\hbar}\sigma}$, the first step of
the WKB approximation. We will further consider these features later on.

\subsection{Wave--function and Copenhagen School}

Let us now consider the relationship between the solutions of (\ref{ES}) and the
reduced action. By (\ref{dfgtp}) it follows that any solution of (\ref{ES}) has
the form
\be
\psi={1\over\sqrt{\S_0'}}\left(Ae^{-{i\over\hbar}\S_0}+Be^{{i\over\hbar}\S_0}
\right).
\l{popca}\ee
In particular, by (\ref{dfgtp}) the transformation $\tilde\psi^D
=A\psi^D+B\psi$, $\tilde\psi=C\psi^D+D\psi$, is equivalent to
\be
e^{{2i\over\hbar}\S_0}\longrightarrow{Ae^{{2i\over\hbar}\S_0}+B\over C
e^{{2i\over\hbar}\S_0}+D},
\l{qsdfgtp}\ee
which is an invariance of $\W=-{\hbar^2\over4m}\{e^{{2i\over\hbar}\S_0},q\}$.
Note that since $\psi^D$ and $\psi$ are arbitrary linearly independent solutions
of (\ref{ES}), we have by (\ref{dfgtp}) that Eq.(\ref{qsdfgtp}) is equivalent to
\be
e^{{2i\over\hbar}\S_0}={A\psi^D+B\psi\over C\psi^D+D\psi}.
\l{qqsdfSgtpA}\ee
This expression shows that a distinguished feature of the present formulation
concerns the appearance of both $\psi^D$ and $\psi$. This is the signal of the
underlying Legendre duality which reflects in the $\psi^D$--$\psi$ duality, a
consequence of the fact that the SE is a second--order linear differential
equation. The appearance of both solutions in relevant formulas is unusual in
the standard approach. So, since in the standard approach the solution of the SE
one considers is the wave--function, it follows that for bound states, only the
$L^2(\RR)$ solution is considered. On the other hand, as we will see in
sect.\ref{epqsat}, Wronskian arguments show that if the SE has an $L^2(\RR)$
solution, then any other linearly independent solution of the SE cannot belong
to $L^2(\RR)$.\footnote{Uniqueness of the wave--function for bound states, may
induce to believe that divergent solutions of the SE appear only in the case
of a non--physical energy value. Thus, one may erroneously understand that in
this context, by linearly independent solution it is meant a solution of the
same SE but with a different value of the energy.} Therefore, the fact that for
some system the wave--function belongs to the $L^2(\RR)$ space, does not mean
that for the same energy level the SE has no divergent solutions. In other
words, one should not confuse the fact that the wave--function for bound states
is unique, corresponding to the non--degeneration theorem of the spectrum for
bound states, with the wrong sentence: ``the SE for bound states has only one
solution". Also note that the fact that the divergent solution cannot be the
wave--function does not imply that it cannot appear in relevant expressions
which in fact characterize the present formulation. In this context it is
important to observe that our approach derives from the EP and, as such, we are
deriving not only QM from a first principle, which looks different from the
axioms of the Copenhagen School, but, as we will see, also its interpretation
is quite different. On the other hand, to be consistent, we have to reproduce
the basic experimental predictions of QM. As we will see, two basic phenomena,
such as the tunnel effect and energy quantization, are in fact a direct
consequence of the EP. It is well--known that in the usual approach these
phenomena are a consequence of the axiomatic interpretation of the
wave--function in terms of probability amplitude. In particular, since in the
classically forbidden regions the wave--function does not vanish identically,
this interpretation implies a non--vanishing probability of finding a particle
in such regions. The probabilistic interpretation also implies that for
suitable potentials the wave--function must belong to $L^2(\RR)$.

At this stage one may wonder how it is possible that tunnel effect and energy
quantization also arise in our approach in which divergent solutions appear.
Let us denote by $\Sigma_E$ the space of solutions of the SE (\ref{ES})
corresponding to the energy level $E$ and consider the case in which
\be
V(q)-E\geq\left\{\begin{array}{ll}P_-^2>0,&q<q_-,\\
P_+^2 >0, & q> q_+,\end{array}\right.
\l{unoaperintroasintoticopiumenox}\ee
where $q_-$ ($q_+$) is the lowest (highest) value of $q$ for which $V(q)-E$
changes sign. We denote by $\sigma^{BD}$ the associated physical discrete
spectrum. Let us then explain in which sense, in the usual approach, the
divergent solutions associated to (\ref{unoaperintroasintoticopiumenox})
cannot be considered physical ones. First note that the precise sentence is
that in the case (\ref{unoaperintroasintoticopiumenox}) the wave--function,
and therefore a particular solution of the SE, must belong to $L^2(\RR)$. From
a physical point of view, what is relevant is that this request selects the
energy spectrum. In particular, the content of the ``Copenhagen axiom'' that
the physical spectrum associated to (\ref{unoaperintroasintoticopiumenox}) is
determined by the condition
\be
The\;wave-function\;belongs\;to\;L^2(\RR),
\l{funzionedonda}\ee
can be separated in two parts. The physically relevant part, in which the
concept of wave--function does not appear, is
\be
The\;SE\;admits\;an\;L^2(\RR)\;solution,
\l{funzionedonda2}\ee
which by itself implies the quantized spectrum, that is
\be
E\in\sigma^{BD}\qquad\Longleftrightarrow\qquad\psi\in L^2(\RR),
\l{funzionedonda3}\ee
for some $\psi\in\Sigma_E$. Therefore, in the context of the energy spectrum,
the meaning of the wave--function may be seen as a possible way to impose the
physical request (\ref{funzionedonda2}), with the remnant part of
(\ref{funzionedonda}) reducing to the meaning of the wave--function and
therefore to its probabilistic interpretation. In other words, in order to
select the physical spectrum we only need (\ref{funzionedonda2}) which contains
less assumptions than those in (\ref{funzionedonda}). This is a first signal
that the Copenhagen interpretation is sufficiently sophisticated to well hide
the underlying nature of QM. In this respect we will see that the
physical request (\ref{funzionedonda2}) directly follows from the EP without
using the axiomatic interpretation of the wave--function! Therefore, as far as
the problem of selecting the physical spectrum is concerned, the role of
(\ref{funzionedonda}) reduces to selecting the values of $E$ for which the SE
admits an $L^2(\RR)$ solution. Since if for a fixed $E$ the SE has an $L^2(\RR)$
solution, then any other linearly independent solution of the SE cannot belong
to $L^2(\RR)$, we have that the effect of (\ref{funzionedonda}) on the selection
of the spectrum is equivalent to
\be
E\notin\sigma^{BD}\qquad\Longleftrightarrow\qquad\psi\notin L^2(\RR),
\l{funzionedonda4}\ee
$\forall\psi\in\Sigma_E$. As we said, the fact that the wave--function must
belong to $L^2(\RR)$ does not mean that for the same $E$ all solutions of
the SE belong to $L^2(\RR)$. It is true the opposite as, if for a fixed
$E$ the SE has an $L^2(\RR)$ solution, then any other linearly independent
solution cannot belong to $L^2(\RR)$. So, for example, if for a potential well
there is a value of $E$ for which the SE has a solution $\psi$ which decreases
exponentially at both spatial infinities, then any other linearly independent
solution $\psi^D$, therefore solution of the same SE with the same $E$ but with
$\psi^D\not\propto\psi$, grows exponentially at $q=\pm\infty$.

The fact that in our formulation both $\psi^D$ and $\psi$ appear, is a
consequence of the underlying Legendre duality implied by the EP. In fact, in
the same way we obtained $p$--$q$ duality in considering the canonical
equations, here we have a manifest $\psi^D$--$\psi$ duality. This can be
nicely expressed by means of the Legendre brackets introduced in \cite{BOMA}.
In particular, repeating the construction in (\ref{dkqp99})--(\ref{2bomas}),
we see that considering $\F(\psi^2)$ as generating function, defined by
$\psi^D=\partial\F/\partial\psi$, we have
\be
\{\psi,\psi\}_{(q)}=0=\{\psi^D,\psi^D\}_{(q)},\qquad\{\psi,\psi^D\}_{(q)}=1.
\l{2bracketsss3}\ee
We stress that this duality structure, which, due to the preferred nature
played by the wave--function, is missing in the usual formulation of QM,
is a basic feature which distinguishes the present formulation from the one
of the Copenhagen School.

\subsection{$\W\longleftrightarrow Q$ and the dual Schr\"odinger equation}

An interesting question is to seek a direct relation between $\W$ and $Q$. To
be more specific, note that while the QSHJE (\ref{1Q}) connects $\W$ and $Q$
and $\S'_0$, it would be desirable to find a differential equation involving
$\W$ and $Q$ only. To this end we observe that (\ref{expoid}) implies
\be
\S_0=\alpha\int^q_{q_0}dx(\W+Q)^{1/2},
\l{IY1}\ee
where $\alpha^2=-2m$. As $Q=\hbar^2\{\S_0,q\}/4m$, we see that the equation
relating $\W$ and the quantum potential is $Q=\hbar^2\{\int^q_{q_0}dx(\W+Q
)^{1/2},q\}/4m$, that is
\be
Q={\hbar^2\over8m}\left({\W''+Q''\over\W+Q}-{5\over4}{(\W'+Q')^2
\over(\W+Q)^2}\right).
\l{YI4}\ee
Note that adding $\W$ to both sides of (\ref{YI4}) we obtain a similar
expression with now $\W$ expressed in terms of $\W+Q$ and its derivatives. This
complementary role between $\W$ and $Q$ is not an accident. To show this, we
write down again Eq.(\ref{ESQ1}) we derived on the basis of duality arguments
\be
\left({\hbar^2\over2m}{\partial^2\over\partial q^2}+Q\right)\phi=0.
\l{IY2}\ee
This shows that $Q$ plays a similar role to that played by $\W$ in the SE. It
is easy to see that if $\phi^D$ and $\phi$ are solutions of (\ref{IY2}), then
\be
\S_0={A\phi^D+B\phi\over C\phi^D+D\phi}.
\l{IY3}\ee
It follows by (\ref{qqsdfSgtpA}) that the solutions of (\ref{IY3}) are related
to the solutions of the SE by
\be
\gamma_\psi(\psi^D/\psi)=e^{{2i\over\hbar}\gamma_\phi(\phi^D/\phi)},
\l{IY4}\ee
where $\gamma_\psi$ and $\gamma_\phi$ denote two M\"obius transformations.

\subsection{$\W^{sd}=\W^0$ and existence of the Legendre
transformation}\l{sbsx}

We now explicitly show how the solution, that is the QSHJE (\ref{1Q}), solves
the problem of defining the Legendre transformation for any state, so that
the $\S_0$ and $\T_0$ descriptions exist for {\it any} state.
Let us first consider the state $\W^0$. In this case we have
\be
{1\over2m}\left({\partial\S_0^0\over\partial q^0}\right)^2+{\hbar^2\over4m}
\{\S_0^0,q^0\}=0,
\l{00}\ee
that by (\ref{expoid}) is equivalent to
\be
\{e^{{2i\over\hbar}\S_0^0},q^0\}=0.
\l{coc}\ee
A possible solution is
\be
e^{{2i\over\hbar}\S_0^0}=\gamma_qq^0.
\l{doif}\ee
Note that the Legendre transform of $\S_0^0$ in (\ref{doif}) is
well--defined. Therefore, the QSHJE solves the problem of defining the Legendre
transform of the reduced action one meets in CM. Now observe that the
self--dual states (\ref{selfduals}), which arose as the highest symmetric points
of the Legendre pair $\S_0$--$\T_0$, have the structure of (\ref{doif}). Hence,
we can identify the reduced action $\S_0^0$ associated to the state $\W^0$ with
the distinguished self--dual state (\ref{sxc1})! In other words, we can set
\be
\gamma_{sd}=\pm{\hbar\over2i},
\l{gammabeta}\ee
so that
\be
\W^{sd}=\W^0,
\l{iddd}\ee
and
\be
\S_0^0(q^0)=\S_0^{sd}(q^{sd})=\pm{\hbar\over2i}\ln\gamma_qq^0,\qquad q^{sd}=
q^0.
\l{1Xg}\ee
We note that the general solution of (\ref{coc}) is
\be
\S_0^0={\hbar\over2i}\ln\left({Aq^0+B\over Cq^0+D}\right),
\l{aa1Xg}\ee
which will be useful in considering both the properties of the physical
solutions with real $p$ and the structure of the trivializing map.

The relevant case in which the Legendre transformation is not defined is
for $\S_0=Aq+B$. According to Eq.(\ref{sothat}), this corresponds to $\W=-A^2/2m
$, that for $V=0$ corresponds to a free system with energy $E=A^2/2m$.
On the other hand, by the M\"obius invariance of the Schwarzian derivative
\be
\W(q)=-{\hbar^2\over4m}\{e^{{2i\over\hbar}\S_0},q\}=-{\hbar^2\over4m}\left
\{{Ae^{{2i\over\hbar}\S_0}+B\over Ce^{{2i\over\hbar}\S_0}+D},q\right\},
\l{gaa11Xy}\ee
$AD-BC\ne 0$. It follows that if $\S_0$ is a solution of Eq.(\ref{1Q}), we have
that also
\be
\tilde\S_0={\hbar\over2i}\ln\left({Ae^{{2i\over\hbar}\S_0}+B\over
Ce^{{2i\over\hbar}\S_0}+D}\right),
\l{hgy}\ee
is a solution in agreement with (\ref{qqsdfSgtpA}). Therefore, we see that in
QM the kinetic term and the quantum term $Q(q)$ are intimately
related: they mix under the symmetry (\ref{gaa11Xy}) of $\W$. This allows us to
solve the problem of defining the Legendre transformation in the case in which
$\W$ is a constant. Indeed, besides $\S_0=\pm\sqrt{2mE}q$, the equation
\be
{1\over2m}\left({\partial\S_0\over\partial q}\right)^2-E+{\hbar^2\over4m}
\{\S_0,q\}=0,
\l{convugualezero}\ee
has solutions
\be
\S_0={\hbar\over2i}\ln\left({Ae^{{2i\over\hbar}\sqrt{2mE}q}+B\over
Ce^{{2i\over\hbar}\sqrt{2mE}q}+D}\right),
\l{alluce2}\ee
where the constants are chosen in such a way that $\S_0\not\propto q$.
Since $\S_0^{cl}=\pm\sqrt{2mE}q$, we have
\be
\lim_{\hbar\longrightarrow0}\ln\left({Ae^{{2i\over\hbar}\sqrt{2mE}q}+B\over
Ce^{{2i\over\hbar}\sqrt{2mE}q}+D}\right)^{\hbar\over2i}={\pm\sqrt{2mE}q},
\l{atoppe}\ee
implying that $A,B,C,D$ should depend on $\hbar$. This important point will be
discussed in detail later on (see subsection \ref{apl}). For the time being
we just observe that the standard solution $\S_0=\pm\sqrt{2mE}q$ considered in
literature has the property that in the $E\longrightarrow0$ limit one obtains
$\S_0=0$ for which both the Schwarzian derivative and the Legendre
transformation are not defined.

\subsection{The trivializing map}

We now derive the VT $q\longrightarrow q^0$ such that $\W\longrightarrow\W^0$.
Although, as we will see, there is a direct way to find it, it is instructive to
first consider the relevant transformation properties with the derived explicit
expression for $\W$. First note that under the $v$--map $q\longrightarrow q^b=
v^b(q)$, we have
\be
\{e^{{2i\over\hbar}\S_0^b(q^b)},q^b\}=\{e^{{2i\over\hbar}\S_0(q)},q^b\}=\left(
\partial_{q^b}q\right)^2\left[\{e^{{2i\over\hbar}\S_0(q)},q\}-\{q^b,q\}\right],
\l{cJi2}\ee
that is
\be
\W^b(q^b)=\left(\partial_{q^b}q\right)^2\left[\W(q)+{\hbar^2\over4m}\{q^b,q\}
\right].
\l{iudvh}\ee
It follows that if
\be
q^b={Ae^{{2i\over\hbar}\S_0(q)}+B\over Ce^{{2i\over\hbar}\S_0(q)}+D},
\l{dsoi2}\ee
then, according to (\ref{cJi2}), we have $\{e^{{2i\over\hbar}\S_0^b(q^b)},
q^b\}=0$, that is $q^b$ is a M\"obius transformation of $q^0$ and
\be
\W^b(q^b)=\W^0(q^0).
\l{awdvbiudvh}\ee
It follows that
\be
q\longrightarrow q^0={Ae^{{2i\over\hbar}\S_0(q)}+B\over Ce^{{2i\over\hbar}\S_0(
q)}+D},
\l{doiqjwdw}\ee
is the solution of the inversion problem Eq.(\ref{9thebasicidea}). Note that
this result can be also directly derived by comparing (\ref{9thebasicidea}) and
(\ref{aa1Xg}).

\mysection{Quantum HJ equation and the reality condition}\l{tqshjeatcl}

We saw that the formulation exhibits $\S_0$--$\T_0$ duality as $\S_0=Aq+B$
does not belong to the space $\K$ of all possible $\S_0$'s. Due to the M\"obius
invariance of the Schwarzian derivative, instead of $\S_0=\sqrt{2mE}q$, which
corresponds to $\W=-E$, we can choose for $\S_0$ the expression given in
(\ref{alluce2}). Similarly, we have seen that the state $\W^0$ corresponds to
$\S_0^0={\hbar\over2i}\ln(Aq^0+B)/(Cq^0+D)$. Therefore, while setting $\psi=R
e^{{i\over\hbar}\hat\S_0}$ the solutions corresponding to the states with $\W=
cnst$ coincide with the classical ones, here we have a basic difference related
to the existence of the Legendre transform of $\S_0$ for any system. Here,
we first discuss these features in connection with the standard version of the
quantum HJ equation, and then we will consider the classical limit. We will see
that, while there are some subtleties involved in considering such a limit for
the standard versions of the quantum HJ equation, in the case of Eq.(\ref{1Q})
there is a natural criterion for defining it. We will conclude this section
by considering the version of the quantum HJ equation used in the WKB
approximation. We will show that also in this case the phase cannot be
considered as the quantum reduced action.

\subsection{Quantum HJ equation and reality of $R$ and $\hat\S_0$}

We now compare the QSHJE (\ref{1Q}) with a more familiar version. Let us denote
by $\psi$ the Schr\"odinger wave--function. While Eq.(\ref{1Q}) is written in
terms of $\S_0$ only, setting
\be
\psi(q)=Re^{{i\over\hbar}{\hat\S_0}},
\l{rs}\ee
with the condition
\be
(R,\hat\S_0)\in\RR^2,
\l{copenhagen}\ee
in Eq.(\ref{ES}), leads to the system of real equations
\be
{1\over2m}\left({\partial\hat\S_0\over\partial q}\right)^2+V-E-{\hbar^2\over
2mR}{\partial^2R\over\partial q^2}=0,
\l{yz11}\ee
\be
{\partial\over\partial q}\left(R^2{\partial\hat\S_0\over\partial q}\right)=0.
\l{yz12}\ee
Note that to identify the wave--function with $Re^{{i\over\hbar}{\hat\S_0}}$ is
an assumption. We will see that in general it should be identified with the
linear combination $R(Ae^{-{i\over\hbar}\hat\S_0}+Be^{{i\over\hbar}\hat\S_0})$.

We now show that (\ref{1Q}) is the natural quantum analogue of the CSHJE. First
of all, when $\hat\S_0$ is constant, (\ref{yz12}) degenerates and (\ref{yz11})
reduces to $\W={\hbar^2}{\partial^2_qR/2mR}$. Since $\hat\S_0$ does not appear
in this equation, its classical limit does not arise in a natural way as for
(\ref{1Q}). In this respect it is worth noticing that there are many cases with
$\hat\S_0=cnst$ despite the fact that $\S_0^{cl}$, as much as $\S_0$, is
non--trivial, most notably the harmonic oscillator. More generally, all the
states for which the wave--function $\psi$ is proportional to a real function
have a trivial $\hat\S_0$ (except for possible nodes of the wave--function where
$R=0$). It follows that for bound states the conjugate momentum would vanish.

In going from the QSHJE to the SE we lose some information about $\S_0$. This is
a consequence of the invariance of $\W$ under a M\"obius transformation of $e^{
{2i\over\hbar}\S_0}$. We also note that with the condition (\ref{copenhagen}),
which implies the existence of states with ${\hat{\S_0}}=cnst$, both the EP and
the $\S_0$--$\T_0$ Legendre duality are not naturally implemented. Later on we
will discuss the reality condition for $\S_0$ and the identification of the
wave--function in terms of $\S_0$. For the time being we consider (\ref{dfgtp})
and do not fix any reality restriction on $\S_0$. The derivation of the
connection between (\ref{yz11})(\ref{yz12}) and (\ref{1Q}) implies a distinction
between two cases. Let us denote by $\psi^D$ a solution of the SE linearly
independent from $\psi$. We may have either $\bar\psi\not\propto\psi$ or $\bar
\psi\propto\psi$. In the first case, since $\W$ is real, we can set
\be
\psi^D=\bar\psi,
\l{sdoiqh}\ee
that is $\psi^D=Re^{{i\over\hbar}\hat\S_0}$, $\psi=Re^{-{i\over\hbar}\hat\S_0}$
and by (\ref{dfgtp})
\be
\S_0=\hat\S_0+\pi k\hbar,
\l{sdooiq}\ee
$k\in\ZZ$. The continuity equation (\ref{yz12}) gives $R\propto 1/\sqrt{\S_0'}$.
Therefore, by (\ref{dfgtp}), that together with the condition (\ref{sdoiqh})
fixes some of the arbitrariness of $\S_0$, we have
\be
Q={\hbar^2\over4m}\{\S_0,q\}=-{\hbar^2\over2mR}{\partial^2R\over\partial q^2},
\l{pdoijq}\ee
and (\ref{yz11}) corresponds to (\ref{1Q}). The difference between (\ref{1Q})
and (\ref{yz11})(\ref{yz12}) becomes relevant when $\bar\psi\propto\psi$. In
particular, (\ref{rs}) and the condition (\ref{copenhagen}) imply that in this
case $\hat\S_0$ must be a constant. Since this constant can be absorbed by a
normalization of $\psi$, we can choose $\hat\S_0=0$. Therefore, up to a
normalization, we have
\be
\psi=R,
\l{oihkl}\ee
which is real. Since also
\be
\psi^D=R\int^q_{q_0}dxR^{-2},
\l{spoqj}\ee
is real, we have that if we identify $\psi$ and $\psi^D$ in
(\ref{oihkl})(\ref{spoqj}) with those in (\ref{dfgtp}), then $\S_0$ would be
purely imaginary, that is
\be
\S_0={\hbar\over2i}\ln\int^q_{q_0}dxR^{-2},
\l{dioupo}\ee
and
\be
\left({\partial\S_0\over\partial q}\right)^2+{\hbar^2\over2}\{\S_0,q\}=
-{\hbar^2\over R}{\partial^2 R\over\partial q^2}.
\l{pojdx}\ee
It follows that (\ref{1Q}) is equivalent to
\be
{\hbar^2\over2mR}{\partial^2R\over\partial q^2}=V-E.
\l{iodx}\ee
On the other hand, for $\hat\S_0=cnst$, (\ref{yz12}) degenerates and
(\ref{yz11}) corresponds to (\ref{iodx}) (and by (\ref{pojdx}) to
(\ref{1Q})). Later on we will derive the real physical solutions of $\S_0$.

\subsection{The classical limit}

We saw that if $\bar\psi\propto\psi$, then $\hat\S_0=cnst$. It follows that
$\S_0^{cl}$, which is never a constant unless $\W=0$, should arise in a rather
involved way from (\ref{yz11}). In particular, the classical limit of
(\ref{yz11}) manifestly does not correspond to the CSHJE. Usually, by quantum
potential it is meant the term
\be
\hat Q=-{\hbar^2\over2mR}{\partial^2 R\over\partial q^2}.
\l{hatQ}\ee
However, when $\bar\psi\propto\psi$, $Q$ and
$\hat Q$ have different structures. In particular, by (\ref{pojdx}) we have
\be
\hat Q=Q+{1\over2m}\left({\partial\S_0\over\partial q}\right)^2.
\l{z1pojdx}\ee
We will see that by itself the reality condition (\ref{copenhagen}) is not an
obstacle to getting the classical limit. The point is that this condition,
together with the {\it assumption} that the wave--function takes the form
(\ref{rs}), causes difficulties in getting the classical limit of
(\ref{yz11})(\ref{yz12}). Thus, since if (\ref{rs}) is a
solution of the SE, then also
\be
\psi=R\left(Ae^{-{i\over\hbar}\hat\S_0}+Be^{{i\over\hbar}\hat\S_0}\right),
\l{LfTx}\ee
is a solution, we should investigate whether there are suitable
$A$ and $B$ leading to a consistent classical limit for
(\ref{yz11})(\ref{yz12}). Suppose that the wave--function is proportional to
a real function, {\it i.e.} $\bar\psi\propto\psi$. In the case of
(\ref{rs}) this would imply $\hat\S_0=cnst$, so that, unless
$\W=0$, one has $\lim_{\hbar\rightarrow0}\hat\S_0\ne\S_0^{cl}$. With the
representation (\ref{LfTx}) the reality condition $\bar\psi\propto\psi$ would
translate to a condition on the constants $A$ and $B$ rather than on $\hat\S_0$.
In this way one can obtain a non--trivial $\hat\S_0$ and then, as we will see,
a consistent classical limit for $\hat\S_0$. We note that the possibility of
taking the linear combination (\ref{LfTx}), from which the classical limit may
be defined, is a consequence of the linearity of the SE. On the other hand, this
linearity is nothing but the manifestation of the invariance of $\W$ under
M\"obius transformations of $\exp(2i\S_0/\hbar)$. Therefore, we see once again
that the M\"obius symmetry, which we already met in considering Legendre
duality and the related canonical equations, plays a basic role.

Let us further discuss the issue of the classical limit for the version
(\ref{yz11})(\ref{yz12}) of the quantum HJ equation. As we observed, the case
$\hat\S_0=0$ is quite different from the classical situation. For example,
for the classical harmonic oscillator of energy $E$ one has
\be
{\partial\S_0^{cl}\over\partial q}=\pm m\omega(a^2-q^2)^{1/2},
\l{classosc}\ee
where $a=(2E/m\omega^2)^{1/2}$ is the amplitude of motion. In the quantum case
we have
\be
\hat\S_0=0,
\l{hatsosc}\ee
and by (\ref{iodx})
\be
\hat Q=E_n-{1\over2}m\omega^2q^2,
\l{hatqosc}\ee
where $E_n=(n+1/2)\hbar\omega$ denotes the energy level of the quantum
oscillator.\footnote{See \cite{Holland} for a discussion of the quantum
harmonic oscillator in the framework of Eqs.(\ref{yz11})(\ref{yz12}).} This
example shows that the condition (\ref{copenhagen}) leads to a
description which has nothing in common with the classical one.

We have seen that these features, related to the condition (\ref{copenhagen}),
do not arise in considering (\ref{1Q}). In particular, while $Q$ is the
universal quantum correction to the CSHJE, this is not the case for $\hat Q$,
as there are physical systems for which $\hat Q$ cannot be seen as quantum
correction to the CSHJE. In fact, we saw that for all the states for which
$\bar\psi\propto\psi$, one has
$\hat\S_0=cnst$. Therefore, troubles in considering the classical limit of
(\ref{yz11})(\ref{yz12}) should arise for all bound states. See
\cite{Holland}\cite{RosenA}\cite{BS}\cite{HA} for related topics.

As observed in \cite{Holland}, in some textbooks the classical limit of
(\ref{yz11})(\ref{yz12}) is understood in a naive sense. The standard argument
is that due to the $\hbar^2$ factor in $\hat Q$ one has that in the $\hbar
\longrightarrow0$ limit $\hat Q$ vanishes. We have explicitly seen that this is
not generally the case. This can be also directly seen by observing that the
fact that $R$ and $\hat\S_0$ in (\ref{yz11})(\ref{yz12}) depend on $\hbar$
implies that taking the classical limit requires a rather different procedure.
Furthermore, even admitting that $\lim_{\hbar\rightarrow0}\hat Q=0$, it remains
to understand what the meaning of Eq.(\ref{yz12}) is in the
$\hbar\longrightarrow0$ limit. We saw that all these questions are naturally
solved in the version (\ref{1Q}) of the quantum HJ equation. In particular,
while in general $\lim_{\hbar\rightarrow0}\hat Q\ne 0$, in our formulation we
have by construction that
\be
\lim_{\hbar\longrightarrow0}Q=0,
\l{JiWXXXB}\ee
so that
\be
\lim_{\hbar\longrightarrow0}\S_0=\S_0^{cl}.
\l{byconstruction}\ee
In particular, this means that CM is seen as an approximate description of QM.
In other words, there are no two worlds, a classical and a quantum one, but only
the quantum one with the classical phase $\hbar\longrightarrow0$ corresponding
to an approximation. In this framework, it is relevant that there exists a QSHJE
different from that considered in Bohm's theory \cite{Bohm}. In particular,
while the quantum and classical harmonic oscillators, as any other system with
real wave--function, seem to belong to two different worlds, this is not the
case for the QSHJE. We also observe that considering CM as an approximation of
QM would suggest a reconsideration of the collapse of the wave--function.

Concerning $\hat Q$, we note that by (\ref{iodx}) it follows that in
the case in which $\bar\psi\propto\psi$, one has
\be
\lim_{\hbar\longrightarrow0}\hat Q=-V+\lim_{\hbar\longrightarrow0}E.
\l{JiWXXXccd}\ee
In this context we observe that in our approach $\W$ is kept fixed in the
classical limit. In this respect the difference between CM and QM resides in
the fact that in the quantum case there are possible constraints on the
structure of the spectrum $\sigma^{qu}$, so that in general
\be
\sigma^{qu}\subset\sigma^{cl}.
\l{qucl}\ee
This means that any quantum energy level is an admissible one in CM. Of course,
for a given $E$ one in general has $\S_0^{cl}\ne\S_0$, so that for a given $E$
one may have regions of space which are classically forbidden, but this does not
exclude that the associated CSHJE makes sense. It is sufficient that $\W\leq0$
in some region, in order for the CSHJE to admit solutions. On the other hand, if
$\W>0$, $\forall q$, then $E$ belongs neither to $\sigma^{qu}$ nor $\sigma^{cl}
$. Therefore, Eq.(\ref{qucl}) is satisfied for any system. We also note that, as
CM is an approximation, the actual observable physical spectrum is $\sigma^{qu}
$. Therefore, from a fixed $\W$ in the QSHJE, it makes sense to consider the
classical limit with $\W$, and therefore $E$, fixed. In particular, the CSHJE is
naturally obtained from Eq.(\ref{1Q}) with potential $V$ and energy $E\in
\sigma^{qu}\subset\sigma^{cl}$. This makes the classical limit (\ref{JiWXXXB})
consistent and well--defined. Therefore, the QSHJE (\ref{1Q}) admits a universal
criterion for defining the classical limit. Note that while keeping $E$ fixed is
the natural procedure to get the classical limit, this does not eliminate the
difficulties in considering the classical limit of the standard version of the
quantum HJ equation (\ref{yz11})(\ref{yz12}). In particular, in the case in
which $\bar\psi\propto\psi$ we have that the classical limit of $\hat Q$,
keeping $E$ fixed, is
\be
\lim_{\hbar\longrightarrow0}\hat Q=E-V=\hat Q,
\l{hatQPP}\ee
that is $\hat Q$ and its classical limit coincide. Since $\hat Q$ does not
vanish in the classical limit, unless $V=E$, it seems more appropriate to call
$Q$ ``quantum potential" rather than $\hat Q$.

\subsection{The WKB approximation and the quantum HJ equation}

It is well--known that another version of the quantum HJ equation arises
in considering the WKB approximation. This is obtained by setting
$\psi=e^{{i\over\hbar}\sigma}$, in the SE, so that ($'\equiv\partial_q$)
\be
{1\over2m}{\sigma'}^2-{i\hbar\over2m}\sigma''=E-V(q),
\l{ll2}\ee
and $\sigma=\sigma_0+{\hbar\over i}\sigma_1+\left({\hbar\over i}\right)^2
\sigma_2+\ldots$. The first three terms are $\sigma_0=\pm\int^q_{q_0}
dxp^{cl}$, $\sigma_1=-{1\over2}\ln p^{cl}$, $\sigma_2={m\over4}{F\over{p^{cl}
}^3}+{m^2\over 8}\int^qdx {F^2\over{p^{cl}}^5}$, where $p^{cl}=\sqrt{2m(E-V)
}$, denotes the modulus of the classical momentum and $F=-V'=p^{cl}{p^{cl}}'
/m$. We have seen that the Schwarzian derivative, which determines the quantum
correction in (\ref{1Q}), arises quite naturally in our formulation. We note
that it also appears in considering the quantum correction $\sigma_2$, that
is $\sigma_2'=\{\sigma_0,q\}/4p^{cl}$.

Also in the case of (\ref{ll2}) one has to consider non--trivial questions
concerning the classical limit. Similarly, if one
considers the wave--function in the WKB approximation $\psi_{WKB}$, and then
defines the quantum potential in a similar way to $\hat Q$, that is with $R=|
\psi_{WKB}|$, one still meets similar problems to those considered with the
version Eqs.(\ref{yz11})(\ref{yz12}). We refer to \cite{Holland} for a
discussion concerning similar features. Here we just observe that a property
of (\ref{ll2}) is that it implies ${\rm Im}\,\sigma\ne 0$ unless $\sigma=Aq+B$.
On the contrary, we will see that the QSHJE (\ref{1Q}) admits real solutions.

\mysection{The wave equation of Classical Mechanics}\l{tweocm}

In this section we will see that the relationship between the QSHJE and the SE
indicates the way to associate to the CSHJE a wave equation. The latter differs
{}from the SE for the addition to $V$ of a non--linear term whose structure is
the same of the quantum potential but with $\S_0$ replaced by $\S_0^{cl}$.
Remarkably, this equation permits an interpretation of the ``classical
wave--function'' in terms of probability amplitude. We will conclude this
section by discussing the different properties of the $\W$ states in CM and QM.

\subsection{The wave--function of Classical Mechanics}

A feature of our formulation is that the existence of both the classical and
quantum versions of the HJ equation suggests considering a wave--function for
CM. A similar problem was considered by Schiller and Rosen \cite{SchillerRosen}.
However, they derived the SE for Classical Statistical Mechanics, using a
suitable modification of Eqs.(\ref{yz11})(\ref{yz12}) (see also sect.2.6 in
\cite{Holland}). Here, we consider a different approach. The idea is to add
the vanishing term
\be
0={\delta^2\over4m}\{\S_0^{cl},q\}-{\delta^2\over4m}\{\S_0^{cl},q\},
\l{zeronellaclassica}\ee
in the CSHJE (\ref{012}), that is
\be
{1\over2m}\left({\partial\S_0^{cl}\over\partial q}\right)^2+\W(q)-{\delta^2
\over4m}\{\S_0^{cl},q\}+{\delta^2\over4m}\{\S_0^{cl},q\}=0.
\l{012quanclass}\ee
Similarly to the way one derives the SE from QSHJE (\ref{1Q}), one can derive
the SE for CM by considering $\W(q)-\delta^2\{\S_0^{cl},q\}/4m$ as an
effective potential. We have
\be
e^{{2i\over\delta}\S_0^{cl}}={\psi^D_{cl}\over\psi_{cl}},
\l{dfgtpD2}\ee
where $\psi^D_{cl}$ and $\psi_{cl}$ are linearly independent solutions of
\be
\left(-{\delta^2\over2m}{\partial^2\over\partial q^2}+V(q)-{\delta^2\over4m}
\{\S_0^{cl},q\}\right)\psi_{cl}=E\psi_{cl}.
\l{ESclass}\ee
This equation can be interpreted as the wave--equation of CM.

\subsection{Probabilistic interpretation of the classical wave--function}

Observe that since for the free particle at rest one has $\S_0^{0\,cl}=cnst$,
it follows that the Schwarzian derivative $\{\S_0^{0\,cl},q\}$ is not defined
in this case. This is solved in QM since the constant function does not belong
to $\K$. Another property of Eq.(\ref{ESclass}) is that since an arbitrary
solution has the form
\be
\psi_{cl}={1\over\sqrt{{\S_0^{cl}}'}}\left(Ae^{-{i\over\delta}\S_0^{cl}}+
Be^{{i\over\delta}\S_0^{cl}}\right),
\l{popclass}\ee
one has that the solutions (\ref{popclass}) with either $A=0$ or $B=0$, that is
\be
\psi_{cl}={A\over\sqrt{{\S_0^{cl}}'}}e^{\pm{i\over\delta}\S_0^{cl}},
\l{popclass3}\ee
satisfy the differential equation
\be
\left(-{\delta^2\over2m}{\partial^2\over\partial q^2}+V(q)+{\delta^2\over2m}
{\partial^2_q|\psi_{cl}|\over|\psi_{cl}|}\right)\psi_{cl}=E\psi_{cl},
\l{ESclass2}\ee
which makes clear the non--linearity of the wave equation of CM. The classical
wave--function solution of (\ref{ESclass2}) has an obvious interpretation.
First observe that
\be
|\psi_{cl}|^2={|A|^2\over p},
\l{zummm}\ee
and note that the probability of finding a particle in an interval is
proportional to the time the particle stays in it. On the other hand, this time
is inversely proportional to the velocity of the particle. It follows from
(\ref{zummm}) that $|\psi_{cl}(q)|^2dq$ is proportional to the probability of
finding the particle in the interval $[q,q+dq]$. Therefore, in this respect, we
have an analogy between the quantum wave--function and the classical one.
A feature of (\ref{012quanclass}) is that it is independent of $\delta$. As
such, we can fix for $\delta$ an arbitrary value. Let us set
\be
\delta=\hbar,
\l{MMRX}\ee
so that
\be
{1\over2m}\left({\partial\S_0^{cl}\over\partial q}\right)^2+\W(q)+{\hbar^2
\over4m}\{\S_0^{cl},q\}-{\hbar^2\over4m}\{\S_0^{cl},q\}=0,
\l{013quanclass}\ee
and
\be
\left(-{\hbar^2\over2m}{\partial^2\over\partial q^2}+V(q)-{\hbar^2\over4m}
\{\S_0^{cl},q\}\right)\psi_{cl}=E\psi_{cl}.
\l{MMESclass}\ee
These classical equations are the dual versions of (\ref{1Q}) and (\ref{ES}). In
particular, in the same way in which (\ref{1Q}) was seen as the deformation of
the CSHJE by the term $Q=\hbar^2\{\S_0,q\}/4m$, one may consider the CSHJE
(\ref{013quanclass}) to be the deformation of the QSHJE by the term $Q^{cl}=-
\hbar^2\{\S_0^{cl},q\}/4m$. In other words, the (\ref{1Q})(\ref{ES}) and
(\ref{013quanclass})(\ref{MMESclass}) show a complementarity between CM and QM.

\subsection{$\W$ states in Classical and Quantum Mechanics}

There is another aspect related to the M\"obius transformations and to the
properties of the $\W$ states. Namely, we have seen that the EP implies that
$\W$ transforms with an inhomogeneous term under VTs. Now, it is natural
to investigate whether, despite the difference between classical and quantum
$\W$ states, there are $v$--maps under which $\W$ transforms in the same way
both in CM and QM. To answer this question one simply note that comparing
(\ref{natura2}) and (\ref{iudvh}) it follows that the only VTs under which
the classical and quantum transformation properties of $\W$ coincide are the
transformations $q^a\longrightarrow q^b$ such that the inhomogeneous term in the
right hand side of (\ref{iudvh}) vanishes, that is
\be
\{q^b,q^a\}=0,
\l{CaSa}\ee
whose solution is
\be
q^b={Aq^a+B\over Cq^a+D}.
\l{CaSa2}\ee
Hence, under VTs corresponding to M\"obius transformations, the $\W$ states
transform in the same way in CM and QM. In particular, under $v$--maps of the
kind (\ref{CaSa2}) we have ($AD-BC=1$)
\be
\W^b(q^b)=\left(Cq^a+D\right)^4\W^a(q^a),\qquad Q^b(q^b)=\left(Cq^a+D
\right)^4Q^a(q^a).
\l{CaSa4}\ee
It follows that
\be
\M(q)={\W(q)\over Q(q)},
\l{IuYh4}\ee
is a M\"obius invariant function. That is, under (\ref{CaSa}) we have
\be
\M^b(q^b)=\M^a(q^a).
\l{CaSa4bis}\ee

\mysection{The trivializing map and quantum transformations}\l{tseattm}

We have seen that the trivializing coordinate plays a crucial role in our
formulation. In this section we will further investigate its properties in the
context of classical and quantum canonical transformations. Furthermore, we will
see that there is an analogy between the trivializing map and the map associated
to the universal covering in the uniformization theory of Riemann surfaces.

Another topic of this section, concerns the relations between the CSHJE and
QSHJE. In particular, we will show that there is a transformation reducing the
QSHJE to the CSHJE. Next, we will show that both the QSHJE and the SE can be
considered in the framework of projective geometry. In particular, we will use
the fact that the Schwarzian derivative can be interpreted as an invariant of
an equivalence problem for curves in $\PP^1$. This fact will shed light on
the geometrical nature of the trivializing map. We will conclude this section
by showing that Heisenberg commutation relations actually arise from the so
called area function of projective geometry.

A central point of our investigation is that the existence of the trivializing
coordinate is closely related to the existence of the self--dual states. A
related issue concerns the structure of the QSHJE. We have seen that this
equation is equivalent to the Schwarzian equation $\{e^{{2i\over\hbar}\S_0},
q\}=-4m\W/\hbar^2$. On the other hand, since $e^{{2i\over\hbar}\S_0}$ is
a M\"obius transformation of the trivializing coordinate, we see that the QSHJE
(\ref{1Q}) can be interpreted as the equation determining $q^0$, namely
\be
\{q^0,q\}=-{4m\over\hbar^2}(V(q)-E).
\l{nonlinearschw}\ee
In this context the SE can be seen as the linearization of
the problem of finding the trivializing map. Since under $q\longrightarrow q^0$
we have $\W\longrightarrow\W^0$, there is the following correspondence between
the derivation of the classical and quantum HJ equations

\vspace{.333cm}

\noindent
{\it As the existence of the classical trivializing
conjugate variables $(Q,P)$, defined by the canonical transformation}
\be
q\longrightarrow Q,\qquad\qquad p\longrightarrow P=cnst=-\partial_Q\S_0^{cl}(q,
Q)|_{Q=cnst},
\l{canonicalusual}\ee
{\it implies the CSHJE} $H(q,p=\partial_q\S_0^{cl})-E=\tilde H(Q,P)=0,$ {\it
also Eq.(\ref{nonlinearschw}) is a consequence of the existence of the
trivializing map}
\be
q\longrightarrow q^0={\gamma_q}^{-1}e^{{2i\over\hbar}\S_0(q)}={i\hbar\over2p_0
},\qquad\qquad p\longrightarrow p_0=(\partial_qq^0)^{-1}p={i\hbar\over2q^0},
\l{duyqg}\ee
{\it under which $\W\longrightarrow\W^0$}.

\vspace{.333cm}

\noindent
Even if in (\ref{duyqg}) we considered non--real functions, later on we will see
that the reality condition will fix a different form which is related to
(\ref{duyqg}) by a M\"obius transformation. We now observe that there is a
property of the approach which allows to make the correspondence between the
classical and quantum cases rather stringent. The point is that due to the
fact that $\S_0^v(q^v)=\S_0(q^0)$, the relation $p_0=\partial_{q^0}\S_0^0(q^0)$
can be equivalently written in the form
\be
p_0={\partial\S_0(q)\over\partial q^0}.
\l{GZS}\ee
As a consequence there is the classical--quantum correspondence
\begin{eqnarray}
p={\partial\S_0^{cl}\over\partial q}\qquad &\Longleftrightarrow &\qquad p=
{\partial\S_0\over\partial q},\\ P=cnst=-{\partial\S_0^{cl}\over\partial Q}|_{Q
=cnst}\qquad &\Longleftrightarrow &\qquad p_0={\partial\S_0\over\partial q^0}.
\end{eqnarray}
Let us show the effect of this transformation on the SE. First of all note that
as $\S_0^v(q^v)=\S_0(q)$, it follows by (\ref{popca}) that if $\psi^v(q^v)$
solves the SE with $\W^v(q^v)=V^v(q^v)-E^v$, then $\psi(q)$ defined by
\be
\psi^v(q^v)(dq^v)^{-1/2}=\psi(q)(dq)^{-1/2},
\l{23}\ee
is solution of the SE with $\W(q)=V(q)-E$. In other words, under $v$--maps the
solutions of the SE transform as $-1/2$--differentials. In particular
\be
\psi^0(q^0)=(\partial_qq^0)^{1/2}\psi(q).
\l{23conzero1}\ee
The SE for $\W^0$ is $\partial_{q^0}^2\psi^0(q^0)=0$, which is equivalent to
\be
{q^{0'}}^{3/2}{\partial^2\psi^0(q^0)\over\partial {q^0}^2}=0,
\l{po2}\ee
where $q^{0'}\equiv\partial_qq^0$. Therefore, it follows from
Eqs.(\ref{doiqjwdw})(\ref{23conzero1})(\ref{po2}) that
\be
{q^{0'}}^{3/2}{\partial^2\psi^0(q^0)\over\partial{q^0}^2}={q^{0'}}^{1/2}
{\partial\over\partial q}{q^{0'}}^{-1}{\partial\over\partial q}{q^{0'}}^{1/2}
\psi(q)=\left({\partial^2\over\partial q^2}+{1\over2}\{e^{{2i\over\hbar}\S_0}
,q\}\right)\psi(q)=0,
\l{po2Xs}\ee
that by (\ref{sothat}) is equivalent to the SE (\ref{ES}).

\subsection{On the quantum canonical transformations}

It follows by (\ref{23}) that the VTs do not preserve the transition amplitudes.
This was expected as the VTs, which connect different physical systems, are
quite different from the transformations one usually considers in QM. In this
context it is useful to recall some facts concerning quantum canonical
transformations (we refer to \cite{Anderson} for further interesting
observations). A first important step was made by Dirac \cite{Dirac} and Weyl
\cite{Weyl} who observed that unitary transformations are canonical. In
\cite{BHJ} it was proposed to define the quantum canonical transformations as
the transformations in the quantum phase space such that the commutator remains
invariant
\be
[q,p]=i\hbar=[Q(q,p),P(q,p)].
\l{DiracB}\ee
This implies that $Q(q,p)=CqC^{-1}$, $P(q,p)=CpC^{-1}$ with $C$ dependent on $q$
and $p$. Hence, the quantum canonical transformations may be non--unitary. As
observed in \cite{Anderson}, progress on the quantum canonical transformations
has been inhibited because of the mistaken belief that such transformations must
be unitary. In this context we recall that we derived the QSHJE just by posing
a similar question to that considered in deriving the CSHJE from a canonical
transformation. Then, the VTs seem closely related to a quantum analogue of the
canonical transformations. In particular, observe that in order to closely
follow the analogy with the classical case, in QM one should start with $q$ and
$p$ considered as {\it independent} variables and then perform a transformation
to a new set of {\it independent} variables $Q$ and $P$. Suppose that the
formalism has been defined. Then, in the case in which $Q$ and $P$ correspond to
the state $\W^0$, the corresponding equation would relate $q$ and $p$ so that
they will become {\it dependent} variables, that is $p=\partial_q\S_0$ with
$\S_0$ solution of the QSHJE (\ref{1Q}) which we obtained from the EP.

\subsection{Analogy with uniformization theory}

As we said, the main feature distinguishing the QSHJE from the CSHJE is that
while the latter is a first--order non--linear differential equation, the QSHJE
is a third--order one. This implies that in the quantum case one should specify
more initial conditions than in the classical case. Let us show how this
reflects in finding the trivializing coordinate.

The concept of trivializing coordinate is reminiscent of that arising in
uniformization theory of Riemann surfaces. We now show the analogy between our
formulation and uniformization theory (see
\cite{FarkasKra}\cite{JonesSingerman}\cite{Uniformization} for an introduction
to uniformization theory, including some physical applications.).
According to uniformization theory, a negatively curved Riemann surface
$\Sigma$ can be represented as
\be
\Sigma\cong H/\Gamma,
\l{cong}\ee
\be
J_\HH:\HH\longrightarrow\Sigma,
\l{JH}\ee
where $\HH=\{w|{\rm Im}\,w>0\}$ is the upper half plane, $\Gamma$ is a
discrete subgroup of $SL(2,\RR)$ (more precisely a Fuchsian group), and
$J_\HH$ is the uniformizing map satisfying
\be
J_\HH\left({Aw+B\over Cw+D}\right)=J_\HH(w),
\l{Xgl2c}\ee
$\left(\begin{array}{c}A\\C\end{array}\begin{array}{cc}B\\D\end{array}\right)
\in\Gamma$. Let us consider the trivial equation on $\HH$
\be
\partial_w^2\phi^0=0,
\l{ttrr}\ee
and observe that since the ratio of two linearly independent solutions of
(\ref{ttrr}) corresponds to a M\"obius transformation of $w$, we have
\be
w={A\phi^0_2+B\phi_1^0\over C\phi^0_2+D\phi_1^0}.
\l{Aphi}\ee
Eq.(\ref{ttrr}) can be seen as the analogue of Eq.(\ref{po2}). Let us set
\be
w=J_\HH^{-1}(z),
\l{JHmeno1}\ee
where $J_\HH^{-1}:\Sigma\longrightarrow\HH$ is the inverse of the uniformizing
map (\ref{JH}). Then, since $\partial_w={1\over {J_\HH^{-1}}'}\partial_z$, we
have that Eq.(\ref{ttrr}) is equivalent to
\be
\left({\partial^2\over\partial z^2}+{1\over2}\{J_\HH^{-1},z\}\right)
(\partial_zJ_\HH^{-1})^{-1/2}\phi^0=0.
\l{fgth}\ee
Suppose that $T(z)={1\over2}\{J_\HH^{-1},z\}$, known as Liouville stress tensor
or Fuchsian projective connection, is given. Then, taking the ratio of two
linearly independent solutions of the {\it uniformizing equation}
\be
\left({\partial^2\over\partial z^2}+T(z)\right)\phi(z)=0,
\l{fgthXd}\ee
would give the inverse of the uniformizing map. Actually, first observe that
by (\ref{fgth}) and (\ref{fgthXd})
\be
\phi^0(w)=(\partial_zw)^{1/2}\phi(z),
\l{sdcxt}\ee
which is the analogue of (\ref{23conzero1}). Then (\ref{ttrr})(\ref{Aphi}) and
(\ref{sdcxt}) yield
\be
J_\HH^{-1}(z)={A\phi_2+B\phi_1\over C\phi_2+D\phi_1}.
\l{gfthxv}\ee
Therefore, the trivializing coordinate of QM is the analogue of
the inverse map of uniformization
\be
q^0\sim J_\HH^{-1}.
\l{drtZ}\ee
While the map $q\longrightarrow q^0$ corresponds to $\W\longrightarrow\W^0$,
in the case of uniformization theory we have that $z\longrightarrow J_\HH^{-1}$
maps a Riemann surface to a simply connected domain (its universal covering).

Going further with the analogy between uniformization theory and our approach,
we note that also in uniformization theory the manifest covariance is due to
the conventional choice of considering $\phi$ as a $-1/2$--differential. This
is exactly the same transformation property of the wave--function: a direct
consequence of choosing the VTs to represent the functional change on the
reduced action induced by the coordinate transformations. As in our formulation,
also in the case of uniformization theory, choosing a different convention would
make the formalism far more cumbersome.

There is another analogy between our formulation and uniformization theory.
Namely, we saw that $e^{{2i\over\hbar}\S_0}$ may be chosen real in the case
in which $\bar\psi\propto\psi$, while when $\bar\psi\not\propto\psi$ we can
first choose $e^{{2i\over\hbar}\S_0}$ to be a phase and then transform it to
a real trivializing map $q^0$ by considering the Cayley map. In this case we
can use the fact that a phase can be transformed to the real axis by any real
M\"obius transformation of the Cayley map\footnote{The Cayley transformation
maps $\Delta=\{z||z|<1\}$ into $\HH$. Then, if $J_\Delta:
\Delta\longrightarrow\Sigma$ is the uniformizing map, we have
$$
J_\HH^{-1}={J_\Delta^{-1}+i\over iJ_\Delta^{-1}+1}.
$$}
\be
q^0=l{e^{{2i\over\hbar}\S_0}+i\over ie^{{2i\over\hbar}\S_0}+1}=
l{\cos(2\S_0/\hbar)\over1-\sin(2\S_0/\hbar)},
\l{oixq}\ee
where $l$ is a constant with the dimension of a length. This means that in
general, given two linearly independent solutions of the SE, we have to solve
the constraint on $A,B,C,D$
\be
{\rm Im}\,X=0,
\l{FtGy}\ee
where
\be
X\equiv{A\psi^D+B\psi\over C\psi^D+D\psi}.
\l{Xequiv}\ee
Once the constraint is solved we have
\be
q^0=\gamma(X),
\l{gZx}\ee
where $\gamma(X)\in\hat\RR$ is a $PGL(2,\RR)$ M\"obius transformation of $X$.
Note that in writing (\ref{gfthxv}) we did not specify the possible values of
$A,B,C,D$. On the other hand, by definition we should require
\be
{\rm Im}\,J_\HH^{-1}>0.
\l{rewqq}\ee
Therefore, similarly to the identification of the trivializing coordinate $q^0$,
also in this case, for a given pair of linearly independent solutions of the
uniformizing equation, we have to find the constraint on the generally {\it
complex} coefficients $A,B,C,D$, such that Eq.(\ref{rewqq}) is satisfied. Since
$PSL(2,\RR)$ maps $\HH$ into itself, once that $A,B,C,D$ are determined, any
other $PSL(2,\RR)$--transformation would correspond to a possible determination
of $J_\HH^{-1}$. As in Eq.(\ref{gZx}), which follows from the reality of the VTs
and corresponds to a residual symmetry of the original M\"obius one, also in the
case of uniformization we have that the original complex M\"obius symmetry is
restricted by (\ref{rewqq}) to a residual $PSL(2,\RR)$--symmetry. However, this
symmetry corresponds to the fact that $PSL(2,\RR)$ maps $\HH$ into itself.
Acting with $\Gamma\subset PSL(2,\RR)$ on $J_\HH^{-1}$, one obtains points in
$\HH$ with the same image in $\Sigma$.

Similarly to the case of QM, where the trivializing map reduces the system to
the state $\W^0$, in the case of uniformization theory, the polymorphic
function $J_\HH^{-1}$ maps a non--trivial topology to a simply connected domain.
More explicitly, similarly to the reduction $\W\longrightarrow\W^0$, also the
Liouville stress tensor $T$ in the uniformizing equation for $\Sigma$ reduces
to the trivial one under the $J_\HH^{-1}$ map. In particular, in the context of
this analogy, connecting different $\W$ states corresponds to connecting
Riemann surfaces that may be different not only in their complex structures but
already at the topological level. A suitable space to describe any possible
Riemann surface is the famous Bers universal Teichm\"uller space $T(1)$
\cite{Bers1}. This can be represented in the form
\be
T(1)=QS(S^1)/M\ddot{o}b (S^1),
\l{oisxp7}\ee
where $QS(S^1)$ is the space of {\it quasisymmetric} maps of the unit circle and
$M\ddot{o}b (S^1)$ denotes the boundary transformations induced by the conformal
automorphisms of the Poincar\'e disc. A quasisymmetric map $f$ is an increasing
self--homeomorphism of the real axis that can be extended to a quasiconformal
mapping of $\HH$ that fixes the point at infinity. According to a
basic result by Beurling and Ahlfors \cite{BeurlingAhlfors} $f$ is
quasisymmetric if for some constant $K$, $1\leq K <\infty$
\be
{1\over K}\leq{f(q+t)-f(q)\over f(q)-f(q-t)}\leq K,
\l{BeurlAhlfo}\ee
where $q\in\RR$ and $t>0$ (for details see {\it e.g.} \cite{Pekonen}).
Eq.(\ref{BeurlAhlfo}) and its duality $K\longleftrightarrow 1/K$ indicate that
these aspects are related to the theory of univalent functions and to distortion
theorems. The role of this fascinating field in basic physical topics, and the
related complex dynamics theory, is still to be developed. For example, the
Koebe $1/4$--theorem and the Schwarz lemma determine inequalities both in
two--dimensional quantum gravity \cite{Uniformization} and $N=2$ SYM theories
\cite{PRDM}. This would suggest that the theory of univalent functions,
including related topics such as uniformization theory and Teichm\"uller theory,
should play a role in considering basic quantum field theoretical aspects such
as possible dualities between infrared and ultraviolet behavior. A related
topic has been investigated by Callan and Wilczek \cite{CallanWilczek}. We also
observe that the extension to $\HH$ of maps defined on $\RR$ is reminiscent of
the Holographic Principle \cite{tHooftSusskind}. Actually, recently ultraviolet
-- infrared dualities have been investigated in the framework of the Holographic
Principle and AdS dualities \cite{Maldacena} by Susskind and Witten
\cite{SusskindWitten}. The physical relevance of the theory of univalent
functions is related to the role of the M\"obius transformations in
uniformization theory. The deepest one being Poincar\'e's discovery that $PSL(2,
\RR)$ M\"obius transformations are both automorphisms of $\HH$ and isometries of
the Poincar\'e metric. The fact that the theory of univalent functions, and in
particular $T(1)$, is of physical interest, was first realized by Bers in
\cite{Bers2} who noticed its relation with the DeWitt superspace. It is worth
noting that the universal Teichm\"uller space $T(1)$ also appears in string
theory.

An interesting point concerning the VTs is that these correspond to real maps.
We have seen that, thanks to the M\"obius symmetry of the Schwarzian derivative,
for any state one can define a real trivializing map. Now, suppose we are
interested in finding the map
\be
q^a\longrightarrow q^b=v^{ba}(q^a),
\l{abconnection}\ee
such that $\W^a(q^a)$ maps to $\W^b(q^b)$. To find it we can choose the pattern
\be
\W^a\longrightarrow\W^0\longrightarrow\W^b.
\l{WBW0WA}\ee
Denoting by $q^0=v^x(q^x)$ the trivializing map reducing $\W^x(q^x)$ to $\W^0(
q^0)$, we have
\be
v^{ba}=v^{b^{-1}}\circ v^a,
\l{ojKK}\ee
so that, since the trivializing maps can be chosen to be real, we have that
$v^{ba}$ is a real map.

\subsection{Quantum transformations}

A property of the QSHJE is that it can be seen as a deformation of the CSHJE
(\ref{012}) by a ``conformal factor''. In fact, noting that
\be
\{\S_0,q\}=-(\partial_q\S_0)^2\{q,\S_0\},
\l{inversoot}\ee
we have that the QSHJE (\ref{1Q}) is equivalent to
\be
{1\over2m}\left({\partial\S_0\over\partial q}\right)^2
\left[1-\hbar^2\U(\S_0)\right]+V(q)-E=0,
\l{hsxdgyij}\ee
where $\U(\S_0)$ is just the canonical potential (\ref{scharz}) we introduced in
the framework of $p$--$q$ duality. Note that (\ref{inversoot}) also implies
\be
2mQ+\hbar^2p^2\U=0.
\l{deppiununpoicapi}\ee
In the case of the state $\W^0(q^0)$, Eq.(\ref{hsxdgyij}) becomes
\be
{1\over2m}\left({\partial\S_0^0\over\partial q^0}\right)^2\left[1-\hbar^2
\U(\S_0)\right]=0,
\l{hsxdgyij0}\ee
whose solution is
\be
\U\left({\hbar\over2i}\ln\gamma(q^0)\right)={1\over\hbar^2},
\l{poidjq}\ee
where
\be
\gamma(q^0)={Aq^0+B\over Cq^0+D},
\l{gammaqzzeroo}\ee
$AD-BC\ne 0$. Eq.(\ref{poidjq}) shows the important role of the purely
quantum mechanical self--dual state (\ref{1Xg}). Eq.(\ref{hsxdgyij}) can be
written in the form
\be
{1\over2m}\left({\partial\S_0\over\partial\hat q}\right)^2+V(q)-E=0,
\l{hsxdgyijccc}\ee
where
\be
\left({\partial q\over\partial\hat q}\right)^2=\left[1-\hbar^2\U(\S_0)\right],
\l{5yt}\ee
or equivalently (we omit the solution with the minus sign)
\be
d\hat q={dq\over\sqrt{1-\beta^2(q)}},
\l{6yt}\ee
with $\beta^2(q)=\hbar^2\U(\S_0)=\hbar^2\{q,\S_0\}/2$. Integrating
(\ref{5yt}) yields
\be
\hat q=\int^q{dx\over\sqrt{1-\beta^2(x)}}.
\l{quantumcoordinate}\ee
Observe that
\be
\lim_{\hbar\longrightarrow0}\hat q=q.
\l{333}\ee

Eq.(\ref{quantumcoordinate}) indicates that in considering the
differential structure one should take into account the effect of the potential
on space geometry. In this context the deformation of the classical HJ equation
amounts to replacing the standard derivative with respect to the classical
coordinate $q$ with the derivative with respect to the quantum coordinate $\hat
q$. This allows to put the QSHJE in the classical form. Namely, setting
$\hat\W(\hat q)=\W(q(\hat q))$ and $\hat\S_0(\hat q)=\S_0(q(\hat q))$,
we have that Eq.(\ref{hsxdgyij}) is equivalent to
\be
{1\over2m}\left({\partial\hat\S_0(\hat q)\over\partial\hat q}\right)^2+
\hat\W(\hat q)=0.
\l{ZmK}\ee
The fact that the QSHJE admits the classical representation (\ref{ZmK})
suggests that classically forbidden regions correspond to critical regions
for the quantum coordinate. Actually, writing Eq.(\ref{quantumcoordinate}) in
the equivalent form ($\s=\S_0(q)$)
\be
\hat q=\int^qdx{\partial_x\S_0\over\sqrt{-2m\W}}=
\int^{\S_0(q)}{d\s\over\sqrt{-2m\W}},
\l{quantumcoordinatebbb}\ee
we see that the integrand is purely imaginary in the classically forbidden
regions $\W>0$. Furthermore, since according to (\ref{hsxdgyij0}), for the state
$\W^0$ the conformal factor vanishes, it follows by (\ref{quantumcoordinate})
that the quantum coordinate for the free particle state with vanishing energy is
divergent.

\subsection{Classical and quantum potentials and equivalence of curves}

We now consider a result obtained by Flanders \cite{Flanders} who showed that
the Schwarzian derivative can be interpreted as an invariant (curvature) of an
equivalence problem for curves in ${\PP}^1$. Let us introduce a frame for
${\PP}^1$, that is a pair ${\bf x},{\bf y}$ of points in affine space ${\bf A}^2
$ such that $[{\bf x},{\bf y}]=1$, where
\be
[{\bf x},{\bf y}]={\bf x}^t\left(\begin{array}{c} 0\\ -1
\end{array}\begin{array}{cc}1\\ 0\end{array}\right){\bf y}=x_1y_2-x_2y_1,
\l{funzionediareuccettina}\ee
is the area function. Observe that the area function has the
$SL(2,\RR)$--symmetry
\be
[\tilde{\bf x},\tilde{\bf y}]=[{\bf x},{\bf y}],
\l{njdi97w}\ee
where $\tilde{\bf x}=R{\bf x}$ and $\tilde{\bf y}=R{\bf y}$ with $R\in SL(2,
\RR)$. Considering the moving frame $s\longrightarrow\{{\bf x}(s),{\bf y}(s)
\}$ and differentiating $[{\bf x},{\bf y}]=1$ yields the structure equations
\be
{\bf x}'=a{\bf x}+b{\bf y},\qquad{\bf y}'=c{\bf x}-a{\bf y},
\l{equazionistrtturali}\ee
where $a,b$, and $c$ depend on $s$. Given a map $\phi=\phi(s)$ from a domain to
$\PP^1$, one can choose a moving frame ${\bf x}(s),{\bf y}(s)$ in such way that
$\phi(s)$ is represented by ${\bf x}(s)$. Observe that this map can be seen as a
curve in ${\PP}^1$. Two mappings $\phi$ and $\psi$ are said to be equivalent if
$\psi=\pi\circ\phi$, with $\pi$ a projective transformation on $\PP^1$.

Flanders considered two extreme situations. The first case corresponds to
$b(s)=0$, $\forall s$. In this case $\phi$ is constant. To see this observe that
taking the derivative of $\lambda{\bf x}$, for some $\lambda(s)\ne 0$, we have
by (\ref{equazionistrtturali}) that $(\lambda{\bf x})'=(\lambda'+a\lambda){\bf x
}$. Choosing $\lambda\propto\exp[{-\int^s_{s_0}dta(t)}]\ne 0$, we have $(\lambda
{\bf x})'=0$, so that $\lambda {\bf x}$ is a constant representative of $\phi$.

The other case is for $b$ never vanishing. There are only two inequivalent
situations. The first one is when $b$ is either complex or positive. It turns
out that it is always possible to choose the following ``natural moving frame"
for $\phi$ \cite{Flanders}
\be
{\bf x}'={\bf y},\qquad{\bf y}'=-k{\bf x}.
\l{primo}\ee
In the other case, corresponding to $b$ real and negative, the natural
moving frame for $\phi$ is
\be
{\bf x}'=-{\bf y},\qquad{\bf y}'=k{\bf x}.
\l{secondo}\ee

A characterizing property of the natural moving frame is that it is determined
up to a sign with $k$ an invariant. Thus, for example, suppose that for a given
$\phi$ there is, besides (\ref{primo}), the natural moving frame ${\bf x}'_1={
\bf y}_1$, ${\bf y}'_1=-k_1{\bf x}_1$. Since both ${\bf x}$ and ${\bf x}_1$ are
representatives of $\phi$, we have ${\bf x}=\lambda{\bf x}_1$, so that ${\bf y}=
{\bf x}'=\lambda'{\bf x}_1+\lambda{\bf y}_1$ and $1=[{\bf x},{\bf y}]=\lambda^2
$. Therefore, ${\bf x}_1=\pm{\bf x}$, ${\bf y}_1=\pm{\bf y}$ and $k_1=k$
\cite{Flanders}.

Let us now review the derivation of Flanders formula for $k$. Consider
$s\longrightarrow {\bf z}(s)$ to be an affine representative of $\phi$ and let
${\bf x}(s),{\bf y}(s)$ be a natural frame. Then ${\bf z}=\lambda{\bf x}$ where
$\lambda(s)$ is never vanishing. Now note that, since ${\bf z}'=\lambda'{\bf x}
+\lambda{\bf y}$, we have that $\lambda$ can be written in terms of the area
function $[{\bf z},{\bf z}']=\lambda^2$. Computing the relevant area functions,
one can check that $k$ has the following expression
\be
2k={[{\bf z},{\bf z}''']+3[{\bf z}',{\bf z}'']\over[{\bf z},{\bf z}']}-
{3\over2}\left({[{\bf z},{\bf z}'']\over[{\bf z},{\bf z}']}\right)^2.
\l{kappassi}\ee

Given a function $z(s)$, this can be seen as the non--homogeneous coordinate of
a point in ${\PP}^1$. Therefore, we can associate to $z$ the map $\phi$ defined
by $s\longrightarrow (1,z(s))={\bf z}(s)$. In this case we have $[{\bf z},{\bf
z}']=z'$, $[{\bf z},{\bf z}'']=z''$, $[{\bf z},{\bf z}''']=z'''$, $[{\bf z}',
{\bf z}'']=0$, and the curvature becomes \cite{Flanders}
\be
k={1\over2}\{z,s\}.
\l{rimarchevole}\ee
Let us now consider an arbitrary state $\W$. We have
\be
\W=-{\hbar^2\over4m}\{e^{{2i\over\hbar}\S_0},q\}=-{\hbar^2\over2m}k_\W.
\l{sothatW}\ee
Similarly, for the quantum potential
\be
Q={\hbar^2\over4m}\{\S_0,q\}={\hbar^2\over2m}k_Q,
\l{sothatQ}\ee
where $k_\W$ is the curvature associated to the map
\be
q\longrightarrow(1,e^{{2i\over\hbar}\S_0(q)}),
\l{splendido}\ee
while the curvature $k_Q$ is associated to the map
\be
q\longrightarrow (1,\S_0(q)).
\l{splendidoDemetrioStratosArea}\ee
Remarkably, the function defining the map (\ref{splendido}) coincides with the
trivializing map. Observe that the SE takes the geometrical form
\be
\left({\partial^2\over\partial q^2}+k_\W\right)\psi=0.
\l{polbuabuicina}\ee
Furthermore, the identity (\ref{expoid}) can be now seen as difference of
curvatures
\be
\left({\partial_q\S_0}\right)^2={\hbar^2}k_\W-{\hbar^2}k_Q,
\l{expoidQW}\ee
and the QSHJE (\ref{1Q}) can be written in the form
\be
{1\over2m}\left({\partial\S_0(q)\over\partial q}\right)^2+\W(q)
+{\hbar^2\over2m}k_Q=0.
\l{aa10bbbxxxbcurv}\ee
Let us now consider the meaning of the natural moving frame in the framework
of the QSHJE. First observe that the structure equations imply that
\be
{\bf x}''=-k{\bf x}.
\l{movingff}\ee
In the case of $k=k_\W$, this equation is the SE, so that
\be
{\bf x}=(\psi^D,\psi),\qquad{\bf y}=(\psi^{D'},\psi'),
\l{xypsidpsi}\ee
and the frame condition is the statement that the Wronskian $W$ of the SE is a
constant
\be
[{\bf x},{\bf y}]=\psi'\psi^D-\psi^{D'}\psi=W=1.
\l{wronschiano}\ee
Hence, the SE determines the natural moving frame associated to the curve in
${\PP}^1$ given by the representative (\ref{splendido}) with $-2m\W/\hbar^2$
denoting the invariant associated to the map. In other words, the Schr\"odinger
problem corresponds to finding the natural moving frame such that $-2m\W/\hbar^2
$ be the invariant curvature. In the $k=k_Q$ case Eq.(\ref{movingff}) becomes
\be
\left({\hbar^2\over2m}{\partial^2\over\partial q^2}+Q\right)\phi=0,
\l{IY2ancora}\ee
so that if $\phi^D$ and $\phi$ are solutions of (\ref{IY2ancora}), then
\be
\S_0={A\phi^D+B\phi\over C\phi^D+D\phi}.
\l{IY3ancora}\ee

\subsection{Area function and commutator}

The area function has a structure which is reminiscent of the commutator. This
feature is emphasized in Eq.(\ref{wronschiano}) that contains the derivative
operator. In fact, Eq.(\ref{wronschiano}) can be written using the commutator
between $\partial_q$ and the ratio of solutions $\psi^D/\psi$. We have
\be
[{\bf x},{\bf y}]=[\partial_q,\psi^D/\psi]_c\psi^2=1,
\l{wronschiano2}\ee
where $[A,B]_c\equiv AB-BA$. Introducing the momentum operator $p=-i\hbar
\partial_q$, (\ref{wronschiano2}) becomes
\be
[\psi^D/\psi,p]_c\psi^2=i\hbar.
\l{oipTw590i}\ee
Note that the $SL(2,\RR)$--symmetry (\ref{njdi97w}) implies
\be
\left[{A\psi^D/\psi+B\over C\psi^D/\psi+D},p\right]_c(C\psi^D+D\psi)^2=i\hbar.
\l{oipTw5902i}\ee
Since, as we will see, the trivializing map is a M\"obius transformation of the
$\psi^D/\psi$, we have that the area function is equivalent to
\be
\left[q^0,p\right]_c(C\psi^D+D\psi)^2=i\hbar.
\l{oipTw5903i}\ee
This reproduces the Heisenberg commutation relation
\be
\left[q^0,p_0\right]_c=i\hbar,
\l{oipTw5904i}\ee
$p_0=-i\hbar\partial_{q^0}$, with (\ref{oipTw5903i}) representing the Jacobian
\be
\partial_qq^0=(C\psi^D+D\psi)^{-2},
\l{iacobianodellatrasfo}\ee
of the trivializing transformation.

\mysection{Canonical variables and M\"obius transformations}\l{cvamt}

We now discuss the role of M\"obius transformations in considering the
canonical variables $p$ and $q$. Once again, we will use the fact that,
according to (\ref{gaa11Xy}), $\W$ remains invariant under a M\"obius
transformation of $e^{{2i\over\hbar}\S_0}$. While under this transformation both
$(\partial_q\S_0)^2$ and $Q$ get transformed, we have the invariance
\be
\left({\partial\tilde\S_0(q)\over\partial q}\right)^2+{\hbar^2\over2}\{\tilde
\S_0,q\}=\left({\partial\S_0(q)\over\partial q}\right)^2+{\hbar^2\over2}\{\S_0,
q\},
\l{1QxXYZ}\ee
where $\tilde\S_0$ is given in (\ref{hgy}). We will begin this section by
considering the explicit dependence of the conjugate momentum on the coordinate
in the case of the state $\W^0$. Next, we will make some preliminary observation
on the possible definition of time parameterization. Then the general expression
for $p$ in terms of $q$ will be derived for an arbitrary state $\W$. We will
conclude this section by a detailed investigation of the symmetries of the
wave--function. By this, we mean the transformations on the initial conditions
for the QSHJE which leave invariant a given eigenfunction of the Hamiltonian.

\subsection{The canonical variables of the state $\W^0$}

Let us consider the conjugate momentum in the case of the state $\W^0$. By
(\ref{aa1Xg}) we have
\be
p_0={i(BC-AD)\hbar\over2(Aq^0+B)(Cq^0+D)}.
\l{pzzzero2}\ee
On the other hand, since $AD-BC\ne 0$, by a common rescaling of $A,B,C,D$, that
by projectivity of the M\"obius transformations do not change (\ref{aa1Xg}), we
can set
\be
AD-BC=\pm i,
\l{determinante}\ee
so that
\be
p_0=\pm {\hbar\over2[AC{q^0}^2+(AD+BC)q^0+BD]}.
\l{pzzwero2}\ee
Observe that if $AC=0$, then either $BC$ or $AD$ vanish and by
(\ref{determinante}) $p_0=\pm\hbar/2(\pm iq^0+BD)$. Similarly, if $BD=0$, then
$p_0=\pm\hbar/2(AC{q^0}^2\pm iq^0)$. It follows that

\vspace{.333cm}

\noindent
{\it $p_0$ can take real values only if $AC\ne0$ and $BD\ne0$, that is $p_0=
\pm\hbar(\alpha{q^0}^2+\beta q^0+\gamma)^{-1}$, $4\alpha\gamma-\beta^2=4$,
$\beta\in\RR$ with both $\alpha$ and $\gamma$ taking values in
$\RR\backslash\{0\}$.}

\vspace{.333cm}

\noindent
By (\ref{determinante}) it follows that the poles of $p_0$
are never on the real $q^0$--axis. Actually, the poles of (\ref{pzzwero2}) are
\be
q^0=-{BC+AD\pm i\over2AC}.
\l{qqs}\ee
Since reality of (\ref{pzzwero2}) implies $(BC+AD)\in\RR$ and $AC\in\RR
\backslash\{0\}$, it follows that the solutions (\ref{qqs}) cannot be real.
By (\ref{determinante}) we can rewrite Eq.(\ref{pzzwero2}) in the form
\be
p_0=\pm{\hbar\ell_1\over({q^0}+\ell_2)^2+\ell_1^2},
\l{pzzweWro2}\ee
where
\be
\ell_1={1\over2AC}\ne0,\qquad\ell_2={BC+AD\over2AC}.
\l{leelles}\ee
Observe that both $\ell_1$ and $\ell_2$ are real constants. A property of $p_0$
is that it vanishes only at infinity
\be
\lim_{q^0\longrightarrow\pm\infty}p_0=0.
\l{XzW}\ee
Furthermore, for $q^0=-\ell_2$, $|p_0|$ reaches its maximum
\be
|p_0(-\ell_2)|={\hbar\over\ell_1}.
\l{WsX4}\ee
Since $\ell_1\ne 0$, we have
\be
|p_0|<\infty.
\l{momentolimitato}\ee

\subsection{Time and elliptic curve as moduli of the state $\W^0$}

Let us consider time parameterization. In CM, this is
derived from the CSHJE by identifying the conjugate momentum with $m\dot q^0$.
Since $p_0=\partial_{q^0}\S_0^0=m\dot q^0$, this identification
would imply a non--vanishing velocity associated to the state $\W^0$. Later on,
we will see that a natural derivation of time parameterization, suggested by
Floyd \cite{Floyd82b}, implies that, generally, one has $\partial_q\S_0\ne m\dot
q$. We will see that with Floyd's time parameterization one has $\dot q^0=0$.

Let us then investigate the structure of $q^0(\tau)$ with $\tau$ defined by
$\partial_q\S_0=m\partial_\tau q$, that is
\be
\tau-\tau_0=m\int^q_{q_0}dx{1\over\partial_x\S_0(x)},
\l{etirutirutauuuuuu}\ee
so that
\be
p_0=m{dq^0\over d\tau}.
\l{velocitacontau}\ee
Eq.(\ref{pzzweWro2}) becomes $m\dot q^0[(q^0+\ell_2)^2+\ell_1^2]\pm\hbar
\ell_1=0$, which can be integrated to
\be
{m\over3}{q^0}^3+m\ell_2{q^0}^2+m(\ell_1^2+\ell_2^2)q^0+c\pm\hbar\ell_1\tau=0,
\l{WxE}\ee
where $c$ is an integration constant. Setting $y^2=\pm 3\hbar\ell_1\tau/m$,
and $\alpha_0=3\ell_2$, $\alpha_1=3(\ell_1^2+\ell_2^2)$, $\alpha_2=3c/m$,
we have that Eq.(\ref{WxE}) becomes the equation of an elliptic curve
\be
y^2={q^0}^3+\alpha_0{q^0}^2+\alpha_1q^0+\alpha_2.
\l{FcT2}\ee
As we noted, this time parameterization, made by identifying the conjugate
momentum with the mechanical one, gives $\dot q^0\ne 0$. We will then consider
Floyd's time parameterization which will imply the expected result $\dot q^0=0$.
However, let us note that it is intriguing that the above identification of
time, borrowed from CM, leads to an underlying non--trivial geometrical
structure. This suggests that time defined through $\partial_q\S_0=m\dot q$
plays some non--trivial role. Even if we will not consider further this
``$\tau$--parameterization" here, its analogy with the classical case deserves
further exploration.

\subsection{The canonical variables of arbitrary states}

The above investigation indicates that there is a structure which generalizes
to arbitrary $\W$ states. In order to determine $p$ as function of $q$, we use
the fact that the SE can be seen as the linearization of the Schwarzian equation
\be
\{e^{{2i\over\hbar}\S_0},q\}=-{4m\over\hbar^2}\W(q).
\l{Ga3}\ee
We have seen that this implies that
\be
e^{{2i\over\hbar}\S_0}={A\psi^D+B\psi\over C\psi^D+D\psi},
\l{0pA}\ee
so that
\be
p={i\over2}{\hbar(AD-BC)W\over AC{\psi^D}^2+(AD+BC)\psi^D\psi+BD\psi^2},
\l{Gh1}\ee
where $W$ denotes the Wronskian
\be
W=\psi'\psi^D-{\psi^D}'\psi,
\l{Wronskian}\ee
which is a constant. Note that $p$ can be written in the form
\be
p={i\over2}{\hbar\tilde W\over\tilde\psi^D\tilde\psi},
\l{pi2}\ee
where
\be
\left(\begin{array}{c}{\tilde\psi}^D\\ \tilde\psi\end{array}\right)=
\left(\begin{array}{c}A\\ C\end{array}\begin{array}{cc}B\\ D
\end{array}\right)\left(\begin{array}{c}\psi^D\\ \psi\end{array}\right),
\l{arrra}\ee
and
$\tilde W={\tilde\psi}'{\tilde\psi}^D-{\tilde\psi}^{D'}\tilde\psi=(AD-BC)W$.
Setting
\be
\alpha={2iAC\over(BC-AD)W},\qquad\beta={2i(AD+BC)\over(BC-AD)W},\qquad\gamma=
{2iBD\over(BC-AD)W},
\l{Xnew1}\ee
Eq.(\ref{Gh1}) becomes
\be
p={\hbar\over\alpha{\psi^D}^2+\beta\psi^D\psi+\gamma\psi^2}.
\l{Gh2}\ee
Since $(AD+BC)^2-4(AC)(BD)=(AD-BC)^2$, it follows by (\ref{Xnew1}) that
\be
\beta^2-4\alpha\gamma=-{4\over W^2}.
\l{Xc}\ee
Note that since $\psi^D$ and $\psi$ are two arbitrary linearly independent
solutions of the SE, we can always choose them to take real values
\be
(\psi^D,\psi)\in\RR^2.
\l{realless}\ee
Actually, since $\W$ is real, it follows that if $\phi$ is a solution of the SE,
then the real function $\psi=\phi+\bar\phi$ is still a solution. Any other
solution has the form
\be
\psi^D(q)=c\psi(q)\int^q_{q_0}dx\psi^{-2}(x)+d\psi(q),
\l{psiDpsi}\ee
for some constants $c$ and $d$. Observe that $W=-c$. Note that a rescaling of
$\psi$ and a change of $c$ and $d$ in (\ref{psiDpsi}) is equivalent to a
transformation of the coefficients $A,B,C,D$ in (\ref{0pA}). By (\ref{Gh2}) it
follows that reality of $(\psi^D,\psi)$ and $(p,q)$ implies $W\in\RR\backslash
\{0\}$, and $\alpha,\beta,\gamma\in\RR$, which in turn, by (\ref{Xc}), imply
$\alpha\ne 0$, $\gamma\ne 0$. Since $\alpha{\psi^D}^2+\beta\psi^D\psi+\gamma
\psi^2$ vanishes for
\be
\psi^D(q)=-{\psi(q)\over2\alpha}\left(\beta\pm{2i\over W}\right),
\l{CarlosSantana3}\ee
and being $W\in\RR\backslash\{0\}$, it follows that $p$ has no poles for real
$q$. The possibility that $\psi^D$ and $\psi$ vanish at the same point is easily
ruled out. For, if $\psi^D(q_0)=0=\psi(q_0)$ then $W=\psi'(q_0)\psi^D(q_0)-
\psi^{D'}(q_0)\psi(q_0)=0$, which is not the case as $\psi^D$ and $\psi$ are
linearly independent so that $W\ne 0$.

Let us consider again Eq.(\ref{Gh2}). Note that writing it in the form
\be
p={\hbar\alpha^{-1}\over\left(\psi^D+{\beta\over2\alpha}\psi\right)^2+
{1\over\alpha^2W^2}\psi^2},
\l{CarlosSantana4}\ee
and setting
\be
\ell_1={1\over\alpha W}=i{AD-BC\over2AC},\qquad\ell_2={\beta\over2\alpha}=
{AD+BC\over2AC},
\l{New2d}\ee
we have
\be
p=\pm{\hbar W\ell_1\over\left(\psi^D+\ell_2\psi\right)^2+{\ell_1^2}\psi^2}.
\l{CarlosSantana5}\ee
Since $AD-BC\ne 0$ we have\footnote{The $\pm 1$ factor is due to the
invariance of the QSHJE (\ref{1Q}) under a change of sign of $\S_0$. This
reflects the fact that the motion can be in two directions. Observe that one can
fix $\ell_1$ to be either positive or negative. Alternatively, one can consider
$\ell_1\in\RR\backslash\{0\}$ and then dropping the $\pm 1$ factor in
(\ref{CarlosSantana5}).}
\be
\ell_1\in\RR\backslash\{0\}.
\l{CarlosSantana6}\ee
{}From the general expression (\ref{CarlosSantana5}) we can directly derive the
solution Eq.(\ref{pzzweWro2}) for the state $\W^0$. Namely, since two linearly
independent solutions of the SE associated with the state $\W^0$ are
\be
\psi^{D^0}=q^0,\qquad\psi^0=1,
\l{psiddpsi}\ee
we have $W=-1$ and Eq.(\ref{CarlosSantana5}) reproduces Eq.(\ref{pzzweWro2}).
Note that Eq.(\ref{CarlosSantana5}) is invariant under a common rescaling of
$\psi^D$ and $\psi$. More generally, any $GL(2,\RR)$--transformation of the
vector
\be
v=\left(\begin{array}{c}\psi^D\\ \psi\end{array}\right),
\l{dF5x}\ee
which in turn may imply also a rescaling of the Wronskian $W$, is always
equivalent to a transformation of the $\ell_k$'s. This can be seen also by
expressing $p$ in a different form. Let us introduce the matrix
\be
\gamma=\left(\begin{array}{c}a_{11}\\a_{21}\end{array}
\begin{array}{cc}a_{12}\\a_{22}\end{array}\right)\in GL(2,\RR).
\l{gammmma}\ee
Setting
\be
a_{11}={c\over2\ell_1\hbar W},\qquad a_{12}={c\ell_2\over2\ell_1\hbar W},
\qquad a_{21}={1\over c},\qquad a_{22}={\ell_1^2+\ell_2^2\over c\ell_2},
\l{sett3}\ee
where $c$ is an arbitrary non--vanishing constant, we have
\be
p^{-1}=\pm v^t\left[\gamma^t\left(\begin{array}{c}0\\1\end{array}
\begin{array}{cc}1\\0\end{array}\right)\gamma-\left(\begin{array}{c}0\\ \sigma
\end{array}\begin{array}{cc}\sigma\\0\end{array}\right)\right]v,
\l{b7tris}\ee
where $\sigma\equiv\det\gamma$, that by (\ref{sett3}) is
\be
\sigma={\ell_1\over2\ell_2\hbar W}.
\l{isWd}\ee
Observe that
\be
\gamma_p\equiv\gamma^t\left(\begin{array}{c}0\\1\end{array}\begin{array}{cc}1\\
0\end{array}\right)\gamma-\left(\begin{array}{c}0\\ \sigma\end{array}
\begin{array}{cc}\sigma\\0\end{array}\right)=2\left(\begin{array}{c}a_{11}a_{21}
\\a_{12}a_{21}\end{array}\begin{array}{cc}a_{12}a_{21}\\a_{12}a_{22}\end{array}
\right),
\l{dfct3E}\ee
so that with the identification (\ref{sett3}), we have
\be
\gamma_p={1\over\ell_1\hbar W}\left(\begin{array}{c}1\\ \ell_2\end{array}
\begin{array}{cc}\ell_2\\ \ell_1^2+\ell_2^2\end{array}\right).
\l{dfct3F}\ee
Note that $\det\gamma_p=1/\hbar^2W^2$ does not depend on the
$\ell_k$'s. Let us write down $p$ in terms of $\gamma_p$
\be
p^{-1}=\pm v^t\gamma_p v.
\l{ReScxD}\ee
Since under a $GL(2,\RR)$--transformation of $v$
\be
v\longrightarrow v_R=R v,
\l{vRv}\ee
\be
R=\left(\begin{array}{c}
a\\ c\end{array}\begin{array}{cc}b\\d\end{array}\right)\in GL(2,\RR),
\l{dt3F34}\ee
the Wronskian gets rescaled by $\rho\equiv\det R$, it follows that
under (\ref{vRv}) $p$ transforms as
\be
p^{-1}\longrightarrow p_R^{-1}=\pm v^tR^t\gamma_{p\rho}Rv,
\l{ReSc}\ee
where $\gamma_{p\rho}=\rho^{-1}\gamma_p$. In particular, as we already noticed,
since $\det\gamma_{p\rho}=(\det R)^{-2}\det\gamma_p$, it follows that $p$ is
invariant under dilatations of the vector $v$. Observe that the effect on $p$ of
any linear transformation of $v$ is equivalent to the following transformation
of $\gamma_p$
\be
\gamma_p\longrightarrow\gamma_p'=R^t\gamma_{p\rho}R.
\l{HyT}\ee
Since $\gamma_p^t=\gamma_p$ and $\det\gamma_p'=\det\gamma$, we have
\be
\gamma_p'={1\over\ell_1'\hbar W}\left(\begin{array}{c}1\\ \ell_2'\end{array}
\begin{array}{cc}\ell_2'\\ {\ell_1'}^2+{\ell_2'}^2\end{array}\right),
\l{dfcU7}\ee
where
\be
\ell_1'=\rho{\ell_1\over(a+\ell_2c)^2+\ell_1^2c^2},\qquad\ell_2'={ab+(bc+
ad)\ell_2+(\ell_1^2+\ell_2^2)cd\over(a+\ell_2c)^2+\ell_1^2c^2}.
\l{ell1ell2}\ee
Any linear transformation of $v$ is equivalent to keeping $v$ fixed and making
a transformation of $\ell_1$ and $\ell_2$.\footnote{The transformed quantity
$\ell_{kR}$ have some interesting properties. For example, there is the
following ``formal coincidence''. Namely, relaxing the condition for the
elements of the $R$--matrix to be constants and setting
$$
R=\left(\begin{array}{c}\psi^D\\ \psi\end{array}\begin{array}{cc}\hbar{\psi^D}'
\\ \hbar {\psi}'\end{array}\right),
$$
it follows by (\ref{ell1ell2}) that $\ell_1\longrightarrow\ell_{1 R}=p$.}
In particular, since $W$ can be chosen to be an arbitrary fixed value, we have
that the degrees of freedom of $p$ are described by the two real parameters
$\ell_1$ and $\ell_2$. Note that setting $W=\pm 1/\hbar$, yields
\be
\gamma_p(\ell_1,\ell_2)=\left(\begin{array}{c}\ell_1^{-1}\\ \ell_2\ell_1^{-1}
\end{array}\begin{array}{cc}\ell_2\ell_1^{-1}\\ \ell_1+\ell_2^2\ell_1^{-1}
\end{array}\right),
\l{df3F1}\ee
so that $\gamma_p\in SL(2,\RR)$, and
\be
|{\rm tr}\,\gamma_p|=|\ell_1^{-1}+\ell_1+\ell_2^2\ell_1^{-1}|\geq2.
\l{GvH}\ee
Since $|{\rm tr}\,\gamma_p|=2$ only in the case in which $\ell_1=1$, $\ell_2=0$,
corresponding to the identity matrix, it follows that $\gamma_p$ is a
hyperbolic matrix.

An interesting feature of the above investigation is that we started with the
general expression (\ref{0pA}) in which $e^{{2i\over\hbar}\S_0}$ is given in
terms of M\"obius transformations of the ratio
\be
w={\psi^D\over\psi}\in\RR,
\l{ratio}\ee
and then arrived at hyperbolic matrices. Let us consider the transformation
\be
\gamma_p (w)={\gamma_{p11}w+\gamma_{p12}\over\gamma_{p21}w+
\gamma_{p22}}={w+\ell_2\over\ell_2 w+(\ell_1^2+\ell_2^2)},
\l{Gha}\ee
and its associated fixed point equation
\be
\gamma_p(w)=w,
\l{GvT}\ee
in the case in which $\gamma_p\ne {\II}_{2\times 2}$. We have
\be
w_\pm={1-\ell_1^2-\ell_2^2\pm\sqrt{(\ell_1^2+\ell_2^2-1)^2+4\ell_2^2}\over
2\ell_2}.
\l{fixedp}\ee
Since, according to (\ref{GvH}), we have $(\ell_1^2+\ell_2^2-1)^2+4\ell_2^2>0$,
it follows that $\gamma_p\ne {\II}_{2\times 2}$ has distinct fixed points lying
on the real axis. This structure indicates that the ratio $w=\psi^D/\psi\in\RR$
can be seen as belonging to the boundary of $\HH$. Let us denote by $\Gamma_H$
the set of all matrices of the kind (\ref{df3F1}). A property of $\Gamma_H$ is
that $\gamma\in\Gamma_H\longrightarrow\gamma^{-1}\in\Gamma_H$, in particular
\be
\gamma_p^{-1}(\ell_1,\ell_2)=\gamma_p\left({\ell_1\over\ell_1^2+\ell_2^2},
-{\ell_2\over\ell_1^2+\ell_2^2}\right).
\l{UyJk}\ee
Suppose that there are elements $\gamma_k$ in $\Gamma_H$ such that
\be
\prod_{k=1}^h(\gamma_{2k-1}\gamma_{2k}\gamma_{2k-1}^{-1}\gamma_{2k}^{-1})=
{\II}_{2\times 2}.
\l{HgY}\ee
Doing this is equivalent to selecting a set of $2h$ pairs $(\ell_1,\ell_2)$.
Then, denoting by $\Gamma_{p\{\ell_k\}}$ the group generated by the
$\gamma_k$'s, we have that with the $2h$ values of $p$ determined by the
$\ell_k$'s moduli, there is associated the genus $h$ Riemann surface
\be
\Sigma_{p\{\ell_k\}}\cong H/\Gamma_{p\{\ell_k\}}.
\l{infinitegenus}\ee
Therefore, we are posing the question of determining the possible sets
$\{\ell_k\}$ such that $\Sigma_{p\{\ell_k\}}$ is a Riemann surface. Since
different $\ell_k$'s correspond to different $p$'s, studying these surfaces
would shed light on the geometry of the possible paths. It would be interesting
to understand whether a Riemann surface of infinite genus may appear.

The generators $\gamma_k$'s of a given $\Gamma_{p\{\ell_k\}}$ would represent
the handles of the Riemann surface. These elements can be represented in the
form
\be
{\gamma_k(w)-w_+\over\gamma_k(w)-w_-}=e^{\lambda}{w-w_+\over w-w_-},
\l{wpiumeno}\ee
where
\be
e^{\lambda}={1+\ell_1^2+\ell_2^2-\sqrt{(\ell_1^2+\ell_2^2-1)^2+4\ell_2^2}\over
1+\ell_1^2+\ell_2^2+\sqrt{(\ell_1^2+\ell_2^2-1)^2+4\ell_2^2}}.
\l{llamm}\ee
This means that any hyperbolic element is conjugate to a dilatation. The scale
factor $\lambda$ has the following geometrical meaning. Consider a point $w\in
H$ and its hyperbolic transformed $\gamma_k(w)$. Then $\lambda(\gamma_k)$
corresponds to the minimal hyperbolic distance between $w$ and $\gamma_k(w)$.
This minimum is reached for $w$ lying in the geodesic intersecting the real
axis at $w_-$ and $w_+$.

\subsection{The symmetries of the wave--function}

The fact that the QSHJE (\ref{1Q}) is a third--order differential equation
implies that there are three integration constants which specify $\S_0$. In
particular, we have seen that two real constants specify $p$. Concerning $\S_0$
there is one more constant which arises by integrating (\ref{CarlosSantana5})
($w=\psi^D/\psi\in\RR$)
\be
e^{{2i\over\hbar}\S_0\{\delta\}}=e^{i\alpha}{w+i\bar\ell\over w-i\ell},
\l{KdT3}\ee
where $\delta=\{\alpha,\ell\}$, with $\alpha$ a real integration constant and
$\ell=\ell_1+i\ell_2$. Observe that the condition $\ell_1\ne 0$ is equivalent to
having $\S_0\ne cnst$ which is a necessary condition to define the term
$\{\S_0,q\}$ in the QSHJE. We also note that changing sign of $\ell_1$
corresponds to changing the sign of $\S_0$; more precisely we have
\be
\S_0\{\alpha,-\bar\ell\}=-\S_0\{\alpha,\ell\}+\hbar\alpha=-\S_0\{-\alpha,\ell\}.
\l{oihcv7}\ee
Note that for the conjugate momentum we have
\be
p={\hbar W (\ell+\bar\ell)\over2\left|\psi^D-i\ell\psi\right|^2}.
\l{momentino}\ee
Let $\psi_E$ be the wave--function of a state of energy $E$. Since $\psi_E$
solves the SE, according to (\ref{popca}), for any fixed set of integration
constants $\delta$, there are coefficients $A$ and $B$ such that
\be
\psi_E\{\delta\}={1\over\sqrt{\S_0'\{\delta\}}}\left(A e^{-{i\over\hbar}
\S_0\{\delta\}}+Be^{{i\over\hbar}\S_0\{\delta\}}\right).
\l{popcaSCHR}\ee

An interesting quantity to consider is the Ermakov invariant \cite{Ermakov}.
This has been rediscovered by Lewis \cite{Lewis} and further investigated also
in connection with the Milne equation \cite{Milne} (see \cite{svariati}
for related investigations). In the context of the SE, the
Ermakov invariant has been considered by Floyd in \cite{Floyd86}
\be
I=(2m)^{-1/2}\left[p\psi_E^2+\hbar^2(\psi_E\partial_qp^{-1/2}-p^{-1/2}
\partial_q\psi_E)^2\right].
\l{Ij2}\ee
This invariant can be constructed in terms of any solution of the SE. Once the
form (\ref{popcaSCHR}) of $\psi_E$ is given, any complex M\"obius
transformation of $e^{{2i\over\hbar}\S_0}$, which according to
(\ref{popcaSCHR}) will in general correspond to a linear transformation
$\psi_E\longrightarrow\tilde\psi_E=a\psi_E^D+b\psi_E$, still solution of the
SE, leaves $I$ invariant. The Ermakov invariant was evaluated in
\cite{Floyd86} in terms of the parameters defining $p$ and of the coefficients
in the expression of $\psi_E$ in terms of $\psi^D$ and $\psi$. Here we have
\be
p\psi_E^2-A^2e^{-{2i\over\hbar}\S_0}-B^2e^{{2i\over\hbar}\S_0}=2AB=\hbar^2(
\psi_E\partial_qp^{-1/2}-p^{-1/2}\partial_q\psi_E)^2+A^2e^{-{2i\over\hbar}\S_0
}+B^2e^{{2i\over\hbar}\S_0},
\l{ppji2}\ee
so that
\be
I=(2m)^{-1/2}4AB.
\l{Ij2bi}\ee
In the case in which either $A$ or $B$ vanish, we have $I=0$.
Let us set (recall that $W\in\RR\backslash\{0\}$)
\be
\epsilon={W(\ell+\bar\ell)\over|W(\ell+\bar\ell)|}={\rm sgn}\,[W(\ell+\bar
\ell)],
\l{eepps}\ee
and define
\be
\phi=\sqrt 2{e^{-i{\alpha\over2}}(\psi^D-i\ell\psi)\over\hbar^{1/2}|W(\ell+
\bar\ell)|^{1/2}}.
\l{phi1}\ee
It follows from (\ref{KdT3}) and (\ref{phi1}) that
\be
e^{{2i\over\hbar}\S_0\{\delta\}}={\bar\phi\over\phi},
\l{phi2}\ee
and since $\phi'\bar\phi-\phi{\bar\phi}'=-2i\epsilon/\hbar$,
we have
\be
p={\hbar\over2i}\partial_q\ln{\phi\over\bar\phi}=\epsilon|\phi|^{-2},
\l{phi3}\ee
so that $\epsilon=\pm 1$ fixes the direction of motion.
By (\ref{phi2}) and (\ref{phi3}) we have
\be
\phi={\epsilon^{1/2}\over\sqrt{\S_0'\{\delta\}}}e^{-{i\over\hbar}\S_0
\{\delta\}},
\l{oihxw9}\ee
so that we obtain the expression of the wave--function in terms of $\phi$
\be
\psi_E\{\delta\}=A\epsilon^{-1/2}\phi+B\epsilon^{1/2}\bar\phi,
\l{pSHR}\ee
where we used $\overline{\epsilon^{-1/2}}=\epsilon^{1/2}$. Note that by $\bar
\phi\not\propto\phi$ and $\phi'\bar\phi-\phi{\bar\phi}'\ne 0$, it follows
that $\phi$ never vanishes. Therefore, a necessary condition for a
solution of the SE to have zeroes is that it must be proportional to a real
function. Hence, if $\psi_E$ has zeroes, then $A\epsilon^{-1/2}\phi+B
\epsilon^{1/2}\bar\phi\propto\bar A\epsilon^{1/2}\bar\phi+\bar B\epsilon^{-1/2}
\phi$, that is $|A|=|B|$, in agreement with the fact that $\psi_E'\bar\psi_E
-{\bar\psi_E}'\psi_E=-2i\epsilon(|A|^2-|B|^2)/\hbar$ can be seen as a Wronskian
only if $\bar\psi_E\not\propto\psi_E$. We also note that in the case in which
$\psi_E=\psi$, we have
\be
A=i\left[{e^{i\alpha}\hbar W\over2(\ell+\bar\ell)}\right]^{1/2},
\qquad B=-\epsilon e^{-i\alpha}A=\bar A,
\l{oidhU}\ee
where the identity $-\epsilon e^{-i\alpha}A=\bar A$ follows from ${\rm sgn}\,
[W/(\ell+\bar\ell)]={\rm sgn}\,[W(\ell+\bar\ell)]$. An interesting question is
to find the transformations $\delta\longrightarrow\delta'$ leaving the state
described by $\psi_E\{\delta\}$ invariant. To this end it is useful to write
$e^{{2i\over\hbar}\S_0}$ in a different form. First of all note that the
expression of $e^{{2i\over\hbar}\S_0}$ in (\ref{KdT3}) can be seen as the
composition of two maps. The first one is the Cayley transformation
\be
w\longrightarrow z=\sigma(w)={w+i\over w-i}={\psi^D+i\psi\over\psi^D-i\psi}\in
S^1,
\l{HyJ8}\ee
where $\sigma$ is the matrix
\be
\sigma={1\over(-2i)^{1/2}}\left(\begin{array}{c}1\\ 1\end{array}
\begin{array}{cc}i\\ -i\end{array}\right).
\l{D73F34}\ee
Then $e^{{2i\over\hbar}\S_0}$ is obtained as the M\"obius transformation
\be
e^{{2i\over\hbar}\S_0}=\gamma_{\S_0}(z)={az+b\over cz+d},
\l{ABCDs}\ee
where, by (\ref{KdT3}), the entries $a,b,c,d$ of $\gamma_{\S_0}$ are
\be
\gamma_{\S_0}={1\over2^{1/2}|\ell+\bar\ell|^{1/2}}\left(\begin{array}{c}
e^{{i\over2}\alpha}(1+\bar\ell)\\ e^{-{i\over2}\alpha}(1-\ell)
\end{array}\begin{array}{cc}e^{{i\over2}\alpha}(1-\bar\ell)\\
e^{-{i\over2}\alpha}(1+\ell)\end{array}\right).
\l{dFF3F1}\ee
Note that $d=\bar a$, $c=\bar b$ and $\mu\equiv\det\gamma_{\S_0}={\rm sgn}\,{
\rm Re}\,\ell$, so that, if ${\rm Re}\,\ell>0$, then $\gamma_{\S_0}\in SU(1,1
)$. By (\ref{popcaSCHR}) and (\ref{ABCDs}) we can write $\psi_E$ in the form
\be
\psi_E\{\delta\}=\left({2i\over\hbar\partial_q\gamma_{\S_0}(z)}\right)^{1/2}
\left(A+B\gamma_{\S_0}(z)\right).
\l{Epaa}\ee
Under the transformation
\be
\delta\longrightarrow\delta'=\{\alpha',\ell'\},
\l{deldel1}\ee
we have
\be
e^{{2i\over\hbar}\S_0\{\delta\}}\longrightarrow e^{{2i\over\hbar}\S_0
\{\delta'\}}=\gamma'_{\S_0}(z)=e^{i\alpha'}{w+i{\bar\ell}'\over w -i\ell'},
\l{KdT3primo}\ee
where $\gamma'_{\S_0}$ is the matrix (\ref{dFF3F1}) with $\alpha$ and $\ell$
replaced by $\alpha'$ and $\ell'$ respectively. For our purpose it is useful
to write $e^{{2i\over\hbar}\S_0\{\delta'\}}$ as the M\"obius
transformation of $e^{{2i\over\hbar}\S_0\{\delta\}}$, that is
\be
e^{{2i\over\hbar}\S_0\{\delta'\}}=\tilde\gamma_{\S_0}\left(\gamma_{\S_0}(z)
\right),
\l{ABCDsz}\ee
where
\be
\tilde\gamma_{\S_0}=\gamma'_{\S_0}\gamma_{\S_0}^{-1}.
\l{gammapprr}\ee
We can now determine the transformations (\ref{deldel1}) such that
\be
\psi_E\{\delta'\}=\left({2i\over\hbar\partial_q\gamma_{\S_0}'(z)}
\right)^{1/2}\left(A+B\gamma_{\S_0}'(z)\right),
\l{popcaSCHRf}\ee
describes the same state described by $\psi_E\{\delta\}$. In other
words, we are considering the transformations of the integration constants of
the QSHJE (\ref{1Q}), corresponding to real $p$, such that $\psi_E\{\delta\}$
remains unchanged up to some multiplicative constant $c$, that is
\be
\psi_E\{\delta\}\longrightarrow\psi_E\{\delta'\}=c\psi_E\{\delta\}.
\l{ccooss}\ee
In order to compute $\psi_E\{\delta'\}$ we observe that by (\ref{gammapprr})
\be
\partial_q\gamma_{\S_0}'=\partial_q\tilde\gamma_{\S_0}\left(\gamma_{\S_0}(z)
\right)={\partial\tilde\gamma_{\S_0}\over\partial\gamma_{\S_0}}\partial_q
\gamma_{\S_0}=\tilde\mu{\partial_q\gamma_{\S_0}\over(\tilde c\gamma_{\S_0}
+\tilde d)^2},
\l{FcV6}\ee
where $\tilde\mu\equiv\det\tilde\gamma_{\S_0}={\rm sgn}\,{\rm Re}\,\tilde\ell$
and
\be
\tilde\gamma_{\S_0}(\gamma_{\S_0})={\tilde a\gamma_{\S_0}+\tilde b\over\tilde
c\gamma_{\S_0}+\tilde d},
\l{CxD}\ee
with
\be
\tilde a=\bar{\tilde d}=e^{{i\over2}\tilde\alpha}{1+\bar{\tilde\ell}\over
2^{1/2}|\tilde\ell+\tilde{\bar\ell}|^{1/2}},\qquad\tilde b=\bar{\tilde c}=
e^{{i\over2}\tilde\alpha}{1-\bar{\tilde\ell}\over2^{1/2}|\tilde\ell+\bar
{\tilde\ell}|^{1/2}},
\l{elementidimatrice}\ee
which are the matrix elements of $\tilde\gamma_{\S_0}$ given by (\ref{dFF3F1})
with the $\delta$--moduli replaced by $\tilde\delta=\{\tilde\alpha,\tilde\ell
\}$. Therefore, by (\ref{popcaSCHRf}) and (\ref{FcV6})
\be
\psi_E\{\delta'\}=\left({2i\over\tilde\mu\hbar\partial_q\gamma_{\S_0}}
\right)^{1/2}\left[A\tilde d+B\tilde b+(A\tilde c+B\tilde a)\gamma_{\S_0}
\right],
\l{ppssid}\ee
and Eq.(\ref{ccooss}) gives
\be
A^2\bar{\tilde b}+AB\tilde a=AB\bar{\tilde a}+B^2\tilde b.
\l{Kj9}\ee
Note that if either $A=0$ or $B=0$, then $\tilde b=0$, {\it i.e.} $\tilde\ell=1
$, and the only transformation leaving invariant the unit ray $\Psi_E$
associated to $\psi_E$ corresponds, as obvious, to adding a constant to $\S_0$.
In the case in which $\psi_E$ is proportional to a real function, one has
\be
B=e^{-i\gamma}A,
\l{igammax}\ee
for some $\gamma\in\RR$. Eq.(\ref{Kj9}) and (\ref{igammax}) imply that the
matrix elements of the transformations leaving the unit ray invariant satisfy
\be
{\rm Im}\,\tilde a={\rm Im}\,(e^{-i\gamma}\tilde b).
\l{immaginario}\ee
In the case in which $\psi_E=\psi$, by (\ref{oidhU}) and (\ref{igammax}) we have
$\gamma=\alpha-(\epsilon+1+4k)\pi/2$, $k\in {\ZZ}$.

While $\Psi_E$ is invariant under the transformations satisfying (\ref{Kj9})
\be
\Psi_E\{\delta\}\longrightarrow\Psi_E\{\delta'\}=\Psi_E\{\delta\},
\l{Pj7}\ee
in the case of the conjugate momentum we have
\be
p\{\delta\}={\hbar\over2i}\partial_q\ln\gamma_{\S_0}\longrightarrow p\{\delta'\}
={\hbar\over2i}\partial_q\ln\gamma_{\S_0}'={\tilde\mu\gamma_{\S_0}\over(\tilde a
\gamma_{\S_0}+\tilde b)(\bar{\tilde b}\gamma_{\S_0}+\bar{\tilde a})}p\{\delta\}.
\l{Fth8}\ee

\mysection{Trajectories and the Equivalence Principle}\l{tatep}

We now start considering the conditions on $\S_0$ coming from the EP. In other
words, until now we did not care about the differential properties of $\S_0$; we
just implicitly assumed that $\S_0$ had such properties that the EP could be
implemented. In other words, instead of defining the properties $\S_0$ should
have in order for the relevant equations to be defined, we did not mention them
reserving part of this section to this basic issue. This approach is
particularly convenient as the conditions for the implementation of the EP
reduce to the conditions that $\S_0$ should satisfy in order for the QSHJE to
exist. As we will see, deriving these conditions will lead to the quantization
of the energy without making use of the conventional axiomatic interpretation of
the wave--function. This is an important result of our approach as we derive a
basic fact directly from the EP. In this section we will also investigate in
detail the symmetry properties of the trivializing map. As we will see, the role
of the M\"obius transformation is that of making the generally projective
transformation, connecting the trivializing map and the ratio $w=\psi^D/\psi$,
equivalent to an affine transformation.

The conditions concerning the existence of the QSHJE are closely related to the
nature of the trivializing map. In particular, we will see that this corresponds
to a local homeomorphism of the extended real line into itself. In this context
we will see that there is an underlying M\"obius symmetry which is related to
the EP. Next, we will consider the case of the free particle and will define the
time parameterization by following Floyd's suggestion of using Jacobi's theorem.
This will lead to a dynamical equation corresponding to the quantum analogue of
Newton's law $F=ma$.

\subsection{Differential properties of the reduced action}

We saw that implementing the EP the CSHJE is modified by the
additive term ${\hbar^2}\{\S_0,q\}/4m$. In order to discuss the conditions for
the existence of the QSHJE (\ref{1Q}), it is convenient to consider its form
(\ref{sothat}) that by the M\"obius symmetry of the Schwarzian derivative and
(\ref{KdT3}) is equivalent to
\be
\{w,q\}=-{4m\over\hbar^2}\W(q).
\l{cosicchesothat}\ee
Existence of this equation requires some conditions on the continuity properties
of $w$ and its derivatives. Since the QSHJE is the consequence of the EP, we can
say that the EP imposes some constraints on $w=\psi^D/\psi$. These constraints
are nothing but the existence of QSHJE (\ref{1Q}) or, equivalently, of
Eq.(\ref{cosicchesothat}). That is, implementation of the EP imposes that
$\{w,q\}$ exists, so that
\be
w\ne cnst,\;w\in C^2(\RR),\;and\;\partial_q^2w\;differentiable\;on\;\RR.
\l{ccnndraft}\ee
Note that requiring $w$ to be of class $C^3(\RR)$ would be an unjustified
restriction as the Schwarzian derivative is defined also in the case in which
$\partial_q^3 w$ is discontinuous (in this case $\W$ is discontinuous).

We now show that the conditions (\ref{ccnndraft}) are not complete. The reason
is that, as we have seen, the implementation of the EP requires that the
properties of the Schwarzian derivative be satisfied. Actually, its very
properties, derived from the EP, led to the identification $\Z(q^a;q^b)=-\hbar^2
\{q^a,q^b\}/4m$. Therefore, in order to implement the EP it is of basic
importance that the transformation properties of the Schwarzian derivative and
its symmetries be satisfied. In deriving the transformation properties of
$\Z(q^a;q^b)$ we noticed how, besides dilatations and translations there is a
highly non--trivial symmetry such as that under inversion. Therefore, we have
that (\ref{cosicchesothat}) must be equivalent to
\be
\{w^{-1},q\}=-{4m\over\hbar^2}\W(q).
\l{cosicchesothat2}\ee
A property of the Schwarzian derivative is duality between its entries
\be
\{w,q\}=-\left({\partial w\over\partial q}\right)^2\{q,w\}.
\l{pksxjq}\ee
This shows that the invariance under inversion of $w$ reflects in the
invariance, up to a Jacobian factor, under inversion of $q$. That is
$\{w,q^{-1}\}=q^4\{w,q\}$, so that the QSHJE (\ref{cosicchesothat}) can be
written in the equivalent form
\be
\{w,q^{-1}\}=-{4m\over\hbar^2}q^4\W(q).
\l{oidjwI939}\ee
In other words, starting from the EP one can arrive to either
Eq.(\ref{cosicchesothat}) or Eq.(\ref{oidjwI939}). The consequence of this fact
is that since under
\be
q\longrightarrow {1\over q},
\l{oiqhd0913}\ee
$0^{\pm}$ maps to $\pm\infty$, we have to extend the continuity conditions
(\ref{ccnndraft}) to the point at infinity. In other words, the continuity
conditions (\ref{ccnndraft}) should hold on the extended real line $\hat\RR
=\RR\cup\{\infty\}$. This aspect is related to the fact that the M\"obius
transformations, under which the Schwarzian derivative transforms as a
quadratic differential, map circles to circles. We stress that we are
considering the systems defined on $\RR$ and not $\hat\RR$. What happens is
that the existence of QSHJE forces us to impose smoothly joining conditions
even at $\pm\infty$, that is (\ref{ccnndraft}) must be extended to
\be
w\ne cnst,\;w\in C^2(\hat\RR),\;and\;\partial_q^2w\;differentiable\;on\;\hat\RR.
\l{ccnn}\ee
As we will see, $w$ is a M\"obius transformation of the trivializing map.
Therefore, Eq.(\ref{pksxjq}), which is defined if and only if $w(q)$ can be
inverted, that is if $\partial_q w\ne 0$, $\forall q\in\RR$, is a consequence
of the cocycle condition (\ref{cociclo3}). By (\ref{oidjwI939}) we see that also
local univalence should be extended to $\hat\RR$. This implies the following
joining condition at spatial infinity
\be
w(-\infty)=\left\{\begin{array}{ll} w(+\infty), & for\;w(-\infty)\ne\pm
\infty,\\ -w(+\infty),& for\;w(-\infty)=\pm\infty.\end{array}\right.
\l{specificandoccnn}\ee
As illustrated by the non--univalent function $w=q^2$, the apparently natural
choice $w(-\infty)=w(+\infty)$, one would consider also in the $w(-\infty)=
\pm\infty$ case, does not satisfy local univalence.

\subsection{The Equivalence Principle and the trivializing map}

We can now derive the explicit expression of the trivializing map
$q\longrightarrow q^0$ under which a state $\W$ reduces to $\W^0$. A first
remark is that, as we have seen, for a given $\W$ there are different
possible $\S_0$'s which are parameterized by the $\delta$--moduli. We already
derived the structure of the trivializing map in (\ref{doiqjwdw}). Here we
derive the explicit expression of the trivializing map as function of the
$\delta$--moduli. This expression follows directly from $\S_0^0(q^0)=\S_0(q)$.
We simply observe that for the state $\W^0$ we can choose, as in
(\ref{psiddpsi}), $\psi^{D^0}=q^0$ and $\psi^0=1$. Then by (\ref{KdT3}) and
$\S_0^0(q^0)=\S_0(q)$ we have
\be
e^{i\alpha_0}{q^0+i\bar\ell_0\over q^0-i\ell_0}=
e^{i\alpha}{w+i\bar\ell\over w-i\ell}.
\l{KdT301}\ee
Therefore, the trivializing map transforming the state $\W$ with moduli $\delta
=\{\alpha,\ell\}$ into the state $\W^0$ with moduli $\delta_0=
\{\alpha_0,\ell_0\}$ is given by the real map
\be
q^0={(\ell_0e^{i\beta}+\bar\ell_0e^{-i\beta})w+i\ell_0\bar\ell e^{i\beta}-i\bar
\ell_0\ell e^{-i\beta}\over2 w\sin\beta+\ell e^{-i\beta}+\bar\ell e^{i\beta}},
\l{g123s}\ee
where $\beta=(\alpha-\alpha_0)/2$. Let us consider the case in which the
functional structure of two reduced actions differs only by a constant, that is
\be
\S_0^a(q^a)=\S_0^b(q^a)+\hbar(\alpha_a-\alpha_b)/2.
\l{differisconoperalfa}\ee
It is clear that these two reduced actions lead to the same conjugate momentum,
that is to the same physical state. In fact, by (\ref{differisconoperalfa}) we
have
\be
p_a(q^a)={\partial\S_0^a(q^a)\over\partial q^a}={\partial\S_0^b(q^a)\over
\partial q^a}={\partial\S_0^b(q^b)\over\partial q^b}|_{q^b=q^a}=p_b(q^a).
\l{papb7}\ee
Therefore, while for general reduced actions $\S_0^a$ and $\S_0^b$ one has $p_a(
q)\ne p_b(q)$, in the case (\ref{differisconoperalfa}) the functional dependence
of $p_a$ and $p_b$ on their arguments coincides and the only effect of the
identification $\S_0^a(q^a)=\S_0^b(q^b)$ is a trivial re--labeling of the
coordinate. Thus, even if $p_a(q^a)\ne p_b(q^b)$, we have $p_a(q)=p_b(q)$.
Therefore, since the transformations (\ref{g123s}) that differ only by the value
of $\beta$ connect the same pair of states, we can set $\alpha=\alpha_0+2k\pi$,
and (\ref{g123s}) becomes
\be
q^0={(\ell_0+\bar\ell_0)w+i\ell_0\bar\ell-i\bar\ell_0\ell\over\ell+\bar\ell}.
\l{qzeroo}\ee
We will call M\"obius states the states parameterized by $\ell$ associated with
a given $\W$.

\subsection{Equivalence Principle and local homeomorphicity of $v$--maps}

We now discuss some basic properties of the $v$--maps. First we note that
consistency of the EP requires continuity of the $v$--maps: since both $q$ and
$q^v$ take values continuously on $\RR$, it is clear that full equivalence
between the two systems requires that the $v$--maps should be continuous. Next
note that the EP implies that there always exists the trivializing map
$q\longrightarrow q^0$ under which $\W\longrightarrow\W^0$. Furthermore, since
the transformation should exist for any pair of $\W$ states, we have that
$v$--maps, including the trivializing map, should be locally invertible. This
provides the pseudogroup property.

We saw that all the above continuity properties hold. In fact these are the
properties we derived for $w$ as a condition for the existence of the QSHJE. On
the other hand, it follows by (\ref{g123s}) that the properties of $w$ reflect
on the properties of the trivializing map. As for $w$, even in considering
$v$--maps we should consider $\hat\RR$. In particular, from the properties of
$w$ and by (\ref{g123s}) it follows that the $v$--maps are local
self--homeomorphisms of $\hat\RR$. Let us map $\hat\RR$ to $S^1$ by means of a
Cayley transformation and then consider the case of the trivializing map. While
\be
z={q-i\over1-iq},
\l{zq1}\ee
spans $S^1$ once, we have that
\be
z^0={q^0-i\over1-iq^0},
\l{wq01}\ee
should run continuously around $S^1$. Since the Cayley transformation is a
global univalent transformation, we have that the $v$--maps induce local
self--homeomorphisms of $S^1$. We also note that since local homeomorphisms are
closed under composition, it follows that local homeomorphicity of any $v$--map
also follows from local homeomorphicity of the trivializing map.

\subsection{The M\"obius symmetry of the trivializing map}

Let us further illustrate the properties of the trivializing map. Choosing a
pair of real linearly independent solutions of the SE different from $(\psi^D,
\psi)$, reflects in a real M\"obius transformation of $w$
\be
w\longrightarrow\tilde w={aw+b\over cw+d}.
\l{wwtilde}\ee
By (\ref{KdT3}), the effect of this transformation on $\S_0$ is
\be
e^{{2i\over\hbar}\S_0\{\delta\}}\longrightarrow e^{{2i\over\hbar}\S_0\{\delta'\}
}=e^{i\alpha}\left({a+i\bar\ell c\over a-i\ell c}\right){w+i(\bar\ell d-ib)/(a+i
\bar\ell c)\over w-i(\ell d+ib)/(a-i\ell c)}=e^{i\alpha'}{w+i\bar\ell'\over
w-i\ell'},
\l{KdT3332}\ee
where
\be
e^{i\alpha'}=e^{i\alpha}{a+i\bar\ell c\over a-i\ell c},\qquad
\ell'={\ell d+ib\over a-i\ell c}.
\l{soddisfa}\ee
Considering the fixed point equation
\be
\ell'=\ell,
\l{puntofisso}\ee
we see that the transformations (\ref{wwtilde}) satisfying
\be
\ell=i{d-a\pm\sqrt{(d-a)^2+4bc}\over2c},
\l{elle5t}\ee
correspond to changing only the phase $\alpha$ and do not affect the conjugate
momentum. Since the transformation (\ref{wwtilde}) corresponds to
\be
{\tilde\psi}^D=a\psi^D+b\psi,\qquad\tilde\psi=c\psi^D+d\psi,
\l{tilldde}\ee
we have that replacing $(\psi^D,\psi)$ in (\ref{CarlosSantana5}) with $({\tilde
\psi}^D,\tilde\psi)$, has no effect on $p$ if the transformation (\ref{tilldde})
satisfies (\ref{elle5t}). This analysis also shows that the transformation of
$w$ connecting (\ref{g123s}) and (\ref{qzeroo}) has the structure
(\ref{wwtilde})(\ref{elle5t}). This is an interesting property of the
trivializing map. In particular, (\ref{g123s}) and (\ref{qzeroo}) show that
there is a symmetry underlying the trivializing map, and therefore, by the
pseudogroup property, the $v$--maps. To see this observe that (\ref{puntofisso})
is equivalent to
\be
d=a-i\ell c-i\ell^{-1}b.
\l{realeunoinino}\ee
On the other hand, reality of $a,b,c,d$ implies the condition
\be
c=-{b^2\over|\ell|^2}.
\l{cuguale}\ee
This means that all the transformations of the kind
\be
\left(\begin{array}{c}\psi^D\\ \psi\end{array}\right)\longrightarrow\left(
\begin{array}{c}\tilde\psi^D\\ \tilde\psi\end{array}\right)=\left
(\begin{array}{c}a\\ -b|\ell|^{-2}\end{array}\begin{array}{cc}b\\ a+ib({\bar
\ell}^{-1}-\ell^{-1})\end{array}\right)\left(\begin{array}{c}\psi^D\\ \psi
\end{array}\right),
\l{2presodah}\ee
induce a transformation of $q^0$ under which $\W^0$ remains invariant. Hence,
the trivializing map is defined up to a group of transformations that we now
derive. First note that the effect of (\ref{2presodah}) on $w$ is
\be
w\longrightarrow\tilde w=\tilde\gamma(w)={a|\ell|^2w+b|\ell|^2
\over-bw+a|\ell|^2+ib(\ell-\bar\ell)},
\l{ind8reyu}\ee
where $\tilde\gamma$ is the matrix in (\ref{2presodah}). From Eq.(\ref{qzeroo})
we have
\be
w=\gamma(q^0),
\l{36pertrecento99}\ee
where
\be
\gamma=(\ell_0+\bar\ell_0)^{-1}\left(\begin{array}{c}\ell+\bar\ell\\ 0
\end{array}\begin{array}{cc}i\bar\ell_0\ell-i\ell_0\bar\ell\\ \ell_0+
\bar\ell_0\end{array}\right).
\l{pqoj99999999}\ee
By (\ref{ind8reyu}) and (\ref{36pertrecento99}) we have
\be
\tilde w=\tilde\gamma(w)=\tilde\gamma(\gamma(q^0))=\tilde\gamma\cdot\gamma(q^0).
\l{ioxu659}\ee
The effect on $q^0$ induced by the transformation (\ref{2presodah}) is defined
by
\be
\tilde w=\gamma(\hat\gamma(q^0))=\gamma\cdot\hat\gamma(q^0),
\l{36pertrecento99909}\ee
that compared with (\ref{ioxu659}) gives
\be
\hat\gamma=\gamma^{-1}\cdot\tilde\gamma\cdot\gamma.
\l{xlk998}\ee
Therefore, given the trivializing map $q\longrightarrow q^0=v^0(q)$ connecting
a M\"obius state of $\W$ to a M\"obius state of $\W^0$, we have that any other
map defined by
\be
\hat{q^0}=\hat\gamma(q^0),
\l{eancora98999999}\ee
still connects the same M\"obius states. These transformations can be
parameterized by $\beta$ in (\ref{g123s}). Let us denote $q^0$ in (\ref{g123s})
by $q^0_\beta$, so that $q^0$ in (\ref{qzeroo}) corresponds to $q^0_{k\pi}$. By
(\ref{g123s}) we have
\be
q_\beta^0=\gamma_1(w),
\l{oxiq979}\ee
where
\be
\gamma_1=\left(\begin{array}{c}\ell_0e^{i\beta}+\bar\ell_0e^{-i\beta}\\ 2
\sin\beta\end{array}\begin{array}{cc}i\ell_0\bar\ell e^{i\beta}-i\bar\ell_0
\ell e^{-i\beta}\\ \ell e^{-i\beta}+\bar\ell e^{i\beta}\end{array}\right).
\l{pqoj999559999}\ee
On the other hand, by (\ref{36pertrecento99}) and (\ref{oxiq979}) we have
$q_\beta^0=\gamma_1\cdot\gamma(q^0)$, so that
\be
q_\beta^0=\gamma_\beta(q^0),
\l{oxiq9790f9wu}\ee
where
\be
\gamma_\beta=\gamma_1\cdot\gamma={\ell+\bar\ell\over\ell_0+\bar\ell_0}
\left(\begin{array}{c}\ell_0e^{i\beta}+\bar\ell_0e^{-i\beta}\\ 2
\sin\beta\end{array}\begin{array}{cc}-2|\ell_0|^2\sin\beta
\\ \ell_0e^{-i\beta}+\bar\ell_0e^{i\beta}\end{array}\right).
\l{BondJamesBondBond}\ee
Therefore, we have that the trivializing map is defined up to the M\"obius
transformation
\be
q_\beta^0={(\ell_0e^{i\beta}+\bar\ell_0 e^{-i\beta})q^0-2|\ell_0|^2\sin\beta
\over2 q^0\sin\beta+\ell_0e^{-i\beta}+\bar\ell_0 e^{i\beta}}.
\l{aaaoxiq9790f9wuxql}\ee
Note that for $\ell_0=|\ell_0|\exp-i[\beta+(2k+1)\pi/2]$, $k\in{\ZZ}$, we have
\be
q_\beta^0=-{|\ell_0|^2\over q^0+2e^{k\pi i}|\ell_0|\cos\beta},
\l{rstaaaoxiq9790f9wuxql}\ee
that for $\beta=(2n+1)\pi/2$, $n\in{\ZZ}$, becomes
\be
q_{(2n+1)\pi/2}^0=-{|\ell_0|^2\over q^0}.
\l{qpwosxq5}\ee
Note that Eq.(\ref{aaaoxiq9790f9wuxql}) can be also directly derived from
(\ref{g123s}) by considering the case in which $w=q^0$ and $\ell=\ell_0$,
corresponding to
\be
e^{i\alpha_0}{q^0_\beta+i\bar\ell_0\over q^0_\beta-i\ell_0}=
e^{i\alpha}{q^0+i\bar\ell_0\over q^0-i\ell_0}.
\l{KdT301aridaije6}\ee
This represents the transformation connecting the same M\"obius state but
with the action having a different $\alpha$. Since this constant has no
dynamical effects, the M\"obius transformations (\ref{aaaoxiq9790f9wuxql})
correspond to a one--parameter group of symmetries of the trivializing map.

It is worth stressing another property of the trivializing map. Namely, we have
seen that it is a M\"obius, or projective, transformation (\ref{g123s}) of the
ratio $w$. On the other hand, this can be reduced to the affine mapping
(\ref{qzeroo}). Recall that in proving that the relation Eq.(\ref{cociclo3})
implies that $\Z(q^a,q^b)\propto\{q^a,q^b\}$, we mentioned the fact that any
element $g$ in the family of one--dimensional complex analytic local
homeomorphisms, such that either $g''(x)/g'(x)=0$ or $\{g(x),x\}=0$, are the
unique solutions of a system of differential equations involving only the first
and higher derivatives, satisfying the pseudogroup property
\cite{Cartan}\cite{GuillStern}. Then, in this context it is interesting to
observe that according to (\ref{g123s}) we have $\{w,q^0\}=0$ which by
(\ref{qzeroo}) reduces to $\partial_{q_0}^2 w/\partial_{q^0} w=0$.

Let us denote by $q_a$ and $q_b$ the coordinates of two M\"obius states
associated to the same $\W$. While in the case of the state $\W^0$, $q^0_a$ and
$q^0_b$ are related by the affine transformation
\be
q^0_b={(\ell_{0,b}+\bar\ell_{0,b})q^0_a+i\ell_{0,b}\bar\ell_{0,a}
-i\bar\ell_{0,b}\ell_{0,a}\over\ell_{0,a}+\bar\ell_{0,a}},
\l{02019}\ee
in the case of M\"obius states associated to arbitrary $\W$ states, the
$v$--map $q_a\longrightarrow q_b=v^{ba}(q_a)$ corresponds to the affine
transformation of the $w$'s
\be
w_b={(\ell_b+\bar\ell_b)w_a+i\ell_b\bar\ell_a-i\bar\ell_b\ell_a\over\ell_a
+\bar\ell_a},
\l{020192}\ee
where $w_k\equiv w(q_k)$, $k=a,b$. The $v$--map connecting two different
states, say $\W^a(q^a)$ and $\W^b(q^b)$, is still defined by (\ref{020192})
where now $w_k\equiv w_k(q_k)=\psi^D_k(q_k)/\psi_k(q_k)$, with $(\psi^D_k,
\psi_k)$ a pair of linearly independent real solutions of the SE defined by
$\W^k(q^k)$.

\subsection{The free particle}

A feature of our formulation is that no use of the usual axioms of QM is made.
In particular, the SE arises as mathematical tool to linearize the QSHJE
(\ref{1Q}). We now start studying the physical consequences of the QSHJE which
is seen as an equation for trajectories. While the concept of trajectories in QM
is reminiscent of Bohm's theory, we will see that there are basic differences.
A first important aspect is that the conjugate momentum
$p=\partial_q\S_0$, which is a real quantity even in the classically forbidden
regions, has a well--defined and natural classical limit. Furthermore, besides
the fact that there is no pilot--wave guide, we will also see that $p\ne m\dot
q$. This is a consequence of the definition of time parameterization which
follows from Jacobi's theorem as done by Floyd in \cite{Floyd82b}.

A basic request to formulating a trajectory interpretation of the QSHJE would be
the derivation of the main features of QM without assuming its axioms. Since
tunnelling and energy quantization are basic features distinguishing QM from CM,
we have to understand whether the QSHJE reproduces such phenomena. We will give
a detailed proof that this is in fact the case: both the tunnel effect and
energy quantization are predictions of the EP and in particular of the QSHJE
which follows from it. In this respect let us recall that in the conventional
formulation of QM the tunnel effect is a consequence of the wave--function
interpretation in terms of probability amplitude of finding the particle in a
measurement process. Since in the classically forbidden regions one generally
has $\psi\ne 0$, there is a non--zero probability of finding the particle in the
region where $V-E>0$. On the other hand, we have seen that $p=\epsilon|\phi|^{-
2}$, which is a real quantity even in the classically forbidden regions. As we
will see, the reality property inside the classically forbidden region is a
property which holds also for the velocity field.

Another feature, which is a consequence of the wave--function interpretation,
concerns the quantization of the energy spectrum. In particular, since $|\psi|^2
dq$ is the probability of finding the particle in the interval $[q,q+dq]$, it
follows that there are situations (bound states) in which the SE admits a
solution $\psi$, identified with the wave--function, belonging to $L^2(\RR)$. We
will see that in our formulation this condition follows from the local
homeomorphicity of $v$--maps, a condition implied by the EP! This condition
directly follows from (\ref{ccnn}) which concerns the existence of the QSHJE.

The above aspects will be discussed in detail later on. We now consider the
solutions of the QSHJE for the free particle of energy $E$. In order to find
the conjugate momentum, we first have to choose two real linearly independent
solutions of the SE. Let us set
\be
\psi^D=a\sin(kq),\qquad\psi=b\cos(kq),
\l{CS8}\ee
where $k=\sqrt{2mE}/\hbar$. For the Wronskian we have $W=-abk$. By
(\ref{CarlosSantana5}) and (\ref{CS8}) we have
\be
p_E=\pm{\hbar(\ell_E+\bar\ell_E)abk\over
2\left|a\sin(kq)-i\ell_E b\cos(kq)\right|^2}.
\l{CS11}\ee

An important topic of our approach that will be discussed later, concerns the
$E\longrightarrow0$ and $\hbar\longrightarrow0$ limits. In doing this we have
to consider $\S_0$ or $p$. This implies that these limits should be performed
by considered $\psi^D$, $\psi$ and $\ell$ simultaneously. However, since any
linear transformation of $\psi^D$ and $\psi$ reflects in a transformation of
$\ell$ and $\alpha$, in considering the $E\longrightarrow0$ limit of $\psi^D$
and $\psi$, we can also require that the corresponding functions reduce to the
one of the state $\W^0$
\be
\lim_{E\longrightarrow0}a\psi^D=q,\qquad\lim_{E\longrightarrow0}b\psi=1.
\l{CS12}\ee
A possible solution is $a=k^{-1}$, $b=1$, so that $W=-1$ and Eq.(\ref{CS11})
becomes
\be
p_E=\pm{\hbar(\ell_E+\bar\ell_E)\over2\left|k^{-1}\sin(kq)-i\ell_E\cos(kq)
\right|^2}.
\l{CS15}\ee

\subsection{Time parameterization}

It is well--known that while in CM the regions where $V-E>0$ are forbidden this
is not the case in QM. In our approach the tunnel effect is due to the fact that
$p$ is a real function even in the classically forbidden regions, a consequence
of the quantum potential which, unlike in Bohm theory, is never trivial. This
is the case also for the state $\W^0$. As stressed in \cite{3}, this
property is reminiscent of the relativistic rest energy.\footnote{In a different
context the usual quantum potential was considered in \cite{MugaSalaSnider} as
an internal potential.} As a consequence of the quantum potential, the dynamics
of the system is rather different from the classical one. In order to specify
it, we have to introduce time parameterization. In \cite{Floyd82b} Floyd made
the interesting proposal of using Jacobi's theorem. According to this theorem,
time parameterization is given by the partial derivative of the reduced action
with respect to $E$, that is
\be
t-t_0={\partial\S_0\over\partial E}.
\l{time}\ee
Using the QSHJE (\ref{1Q}), we can write (\ref{time}) in the form
\be
t-t_0={\partial\over\partial E}\int^q_{q_0}dx{\partial\S_0\over\partial x}=
\left({m\over2}\right)^{1/2}\int^q_{q_0}dx{1-\partial_EQ\over (E-V-Q)^{1/2}},
\l{tt0}\ee
where $q_0\equiv q(t_0)$. The velocity and acceleration are \cite{Floyd82b}
\be
{dq\over dt}=\left({dt\over dq}\right)^{-1}={\partial_q\S_0\over m(1-\partial_E
\V)},
\l{Axtt0}\ee
and
\be
{\ddot q}=-{\dot q}^3{d^2 t\over d q^2}={2(E-\V)\partial_q\partial_E\V\over m(1
-\partial_E\V)^3}-{\partial_q\V\over m(1-\partial_E\V)^2},
\l{Axtte4}\ee
where $\V$ denotes the ``effective potential"
\be
\V(q)\equiv V(q)+Q(q).
\l{effpt}\ee

\subsection{Quantum mass field: dynamical equation}

Observe that the deformation of the classical relation between conjugate
momentum and velocity is provided by the quantum potential through the relation
\be
\dot q={p\over m(1-\partial_EQ)}\ne{p\over m}.
\l{ne1}\ee
The relationship between $p$ and $\dot q$ can be written in an alternative way
by using the QSHJE (\ref{1Q}). In fact, the partial derivative of (\ref{1Q})
with respect to $E$ gives
\be
m{\partial Q\over\partial E}=m-p{\partial p\over\partial E},
\l{bjiu1}\ee
so that (\ref{ne1}) becomes
\be
\dot q={1\over\partial_E p},
\l{bijiu2}\ee
which is satisfied also classically. The fact that $\dot q\ne p/m$ is somehow
reminiscent of special relativity in which there is the correction due to the
$\gamma$ factor. In particular, for the free particle we have
\be
E={m\over2}{\dot q}^2(1-\partial_EQ)^2+Q.
\l{relazione3}\ee
This suggests defining the quantum mass field
\be
m_Q=m(1-\partial_EQ),
\l{quantummass}\ee
so that
\be
p=m_Q\dot q.
\l{scrittaalleh0min1618april}\ee
This form of the conjugate momentum allows us to write the QSHJE (\ref{1Q}) in
terms of $\stackrel{.}{q},\stackrel{..}{q},\stackrel{...}{q}$. Using
\be
p'=m_Q'\dot q+m_Q{\ddot q\over\dot q},\qquad p''=m_Q''\dot q+2m_Q'{\ddot q\over
\dot q}+m_Q\left({\stackrel{...}{q}\over{\dot q}^2}-{{\ddot q}^2\over{\dot q}^3
}\right),
\l{primi}\ee
and recalling that
\be
Q={\hbar^2\over4m}\{\S_0,q\}={\hbar^2\over4m}\left({p''\over p}-{3\over2}{{p'
}^2\over p^2}\right),
\l{Qp}\ee
we have that the QSHJE (\ref{1Q}) is equivalent to the dynamical equation
\be
{m_Q^2\over2m}{\dot q}^2+V(q)-E+{\hbar^2\over4m}\left({m_Q''\over m_Q}-{3\over2}
{{m'_Q}^2\over m_Q^2}-{m'_Q\over m_Q}{\ddot q\over{\dot q}^2}+{\stackrel{...}{q}
\over{\dot q}^3}-{5\over2}{{\ddot q}^2\over{\dot q}^4}\right)=0.
\l{QSHJELD}\ee
Note that taking the time derivative of this equation yields the quantum
analogue of Newton's law. We also observe that the solution of the equation of
motion depends on the integration constants ${\rm Re}\,\ell$ and ${\rm Im}\,
\ell$ which play the role of hidden variables.

\mysection{Equivalence Principle and fundamental constants}\l{epafc}

We have seen that the EP implied the QSHJE in which the Planck constant plays
the role of covariantizing parameter. The fact that a fundamental constant
follows from the EP suggests that other fundamental constants as well may be
related to such a principle. One hint in this direction is to observe that the
EP we formulated is reminiscent of Einstein's EP \cite{Einstein}. The similarity
between our formulation and the basis of GR suggests that the long--standing
problem of quantizing gravity is related to a deep relationship between gravity
and the foundations of QM. This indicates that the EP should be seen as a
principle underlying all possible interactions. In
particular, not only gravity but any physical system under the action of an
arbitrary potential should be equivalent to the free system. We have seen that
this implied the introduction of the quantum potential. Unlike in the
conventional approach to QM, this function is never trivial. In particular, by
(\ref{pzzweWro2}) it follows that in the case of the state $\W^0$ we have
\be
Q^0=-{p_0^2\over2m}=-{\hbar^2(\ell_0+\bar\ell_0)^2\over8m|q^0-i\ell_0|^4}.
\l{oix9ui87}\ee
We note that, unlike the relativistic rest energy, which depends on the mass
only, the quantum potential depends on $V-E$ itself. It is just this peculiar
property of $Q$ which makes the tunnel effect possible: the intrinsic
``self--energy'' $Q$ is the particle reaction to any external potential. This
intrinsic effect makes $p$ a real field also in the classically forbidden
regions.

The connection between our formulation and GR would suggest the appearance of
other fundamental constants besides the Planck constant. Since our
formulation concerns QM, it is clear that a connection with GR would suggest
the appearance of the Newton constant, in particular one should expect that
the Planck length should arise in a natural way. We now follow \cite{3} to show
that in considering the classical and $E\longrightarrow0$ limits in the case of
the free particle, one has to introduce fundamental lengths. The first condition
is that in the $\hbar\longrightarrow0$ limit Eq.(\ref{CS15}) reduces to the
classical solution
\be
\lim_{\hbar\longrightarrow0}p_E=\pm\sqrt{2mE}.
\l{bos1S11b}\ee
On the other hand, we should also have
\be
\lim_{E\longrightarrow0}p_E=p_0=\pm{\hbar(\ell_0+\bar\ell_0)\over2|q-i\ell_0|^2
},
\l{bisCS11b}\ee
in agreement with (\ref{pzzweWro2}). Let us first consider Eq.(\ref{bos1S11b}).
Due to the $\ell_E\cos(kq)$ term in (\ref{CS15}), we see that existence of the
classical limit would imply some condition on $\ell_E$. In
particular, in order to reach the classical value $\sqrt{2mE}$ in the $\hbar
\longrightarrow0$ limit, the quantity $\ell_E$ should depend on $E$. Let us set
\be
\ell_E\sim k^{-1}+\lambda_E,
\l{prova1}\ee
for some $\lambda_E$. Recalling that $p=\epsilon|\phi|^{-2}$, we have
\be
p_E=\pm{\hbar(\ell_E+\bar\ell_E)\over2|k^{-1}\sin(kq)-i\ell_E\cos(kq)|^2}=\pm{
\sqrt{2mE}+mE(\lambda_E+\bar\lambda_E)/\hbar\over\left|e^{ikq}+\lambda_E k
\cos(kq)\right|^2}.
\l{prova2}\ee
Hence, if $\lambda_E\sim\hbar^n$, $n>1$, then $p_E$ would satisfy
(\ref{bos1S11b}). However, we now show that the condition (\ref{bisCS11b})
implies that (\ref{prova1}), although correct as $\lambda_E$ is still
undetermined, is not the natural one. In fact, to get the limit (\ref{bisCS11b})
one observes that $q$ in the right hand side arises from the $E\longrightarrow0$
limit of $k^{-1}\sin(kq)$. This shows that to get this limit we cannot
manipulate the expression (\ref{CS15}) by moving $k^{-1}$ from the $k^{-1}\sin(k
q)$ term. On the other hand, with the position (\ref{prova1}), we would have
\be
p_E\;{}_{\stackrel{\sim}{E\longrightarrow0}}{2\hbar^2(2mE)^{-1/2}+\hbar
(\lambda_E+\bar\lambda_E)\over2|q-i\hbar(2mE)^{-1/2}-i\lambda_E|^2},
\l{prova3}\ee
implying a rather involved cancellation of the divergent $E^{-1/2}$ terms
which should come from $\lambda_E$. This suggests considering the position
\be
\ell_E=k^{-1}f(E,\hbar)+\lambda_E,
\l{prova4}\ee
where $f$ is dimensionless. Since $\lambda_E$ is arbitrary, we choose
$f$ to be real. By (\ref{prova2}) and (\ref{prova4}) we have
\be
p_E=\pm{\sqrt{2mE}f(E,\hbar)+mE(\lambda_E+\bar\lambda_E)/\hbar\over
\left|e^{ikq}+(f(E,\hbar)-1+\lambda_Ek)\cos(kq)\right|^2},
\l{prova4cc}\ee
so that, forgetting for a moment $\lambda_E$, in order to recover the
classical result we have that in the $\hbar\longrightarrow0$ limit $f$
has to satisfy the condition
\be
\lim_{\hbar\longrightarrow0} f(E,\hbar)=1.
\l{prova5}\ee
On the other hand, cancellation of the divergent term $E^{-1/2}$ in
\be
p_E\;{}_{\stackrel{\sim}{E\longrightarrow0}}{\hbar^2(2mE)^{-1/2}f(E,\hbar)+\hbar
(\lambda_E+\bar\lambda_E)\over|q-i\hbar(2mE)^{-1/2}f(E,\hbar)-i\lambda_E|^2},
\l{prova3truciolo}\ee
yields
\be
\lim_{E\longrightarrow0}E^{-1/2}f(E,\hbar)=0.
\l{prova6}\ee
This analysis shows that the request that both the $\hbar\longrightarrow0$ and
$E\longrightarrow0$ limits exist, is a strong constraint as
we have to introduce the dimensionless quantity $f$ which depends both on $E$
and $\hbar$ and satisfies the conditions (\ref{prova5}) and (\ref{prova6}). We
now see that this suggests introducing the Planck length.

\subsection{$\hbar\longrightarrow0, E\longrightarrow0$ and Planck length}\l{apl}

The limit (\ref{bisCS11b}) can be seen as the limit in which the trivializing
map reduces to the identity. Actually, the trivializing map connecting
$\W=-E$ and $\W^0$, reduces to the identity map in the
$E\longrightarrow0$ limit. In the above investigation we considered $q$ as an
independent variable, however one can also consider $q_E(q^0)$ so that $\lim_{E
\to0} q_E=q^0$ and in the above formulas one can replace $q$ with $q_E$.

In considering the two limits $E\longrightarrow0$ and $\hbar\longrightarrow0$,
one has to introduce basic lengths. We know from (\ref{prova6}) that $k$ must
enter in the expression of $f(E,\hbar)$. Since $f$ is a dimensionless constant,
we need at least one more constant with the dimension of a length. Two
fundamental lengths one can consider are the Compton length
\be
\lambda_c={\hbar\over mc},
\l{prova7}\ee
and the Planck length
\be
\lambda_p=\sqrt{\hbar G\over c^3}.
\l{prova8}\ee
Two dimensionless quantities depending on $E$ are
\be
x_c=k\lambda_c=\sqrt{2E\over mc^2},
\l{kComptlength}\ee
and
\be
x_p=k\lambda_p=\sqrt{2mEG\over\hbar c^3}.
\l{kPlancklength}\ee
On the other hand, since $x_c$ does not depend on $\hbar$, this combination is
ruled out by (\ref{prova5}). Then we see that a natural expression for $f$ is
a function of the Planck length times $k$. Let us set
\be
f(E,\hbar)=e^{-\alpha(x_p^{-1})},
\l{prova9}\ee
where
\be
\alpha(x_p^{-1})=\sum_{k\geq1}\alpha_kx_p^{-k}.
\l{alphapp}\ee
Eqs.(\ref{prova5})(\ref{prova6}) correspond to conditions on the $\alpha_k$'s.
For example, in the case in which one considers $\alpha$ to be the function
$\alpha(x_p^{-1})=\alpha_1x_p^{-1}$, then by (\ref{prova6}) we have
$\alpha_1>0$. Therefore, we have
\be
\ell_E=k^{-1}e^{-\alpha(x_p^{-1})}+\lambda_E.
\l{prova10}\ee
In order to consider the structure of $\lambda_E$, we note that although $e^{-
\alpha(x_p^{-1})}$ cancelled the $E^{-1/2}$ divergent term, we still have some
conditions to be satisfied. To see this note that
\be
p_E=\pm{\sqrt{2mE}e^{-\alpha(x_p^{-1})}+mE(\lambda_E+\bar\lambda_E)/\hbar
\over\left|e^{ikq}+(e^{-\alpha(x_p^{-1})}-1+k\lambda_E)\cos(kq)\right|^2},
\l{prova20}\ee
so that the condition (\ref{bos1S11b}) implies
\be
\lim_{\hbar\longrightarrow0}{\lambda_E\over\hbar}=0.
\l{prova21}\ee
To discuss this limit, we first note that by (\ref{prova2}) and (\ref{prova10})
\be
p_E=\pm{2\hbar k^{-1}e^{-\alpha(x_p^{-1})}+\hbar(\lambda_E+\bar\lambda_E)\over2
\left|k^{-1}\sin(kq)-i\left(k^{-1}e^{-\alpha(x_p^{-1})}+\lambda_E\right)\cos(kq)
\right|^2}.
\l{prova21bbv}\ee
So that, since
\be
\lim_{E\longrightarrow0}k^{-1}e^{-\alpha(x_p^{-1})}=0,
\l{prova21ccv}\ee
we have by (\ref{bisCS11b}) and (\ref{prova21bbv}) that
\be
\lim_{E\longrightarrow0}\lambda_E=\lim_{E\longrightarrow0}\ell_E=\ell_0.
\l{prova22}\ee
Consider now the limit
\be
\lim_{\hbar\longrightarrow0}p_0=0.
\l{prova23}\ee
Writing $p_0$ in the equivalent form
\be
p_0=\pm{\hbar(\ell_0+\bar\ell_0)\over2|q^0-i\ell_0|^2},
\l{prova24}\ee
we see that to get the classical limit, we have to consider both cases $q^0\ne
0$ and $q^0=0$. In particular, we see that the behavior of $\ell_0$ should be
\be
\ell_0\sim\hbar^\gamma,\qquad-1<\gamma<1.
\l{prova25}\ee
Lengths having the term $\hbar^\gamma$, $-1<\gamma<1$, can be constructed by
means of $\lambda_c$ and $\lambda_p$. We also note that a constant length
independent from $\hbar$ is given by $\lambda_e=e^2/mc^2$ with $e$ the
electric charge.\footnote{Note that the ratio $\lambda_e/\lambda_c$ is the
fine--structure constant $\alpha=e^2/\hbar c$.} Then $\ell_0$ can be considered
as a suitable function of $\lambda_c$, $\lambda_p$ and $\lambda_e$ satisfying
the constraint (\ref{prova25}). Consistency between (\ref{prova21}) and
(\ref{prova25}) implies that the condition (\ref{prova21}) is given by the
$\lambda_E/\lambda_0$ term of $\lambda_E$. In particular, the above
investigation indicates that a natural way to express $\lambda_E$ is
\be
\lambda_E=e^{-\beta(x_p)}\lambda_0,
\l{prova26}\ee
where
\be
\beta(x_p)=\sum_{k\geq1}\beta_k x_p^k.
\l{prova267Y}\ee
Any possible choice should satisfy the conditions (\ref{prova21}) and
(\ref{prova22}). For example, for $\ell_E$ built with $\beta(x_p)=\beta_1x_p$,
one should have $\beta_1>0$. Summarizing, we have
\be
\ell_E=k^{-1}e^{-\alpha(x_p^{-1})}+e^{-\beta(x_p)}\ell_0,
\l{prova27}\ee
where
\be
\ell_0=\ell_0(\lambda_c,\lambda_p,\lambda_e),
\l{prova28}\ee
and for the conjugate momentum of the state $\W=-E$, we have
\be
p_E=\pm{2k^{-1}\hbar e^{-\alpha(x_p^{-1})}+\hbar e^{-\beta(x_p)}(\ell_0+\bar
\ell_0)\over2\left|k^{-1}\sin(kq)-i\left(k^{-1}e^{-\alpha(x_p^{-1})}+e^{-\beta
(x_p)}\ell_0\right)\cos(kq)\right|^2}.
\l{prova21bbvxx}\ee

We observe that the above investigation answers the question posed at the end
of subsect.\ref{sbsx}, where we noticed non--triviality of the classical limit.
In particular, one can check that the coefficients in (\ref{atoppe}), which are
defined up to a global scaling factor, are
\be
A=e^{{i\over2}\alpha}(k^{-1}-\bar\ell_E),\quad B=-e^{{i\over2}\alpha}(\bar
\ell_E+k^{-1}),\quad C=e^{-{i\over2}\alpha}(k^{-1}+\ell_E),\quad D=e^{-{i\over
2}\alpha}(\ell_E-k^{-1}),
\l{ABCD1t}\ee
where we used
\be
{w+i\bar\ell_E\over w-i\ell_E}={k^{-1}\tan kq+i\bar\ell_E\over k^{-1}\tan kq
-i\ell_E}={(k^{-1}-\bar\ell_E)e^{2ikq}-\bar\ell_E-k^{-1}\over(k^{-1}+\ell_E)
e^{2ikq}+\ell_E-k^{-1}}.
\l{ABCD2t}\ee
The limit (\ref{atoppe}) is a direct consequence of the fact that the classical
limit of (\ref{prova21bbvxx}) is $\pm\sqrt{2mE}$.

A feature related to the above investigation concerns the analogue of
de Broglie's wave--length
\be
\lambda(q)={h\over|p|}={h|\phi|^2},
\l{lengthqdependent}\ee
where $h$ denotes the Planck constant. Let us denote by $q_R$ a solution of the
equation
\be
\psi^D(q)=-\psi(q){\rm Im}\,\ell,
\l{qminimo}\ee
that determines the points where $\psi^D-i\ell\psi$ is purely imaginary. We have
\be
\lambda(q_R)=2\pi\psi^2(q_R)|W|^{-1}{\rm Re}\,\ell,
\l{lengthqdependent999}\ee
and
\be
|p(q_R)|={\hbar|W|\over\psi^2(q_R){\rm Re}\,\ell}.
\l{lengthqdependent9999}\ee
In the case of the state $\W^0$, we have
\be
\lambda^0(q^0)=4\pi{|q^0-i\ell_0|^2\over\ell_0+\bar\ell_0},
\l{lengthqdependentinzero}\ee
that reaches its minimum precisely at $q^0=q^0_R=-{\rm Im}\,\ell_0$
\be
\lambda^0(-{\rm Im}\,\ell_0)=2\pi{\rm Re}\,\ell_0.
\l{lengthqdependentinzeromimnimum}\ee
Under the trivializing map all the solutions of (\ref{qminimo})
map to $q^0_R$. In fact, by (\ref{qzeroo}) and (\ref{qminimo})
\be
-{\rm Im}\,\ell_0={-(\ell_0+\bar\ell_0){\rm Im}\,\ell
+i\ell_0\bar\ell-i\bar\ell_0\ell\over\ell+\bar\ell}.
\l{qzerooastratto}\ee
More generally, by (\ref{020192})
\be
-{\rm Im}\,\ell_b={-(\ell_b+\bar\ell_b){\rm Im}\,\ell_a
+i\ell_b\bar\ell_a-i\bar\ell_b\ell_a\over\ell_a+\bar\ell_a},
\l{qzerooastratto2222}\ee
which can be also seen as a consequence of (\ref{qzerooastratto}) and of the
pseudogroup property of $v$--maps. Note that since
\be
p_0dq^0=pdq,
\l{cqko}\ee
we have the transformation property
\be
\lambda^0(q^0)=\left|{\partial q^0\over\partial q}\right|\lambda(q),
\l{wlicyu}\ee
which is equivalent to
\be
\lambda^0(q^0)=\left|{\phi^0(q^0)\over\phi(q)}\right|^2\lambda(q),
\l{wlicyuduetto}\ee
where $\phi^0$ is the function given in (\ref{phi1}) in the case of $\W^0$.

We have seen that in considering the $\hbar\longrightarrow0$ and $E
\longrightarrow0$ limits, one has to introduce some fundamental lengths. The
structure of $\ell_E$ indicates a possible way in which fundamental constants
and related interactions may arise. In particular, Eq.(\ref{prova27}) suggests
that in a suitable context these quantities may play the role of a natural
cut--off. In this respect we observe that in the expression of $\ell_E$ both
$x_p$ and $x_p^{-1}$ appears, a fact suggesting a possible ultraviolet--infrared
duality. In this context we recall that the basic M\"obius symmetry
(\ref{aaaoxiq9790f9wuxql}) is intrinsically related to the structure of the
QSHJE and then of the trivializing map. One may consider the state $\W^0$ as a
sort of vacuum state with all the other states connected by the trivializing
map. Then, thanks to the pseudogroup property of $v$--maps, the M\"obius
symmetry of the state $\W^0$, reflects in symmetry properties of arbitrary
states. In this respect we stress that the symmetry associated to the state
$\W^0$ includes the inversion (\ref{qpwosxq5}). In particular, observe that
after the rescaling of the space coordinate
\be
q^0_\beta\longrightarrow X^0_\beta={q^0_\beta\over|\ell_0|},
\l{spacerescaling}\ee
we have that (\ref{qpwosxq5}) precisely corresponds to the $S$--duality
transformation
\be
X_{(2n+1)\pi/2}^0=-{1\over X^0},
\l{qpwosxcwco91}\ee
which makes a correspondence between long and short distances.

We have seen that the $\ell_E$'s naturally appear in the framework of the
M\"obius transformations whose origin traces back to the $p$--$q$ duality we
studied at the beginning of our investigation. Actually, $p$--$q$ duality is
connected to the $\psi^D$--$\psi$ duality. In particular, while in considering
the self--dual states we were forced to introduce the dimensional constants
$\gamma$, $\gamma_p$ and $\gamma_q$, in the case discussed above we introduced
the dimensional constants $\lambda_p$, $\lambda_c$ and $\lambda_e$. As for
$p$--$q$ duality, in which the dimensionality of the constants is related to the
M\"obius symmetry and to the different dimensional properties of $p$ and $q$,
in the case of $\psi^D$ and $\psi$ we have that any linear combination of
$\psi^D$ and $\psi$ corresponds to a M\"obius transformation of $w=\psi^D/\psi$.
On the other hand, a linear combination between $\psi^D$ and $\psi$ forces us,
as in the case of $p$ and $q$, to introduce dimensional constants. The origin of
this is, once again, the existence of the self--dual state, which actually
coincides with the state $\W^0$. In fact, since for the state $\W^0$ the
functions $\psi^D$ and $\psi$ must solve the equation $\partial_q^2\psi^D=0=
\partial_q^2\psi$, we necessarily have to introduce a constant length in
considering linear combinations of $\psi^D=q$ and $\psi=cnst$. Since these
solutions correspond to $\W^0\equiv 0$, we do not have any constant length
provided by the problem. This leads considering fundamental constants. This is
the basic reason of the appearance of $\lambda_p$, $\lambda_c$ and $\lambda_e$.
Since these constants appear in $\ell$, we have that in our formulation of QM
there are hidden variables depending on the Planck length.

\subsection{Time, energy and trajectories}

We now consider an aspect concerning Floyd's proposal of using Jacobi's
theorem to define time parameterization. The point is that in considering
$\partial_E\S_0$, one has to understand whether the $E$ dependence of $\ell$, as
in the case of $\ell_E$ for the state with $\W=-E$, should be derived. In other
words, one should fix which terms in the expression of $\S_0$ one has to
consider as explicit. For example, in the case of the free particle of energy
$E$, one can absorb the multiplicative $k^{-1}$ factor in $w=k^{-1}\tan kq$ just
by a rescaling of $\ell_E$ in the expression of the reduced action that we
denote by $\S_0^E$. If in evaluating the partial derivative $\partial_E\S_0^E$
one considers the dependence of $\ell_E$ on $E$ as the unique implicit one, then
the value of $\partial_E\S_0^E$ is not well--defined: a common $E$--dependent
rescaling of $w+i\ell_E$ and $w-i\ell_E$ in (\ref{KdT3}) would change
$\partial_Ew$, and therefore $t-t_0=\partial_E\S_0^E$, leaving $\S_0^E$
invariant. In order to better understand these aspects, let us consider time
parameterization in the case of the state $\W^0$. Since this state corresponds
to the $E\longrightarrow0$ limit of the state $\W=-E$, we have
\be
t-t_0=\lim_{E\longrightarrow0}{\partial\S_0^E\over\partial E}.
\l{tempoperw0}\ee
On the other hand, for any fixed $\ell_0$ any different choice of the functions
$\alpha(x_p^{-1})$ and $\beta(x_p)$ corresponds by construction to a set of
$\ell_E$'s with the common property of having the same limit $\ell_0$. This
means that for any choice of the functions $\alpha(x_p^{-1})$ and $\beta(x_p)$
one obtains always $\S_0^0$ with the same $\ell_0$. Hence, different choices of
$\alpha(x_p^{-1})$ and $\beta(x_p)$ will correspond to the same M\"obius state
in the $E\longrightarrow0$ limit. This implies that the limit cannot depend on
$\alpha(x_p^{-1})$ and $\beta(x_p)$. What remains to understand is whether the
$k^{-1}$ factor in $w=k^{-1}\tan kq$ and in $\ell_E=k^{-1}e^{-\alpha(x_p^{-1})}
+e^{-\beta(x_p)}\ell_0$, should be derived. A natural possibility that would
allow a non ambiguous definition of time parameterization, is to assume that all
the terms depending on $E$ that can be absorbed in a redefinition of $\ell$
should not be considered in evaluating $\partial_E\S_0$. In other words, the
only $E$ terms in $\S_0$ that should be considered in evaluating $\partial_E
\S_0$ are those for which the transformation $E\longrightarrow E'\ne E$ does not
correspond to a M\"obius transformation of $e^{{2i\over\hbar}\S_0}$, that is
such that $\W\longrightarrow\W'\ne\W$. Let us consider again the case of the
{}free particle. We have $w=k^{-1}\tan kq$. While changing $E$ in $E'$ in the
$k^{-1}$ factor would give $\{{k'}^{-1}\tan kq,q\}=\{w,q\}=-4mE/\hbar^2$, where
$k'=\sqrt{2mE'}/\hbar$, in the case in which we change the $E$ appearing in
$\tan kq$ we have $\{k^{-1}\tan k'q,q\}=-4mE'/\hbar^2\ne-4mE/\hbar^2$. Hence,
the $k^{-1}$ term in $w$ should not be derived. The constant corresponding to
the initial conditions for $\S_0^E$ are $\alpha$ and $k\ell_E$ and the only term
in $\S_0^E$ to be derived with respect to $E$ is $kw=\tan kq$. It follows that
the equation for the trajectories of the free particle of energy $E$ is
\be
t-t_0={\partial\S_0^E\over\partial E}={k(\ell_E+\bar\ell_E)\over
2|\sin(kq)-ik\ell_E\cos(kq)|^2}\sqrt{m\over2E}q.
\l{tt0se}\ee
Note that since by (\ref{prova4})(\ref{prova5}) and (\ref{prova21})
\be
\lim_{\hbar\longrightarrow0}k\ell_E=1,
\l{limitekelle1}\ee
we have that the trajectories (\ref{tt0se}) precisely collapse
to the classical one in the $\hbar\longrightarrow0$ limit
\be
t-t_0=\sqrt{m\over2E}q.
\l{tt0seclassico}\ee
Hence, the QSHJE correctly leads to trajectories whose classical limit arises
in a natural way. Note that the quantum trajectory (\ref{tt0se}) can be
written as the multiplicative correction of the classical one
\be
t-t_0=F_{qu}(q)\sqrt{m\over2E}q,
\l{tt0se2}\ee
where
\be
F_{qu}(q)={k(\ell_E+\bar\ell_E)\over2|\sin(kq)-ik\ell_E\cos(kq)|^2}.
\l{xDtzxd}\ee

It is easy to see that the right hand side of (\ref{tt0se}) diverges in
the $E\longrightarrow0$ limit. This means that the state $\W^0$ corresponds,
as in the classical limit, to a particle at rest, that is
\be
{\dot q}^0=0.
\l{qdotzero0}\ee
Therefore, in the formulation derived from the EP a free particle with
vanishing energy is at rest, as in the classical case. However, while in
CM one has $\partial_q\S_0^{cl}=m\dot q$, in QM we have $\partial_q\S_0\ne m
\dot q$. In particular, while for the state $\W^0$ we have $\dot q=0$, the
conjugate momentum is non--vanishing. This can be considered as the
effect of the quantum mass field. Let us denote by $m_{Q_E}$ the quantum mass
field corresponding to the free particle of energy $E$. It follows from
(\ref{pzzweWro2})(\ref{scrittaalleh0min1618april}) and (\ref{qdotzero0}) that
\be
\lim_{E\longrightarrow0}m_{Q_E}=\pm\infty.
\l{mqediverge}\ee
Hence, the fact that $p_0\not\equiv0$ is due to the divergence of
$\partial_EQ_E$ in the $E\longrightarrow0$ limit.

While the state $\W^0$ corresponds, as in the classical limit, to a particle at
rest, in the $\W=-E$ case the velocity is not a constant. In fact, deriving
(\ref{tt0se}) with respect to $q$ we obtain
\be
\dot q={2|\sin(kq)-ik\ell_E\cos(kq)|^4\over k(\ell_E+\bar\ell_E)\left\{|\sin(kq)
-ik\ell_E\cos(kq)|^2+kq[(|k\ell_E|^2-1)\sin(2kq)+ik(\ell_E-\bar\ell_E)\cos(2kq)]
\right\}}\sqrt{2E\over m},
\l{dottorediq}\ee
which is a constant in the classical limit only
\be
\lim_{\hbar\longrightarrow0}\dot q=\sqrt{2E/m}.
\l{velocitaclassica}\ee
Therefore, for the trajectory (\ref{tt0se}) we have $\S_0^E\ne\sqrt{2mE}q$
unless $k\ell_E=1$, which holds in the classical limit only. Now recall that in
discussing $p$--$q$ duality, we observed that since this is based on the
existence of the Legendre transform of $\S_0$, it would break down when
either $\S_0=cnst$ or $\S_0\propto q+cnst$. Hence, requiring a formulation with
manifest $p$--$q$ duality for all the possible states forces one to discard CM.
On the other hand, we derived the QSHJE (\ref{1Q}) from the EP without assuming
any $p$--$q$ duality. Then we noticed that, due to the $\{\S_0,q\}$ term, the
QSHJE could not be defined for $\S_0=cnst$. Hence, a condition for the existence
of $p$--$q$ duality was a simple consequence of the implementation of the EP. It
remains to understand whether also the case $\S_0^E\propto q+cnst$, for which
the Legendre transform of $\S_0$ degenerates, could be discarded for reasons
related to the implementation of the EP. This would give an even deeper
relationship between the EP and $p$--$q$ duality. We have seen above that this
is in fact the case: this came out as a consistency condition on the limits. In
particular, the solution $\S_0^E=\sqrt{2mE}q$, would imply
\be
\S_0^0=\lim_{E\longrightarrow0}\S_0^E=0,
\l{bertuccia}\ee
that we already discarded. Therefore, the EP guarantees $p$--$q$ duality.

\mysection{Equivalence Principle, tunnelling and quantized spectra}\l{epqsat}

In this section we show how tunnelling arises in the context of the QSHJE. We
also show, following \cite{1l2}, the basic fact that when $\lim_{q\to\pm\infty}
\W>0$ the QSHJE is defined only in the case in which the
corresponding SE has square summable solutions. This implies the standard result
on the quantized energy spectra without making use of any axiomatic
interpretation of the wave--function. We will also discuss the structure of the
spaces $\H$ and $\K$ of the admissible $\W$'s and $\S_0$'s respectively.

\subsection{Tunnelling}

It is well--known that it may happen that there are space regions which are
forbidden to classical trajectories. From the CSHJE one has
\be
p=\pm\sqrt{2m(E-V)},
\l{wu1classico}\ee
so that the classically forbidden region is $\Omega=\{q\in\RR|V(q)-E>0\}$.
The situation is completely different in the case of the QSHJE where
\be
p=\pm\sqrt{2m(-V-Q+E)}.
\l{wu1}\ee
Hence, due to the $Q$ term, we have that even if $q\in\Omega$, $p$ may be real.
In fact, since $p=\epsilon|\phi|^{-2}$, we have $p\in\RR$, $\forall q\in\RR$.
As we will see the exception arises just for the case of the infinitely deep
potential well. In other words, we have the tunnel effect for the trajectories
described by the QSHJE. In the usual formulation, tunnelling is essentially a
consequence of the axiomatic interpretation of the wave--function in terms of
probability amplitude. Note that we considered reality of $p$. However, to have
the tunnel effect one should check that $\dot q$, like $p$, is a real quantity
for any $q$. It can be seen, for example by (\ref{bijiu2}), that this is in
fact the case. This can be also seen by noticing that
\be
t-t_0={\partial\S_0\over\partial E}={W_E\over W}p,\qquad W_E\equiv\psi^D
\partial_E\psi-\psi\partial_E\psi^D,
\l{azzt5}\ee
implies $t\in\RR$, $\forall q\in\RR$. The reason underlying the tunnel effect
resides in the fact that it is always possible to choose a pair of linearly
independent solutions of the SE which are real for any $q\in\RR$. Deriving
(\ref{azzt5}) with respect to $q$ we obtain the expression of the velocity
\be
\dot q={W\over\partial_q(pW_E)}.
\l{velocity}\ee
Hence, there are no forbidden regions associated to the QSHJE. We note that this
can be also seen as a direct consequence of the EP. In fact, since the EP
implies that (\ref{VIP}) is a local self--homeomorphism of $\hat\RR$, even the
existence of only one $\S_0$, $\forall q\in\RR$, implies that any reduced action
will be real for any real value of the coordinate. On the other hand, if the
reduced actions take real values for all $q\in\RR$, then $p=\partial_q\S_0\in
\RR$, $\forall q\in\RR$. This implies that $t\in\RR$, $\forall q\in\RR$. The
exception arises for the infinitely deep potential well. We will see that this
situation will come as a particular limiting case.

\subsection{Quantized spectra from the Equivalence Principle}

We saw that the EP implied the QSHJE (\ref{1Q}). However, although this equation
implies the SE, we saw that there are aspects concerning the canonical variables
which arise in considering the QSHJE rather than the SE. In this respect a
natural question is whether the basic facts of QM also arise in our formulation.
A basic point concerns a property of many physical systems such as energy
quantization. This is a matter of fact beyond any interpretational aspect of QM.
Then, as we used the EP to get the QSHJE, it is important to understand how
energy quantization arises in our approach. According to the EP, the QSHJE
contains all the possible information on a given system. Then, the QSHJE itself
should be sufficient to recover the energy quantization including its structure.
In the usual approach the quantization of the spectrum arises from the basic
condition that in the case in which $\lim_{q\to\pm\infty}\W>0$, the
wave--function should vanish at infinity. Once the possible solutions are
selected, one also imposes the continuity conditions whose role in determining
the possible spectrum is particularly transparent in the case of discontinuous
potentials. For example, in the case of the potential well, besides the
restriction on the spectrum due to the $L^2(\RR)$ condition for the
wave--function (a consequence of the probabilistic interpretation of the
wave--function), the spectrum is further restricted by the smoothly joining
conditions. Since the SE contains the term $\partial_q^2\psi$, the continuity
conditions correspond to an existence condition for this equation. On the other
hand, also in this case, the physical reason underlying this request is the
interpretation of the wave--function in terms of probability amplitude.
Actually, strictly speaking, the continuity conditions come from the continuity
of the probability density $\rho=|\psi|^2$. This density should also satisfy the
continuity equation $\partial_t\rho+\partial_qj=0$, where $j=i\hbar(\psi
\partial_q\bar\psi-\bar\psi\partial_q\psi)/2m$. Since for stationary states
$\partial_t\rho=0$, it follows that in this case $j=cnst$. Therefore, in the
usual formulation, it is just the
interpretation of the wave--function in terms of probability amplitude, with the
consequent meaning of $\rho$ and $j$, which provides the physical motivation for
imposing the continuity of the wave--function and of its first derivative.

Now observe that in our formulation the continuity conditions arise from the
QSHJE. In fact, (\ref{ccnn}) implies continuity of $\psi^D$, $\psi$,
with $\partial_q\psi^D$ and $\partial_q\psi$ differentiable, that is
\be
Equivalence\;Principle\;\longrightarrow\;continuity\;of\;(\psi^D,\psi),\;and
\;(\partial_q\psi^D,\partial_q\psi)\;differentiable.
\l{equivalenzaederivata}\ee

Let us now consider the condition about the existence of an $L^2(\RR)$ solution
in the case in which $\lim_{q\to\pm\infty}\W>0$. Let us first note that in the
interesting paper \cite{Floyd82b}, the possible values of $E$ for which the SE
has no $L^2(\RR)$ solution were a priori discarded. Apparently, in considering
the QSHJE there are no reasons to make this assumption. Thus, even if in
\cite{Floyd82b} the axiomatic interpretation of the wave--function is refused,
and a basic trajectory interpretation of QM based on the QSHJE is provided, we
note that imposing the $L^2(\RR)$ condition appears as an unjustified
assumption. Actually, this is what essentially happens also in Bohm's theory,
where the $L^2(\RR)$ condition is essentially assumed. To a great extent this
seems to happen also in the stochastic formulation of QM. On the other hand,
in our approach we have a basic natural principle which motivates both the QSHJE
and then the continuity condition. It is therefore of basic importance to
investigate whether the existence of an $L^2(\RR)$ solution is also a
consequence of the EP. We now show that this is in fact the case!

In the following we will derive a result concerning the energy spectra. In this
context we will see that if $V(q)>E$, $\forall q\in\RR$, then there are no
solutions such that the ratio of two real linearly independent solutions of the
SE corresponds to a local self--homeomorphism of $\hat\RR$. The fact that this
is an unphysical situation can be also seen from the fact that the case $V>E$,
$\forall q\in\RR$, has no classical limit. Therefore, if $V>E$ both at $-\infty$
and $+\infty$, a physical situation requires that there are at least two points
in which $V-E=0$. More generally, if the potential is not continuous, we should
have at least two turning points for which $V(q)-E$ changes sign. Let us denote
by $q_-$ ($q_+$) the lowest (highest) turning point. We will prove the following
basic fact

\vspace{.333cm}

\noindent
{\it If}
\be
V(q)-E\geq\left\{\begin{array}{ll}P_-^2>0,& q<q_-,\\ P_+^2>0,&q>q_+,
\end{array}\right.
\l{asintoticopiumeno}\ee
{\it then $w$ is a local self--homeomorphism of $\hat\RR$
if and only if the corresponding SE has an $L^2(\RR)$ solution.}

\vspace{.333cm}

\noindent
Note that by (\ref{asintoticopiumeno}) we have
\be
\int^{-\infty}_{q_-}dx\kappa(x)=-\infty,\quad\int^{+\infty}_{q_+}dx\kappa(x)=+
\infty,
\l{divergono}\ee
where $\kappa=\sqrt{2m(V-E)}/\hbar$. Before going further, let us stress that
what we actually need to prove is that in the case (\ref{asintoticopiumeno}),
the joining condition (\ref{specificandoccnn}) requires that the corresponding
SE has an $L^2(\RR)$ solution. Observe that while (\ref{ccnn}), which however
follows from the EP, can be recognized as the standard condition
(\ref{equivalenzaederivata}), the other condition (\ref{specificandoccnn}),
which still follows from the existence of the QSHJE, and therefore from the EP,
is not directly recognized in the standard formulation. Since this leads to
energy quantization, while in the usual approach one needs one more assumption,
we see that there is a quite fundamental difference between the QSHJE and the
SE. We stress that (\ref{ccnn}) and (\ref{specificandoccnn}) guarantee that $w$
is a local self--homeomorphism of $\hat\RR$.

Let us first show that the request that the corresponding SE has an $L^2(\RR)$
solution is a sufficient condition for $w$ to satisfy (\ref{specificandoccnn}).
Let $\psi\in L^2(\RR)$ and denote by $\psi^D$ a linearly independent solution.
As we will see, the fact that $\psi^D\not\propto\psi$ implies that if $\psi\in
L^2(\RR)$, then $\psi^D\notin L^2(\RR)$. In particular, $\psi^D$ is divergent
both at $q=-\infty$ and $q=+\infty$. Let us consider the real ratio
\be
w={A\psi^D+B\psi\over C\psi^D+D\psi},
\l{arbitraryratio}\ee
where $AD-BC\ne 0$. Since $\psi\in L^2(\RR)$, we have
\be
\lim_{q\longrightarrow\pm\infty}w=\lim_{q\longrightarrow\pm\infty}
{A\psi^D+B\psi\over C\psi^D+D\psi}={A\over C},
\l{arbitygvxy}\ee
that is $w(-\infty)=w(+\infty)$. In the case in which $C=0$ we have
\be
\lim_{q\longrightarrow\pm\infty}w=\lim_{q\longrightarrow\pm\infty}{A\psi^D
\over D\psi}=\pm\epsilon\cdot\infty,
\l{arbitygvxyconczero}\ee
where $\epsilon=\pm1$. The fact that $\lim_{q\to\pm\infty}{A\psi^D/D\psi}$
diverges follows from the mentioned properties of $\psi^D$ and $\psi$. It
remains to check that if $\lim_{q\to-\infty}{A\psi^D/D\psi}=-\infty$, then
$\lim_{q\to+\infty}{A\psi^D/D\psi}=+\infty$, and vice versa. This can be
seen by observing that
\be
\psi^D(q)=c\psi(q)\int^q_{q_0}dx\psi^{-2}(x)+d\psi(q),
\l{oqqw}\ee
$c\in\RR\backslash\{0\}$, $d\in\RR$. Since $\psi\in L^2(\RR)$ we have
$\psi^{-1}\not\in L^2(\RR)$ and $\int^{+\infty}_{q_0}dx\psi^{-2}(x)=+\infty$,
$\int^{-\infty}_{q_0}dx\psi^{-2}(x)=-\infty$, so that $\psi^D(-\infty)/\psi
(-\infty)=-\epsilon\cdot\infty=-\psi^D(+\infty)/\psi(+\infty)$, where
$\epsilon={\rm sgn}\,c$.

We now show that the existence of an $L^2(\RR)$ solution of the SE is a
necessary condition to satisfy the joining condition (\ref{specificandoccnn}).
We give two different proofs of this, one is based on the WKB approximation
while the other one uses Wronskian arguments. In the WKB approximation, we have
\be
\psi={A_-\over\sqrt{\kappa}}e^{-\int^q_{q_-}dx\kappa(x)}
+{B_-\over\sqrt{\kappa}}e^{\int^q_{q_-}dx\kappa(x)},\qquad q\ll q_-,
\l{Pantani1}\ee
and
\be
\psi={A_+\over\sqrt{\kappa}}e^{-\int^q_{q_+}dx\kappa(x)}
+{B_+\over\sqrt{\kappa}}e^{\int^q_{q_+}dx\kappa(x)},\qquad q\gg q_+.
\l{Pantani2}\ee
In the same approximation, a linearly independent solution has the form
\be
\psi^D={A_-^D\over\sqrt{\kappa}}e^{-\int^q_{q_-}dx\kappa(x)}
+{B_-^D\over {\kappa}}e^{\int^q_{q_-}dx\kappa(x)},\qquad q\ll q_-.
\l{Pantani1D}\ee
Similarly, in the $q\gg q_+$ region we have
\be
\psi^D={A_+^D\over\sqrt{\kappa}}e^{-\int^q_{q_+}dx\kappa(x)}
+{B_+^D\over\sqrt{\kappa}}e^{\int^q_{q_+}dx\kappa(x)},\qquad q\gg q_+.
\l{Pantani2D}\ee
Note that (\ref{Pantani1})--(\ref{Pantani2}) are derived by solving the
differential equations corresponding to the WKB approximation for $q\ll q_-$ and
$q\gg q_+$, so that the coefficients of $\kappa^{-1/2}\exp\pm\int^q_{q_-}dx
\kappa(x)$, {\it e.g.} $A_-$ and $B_-$ in (\ref{Pantani1}), cannot be
simultaneously vanishing. In particular, the fact that $\psi^D\not\propto\psi$
yields
\be
A_-B_-^D-A_-^DB_-\ne 0,\qquad A_+B_+^D-A_+^DB_+\ne 0.
\l{PantaniGirodItaliaeTour}\ee
Let us now consider the case in which, for a given $E$ satisfying
(\ref{asintoticopiumeno}), any solution of the corresponding SE diverges at
least at one of the two spatial infinities, that is
\be
\lim_{q\longrightarrow +\infty} (|\psi(-q)|+|\psi(q)|)=+\infty.
\l{caruccioe}\ee
This implies that there is a solution diverging both at $q=-\infty$ and $q=+
\infty$. In fact, if two solutions $\psi_1$ and $\psi_2$ satisfy $\psi_1(-
\infty)=\pm\infty$, $\psi_1(+\infty)\ne\pm\infty$ and $\psi_2(-\infty)\ne\pm
\infty$, $\psi_2(+\infty)=\pm\infty$, then $\psi_1+\psi_2$ diverges at $\pm
\infty$. On the other hand,
(\ref{PantaniGirodItaliaeTour}) rules out the case in which all the solutions in
their WKB approximation are divergent only at one of the two spatial infinities,
say $-\infty$. Since, in the case (\ref{asintoticopiumeno}), a solution which
diverges in the WKB approximation is itself divergent (and vice versa), we have
that in the case (\ref{asintoticopiumeno}), the fact that all the solutions of
the SE diverge only at one of the two spatial infinities cannot occur.

Let us denote by $\psi$ a solution which is divergent both at $-\infty$ and
$+\infty$. In the WKB approximation this means that both $A_-$ and $B_+$ are
non--vanishing, so that
\be
\psi{}_{\;\stackrel{\sim}{q\longrightarrow-\infty}\;}{A_-\over\sqrt\kappa}e^{-
\int^q_{q_-}dx\kappa},\qquad\psi{}_{\;\stackrel{\sim}{q\longrightarrow+\infty}
\;}{B_+\over\sqrt{\kappa}}e^{\int^q_{q_+}
dx\kappa}.
\l{equazione1piumeno}\ee
The asymptotic behavior of the ratio $\psi^D/\psi$ is given by
\be
\lim_{q\longrightarrow-\infty}{\psi^D\over\psi}={A_-^D\over A_-},\qquad
\lim_{q\longrightarrow+\infty}{\psi^D\over\psi}={B_+^D\over B_+}.
\l{osserva1}\ee
Note that since in the case at hand any divergent solution also diverges in the
WKB approximation, we have that (\ref{caruccioe}) rules out the case $A^D_-=
B_+^D=0$. Let us then suppose that either $A_-^D=0$ or $B_+^D=0$. If $A_-^D=0$,
then $w(-\infty)=0\ne w(+\infty)$. Similarly, if $B_+^D=0$, then $w(+\infty)=0
\ne w(-\infty)$. Hence, in this case $w$, and therefore the trivializing map,
cannot satisfy (\ref{specificandoccnn}). On the other hand, also in the case in
which both $A_-^D$ and $B_+^D$ are non--vanishing, $w$ cannot satisfy
Eq.(\ref{specificandoccnn}). For, if $A_-^D/A_-=B_+^D/B_+$,
then
\be
\phi=\psi-{A_-\over A^D_-}\psi^D=\psi-{B_+\over B^D_+}\psi^D,
\l{wouldbesarebbe}\ee
would be a solution of the SE whose WKB approximation has the form
\be
\phi={B_-\over\sqrt{\kappa}}e^{\int^q_{q_-}dx\kappa(x)},\qquad q\ll q_-,
\l{Pantani1Giro}\ee
and
\be
\phi={A_+\over\sqrt{\kappa}}e^{-\int^q_{q_+}dx\kappa(x)},\qquad q\gg q_+.
\l{Pantani2Tour}\ee
Hence, if $A_-^D/A_-=B_+^D/B_+$, then there is a solution whose WKB
approximation vanishes both at $-\infty$ and $+\infty$. On the other hand,
we are considering the values of $E$ satisfying Eq.(\ref{asintoticopiumeno})
and for which any solution of the SE has the property
(\ref{caruccioe}). This implies that no solutions can vanish both at
$-\infty$ and $+\infty$ in the WKB approximation. Hence
\be
{A_-^D\over A_-}\ne{B_+^D\over B_+},
\l{aebconesenzaddoversi}\ee
so that $w(-\infty)\ne w(+\infty)$. We also note that not even the case $w(-
\infty)=\pm\infty=-w(+\infty)$ can occur, as this would imply that $A_-=B_+=0$,
which in turn would imply, against the hypothesis, that there are solutions
vanishing at $q=\pm\infty$. Hence, if for a given $E$ satisfying
(\ref{asintoticopiumeno}), any
solution of the corresponding SE diverges at least at one of the two spatial
infinities, we have that the trivializing map has a discontinuity at $q=\pm
\infty$. As a consequence, the EP cannot be implemented in this case so that
this value $E$ cannot belong to the physical spectrum.

Therefore, the physical values of $E$ satisfying (\ref{asintoticopiumeno}) are
those for which there are solutions which are divergent neither at $-\infty$
nor at $+\infty$. On the other hand, from the WKB approximation and
(\ref{asintoticopiumeno}), it follows that the non--divergent solutions must
vanish both at $-\infty$ and $+\infty$. It follows that the only energy levels
satisfying the property (\ref{asintoticopiumeno}), which are compatible with the
EP, are those for which there exists the solution vanishing both at $\pm\infty$.
On the other hand, solutions vanishing as $\kappa^{-1/2}\exp\int^q_{q_-}dx
\kappa$ at $-\infty$ and $\kappa^{-1/2}\exp-\int^q_{q_+}dx\kappa$ at
$+\infty$, with $P^2_\pm>0$, cannot contribute with an infinite value to
$\int^{+\infty}_{-\infty}dx\psi^2(x)$. The reason is that existence of the
QSHJE requires that $\{e^{{2i\over\hbar}\S_0},q\}$ be
defined and this, in turn, implies that any solution of the SE must be
continuous. On the other hand, since $\psi$ is continuous, and
therefore finite also at finite values of $q$, we have $\int^{q_b}_{q_a}dx\psi^2
(x)<+\infty$ for all finite $q_a$ and $q_b$. In other words, the only
possibility for a continuous function to have a divergent value of $\int^{+
\infty}_{-\infty}dx\psi^2(x)$ comes from its behavior at $\pm\infty$. Therefore,
since the implementation of the EP in the case (\ref{asintoticopiumeno})
requires that the corresponding $E$ should admit a solution with the behavior
\be
\psi{}_{\;\stackrel{\sim}{q\longrightarrow-\infty}\;}{A_-\over\sqrt{\kappa}}
e^{\int^q_{q_-}dx\kappa},\qquad\psi{}_{\;\stackrel{\sim}{q\longrightarrow+
\infty}\;}{B_+\over\sqrt{\kappa}}e^{-\int^q_{q_+}dx\kappa},
\l{equazione1piumenoicse}\ee
we have the following basic fact

\vspace{.333cm}

\noindent
{\it
The values of $E$ satisfying
\be
V(q)-E\geq\left\{\begin{array}{ll}P_-^2>0,&q<q_-,\\ P_+^2>0,&q>q_+,
\end{array}\right.
\l{asintoticopiumenofgt}\ee
are physically admissible if and only if the corresponding SE
has an $L^2(\RR)$ solution.
}

\vspace{.333cm}

We now give another proof of the fact that if $\W$ is of the type
(\ref{asintoticopiumenofgt}), then the corresponding SE must have an
$L^2(\RR)$ solution in order to satisfy (\ref{specificandoccnn}).
In particular, we will show that this is a necessary condition. That this
is sufficient has been already proved above.

Let us start by observing that Wronskian arguments, which can be found
in Messiah's book \cite{Messiah}, imply that if $V(q)-E\geq P_+^2>0$,
$q>q_+$, then as $q\longrightarrow +\infty$, we have ($P_+>0$)

\begin{itemize}
\item[{\bf --}]{There is a solution of the SE that vanishes at least as
$e^{-P_+q}$.}
\item[{\bf --}]{Any other linearly independent solution diverges at least as
$e^{P_+q}$.}
\end{itemize}

\noindent
Similarly, if $V(q)-E\geq P_-^2>0$, $q<q_-$, then as $q\longrightarrow-
\infty$, we have ($P_->0$)

\begin{itemize}
\item[{\bf --}]{There is a solution of the SE that vanishes at least as
$e^{P_-q}$.}
\item[{\bf --}]{Any other linearly independent solution diverges at least as
$e^{-P_-q}$.}
\end{itemize}

\noindent
These properties imply that if there is a solution of the SE in $L^2(\RR)$, then
any solution is either in $L^2(\RR)$ or diverges both at $-\infty$ and
$+\infty$. Let us show that the possibility that a solution vanishes only at one
of the two spatial infinities is ruled out. Suppose that, besides the $L^2(\RR)$
solution, which we denote by $\psi_1$, there is a solution $\psi_2$ which is
divergent only at $+\infty$. On the other hand, the above properties show that
there exists also a solution $\psi_3$ which is divergent at $-\infty$. Since the
number of linearly independent solutions of the SE is two, we have $\psi_3=A
\psi_1+B\psi_2$. However, since $\psi_1$ vanishes both at $-\infty$ and
$+\infty$, we see that $\psi_3=A\psi_1+B\psi_2$ can be satisfied only if
$\psi_2$ and $\psi_3$ are divergent both at $-\infty$ and $+\infty$. This fact
and the above properties imply that

\vspace{.333cm}

\noindent
{\it If the SE has an $L^2(\RR)$ solution, then any solution has two possible
asymptotics}

\begin{itemize}
\item[{\bf --}]{Vanishes both at $-\infty$ and $+\infty$ at least as $e^{P_-q}$
and $e^{-P_+q}$ respectively.}
\item[{\bf --}]{Diverges both at $-\infty$ and $+\infty$ at least as $e^{-P_-q}$
and $e^{P_+q}$ respectively.}
\end{itemize}

\vspace{.333cm}

\noindent
Similarly, we have

\vspace{.333cm}

\noindent
{\it If the SE does not admit an $L^2(\RR)$ solution,
then any solution has three possible asymptotics}

\begin{itemize}
\item[{\bf --}]{Diverges both at $-\infty$ and $+\infty$ at least as $e^{-P_-q}$
and $e^{P_+q}$ respectively.}
\item[{\bf --}]{Diverges at $-\infty$ at least as $e^{-P_-q}$ and vanishes at
$+\infty$ at least as $e^{-P_+q}$.}
\item[{\bf --}]{Vanishes at $-\infty$ at least as $e^{P_-q}$ and diverges at
$+\infty$ at least as $e^{P_+q}$.}
\end{itemize}

\vspace{.333cm}

\noindent
Let us consider the ratio $w=\psi^D/\psi$ in the latter case. Since any
different choice of linearly independent solutions of the SE corresponds to a
M\"obius transformation of $w$, we can choose\footnote{Here by $\sim$ we mean
that $\psi^D$ and $\psi$ either diverge or vanish ``at least as".}
\be
\psi^D_{\;\stackrel{\sim}{q\longrightarrow-\infty}\;}a_-e^{P_-q},\qquad\qquad
\psi^D_{\;\stackrel{\sim}{q\longrightarrow+\infty}\;}a_+e^{P_+q},
\l{psiddi340}\ee
and
\be
\psi_{\;\stackrel{\sim}{q\longrightarrow-\infty}\;}b_-e^{-P_-q},\qquad\qquad
\psi_{\;\stackrel{\sim}{q\longrightarrow+\infty}\;}b_+e^{-P_+q}.
\l{psii340}\ee
Their ratio has the asymptotics
\be
{\psi^D\over\psi}{}_{\;\stackrel{\sim}{q\longrightarrow-\infty}\;}c_-e^{2P_-q}
\longrightarrow0,\qquad\qquad{\psi^D\over\psi}{}_{\;\stackrel{\sim}{q
\longrightarrow+\infty}\;}c_+e^{2P_+q}\longrightarrow\pm\infty,
\l{psiddisupsi340}\ee
so that $w$ cannot satisfy Eq.(\ref{specificandoccnn}). This concludes the
alternative proof of the fact that, in the case (\ref{asintoticopiumenofgt}),
the existence of the $L^2(\RR)$ solution is a necessary condition in order
(\ref{specificandoccnn}) be satisfied. The fact that this is a sufficient
condition has been proved previously in deriving Eq.(\ref{arbitygvxy}).

The above results imply that the usual quantized spectrum arises as a
consequence of the EP. As examples we will consider the potential
well and the simple and double harmonic oscillators.

Let us note that we are considering real solutions of the SE. Thus, apparently,
in requiring the existence of an $L^2(\RR)$ solution, one should specify the
existence of a real $L^2(\RR)$ solution. However, if there is an $L^2(\RR)$
solution $\psi$, this is unique up to a constant, and since also $\bar\psi\in
L^2(\RR)$ solves the SE, we have that an $L^2(\RR)$ solution of the SE is real
up to a phase.

\subsection{The index of the trivializing map}

Let us consider some further properties of the trivializing map. Let $n$ be the
index $I[q^0]$ of the covering associated to the trivializing map (\ref{g123s}).
This is the number of times $q^0$ spans $\hat\RR$ while $q$ spans $\hat\RR$.
Since $q^0$ and $w$ are related by a M\"obius transformation, we have
\be
I[q^0]=I[w].
\l{qzeroew}\ee
Another property of the trivializing map is that its index depends on $\W$ but
not on the specific M\"obius state. Note that since $p$ does not vanish for
finite values of $w$, it follows that $I[q^0]$ coincides with the number of
zeroes of $w$. This fact may be also understood by recalling the Sturm theorem
(see \cite{Arnold2}\cite{Ovsienko1}\cite{Ovsienko2} for a simple geometrical
interpretation),
stating that given two linearly independent solutions $\psi^D$ and $\psi$ of
$\psi''=K\psi$, between any two zeroes of $\psi^D$ there is one zero of
$\psi$.\footnote{This is another manifestation of $\psi^D$--$\psi$ duality. In
this context we stress that, while in the usual approach the wave--function
plays the central role, our description contains both $\psi^D$ and $\psi$.} This
theorem, and the condition that the values of $E$ satisfying
(\ref{asintoticopiumenofgt}) should correspond to a SE having an $L^2(\RR)$
solution, guarantees local homeomorphicity of the trivializing map. It remains
to understand the case in which $V(q)>E$, $\forall q\in\RR$. We already noticed
that, since there is no classical limit in this case, these solutions are not
admissible ones. This also follows, as it should, from the fact that these
solutions do not satisfy (\ref{specificandoccnn}). To see this, it is sufficient
to note that if $\psi$ decreases as $q\longrightarrow-\infty$, then by $\psi''/
\psi=4m\W/\hbar^2>0$, $\forall q\in\RR$, it follows that $\psi$ is always
convex, $\psi\not\in L^2(\RR)$. Therefore, the absence of turning points does
not modify the essence of the conclusions in the previous subsection and if
$V(q)>E$, $\forall q\in\RR$, then (\ref{specificandoccnn}) cannot be satisfied.
We refer to (\ref{introesempioill2}) for an explicit example.

\subsection{Equivalence Principle and admissible potentials}

We can now better define the spaces $\H$ and $\K$ that we introduced as the
spaces of all possible $\W$ and $\S_0$ respectively. We introduced these spaces
and then arrived to the conclusions that the implementation of the EP univocally
implied the QSHJE (\ref{1Q}). The appearance of $\{\S_0,q\}$ then implied that
the trivializing map must be a local self--homeomorphism of $\hat\RR$. This
means that the possible $\W$'s which are compatible with the EP are all the
functions which are $-\hbar^2/4m$ times the Schwarzian derivative of a local
self--homeomorphism of $\hat\RR$. This is nothing but a consequence of the
cocycle condition (\ref{inhomtrans}) and therefore of the EP. Therefore, we have

\vspace{.333cm}

\noindent
{\it The space $\H$ of admissible $\W$ states consists of the functions of the
form}
\be
\W=-{\hbar^2\over4m}\{w,q\},
\l{ssequazionediSchwarz}\ee
{\it where $w$ is an arbitrary local homeomorphism of $\hat\RR$ into itself.}

\vspace{.333cm}

\noindent
Similarly, we have

\vspace{.333cm}

\noindent
{\it The space $\K$ of admissible reduced actions $\S_0$, consists of the
functions of the form}
\be
\S_0(q)={\hbar\over2i}\ln\left({w(q)+i\bar\ell\over w(q)-i\ell}\right)+
{\alpha\hbar\over2},
\l{ssequazidiSchwarz}\ee
{\it where $w$ is an arbitrary local homeomorphism of $\hat\RR$ into itself,
and\footnote{To be precise, note that even if $\S_0$ takes complex values for
$\alpha\in\CC$, there are no specific reasons to forbid this as both $p$ and
$\dot q$ are independent of $\alpha$.} $\alpha\in\RR$, ${\rm Re}\,\ell\ne0$.}

\mysection{The potential well and the harmonic oscillators}\l{pwho}

In this section we will consider some relevant examples of the general fact we
proved in the previous section. In particular, we will consider the cases of the
potential well and of the simple and double harmonic oscillators. We will
explicitly see that in the case one considers the energy values for which the
corresponding SE has not $L^2(\RR)$ solutions, the ratio
$\psi^D/\psi$ has a discontinuity at spatial infinity.

\subsection{The potential well}

We now show how the quantized spectrum and its structure arise in the case
of the potential well
\be
V(q)=\left\{\begin{array}{ll}0,&|q|\leq L,\\ V_0,&|q|>L.\end{array}\right.
\l{Vu1}\ee
According to (\ref{CarlosSantana5}), in order to find the expression for the
conjugate momentum, we have to find two linearly independent solutions of the
SE which should satisfy the continuity conditions. Let us set
\be
k={\sqrt{2mE}\over\hbar},\qquad\kappa={\sqrt{2m(V_0-E)}\over\hbar}.
\l{I91d}\ee
Since $V(q)$ is an even function, we can choose solutions of definite parity.
For $|q|\leq L$ we can choose either $\psi_1^1=\cos(kq)$ or $\psi_2^1=\sin(kq)$.
For $q>L$ we can choose either $\psi_1^2=Ae^{-\kappa q}$ or $\psi_2^2=B
e^{\kappa q}$
(or any their linear combination). Parity of $V(q)$ fixes the solutions for
$q<-L$. In general choosing $\psi_i^1$ and $A\psi_1^2+B\psi_2^2$ will not give
quantized spectra. Before considering this general situation, we consider the
four cases given by the joining conditions $\psi_i^1=\psi_j^2$, $\partial_q
\psi_i^1=\partial_q\psi_j^2$ at $q=L$. We will denote such solutions by $(i,j)$.
We will see that the cases $(1,2)$ and $(2,2)$ correspond to a trivializing map
which is discontinuous at $q=\pm\infty$ so that, according to the EP, these
solutions are not physical and must be discarded. Let us first consider the
$(1,1)$ case.

\vspace{.6333cm}

\centerline{\underline{$i=1,j=1$}}

\vspace{.5333cm}

\noindent
Imposing the continuity conditions (\ref{equivalenzaederivata}), we have
\be
\psi=\left\{\begin{array}{ll}\cos(kL)\exp[\kappa(q+L)],&q<-L,\\ \cos(kq)
,&|q|\leq L,\\ \cos(kL)\exp[-\kappa(q-L)],&q>L,\end{array}\right.
\l{unouno}\ee
and
\be
k\tan(kL)=\kappa.
\l{unounokk}\ee
A linearly independent solution is given by
\be
\psi^D=[2k\sin(kL)]^{-1}\cdot\left\{\begin{array}{ll}\cos(2kL)\exp[\kappa
(q+L)]-\exp[-\kappa(q+L)],&q<-L,\\ 2\sin(kL)\sin(kq),&|q|\leq L,\\
\exp[\kappa(q-L)]-\cos(2kL)\exp[-\kappa(q-L)],&q>L.\end{array}\right.
\l{unounoD}\ee
Note that any other possible solution has the form $A\psi^D+B\psi$. Let us
now consider the trivializing map associated to this solution. To do this it
is sufficient to find $w$. In the $(1,1)$ case we have
\be
{\psi^D\over\psi}=[k\sin(2kL)]^{-1}\cdot\left\{\begin{array}{ll}\cos(2kL)-
\exp[-2\kappa(q+L)],&q<-L,\\ \sin(2kL)\tan(kq),&|q|\leq L,\\
\exp[2\kappa(q-L)]-\cos(2kL),&q>L.\end{array}\right.
\l{mapunouno}\ee
Observe that
\be
\lim_{q\longrightarrow\pm\infty}{\psi^D\over\psi}=\pm\infty.
\l{infinito11}\ee
Hence, in the $(1,1)$ case the trivializing map is a local self--homeomorphism
of $\hat\RR$ as required by the EP, so that the solutions of
(\ref{unounokk}) are physical energy levels. In particular, if $E_n$ is the
$(n+1)^{th}$ solution of (\ref{unounokk}), then it follows from
(\ref{mapunouno}) that the map covers $\hat\RR$ $(n+1)$--times.

\vspace{.6333cm}

\centerline{\underline{$i=2,j=1$}}

\vspace{.5333cm}

\noindent
In the $(2,1)$ case we have
\be
\psi=k^{-1}\cdot\left\{\begin{array}{ll}-\sin(kL)\exp[\kappa(q+L)],&q<-L,\\
\sin(kq),&|q|\leq L,\\ \sin(kL)\exp[-\kappa(q-L)],&q>L.\end{array}\right.
\l{dueuno}\ee
where
\be
k\cot(kL)=-\kappa.
\l{dueunokk}\ee
A linearly independent solution is given by
\be
\psi^D=[2\cos(kL)]^{-1}\cdot\left\{\begin{array}{ll}\exp[-\kappa(q+L)]+\cos(2kL)
\exp[\kappa(q+L)]&q<-L,\\ 2\cos(kL)\cos(kq),&|q|\leq L,\\
\exp[\kappa(q-L)]+\cos(2kL)\exp[-\kappa(q-L)],&q>L.\end{array}\right.
\l{dueunoD}\ee
and the ratio $w=\psi^D/\psi$ is given by
\be
{\psi^D\over\psi}=k[\sin(2kL)]^{-1}\cdot\left\{\begin{array}{ll}
-\cos(2kL)-\exp[-2\kappa(q+L)],&q<-L,\\ \sin(2kL)\cot(kq),&|q|\leq L,\\
\cos(2kL)+\exp[2\kappa(q-L)],&q>L.\end{array}\right.
\l{mapdueuno}\ee
Even in this case we have
\be
\lim_{q\longrightarrow\pm\infty}{\psi^D\over\psi}=\pm\infty,
\l{infinito21}\ee
so that the associated trivializing map is a local self--homeomorphism of
$\hat\RR$ and the corresponding spectrum given by the solutions of
(\ref{dueunokk}) is a physical one.

\vspace{.6333cm}

\centerline{\underline{$i=1,j=2$}}

\vspace{.5333cm}

\noindent
The $(1,2)$ case is the one of the two cases in which there are not solutions of
the SE which are vanishing both at $-\infty$ and $+\infty$. We have
\be
\psi=\left\{\begin{array}{ll}\cos(kL)\exp[-\kappa(q+L)],&q<-L,\\ \cos(kq),&|q|
\leq L,\\ \cos(kL)\exp[\kappa(q-L)],&q>L.\end{array}\right.
\l{unodue}\ee
Also in this case we would have a quantized spectrum. In fact, the continuity
conditions give
\be
k\tan(kL)=-\kappa.
\l{unoduekk}\ee
However, we now show that these solutions do not satisfy the EP. Let us consider
the dual solution
\be
\psi^D=[2k\sin(kL)]^{-1}\cdot\left\{\begin{array}{ll}\cos(2kL)\exp[-\kappa(q+L)]
-\exp[\kappa(q+L)],&q<-L,\\ 2\sin(kL)\sin(kq),&|q|\leq L,\\ \exp[-\kappa(q-L)]-
\cos(2kL)\exp[\kappa(q-L)],&q>L.\end{array}\right.
\l{unodueD}\ee
These solutions provide an explicit example of the general case we discussed in
considering the asymptotic conditions coming from the EP. Namely, observe that
in the $(1,2)$ case, both $\psi$ and $\psi^D$ are divergent at $\pm\infty$. Also
observe that there is no way to have a solution vanishing both at $\pm\infty$.
To see this, note that for a given $E$, solution of (\ref{unoduekk}), any
solution can be written as $\phi=A\psi^D+B\psi$. Let us choose $\phi$ in such a
way that it vanishes at $+\infty$ that is $\phi=k\psi^D+\cot(2kL)\psi$, so that
\be
\phi=[2\sin(kL)]^{-1}\cdot\left\{\begin{array}{ll}2\cos(2kL)\exp[-\kappa(q+L)]-
\exp[\kappa(q+L)],&q<-L,\\ 2\sin(kL)[\sin(kq)+\cot(2kL)\cos(kq)],&|q|\leq L,\\
\exp[\kappa(q-L)],&q>L.\end{array}\right.
\l{phiunodue}\ee
Hence, even if $\phi_{\;\stackrel{\sim}{q\to+\infty}\;}0$, we have that it
diverges at $-\infty$
\be
\phi_{\;\stackrel{\sim}{q\longrightarrow-\infty}\;}
-{\cos(2kL)\over\sin(kL)}\exp[-\kappa(q+L)].
\l{fifomenoinf}\ee
Note that the only case in which $\phi$ would vanish at $-\infty$ is for $\cos(2
kL)=0$. On the other hand, (\ref{unoduekk}) and number theoretical arguments
show that this is never the case. We have seen that, depending on $\W$, there
are cases with a quantized spectrum even if the SE does not have $L^2(\RR)$
solutions. However, as discussed in the general case, in this case the EP cannot
be satisfied and the corresponding values of $E$ cannot be in the physical
spectrum. Let us explicitly see how the $(1,2)$ case would give a discontinuous
trivializing map. To see this it is sufficient to study the behavior of the
ratio $\psi^D/\psi$. By (\ref{unodue}) and (\ref{unodueD}), we have
\be
{\psi^D\over\psi}=[k\sin(2kL)]^{-1}\cdot\left\{\begin{array}{ll}\cos(2kL)-\exp[2
\kappa(q+L)],&q<-L,\\ \sin(2kL)\tan(kq),&|q|\leq L,\\ \exp[2\kappa(q-L)]-\cos(2k
L),&q>L,\end{array}\right.
\l{mapunodue}\ee
whose asymptotic behavior is
\be
\lim_{q\longrightarrow\pm\infty}{\psi^D\over\psi}=\mp k^{-1}\cot(2kL).
\l{limitidoversisplendido}\ee
It follows that the only possibility to have $w(-\infty)=w(+\infty)$
is that $k^{-1}\cot(2kL)=0$. However, this equation is not compatible
with the condition (\ref{unoduekk}). Hence, we have
\be
w(-\infty)\ne w(+\infty).
\l{AsdAds}\ee
It follows that the $E$'s solutions of (\ref{unoduekk}) must be discarded. We
also note that the other possibility which would give a trivializing map,
corresponding to a local self--homeomorphism of $\hat\RR$, would be for $w(-
\infty)=\pm\infty=-w(+\infty)$. However, this case would correspond to $k^{-1}
\cot(2kL)=\pm\infty$, which has not solutions compatible with (\ref{unoduekk}).

\vspace{.6333cm}

\centerline{\underline{$i=2,j=2$}}

\vspace{.5333cm}

\noindent
In this case we have
\be
\psi=k^{-1}\cdot\left\{\begin{array}{ll}-\sin(kL)\exp[-\kappa(q+L)],&q<-L,\\
\sin(kq),&|q|\leq L,\\ \sin(kL)\exp[\kappa(q-L)],&q>L,\end{array}\right.
\l{duedue}\ee
where
\be
k\cot(kL)=\kappa.
\l{dueduekk}\ee
For the dual solution, we have
\be
\psi^D=[2\cos(kL)]^{-1}\cdot\left\{\begin{array}{ll}\exp[\kappa(q+L)]+
\cos(2kL)\exp[-\kappa(q+L)],&q<-L,\\ 2\cos(kL)\cos(kq),&|q|\leq L,\\
\exp[-\kappa(q-L)]+\cos(2kL)\exp[\kappa(q-L)],&q>L,\end{array}\right.
\l{duedueD}\ee
and the ratio of the solutions is
\be
{\psi^D\over\psi}=k[\sin(2kL)]^{-1}\cdot\left\{\begin{array}{ll}
-\cos(2kL)-\exp[2\kappa(q+L)],&q<-L,\\ \sin(2kL)\cot(kq),&|q|\leq L,\\
\cos(2kL)+\exp[-2\kappa(q-L)],&q>L.\end{array}\right.
\l{mapduedue}\ee
The asymptotic behavior is
\be
\lim_{q\longrightarrow\pm\infty}{\psi^D\over\psi}=\pm k\cot(2kL).
\l{limitidoversisplendido2}\ee
It follows that the only possibility to have either $w(-\infty)=w(+\infty)$ or
$w(-\infty)=\pm\infty=-w(+\infty)$ is that either $k\cot(2kL)=0$ or $k\cot(2k
L)=\pm\infty$. However, also in this case this equation is not compatible with
the condition (\ref{dueduekk}). Hence, we have
\be
w(-\infty)\ne w(+\infty),
\l{AsdAds22}\ee
and the values of $E$ solutions of (\ref{dueduekk}) must be discarded.

Let us make some remarks. First, recall that even in this case, for any
admissible $E$, there are M\"obius states parameterized by the values of $\ell$.
Secondly, we note once again the crucial role of the dual solution: while $\psi$
is vanishing at $\pm\infty$, the fact that $\psi^D$ is divergent implies that
$p$ and $\dot q$ vanish exponentially at infinity. This makes evident the
crucial role of the dual solution, an aspect which is peculiar of the present
formulation. We also note that even in this case the classical limit is
completely under control. For example, in the $i=j=1$ case by
(\ref{momentino})(\ref{unouno}) and (\ref{unounoD}) we have
$$
p=-{\hbar\over2}(\ell+\bar\ell)k^2\sin^2(kL)\cdot
$$
\be
\cdot\left\{\begin{array}{ll}4\left|-\exp[-\kappa(q+L)]+[\cos(2kL)-ik\ell\sin
(2kL)]\exp[\kappa(q+L)]\right|^{-2},&q<-L,\\ \sin^{-2}(kL)\left|\sin(kq)-ik
\ell\cos(kq)\right|^{-2},&|q|\leq L,\\ 4\left|\exp[\kappa(q-L)]-[\cos(2kL)+i
k\ell\sin(2kL)]\exp[-\kappa(q-L)]\right|^{-2},&q>L.\end{array}\right.
\l{ppppunounoD}\ee
While in the region $|q|\leq L$ the classical limit parallels the analysis
we considered in the case of the free particle, in the regions $|q|>L$ we have
\be
\lim_{\hbar\longrightarrow0}p=0,
\l{dovutoaexppiumenokappaq}\ee
which is a consequence of the $\exp(\pm kq)$ terms.

Above we considered the four cases in which $\psi_i^1$ is smoothly joined to
$\psi_j^2$ at $q=L$. However, one can equivalently consider $\psi_i^1$ and an
arbitrary linear combination $A\psi_1^2+B\psi_2^2$. This is the example in the
Introduction. In particular, by (\ref{introcasogeneraledue}) we see that
(\ref{introariprovace}) is the $(2,1)$ case, with (\ref{dueuno}) corresponding
to (\ref{introcasogenerale1}) and (\ref{dueunoD}) to (\ref{introcasogenerale3}).
Similarly, by (\ref{introcasogeneralequattro}) the case (\ref{introtoprocnc})
is the $(1,1)$ case, with (\ref{unouno}) corresponding to
(\ref{introcasogenerale3}) and (\ref{unounoD}) to (\ref{introcasogenerale1}).

In conclusion, we note that the joining condition (\ref{specificandoccnn})
implied by the EP, leads, as follows from the general results we proved, to the
usual spectrum. We also saw that there is a quantized spectrum associated to the
case in which there are divergent solutions only. Even if these are not
admissible solutions, we note that this implied a doubling of the relevant
transcendental equations.

\subsection{The infinitely deep potential well and the free particle}

The case of the infinitely potential well can be studied as limiting case in
which $V_0\longrightarrow\infty$. However, we observe that for $V_0=\infty$,
there is a peculiar situation. The reason is that, according to the EP the
reduced action cannot be a constant in a finite (or infinite) region. In the
case of the infinitely deep potential well, we have $\S_0=cnst$ and $p=0$ in
the region $|q|>L$. However, according to the EP, we have that $\S_0$, and
therefore $p$, spreads on all the existent space. In this sense, the infinitely
deep potential well can be seen as a restriction of $\RR$ to a finite interval.
Then, in this case, the boundary at $q=\pm L$ plays the role of the points
$\pm\infty$ and the joining condition (\ref{specificandoccnn}) is replaced by
\be
w(-L)=\left\{\begin{array}{ll}w(+L),&for\,w(-L)\ne\pm\infty,\\ -w(+L),
&for\,w(-L)=\pm\infty,\end{array}\right.
\l{continuitabucainfinita}\ee
which is in fact satisfied as\footnote{Note that with this choice we have $w(-L)
=w(+L)$. Interchanging $\psi^D$ and $\psi$, this becomes $w(-L)=-w(+L)$.}
\be
{\psi^D\over\psi}=\left\{\begin{array}{ll}k^{-1}\tan({n\pi\over2L}q),&n
\,{\rm even},\\ k\cot({n\pi\over2L}q),&n\,{\rm odd}.\end{array}\right.
\l{continuitabucainfinita2}\ee
In this way, the discontinuity at $q=\pm L$ of $\psi^D$, and therefore of
$\partial_q\psi$, which unavoidably arises when $V_0=+\infty$, naturally
disappears. We note that, while this solution is a consequence of the EP, in the
conventional approach to QM one considers the self--adjoint extension of the
operators (see {\it e.g.} \cite{Thirring}). Observe that the solutions
(\ref{continuitabucainfinita2}) correspond to the cases $i=1,j=1$ and $i=2,j=1$
in the limit $V_0\longrightarrow +\infty$. For the energy levels we have
\be
E_n={\hbar^2\pi^2\over8mL^2}n^2,\qquad n=1,2,3,\ldots.
\l{Eenne}\ee
This way of considering the infinitely deep well is also consistent with the
limit case $L\longrightarrow+\infty$ under which $\S_0$ reduces to the one of
the free particle. In doing this one has to consider the double scaling limit
$L\longrightarrow+\infty$ and $n\longrightarrow+\infty$, with the ratio $n/L$
taking continuous finite values
\be
\lim_{(n,L)\longrightarrow(+\infty,+\infty)}{n\over L}={\sqrt{8mE}\over
\hbar\pi}.
\l{doppiolimite}\ee
Observe that this way of obtaining the free particle shows that in this case
the trivializing map satisfies the continuity at $\pm\infty$ obtained from
(\ref{continuitabucainfinita}) in the $L\longrightarrow +\infty$ limit.
Furthermore, the discontinuity of the ratio $w=\psi^D/\psi$ at $\pm\infty$
of the non--admissible solutions disappears in the $L\longrightarrow+\infty$
limit with the consequence that these become physical solutions. Let us also
note that in the case of the free particle, the ratio of solutions is a M\"obius
transformation of $w_E=k^{-1}\tan kq$, so that the index of the trivializing map
is infinite in this case
\be
I[w_E]=\infty.
\l{indicewE}\ee
Note that $\lim_{E\to 0}w_E=q^0$, so that in this limit the right hand side of
(\ref{indicewE}) collapses to $1$. Let $n$ be the maximum number of zeroes that
a solution of the SE associated to $\W$ may have and denote by $w_n$ the ratio
of two linearly independent solutions of the SE. Local homeomorphicity of the
trivializing map and Sturm's theorem on zeroes of solutions of the SE, yield
\be
I[w_n]=n.
\l{indicewn}\ee

\subsection{The simple and double harmonic oscillators}

Another relevant system with quantized spectrum is the harmonic oscillator.
We now consider the case of the double harmonic oscillator, with the simple one
corresponding to a particular case. The Hamiltonian describing the relative
motion of the reduced mass of the double harmonic oscillator is
\be
H={p^2\over2m}+{1\over2}m\omega^2(|q|-q_0)^2.
\l{doppiooscarm}\ee
The reduced action is given by (\ref{KdT3}) with $\psi^D$ and
$\psi$ real linearly independent solutions of the SE
\be
\left(-{\hbar^2\over2m}{\partial^2\over\partial q^2}+
{1\over2}m\omega^2(|q|-q_0)^2\right)\psi=E\psi.
\l{schrdoparosc}\ee
Let us set 
\be
E=\left(\mu+{1\over2}\right)\hbar\omega,
\l{muiota}\ee
and
\be
z'=\alpha(q+q_0),\quad q\leq0,\qquad z=\alpha(q-q_0),\quad q\geq0,
\l{zzprimo}\ee
where $\alpha=\sqrt{2m\omega/\hbar}$. Note that $z'(-q)=-z(q)$ and that for $q_0
=0$ the system reduces to the simple harmonic oscillator. We stress that since
we consider (\ref{schrdoparosc}) for any real $E$, at this stage $\mu$ is an
arbitrary real number. The SE (\ref{schrdoparosc}) is equivalent to
\be
{\partial^2\psi\over\partial{z'}^2}+\left(\mu+{1\over2}-{{z'}^2\over4}
\right)\psi=0,\qquad q\leq0,
\l{schrdoparosc1}\ee
and
\be
{\partial^2\psi\over\partial z^2}+
\left(\mu+{1\over2}-{z^2\over4}\right)\psi=0,\qquad q\geq0.
\l{schrdoparosc2}\ee
A solution of (\ref{schrdoparosc2}) is given for any $\mu$ by the parabolic
cylinder function (see {\it e.g.} \cite{MagnusOberhettinger})
\be
D_\mu(z)=2^{\mu\over2}e^{-{z^2\over4}}\left[{\Gamma({1\over2})\over\Gamma
[{1-\mu\over2}]}{}_1F_1(-{\mu\over2};{1\over2};{z^2\over2})+{z\over\sqrt 2}
{\Gamma(-{1\over2})\over\Gamma(-{\mu\over2})}{}_1F_1[{1-\mu\over2};{3\over2}
;{z^2\over2}]\right],
\l{cilindroparabolico}\ee
where ${}_1F_1$ is the confluent hypergeometric function
\be
{}_1F_1(a;c;z)=1+{a\over c}{z\over1!}+{a(a+1)\over c(c+1)}{z^2\over2!}+\ldots.
\l{ipergconfl}\ee
We are interested in the behavior of $w$ at $\pm\infty$. If $w(-\infty)\ne\pm
\infty$ we have to impose $w(-\infty)=w(+\infty)$, while if $w(-\infty)=\pm
\infty$, then we should have $w(-\infty)=-w(+\infty)$. To consider this aspect
we need the behavior of $D_\mu$ for $|z|\gg 1$, $|z|\gg |\mu|$. For
$\pi/4<\arg z<5\pi/4$ we have
$$
D_\mu(z){}_{\;\stackrel{\sim}{|z|\gg 1}\;}-{\sqrt{2\pi}\over\Gamma(-\mu)}
e^{\mu\pi i}e^{z^2/4}z^{-\mu-1}\left[1+{(\mu+1)(\mu+2)\over2z^2}
+{(\mu+1)(\mu+2)(\mu+3)(\mu+4)\over2\cdot4z^4}+\ldots\right]
$$
\be
+e^{-z^2/4}z^\mu\left[1-{\mu(\mu-1)\over2z^2}
+{\mu(\mu-1)(\mu-2)(\mu-3)\over2\cdot4z^4}-\ldots\right],
\l{menoinf}\ee
while for $|\arg z|<3\pi/4$
\be
D_\mu(z){}_{\;\stackrel{\sim}{|z|\gg 1}\;}e^{-z^2/4}z^\mu\left[1-{\mu(\mu-1)
\over2z^2}+{\mu(\mu-1)(\mu-2)(\mu-3)\over2\cdot4z^4}-\ldots\right].
\l{piuinf}\ee
A property of $D_\mu(z)$ is that also $D_\mu(-z)$ solves (\ref{schrdoparosc2}).
If $\mu$ is a non--negative integer, then $D_\mu(z)$ and $D_\mu(-z)$
coincide.\footnote{Note that if $\mu$ is a non--negative integer, then
$\Gamma^{-1}(-\mu)=0$, so that in this case the first term in (\ref{menoinf})
cancels.} Let us then consider the case in which $\mu$ is not a non--negative
integer. Two sets of solutions of (\ref{schrdoparosc}), which can be written in
the form (\ref{schrdoparosc1})(\ref{schrdoparosc2}), are
\be
\psi_\pm=\left\{\begin{array}{ll}\pm D_\mu\left(-\alpha (q+q_0)\right),& q\leq
0,\\{}&{}\\ D_\mu\left(\alpha(q-q_0)\right),&q\geq0,\end{array}\right.
\l{paridmu}\ee
where $\mu$ is solution of the continuity condition at $q=0$. For the even
solution $\psi_+$, this gives
\be
D_\mu'\left(-\alpha q_0\right)=0,
\l{paridmucond}\ee
whereas for the odd one $\psi_-$, we have
\be
D_\mu\left(-\alpha q_0\right)=0.
\l{disparidmucond}\ee
For each $\mu$, solution of Eq.(\ref{paridmucond}) (Eq.(\ref{disparidmucond})),
we have that a solution $\psi^D_-$ ($\psi^D_+$) of the SE
(\ref{schrdoparosc}), linearly independent from $\psi_+$ ($\psi_-$), is given by
\be
\psi^D_\mp=\left\{\begin{array}{ll}\mp D_\mu\left(\alpha(q+q_0)\right)\mp a_\mp
D_\mu\left(-\alpha(q+q_0)\right),&q\leq0,\\{}&{}\\ D_\mu\left(-\alpha(q-
q_0)\right)+a_\mp D_\mu\left(\alpha(q-q_0)\right),&q\geq0.\end{array}\right.
\l{paridmud}\ee
We have
\be
a_-=-{D_\mu(\alpha q_0)\over D_\mu(-\alpha q_0)},
\l{amenooo1}\ee
with $\mu$ solution of Eq.(\ref{paridmucond}). Similarly
\be
a_+={D_\mu'(\alpha q_0)\over D_\mu'(-\alpha q_0)},
\l{apiuuu1}\ee
where $\mu$ is solution of Eq.(\ref{disparidmucond}). Therefore, for a given
$\mu$ solution of Eq.(\ref{paridmucond}) (Eq.(\ref{disparidmucond})), $\psi_+$
and $\psi^D_-$ ($\psi_-$ and $\psi^D_+$) are linearly independent solutions of
the SE (\ref{schrdoparosc}). The transcendental equations (\ref{paridmucond})
and (\ref{disparidmucond}) determine, each one, a set of possible energy
eigenvalues. Let us now consider the asymptotic behavior of ${\psi_-^D/\psi_+}$
and ${\psi_+^D/\psi_-}$. By (\ref{paridmu}) and (\ref{paridmud}), we have
\be
{\psi^D_-\over\psi_+}=\left\{\begin{array}{ll}-a_--{D_\mu\left(\alpha(q+q_0)
\right)/D_\mu\left(-\alpha(q+q_0)\right)},&q\leq0,\\{}&{}\\ a_-+D_\mu\left(-
\alpha(q-q_0)\right)/D_\mu\left(\alpha(q-q_0)\right),& q\geq0,\end{array}\right.
\l{rapportosolllx}\ee
and
\be
{\psi^D_+\over\psi_-}=\left\{\begin{array}{ll}-a_+-D_\mu\left(\alpha(q+q_0)
\right)/D_\mu\left(-\alpha(q+q_0)\right),&q\leq0,\\{}&{}\\ a_++D_\mu\left(-
\alpha(q-q_0)\right)/D_\mu\left(\alpha(q-q_0)\right),&q\geq0.\end{array}\right.
\l{rapportosollly}\ee
By (\ref{menoinf}) and (\ref{piuinf}), we have the asymptotics\footnote{The
asymptotics (\ref{menoinf}) and (\ref{piuinf}) include $\arg z'=\pi$ and $\arg
z=0$ respectively, so that
$$
{z'}^{-\mu-1}(-z')^{-\mu}=-e^{-\mu\pi i}|z'|^{-2\mu-1},\qquad
z^{-\mu}(-z)^{-\mu-1}=-e^{-\mu\pi i}|z|^{-2\mu-1}.
$$}
\be
{\psi^D_-\over\psi_+}{}_{\;\stackrel{\sim}{q\longrightarrow-\infty}\;}{\sqrt{2
\pi}\over\Gamma(-\mu)}e^{\mu\pi i}e^{{z'}^2/2}{z'}^{-\mu-1}(-z')^{-\mu}
\longrightarrow-\infty,
\l{asssymptmeno1}\ee
and
\be
{\psi^D_-\over\psi_+}{}_{\;\stackrel{\sim}{q\longrightarrow+\infty}\;}-{\sqrt{2
\pi}\over\Gamma(-\mu)}e^{\mu\pi i}e^{z^2/2}z^{-\mu}(-z)^{-\mu-1}
\longrightarrow+\infty.
\l{asssymptpiu1}\ee
By (\ref{rapportosolllx})(\ref{rapportosollly}) we see that $\psi^D_+/\psi_-$
and $\psi^D_-/\psi_+$ have the same asymptotic limits. Hence, since
\be
\lim_{q\longrightarrow-\infty}{\psi^D_-\over\psi_+}=-\infty=-\lim_{q
\longrightarrow +\infty}{\psi^D_-\over\psi_+},
\l{caruccetinuccettinettina}\ee
and similarly for $\psi^D_+/\psi_-$, we have that the trivializing map
associated to the double harmonic oscillator, with energy levels given by
(\ref{paridmucond}) and (\ref{disparidmucond}), is a local homeomorphism of
$\hat\RR$ into itself. Other possible solutions of the SE (\ref{schrdoparosc})
are
\be
\psi_\pm=\left\{\begin{array}{ll}\pm D_\mu\left(\alpha(q+q_0)\right),&q\leq
0,\\{}&{}\\ D_\mu\left(-\alpha(q-q_0)\right),&q\geq0,\end{array}\right.
\l{paridmudiv}\ee
where $\mu$ is solution of the continuity condition at $q=0$. For the
even solution $\psi_+$, this condition gives
\be
D_\mu'\left(\alpha q_0\right)=0,
\l{paridmuconddiv}\ee
whereas for the odd one $\psi_-$, we have
\be
D_\mu\left(\alpha q_0\right)=0.
\l{disparidmuconddiv}\ee
For each $\mu$ solution of Eq.(\ref{paridmuconddiv})
(Eq.(\ref{disparidmuconddiv})) we have that a solution $\psi^D_-$ ($\psi^D_+$)
of (\ref{schrdoparosc}), linearly independent from $\psi_+$ ($\psi_-$), is
given by
\be
\psi^D_\mp=\left\{\begin{array}{ll}\mp D_\mu\left(-\alpha(q+q_0)\right)\mp
b_\mp D_\mu\left(\alpha(q+q_0)\right),&q\leq0,\\{}&{}\\ D_\mu\left(\alpha(q-
q_0)\right)+b_\mp D_\mu\left(-\alpha(q-q_0)\right),&q\geq0,\end{array}\right.
\l{paridmuddiv}\ee
where
\be
b_-=-{D_\mu(-\alpha q_0)\over D_\mu(\alpha q_0)},
\l{amenooo1div}\ee
with $\mu$ solution of Eq.(\ref{paridmuconddiv}). Similarly
\be
b_+={D_\mu'(-\alpha q_0)\over D_\mu'(\alpha q_0)},
\l{apiuuu1dvi}\ee
where $\mu$ is solution of Eq.(\ref{disparidmuconddiv}).
By (\ref{paridmudiv}) and (\ref{paridmuddiv}), we have
\be
{\psi^D_-\over\psi_+}=\left\{\begin{array}{ll}-b_--{D_\mu\left(-\alpha(q+q_0)
\right)/D_\mu\left(\alpha(q+q_0)\right)},&q\leq0,\\{}&{}\\ b_-+D_\mu\left(
\alpha(q-q_0)\right)/D_\mu\left(-\alpha(q-q_0)\right),& q\geq0,\end{array}
\right.
\l{rapportosolllxdiv}\ee
and
\be
{\psi^D_+\over\psi_-}=\left\{\begin{array}{ll}-b_+- D_\mu\left(-\alpha(q+q_0
)\right)/D_\mu\left(\alpha(q+q_0)\right),&q\leq0,\\{}&{}\\ b_++D_\mu
\left(\alpha(q-q_0)\right)/D_\mu\left(-\alpha(q-q_0)\right),&q\geq0.
\end{array}\right.
\l{rapportosolllydiv}\ee
By (\ref{menoinf}) and (\ref{piuinf}), we have the asymptotics
\be
{\psi^D_-\over\psi_+}{}_{\;\stackrel{\sim}{q\longrightarrow\pm\infty}\;}\pm b_-,
\l{asssymptmeno1div}\ee
and
\be
{\psi^D_+\over\psi_-}{}_{\;\stackrel{\sim}{q\longrightarrow\pm\infty}\;}\pm b_+.
\l{asssymptmeno2div}\ee
Since both $b_-$ and $b_+$ are finite non--vanishing, we have that the ratios
$\psi^D_-/\psi_+$ and $\psi^D_+/\psi_-$ associated to the solutions
(\ref{paridmuconddiv}) and (\ref{disparidmuconddiv}) respectively, do not
satisfy the joining condition (\ref{specificandoccnn}). Therefore, according to
the EP, the energy eigenvalues defined by (\ref{paridmuconddiv})
and (\ref{disparidmuconddiv}) are not physical ones.

Thus, besides the potential well, also in this case, in agreement with the
general theorem (\ref{asintoticopiumenofgt}), the only physical solutions are
(\ref{paridmucond})(\ref{disparidmucond}), with the $L^2(\RR)$ solutions given
in (\ref{paridmu}).

Let us now consider the simple harmonic oscillator. In this case $q_0=0$, that
is $z'=z$. By (\ref{paridmucond}) and (\ref{disparidmucond}) it follows that
the continuity conditions in the case of the harmonic oscillator are
\be
D_\mu(0)=0,
\l{laprimachehaidetto}\ee
and
\be
D_\mu'(0)=0.
\l{lasecondachehaidetto}\ee
Let us write down the initial values of $D_\mu$
\be
D_\mu(0)=2^{\mu/2}{\Gamma(1/2)\over\Gamma[(1-\mu)/2]},\qquad
D_\mu'(0)=2^{(\mu-1)/2}{\Gamma(-1/2)\over\Gamma(-\mu/2)}.
\l{tutteedue}\ee
Since the poles of $\Gamma(\mu)$ are at the non--positive integer values
of $\mu$, the solutions of (\ref{laprimachehaidetto}) are
\be
\mu=2n+1,\qquad n=0,1,2,\ldots,
\l{muduenpiu1}\ee
while for Eq.(\ref{lasecondachehaidetto}) we have
\be
\mu=2n,\qquad n=0,1,2,\ldots.
\l{muduen}\ee
For non--negative integer values of $n$ we have
\be
D_n(z)=(-1)^ne^{z^2/4}{d^n\over dz^n}e^{-z^2/2}=e^{-z^2/4}H_n(z),
\l{parabolacilindroHermite}\ee
where the $H_n$'s are the Hermite polynomials. By
(\ref{muiota})(\ref{muduenpiu1}) and (\ref{muduen}), we obtain the spectrum
of the harmonic oscillator
\be
E=\left(n+{1\over2}\right)\hbar\omega,\qquad n=0,1,2,\ldots.
\l{muiota2}\ee

We already observed that since $\Gamma^{-1}(-n)=0$ for $n=0,1,2,\ldots$, it
follows that $D_n(z)$ and $D_n(-z)$ are linearly dependent. Besides
Eq.(\ref{parabolacilindroHermite}), this can be also seen from the fact that
$D_n(z)$ and $D_n(-z)$ are both convergent with the same leading terms, implies
that $D_n(z)=(-1)^nD_n(-z)$ (where the $(-1)^n$ factor follows from
(\ref{menoinf}) and (\ref{piuinf})). This also follows from the relation
\be
D_\mu(z)=e^{\mu\pi i}D_\mu(-z)+{\sqrt{2\pi}\over\Gamma(-\mu)}
e^{(\mu+1)\pi i/2}D_{-\mu-1}(-iz),
\l{didXXX}\ee
which implies that $D_{-n-1}(-iz)$ is a solution of the SE of the simple
harmonic oscillator, which is linearly independent from $D_n(z)$. Then, in this
case one has
\be
\psi=D_n(\alpha q),\qquad\psi^D=e^{-(n+1)\pi i/2}D_{-n-1}(-i\alpha q),
\l{dualioscillatore}\ee
where the role of the factor $e^{-(n+1)\pi i/2}$ is that of making
$\psi^D$ real. The ratio
\be
{\psi^D\over\psi}=e^{-(n+1)\pi i/2}{D_{-n-1}(-i\alpha q)\over D_n(\alpha q)},
\l{rapportooscc}\ee
has the asymptotics
\be
{\psi^D\over\psi}{}_{\;\stackrel{\sim}{q\longrightarrow -\infty}\;}
e^{\alpha^2 q^2/2}(\alpha q)^{-2n-1}\longrightarrow-\infty,
\l{rapportlimitec1}\ee
\be
{\psi^D\over\psi}{}_{\;\stackrel{\sim}{q\longrightarrow +\infty}\;}
e^{\alpha^2 q^2/2}(\alpha q)^{-2n-1}\longrightarrow+\infty,
\l{rapportlimitec2}\ee
showing that, according to the general result we derived previously, also the
standard spectrum of the harmonic oscillator defines a local homeomorphism of
$\hat\RR$ into itself.

\subsection{The general case}

In the previous investigation we considered solutions of the SE for which $\psi$
contains either the vanishing or divergent solution. However, one may consider
the following general form
\be
\psi=\left\{\begin{array}{ll}D_\mu(-z')+cD_\mu(z'),&q\leq0,\\
{}&{}\\ D_\mu(z)+cD_\mu(-z),&q\geq0,\end{array}\right.
\l{zummmparidmudiv}\ee
where
\be
c={D_\mu'(-\alpha q_0)\over D_\mu'(\alpha q_0)}.
\l{FrankZappa56}\ee
For any given $\mu$, a linearly independent solution is given by
\be
\psi^D=\left\{\begin{array}{ll}-D_\mu(-z')-dD_\mu(z'),&q\leq0,\\
{}&{}\\ D_\mu(z)+dD_\mu(-z),&q\geq0,\end{array}\right.
\l{zummmdisparidmudiv}\ee
where
\be
d=-{D_\mu(-\alpha q_0)\over D_\mu(\alpha q_0)}.
\l{FrankZappa562}\ee
The ratio
\be
{\psi^D_+\over\psi_-}=\left\{\begin{array}{ll}-(D_\mu(-z')+dD_\mu(z'))/(
D_\mu(-z')+cD_\mu(z')),&q\leq0,\\{}&{}\\ (D_\mu(z)+dD_\mu(-z))/
(D_\mu(z)+cD_\mu(-z)),&q\geq0,\end{array}\right.
\l{zummmdsjhdoisparidmudiv}\ee
has the asymptotics behavior
\be
\lim_{q\longrightarrow\pm\infty}{\psi^D\over\psi}=\pm{d\over c}.
\l{dsucabbastanza}\ee
This shows that Eq.(\ref{specificandoccnn}) is satisfied in the case in
which either
\be
c=0,
\l{cuguale0}\ee
or
\be
d=0.
\l{duguale0}\ee
These cases correspond to the ones discussed in
(\ref{paridmu})--(\ref{caruccetinuccettinettina}).
Note that if $\mu=0,1,2,\ldots$, then $\psi$ in (\ref{zummmparidmudiv}) and
its dual (\ref{zummmdisparidmudiv}) are not linearly independent. In this case
there is always a solution vanishing both at $-\infty$ and $+\infty$. If
$q_0=0$, then this situation corresponds to the harmonic oscillator. Generally,
for an arbitrary $q_0$ and for $\mu=0,1,2,\ldots$, the solution $\psi^D$ in
(\ref{zummmdisparidmudiv}) is replaced by
\be
\psi^D=\left\{\begin{array}{ll}-D_n(-z')-d'D_{-n-1}(-iz'),&q\leq0,\\
{}&{}\\ D_n(z)+d'D_{-n-1}(iz),&q\geq0,\end{array}\right.
\l{zummmdisparidmudiv3456}\ee
where now
\be
d'=-{D_n(-\alpha q_0)\over D_n(-i\alpha q_0)}.
\l{FrankZappa28ottobre56}\ee
We note that above we used the parabolic cylinder functions with real argument.
Then, the fact that for $\mu=0,1,2,\ldots$, $D_\mu(z)$ and $D_\mu(-z)$ are not
linearly independent, forced us to use $D_\mu(-iz)$. In this context we observe
that $D_\mu(z)$ and $D_\mu(-iz)$ are always linearly independent so that the
dual solution (\ref{zummmdisparidmudiv3456}) can be extended to arbitrary
values of $\mu$.

\mysection{Generalizations}\l{tdc}

{\it This section has been written in collaboration with Gaetano Bertoldi.}

\vspace{.333cm}

\noindent
A property of the SE, which can be seen as a consequence of the EP, is that the
reduced action associated to the state $\W^0$ is not constant. Actually, the
structure of this solution is rather peculiar and turns out to be sufficient to
fix the SE both in any dimension and for time--dependent potentials. This is not
a surprise since basic features of the usual approach to QM, such as uncertainty
relations, energy quantization, tunnel effect, Hilbert spaces etc. already arise
in the one--dimensional stationary case. In this section we derive the quantum
HJ equation for time--dependent potentials. Next, we will consider the higher
dimensional case. Finally, we will derive the relativistic quantum HJ equation.

\subsection{Time--dependent Schr\"odinger equation}

Let us start by noticing that since for stationary states one has
\be
\S(q,t)=\S_0(q)-Et,
\l{unohha}\ee
it follows that we can rewrite the QSHJE (\ref{1Q}) in the form
\be
{\partial\S\over\partial t}+{1\over2m}\left({\partial\S(q)\over\partial q}
\right)^2+V(q)+{\hbar^2\over4m}\{\S,q\}=0.
\l{1Q2bis}\ee
This equation and Eq.(\ref{ES}) are equivalent to
\be
i\hbar{\partial\psi(q,t)\over\partial t}=\left(-{\hbar^2\over2m}{\partial^2
\over\partial q^2}+V(q)\right)\psi(q,t),
\l{z758xx}\ee
with
\be
\psi(q,t)={1\over\sqrt{\S'}}e^{{i\over\hbar}\S}.
\l{45yi}\ee
A peculiar property of Eq.(\ref{z758xx}) is that it is satisfied by all
possible solutions of the SE irrespectively of the specific
energy eigenvalue. Let us set\footnote{Here we consider the case of discrete
spectra, however similar arguments extend to the continuous ones.}
\be
\Phi_k={1\over\sqrt{\S_k'}}e^{{i\over\hbar}\S_k}=e^{-{i\over\hbar}E_kt}\bar
\phi_k,
\l{Gh2B}\ee
where $\S_k=\S_{0,k}-E_kt$, and $\S_{0,k}$ is the reduced
action corresponding to the eigenvalue $E_k$ and
\be
\phi_k={1\over\sqrt{\S_{0,k}'}}e^{-{i\over\hbar}\S_{0,k}}.
\l{Hyubbis}\ee
Since
\be
\left(-{\hbar^2\over2m}{\partial^2\over\partial q^2}+V(q)\right)\bar\phi_k=E_k
\bar\phi_k,
\l{Gh3B}\ee
we have that Eq.(\ref{z758xx}) is solved by
\be
\Psi(q,t)=\sum_kc_k\Phi_k,
\l{Gh4}\ee
with $c_k$'s arbitrary constants. While $\bar\Psi$
satisfies the complex conjugated of Eq.(\ref{z758xx}), the function
\be
\Psi^D(q,t)=\sum_kd_k\Phi_k^D,
\l{Gh5}\ee
where $\Phi^D_k=e^{-{i\over\hbar} E_kt}\phi_k$, is still solution of
(\ref{z758xx}). In particular, the most general solution of (\ref{z758xx}) is
\be
\Phi=\sum_k(A_k\Phi_k^D+B_k\Phi_k).
\l{Gh7}\ee
An interesting aspect of Eq.(\ref{z758xx}) is that, like (\ref{1Q2bis}), the
energy level does not appear: it is not specified, it appears, as in the
classical HJ equation, as an integration constant. On the other hand, this is
precisely the same role played by $\alpha$ and $\ell$. As in the SE (\ref{ES})
there is no information about the initial conditions determining $\alpha$ and
$\ell$, in the SE (\ref{z758xx}) this lack of information is extended to the
energy level. This aspect is evident in Eq.(\ref{z758xx}) where in order to
determine $\S$ we need to know the initial conditions fixing the energy level
$E$ and the constant $\alpha$ and $\ell$. Therefore, in the framework of the
Quantum HJ Equation associated to Eq.(\ref{z758xx}) and that will be considered
later on, the lack of information on the energy level is similar in nature to
that associated to the integration constants $\alpha$ and $\ell$. As a
consequence, while a given solution of the SE (\ref{ES}) is not sufficient to
specify $\S_0$, in the case of Eq.(\ref{z758xx}) the energy level $E$ should be
specified as one more integration constant. Fixing the value of $E$ would be
equivalent to specify which is the non--vanishing pair $(A_k,B_k)$ in
(\ref{Gh7}). Once this is done, one needs further conditions to
completely fix $\S$. This analogy has the following natural consequence: as in
the case of $\psi_E$, whose knowledge in general does not completely fix $\S_0$,
the knowledge of $\Phi$ in (\ref{Gh7}) is not sufficient to completely fix $\S$.

Let us now consider the case of a system that at the time $t<t_1$ is described
by $\S_{0,k}$, corresponding to the energy level $E_k$, associated to a given
potential $V(q)$. Subsequently a perturbation, described by some time--dependent
potential, is turned on in such a way that the system at the time $t>t_2>t_1$
has a reduced action $\S_{0,n}$ corresponding to the energy level $E_n$.
Although $\Phi_k$ and $\Phi_n$ satisfy the SE (\ref{Gh3B}) with different energy
level, they satisfy the same SE in the form of Eq.(\ref{z758xx}). In other
words, (\ref{z758xx}) is the equation which remains form invariant for $t<t_1$
and $t>t_2$, and of which both $\Phi_k$ and $\Phi_n$ are solutions. As we have
seen, this property is related to the linearity of Eq.(\ref{z758xx}). We now
show that linearity and consistency imply that in the case of time--dependent
potentials we have
\be
i\hbar{\partial\psi(q,t)\over\partial t}=\left(-{\hbar^2\over2m}
{\partial^2\over\partial q^2}+V(q,t)\right)\psi(q,t).
\l{Az1a}\ee
To show this we first observe that to find the generalization of
Eq.(\ref{z758xx}) in the case of time--dependent potentials the equation should
have the following properties
\begin{itemize}
\item[{\it a})]{Linearity.}
\item[{\it b})]{In the time--independent case it should reduce to
Eq.(\ref{z758xx}).}
\item[{\it c})]{In the $\hbar\longrightarrow0$ limit it should reduce to the
CHJE (\ref{09BV7U3}).}
\end{itemize}
These three conditions imply that the generalization of Eq.(\ref{z758xx}) to the
case in which the potential depends on time has the form
\be
\sum_{\alpha,\beta,\gamma,\delta}c_{\alpha\beta\gamma\delta}m^\alpha
\hbar^\beta{\partial^{\gamma+\delta}\psi\over\partial t^\gamma\partial q^\delta}
=\left(-{\hbar^2\over2m}{\partial^2\over\partial q^2}+V(q,t)\right)\psi,
\l{758xx}\ee
with $c_{\alpha\beta\gamma\delta}$ dimensionless constants. Even if the case of
the time--dependent perturbation, considered previously, already shows that the
only non--vanishing coefficient in (\ref{758xx}) is $c_{0110}=i$, we show how
(\ref{Az1a}) can be derived from (\ref{758xx}) as a consequence of the points
{\it a} and {\it b} above. The fact that in the stationary case Eq.(\ref{758xx})
should correspond to (\ref{ES}), which is equivalent to (\ref{z758xx}), implies
\be
\gamma\geq1.
\l{gammageq1}\ee
Let us now perform the dimensional analysis. A comparison between the left-- and
right--hand sides of Eq.(\ref{758xx}) shows that the operator $m^\alpha
\hbar^\beta\partial^\gamma_t\partial^\delta_q$ has the dimension of the energy,
so that
\be
\beta=1-\alpha,\qquad\gamma=1+\alpha,\qquad\delta=-2\alpha.
\l{dimensioanleanalisi}\ee
Eqs.(\ref{gammageq1}) and (\ref{dimensioanleanalisi}) yield
\be
\gamma\geq1\longrightarrow\alpha\geq0\longrightarrow\left\{\begin{array}{ll}
\beta\leq1,\\ \delta\leq0.\end{array}\right.
\l{laiesaoioiaoimmeia}\ee
On the other hand, we have $\delta\geq0$, so that by
(\ref{dimensioanleanalisi}) and (\ref{laiesaoioiaoimmeia}) one obtains
\be
\alpha=0,\qquad\beta=1,\qquad\gamma=1,\qquad\delta=0.
\l{abcd}\ee
Therefore, the only surviving term in the left hand side of (\ref{758xx}) is
$c_{0110}$. Comparing with the stationary case we get $c_{0110}=i$, that is
Eq.(\ref{758xx}) reduces to (\ref{Az1a}). Making the identification
\be
\psi(q,t)=R(q,t)e^{{i\over\hbar}\S(q,t)},
\l{45yighx}\ee
and using (\ref{Az1a}) we obtain the Quantum HJ Equation
\be
{\partial\S\over\partial t}+{1\over2m}\left({\partial\S\over\partial q}
\right)^2+V(q,t)-{\hbar^2\over2mR}{\partial^2R\over\partial q^2}=0,
\l{U1a}\ee
\be
{\partial R^2\over\partial t}+{1\over m}{\partial\over\partial q}
\left(R^2{\partial\S\over\partial q}\right)=0.
\l{U2a}\ee
Note that in general the quantum potential $-\hbar^2{\partial^2_q R}/2mR$, does
not correspond to the one considered in the literature ({\it e.g.} the Bohmian
one). The reason is that in general $\psi$ in (\ref{45yighx}), solution of the
SE, does not correspond to the wave--function. Thus, for example, in the
stationary case, in which $\S(q,t)=\S_0(q)-Et$, we always have to satisfy the
conditions (\ref{ccnn}), in particular $\S_0\ne cnst$.

\subsection{Higher dimension}

Similarly to the case of the Heisenberg uncertainty relations
\be
\Delta p_k\Delta q_k\geq\hbar/2,
\l{pov3}\ee
in spite of being intrinsically one--dimensional, also our formulation
essentially implies QM in higher dimension. In fact, it turns out that the EP
implies the SE also in higher dimension. A detailed analysis of this argument
is the subject of the forthcoming paper \cite{BFM}. Here, we first outline
the basic steps in \cite{BFM} and then propose an alternative approach based
on space compactification. Let us consider the $D$--dimensional CSHJE
\be
{1\over2m}\sum_{k=1}^D\left({\partial\S_0^{cl}(q)\over\partial q_k}
\right)^2+\W(q)=0.
\l{0xy12}\ee
Also in this case, given another system with reduced action $\S_0^{cl\,v}$,
we set
\be
\S_0^{cl\,v}(q^v)=\S_0^{cl}(q).
\l{wesette}\ee
Since $\S_0^{cl\,v}(q^v)$ must satisfy the CSHJE
\be
{1\over2m}\sum_{k=1}^D\left({\partial\S_0^{cl\,v}(q^v)
\over\partial q^v_k}\right)^2+\W^v(q^v)=0,
\l{odiqw9Tghy}\ee
it follows by (\ref{wesette}) that
\be
p_k\longrightarrow p^v_k={\partial\S_0^{cl\,v}(q^v)\over\partial q^v_k}=
\sum_{i=1}^D{\partial q_i\over\partial q^v_k}{\partial\S_0^{cl}(q)\over
\partial q_i}=\sum_{i=1}^DJ_{ki}p_i,
\l{insomma}\ee
where $J_{ki}$ denotes the Jacobian matrix
\be
J_{ki}={\partial q_i\over\partial q^v_k}.
\l{hfgt1}\ee
Let us set
\be
(p^v|p)={\sum_kp_k^{v^2}\over\sum_kp_k^2}={p^tJ^tJp\over p^tp}.
\l{natura2xy}\ee
Note that in the one--dimensional case
\be
(p^v|p)=\left({p^v\over p}\right)^2=\left({\partial\S_0^{cl}\over
\partial q^v}{\partial q\over\partial\S_0^{cl}}\right)^2=\left({\partial
q^v\over\partial q}\right)^{-2},
\l{onedimensionalcase}\ee
so that the Jacobian of the coordinate transformation is the ratio of
momenta. By (\ref{0xy12}) we have
\be
\W(q)\longrightarrow\W^v(q^v)=(p^v|p)\W(q),
\l{piccolo}\ee
that for the state $\W^0$ gives
\be
\W^0(q^0)\longrightarrow\W^v(q^v)=(p^v|p^0)\W^0(q^0)=0.
\l{qdddppp}\ee
Thus we have that in any dimension the EP cannot be consistently implemented
in CM. It is therefore clear that also in the higher dimensional case the
implementation of the EP requires the deformation of the CSHJE. Therefore
\be
{1\over2m}\sum_{k=1}^D\left({\partial\S_0(q)\over\partial q_k}\right)^2+\W(q)
+Q(q)=0.
\l{1Xvzs}\ee
The properties of $\W+Q$ under the VT (\ref{wesette}) are determined by the
transformed equation
\be
{1\over2m}\sum_{k=1}^D\left({\partial\S_0^{cl\,v}(q^v)\over\partial q_k^v}
\right)^2+\W^v(q^v)+Q^v(q^v)=0,
\l{ocwi0ml}\ee
which by (\ref{wesette}) and (\ref{1Xvzs}) yields
\be
\W^v(q^v)+Q^v(q^v)=(p^v|p)\left(\W(q)+Q(q)\right).
\l{1Tdzs}\ee
The only possibility to reach any state $\W^v$ from $\W^0$ is that it transforms
with an inhomogeneous term. Namely, as $\W^0(q^0)\longrightarrow\W^v(q^v)\ne 0$,
it follows by (\ref{1Tdzs}) that for each pair of states we have
\be
\W^v(q^v)=(p^v|p^a)\W^a(q^a)+\Z(q^a;q^v),
\l{1Z4}\ee
and
\be
Q^v(q^v)=(p^v|p^a)Q^a(q^a)-\Z(q^a;q^v).
\l{BtD21Tdzs}\ee
Setting $\W^a(q^a)=\W^0(q^0)$ in Eq.(\ref{1Z4}) yields
\be
\W^v(q^v)=\Z(q^0;q^v),
\l{ddBtD1Tdzs}\ee
so that, according to the EP, even in higher dimension all the states
correspond to the inhomogeneous part in the transformation of the state $\W^0$
induced by $v$--maps. Comparing
\be
\W^b(q^b)=(p^b|p^a)\W^a(q^a)+\Z(q^a;q^b)=\Z(q^0;q^b),
\l{ganzatezs}\ee
with the same formula with $q^a$ and $q^b$ interchanged, we have
\be
\Z(q^b;q^a)=-(p^a|p^b)\Z(q^a;q^b),
\l{inparticolarezs}\ee
in particular
\be
\Z(q;q)=0.
\l{inparticolarebzs}\ee
More generally, comparing
\be
\W^b(q^b)=(p^b|p^c)\W^c(q^c)+\Z(q^c;q^b)=(p^b|p^a)\W^a(q^a)+
(p^b|p^c)\Z(q^a;q^c)+\Z(q^c;q^b),
\l{associatidoczs}\ee
with (\ref{ganzatezs}) we obtain the basic cocycle condition
\be
\Z(q^a;q^c)=(p^c|p^b)\left[\Z(q^a;q^b)-\Z(q^c;q^b)\right],
\l{cociclo3xxszs}\ee
which is a direct consequence of the EP. Eq.(\ref{cociclo3xxszs}) is the
higher dimensional generalization of Eq.(\ref{cociclo3}). Similarly to the
one--dimensional case, also Eq.(\ref{cociclo3xxszs}) univocally leads to the
higher dimensional SE. The details of the derivation will be given in
\cite{BFM}. Let us just mention that the higher dimensional version of the
identity (\ref{expoid}) is \cite{BFM}
\be
\alpha^2(\nabla\S_0)\cdot(\nabla\S_0)={\Delta(Re^{\alpha\S_0})\over Re^{\alpha
\S_0}}-{\Delta R\over R}-\alpha\left(2{\nabla R\cdot\nabla\S_0\over R}+\Delta
\S_0\right),
\l{identity}\ee
for any constant $\alpha$. Therefore, if $R$ satisfies the continuity equation
\be
\nabla R^2\cdot\nabla\S_0+R^2\Delta\S_0=0,
\l{conteq}\ee
we have
\be
(\nabla\S_0)\cdot(\nabla\S_0)=\hbar^2\left({\Delta R\over R}-{\Delta(Re^{\pm{i
\S_0\over\hbar}})\over Re^{\pm{i\S_0\over\hbar}}}\right),
\l{identity2}\ee
and the general form of the QHJE in $D$--dimensions is provided by
(\ref{conteq}) and
\be
{1\over2m}(\nabla\S_0)\cdot(\nabla\S_0)+\W+Q=0,
\l{QHJE}\ee
where $Q$ is the quantum potential
\be
Q(q)=-{\hbar^2\over2m}{\Delta R\over R}.
\l{identif}\ee
We stress the important fact that in general, as in the case of the other
Quantum HJ Equations we derived, $Q$ does not correspond to the usual quantum
potential. The reason for this is always the same. Namely, while in the
standard approach Eqs.(\ref{QHJE})(\ref{identif}) are derived from the
Schr\"odinger equation by identifying $Re^{{i\over\hbar}\S_0}$ with the
wave--function, here the identification is between $Re^{{i\over\hbar}\S_0}$
and a solution of the Schr\"odinger equation which in general does not
correspond to the wave--function.

Here we show an alternative derivation with respect to \cite{BFM} which is based
on space compactification. As we will see, the approach is reminiscent of that
in string theory and sets in a natural way a correspondence between potentials
and space compactification. We will see that this approach poses a mathematical
problem and will lead to the suggestion that potentials (forces) present in
Nature have a geometrical origin, as they are connected with the properties of
compactification. Let us start by observing that for potentials of the form
\be
V(q)=\sum_{k=1}^DV_k(q_k),
\l{separa}\ee
the EP implies the decoupled quantum HJ equations
\be
{1\over2m}\left({\partial\S_{0,k}(q_k)\over\partial q_k}\right)^2+\W_k(q_k)
+{\hbar^2\over4m}\{\S_{0,k},q_k\}=0,
\l{x22x22}\ee
$k=1,\ldots, D$, where $\W_k(q_k)\equiv V_k(q_k)-E_k$. Eq.(\ref{x22x22}) implies
the Schr\"odinger equation
\be
\left(-{\hbar^2\over2m}\Delta_D(q)+\W(q)\right)\psi(q)=0,
\l{yz1xxxx7}\ee
where
\be
\W(q)=\sum_{k=1}^D\W_k(q_k),
\l{qiuh}\ee
and $\Delta_D(q)$ denotes the $D$--dimensional Laplacian
\be
\Delta_D(q)=\sum_{k=1}^D{\partial^2\over\partial q^2_k}.
\l{lapll}\ee
Observe that
\be
\psi(q)=\prod_{k=1}^D\psi_k(q_k),
\l{dftxx}\ee
where each $\psi_k$ is solution of the one--dimensional SE
\be
\left(-{\hbar^2\over2m}{\partial^2\over\partial q^2_k}+\W_k(q_k)\right)
\psi_k(q_k)=0.
\l{yz1dxxxx7}\ee
Also in higher dimension the state $\W^0$ corresponds to the non--trivial
solution
\be
\S_0^0={\hbar\over2i}\sum_{k=1}^D\ln\left(q_k^0+i\bar\ell_{0,k}
\over q_k^0-i\ell_{0,k}\right).
\l{nonttt}\ee
This guarantees that the Legendre transformation
\be
\S_0=\sum_{k=1}^Dp_k{\partial\T_0\over\partial p_k}-\T_0,
\l{podl}\ee
is defined for any physical system. As a consequence of the involutive nature of
the Legendre transformation, we have that $\S_0$--$\T_0$ duality extends to
higher dimension.

A first remark for our construction is that Eq.(\ref{yz1xxxx7}), due to its
structure, holds also for potentials which are more general than (\ref{separa}).
To see this observe that the Laplacian is invariant under rototranslations of
the coordinate (parity transformations can be included in the following
construction)
\be
\Delta_D(\tilde q)=\sum_{k=1}^D{\partial^2\over\partial{\tilde q}^2_k}=
\Delta_D(q)
\l{lapll2}\ee
where
\be
\tilde q_k=\sum_{j=1}^DR_{kj}q_j+B_k,
\l{rottrinv}\ee
\be
R^tR={\II}_D.
\l{unoo}\ee
It follows that Eq.(\ref{yz1dxxxx7}) is equivalent to
\be
\left(-{\hbar^2\over2m}
\Delta_D(\tilde q)+\tilde\W(\tilde q)\right)\tilde\psi(\tilde q)=0,
\l{yz1xxxx71}\ee
where $\tilde\W(\tilde q)=\W(q(\tilde q))$, and $\tilde\psi(\tilde q)=\psi(q(
\tilde q))$. The observation is that in general both $\tilde\W(\tilde q)$ and
$\tilde\psi(\tilde q)$ would not decompose as a sum and product respectively,
namely
\be
\tilde\W(\tilde q)=\sum_{k=1}^D\W_k(q_k)\ne\sum_{k=1}^D\tilde\W_k(\tilde q_k),
\l{FBT2}\ee
\be
\tilde\psi(\tilde q)=\prod_{k=1}^D\psi_k(q_k)\ne\prod_{k=1}^D\tilde
\psi_k(\tilde q_k).
\l{FenderStratocaster}\ee
Hence, removing the tilde from (\ref{yz1xxxx71}), we have that the EP
implies at least the $D$--dimensional SE
\be
\left(-{\hbar^2\over2m}\Delta_D(q)+\W(q)\right)\psi(q)=0,
\l{yz1fffxxaxx7}\ee
where now $\W(q)$ is more general than (\ref{qiuh}).

\subsection{Schr\"odinger equation and space compactification}

The previous construction can be generalized to get a wider class of potentials.
The point is to first consider the SE in $D+N$ dimensions.
Eq.(\ref{x22x22}) implies
\be
\left(-{\hbar^2\over2m}\Delta_{D+N}(q)+\W(q)\right)\psi(q)=0,
\l{yz1xxxx78zd}\ee
where
\be
\W(q)=\sum_{k=1}^{D+N}\W_k(q_k).
\l{qiuhxxx}\ee
Performing a rototranslation we have
\be
\left(-{\hbar^2\over2m}
\Delta_{D+N}(\tilde q)+\tilde\W(\tilde q)\right)\tilde\psi(\tilde q)=0,
\l{yz1xxxx78fff}\ee
which is equivalent to
\be
\left(-{\hbar^2\over2m}\Delta_D(\tilde q)+\tilde\W(\tilde q)-{\hbar^2\over2m}
{\Delta_N(\tilde q)\tilde\psi(\tilde q)\over\tilde\psi(\tilde q)}
\right)\tilde\psi(\tilde q)=0,
\l{3409}\ee
where
\be
\Delta_D(\tilde q)=\sum_{k=1}^D{\partial^2\over\partial{\tilde q}^2_k},\qquad
\Delta_N(\tilde q)=\sum_{k=D+1}^{D+N}{\partial^2\over\partial{\tilde q}^2_k}.
\l{GibsonLesPaulCustom}\ee
Eq.(\ref{3409}) implies the $D$--dimensional SE
\be
\left(-{\hbar^2\over2m}\Delta_D(\tilde q)+\W_{eff}(\tilde q)\right)\tilde
\psi(\tilde q)=0,
\l{3409bisse}\ee
where
\be
\W_{eff}(\tilde q)=\tilde\W(\tilde q)-{\hbar^2\over2m}{\Delta_N(\tilde q)
\tilde\psi(\tilde q)\over\tilde\psi(\tilde q)},
\l{axd3}\ee
with $\tilde q_{D+1},\ldots,\tilde q_{D+N}$ seen as parameters for the
potential.

The above approach seems to be sufficiently powerful to produce a wide range
of potentials from compactification. It is then interesting to understand
whether an arbitrary potential in $D$--dimensions can be obtained by a
rototranslation and then compactifying $N$--dimensions. One should inductively
define the compactification when $N\longrightarrow\infty$. There are many other
questions concerning this approach. For example, it would be interesting to find
in the case of some relevant potentials the associated minimal value of $N$ such
that it can be reproduced by compactification.

\subsection{Time--dependent case}

Let us consider a potential of the form
\be
V(q,t)=\sum_{k=1}^DV_k(q_k,t).
\l{separacont}\ee
Since
\be
{i\hbar}{\partial\psi_k(q_k,t)\over\partial t}=\left(-{\hbar^2\over2m}
{\partial^2\over\partial q_k^2}+V_k(q_k,t)\right)\psi_k(q_k,t),
\l{z758xx33tt}\ee
it follows that
\be
i\hbar{\partial\psi(q,t)\over\partial t}=\left(-{\hbar^2\over2m}
\Delta_D(q)+V(q,t)\right)\psi(q,t),
\l{z758xx33ttc}\ee
where
\be
\psi(q,t)=\prod_{k=1}^D\psi_k(q_k,t).
\l{separacontpsi}\ee
Since space compactification does not affect the time component, we have
that Eq.(\ref{z758xx33ttc}) holds for all the potentials obtained by the
procedure described in the previous subsection.

\subsection{Relativistic extension}

A property of the EP is that it has a universal character. We already saw that
it implies the cocycle condition also in higher dimension. This cocycle
condition holds also in the relativistic case since it arises as consistency
condition on the transformation properties of $\W$ and $Q$. Thus, in the
relativistic case one arrives to the same cocycle condition (\ref{cociclo3}) and
to its higher dimensional version (\ref{cociclo3xxszs}). In this sense the
quantum correction has always the same structure. For example, in the
one--dimensional relativistic case it is still given by the Schwarzian
derivative of the reduced action times $\hbar^2/4m$. Since the classical
stationary HJ equation has the form
\be
{1\over2m}\left({\partial\S_0^{cl}\over\partial q}\right)^2+\W_{rel}=0,
\l{012relativisticobisse}\ee
where
\be
\W_{rel}\equiv{1\over2mc^2}[m^2c^4-(V-E)^2],
\l{wrelativisticoprbisse}\ee
we immediately see that, according to the EP, its quantum version is
\be
{1\over2m}\left({\partial\S_0\over\partial q}\right)^2+\W_{rel}+
{\hbar^2\over4m}\{\S_0,q\}=0,
\l{012relativisticobissequanticodellecreature}\ee
which has been considered also in \cite{FloydKG}.
Similarly to the non--relativistic case, we have
\be
e^{{2i\over\hbar}\S_0}=e^{i\alpha}{w_{rel}+i\bar\ell\over w_{rel}-i\ell},
\l{odxi9Iu}\ee
where $w_{rel}=\phi^D/\phi$, with $\phi^D$ and $\phi$ denoting two
real linearly independent solutions of the Klein--Gordon equation
\be
-c^2\hbar^2{\partial^2\phi\over\partial q^2}+(m^2c^4-E^2+2EV-V^2)\phi=0.
\l{oixhwJn}\ee

\mysection{Conclusions}\l{conclusions}

Let us make some concluding remarks concerning the main results of our
investigation. We started by
discussing our recent formulation of an EP from which the QSHJE was derived
\cite{1}\cite{1l2}\cite{3prima}\cite{3}. We focused on formulating the general
theory by starting from basic concepts leading in a natural way to a new
formulation of QM. As a result we obtained basic characteristics of QM from a
first principle such as the postulated equivalence of states under spatial
coordinate transformations. The initial sections were devoted to a critical
examination of the rest frame and time parameterization as considered in CM.
This analysis was done by a parallel investigation of a basic relationship
between the Legendre transformation and second--order linear differential
equations first observed in \cite{M1} in the framework of Seiberg--Witten
theory \cite{SW}. We saw that the EP univocally leads to the QSHJE, which in
turn implies the SE, which has a natural interpretation in terms of
trajectories depending on initial conditions parameterized by the constant
$\ell$. These conditions are lost in the SE which we introduced as a
mathematical tool to solve the QSHJE. Thus ${\rm Re}\;\ell$ and ${\rm Im}\;
\ell$ can be seen as a sort of hidden variables depending on fundamental
constants, notably on the Planck and Compton lengths. This suggests that
fundamental interactions may in fact arise in the framework of the EP.

An alternative formulation of QM should reproduce the basic experimental
facts, such as tunnel effect and energy quantization. In Bohm's theory
\cite{Bohm} and stochastic quantization \cite{NelsonGuerra}, the $L^2(\RR)$
condition for the wave--function in the case (\ref{asintoticopiumeno}),
which in fact implies tunnelling and energy quantization, still
arises from a re--interpretation of the Copenhagen axioms. In our formulation
we started with a criticism of the distinguished role of time and were then
led to formulate the Equivalence Postulate in which spatial coordinate
transformations are considered. We considered the transformations in
the framework of HJ theory in which dynamics is described in terms of the
functional relation between $p$ and $q$, so that time parameterization does
not appear. While in the Hamilton and Lagrange equations of motion time
derivatives appear also in the stationary case, in HJ theory the time
parameterization is introduced only after one identifies $p$ and $m\dot q$.
In this way one can consider possible relations among different systems
without introducing time parameterization, a concept that, as we saw, is
related to the privileged nature of the rest frame. As a matter of fact,
this property of HJ theory is in fact at the heart of our formulation of QM.

The above discussion illustrates in which sense our EP differs from that
formulated by Einstein \cite{Einstein}. Loosely speaking, one can say that
formulating Einstein's EP in a strong sense, that is for all possible
potentials, and before introducing time parameterization, leads to QM. The
basic fact is that implementation of the EP univocally leads to the QSHJE
and therefore to the SE. Furthermore, it has been shown that the basic
$L^2(\RR)$ condition is derived from the QSHJE itself. Thus, tunnelling and
energy quantization directly follows from the EP and can be seen as
basic tests of our formulation.

Our formulation has manifest $p$--$q$ duality which traces back to the
involutive nature of the Legendre transformation. A fact clearly expressed in
terms of the Legendre brackets. This feature reflects in the appearance in the
formalism of both $\psi^D$ and $\psi$. In particular, while in the standard
formulation of QM one usually considers a particular solution of the SE as the
relevant quantity, here the relevant quantity is the ratio $w=\psi^D/\psi$
whose role is particularly transparent in considering the geometrical concept of
trivializing map where, like in the case of $\S_0$, $p$ and $\dot q$, $\psi^D$
and $\psi$ appear simultaneously.

Obtaining QM from a principle which is reminiscent of Einstein's EP
\cite{Einstein}, suggests that difficulties underlying quantization of gravity
concern the way in which one usually considers the principles underlying the two
theories. In particular, on general grounds, the fact that GR describes
space--time as intrinsically related to matter, and that
this is done in the framework of well--defined trajectories, indicates that the
usual formulation of QM may have basic conceptual obstructions when
gravitation is taken into account. The fact that QM arises from a
simple principle, may in fact open the way to a reformulation of the problem of
quantum gravity from a completely unexpected and somehow surprising perspective.
In this context we note that the appearance of the M\"obius symmetry and the
role of the quantum potential may suggest that interactions present in
Nature are deeply related to it. We saw that different initial conditions are
related by a M\"obius transformation. Furthermore, these conditions, seen as
hidden variables, have an intrinsic dependence on fundamental constants such as
the Planck length. A concrete signal that our formulation of QM is deeply
connected to Gravitation.

Another distinguished feature of the present formulation is the fact that the
quantum potential is never vanishing. This property is somehow reminiscent of
the relativistic rest energy. This supports the old suspicion that special
relativity itself and QM are in fact related. In the usual formulation this
possible connection is suggested in considering the de Broglie wave--length. In
this respect we note that in sect.\ref{epafc} we introduced a similar quantity
expressed in terms of the conjugate momentum.

The above observations indicate that the problems one meets in formulating a
theory in which the interactions appears to be unified, may in fact be connected
to the usual scheme we have about basic concepts. A peculiarity of this picture
is that it somehow reproduces the historical development of Theoretical Physics
research. So, the standard picture is to think of QM as the basic framework in
which interactions should be described. In this way we have that from one side
there is the QM framework and on the other one there are the four distinguished
fundamental interactions. The investigation we performed in the present paper
may be an alternative starting point towards an effective unification of
fundamental interactions. However, in order to formulate it, we have to
completely reconsider basic concepts such as the meaning of time, the role of
trajectories and their connection with the concept of force at distance. In this
context it is exciting to consider the possibility that the structure of
interactions is in fact intimately related to the quantum potential that, as we
have seen, plays the role of intrinsic energy. More precisely, it can be seen as
a sort of self--energy of the particle whose structure depends on the
external potential. In particular, this self--energy should provide the
key to understand how particle interacts with external forces and can be
considered as particle response to external perturbations. These observations
would lead to replace the above scheme by a single theory.

It is not a fortuitous event that the EP we formulated originated from a
critical analysis of the role of time in CM. This was done after noticing the
peculiar nature of the rest frame. It is actually unavoidable that the classical
concept of time loses its nature in the present formulation. This is somehow a
parallel evolution of what happens in GR.

Another feature concerns the similarity of our formulation with the theory of
uniformization of Riemann surfaces. It is even more frequent that this
fascinating theory enters, in different contexts, in basic physical topics. We
have seen that the concept of trivializing map is essentially the same as the
one in Riemann surfaces theory. We believe that many of the geometrical concepts
underlying our approach are related to the fact that the topic
concerning complex structures of Riemann surfaces may be intimately related to
the appearance of complex numbers in QM.

\vspace{.333cm}

\noindent
{\bf Acknowledgements}.
It is a pleasure to thank M. Appleby, D. Bellisai, M. Bochicchio, L. Bonora,
R. Carroll, F. De Felice, G.F. Dell'Antonio, F. Guerra, F. Illuminati,
A. Kholodenko, A. Kitaev, G. Marmo, R. Nobili, R. Onofrio, F. Paccanoni,
P. Sergio, G. Travaglini, G. Vilasi and R. Zucchini for stimulating discussions.
Special thanks are due to E.R. Floyd for interesting discussions on the QSHJE,
to G. Bertoldi for contributions in some aspects of the present formulation
of QM and to G. Bonelli, E. Gozzi, J.M. Isidro, P.A. Marchetti, P. Pasti
and M. Tonin for basic discussions on the foundations of QM.

Work supported in part by DOE Grant No.\ DE--FG--0287ER40328 (AEF)
and by the European Commission TMR programme ERBFMRX--CT96--0045 (MM).


\newpage

\begin{thebibliography}{99}

\bibitem{1} A.E. Faraggi and M. Matone, Phys. Lett. {\bf B450} (1999) 34,
hep-th/9705108; Phys. Lett. {\bf B437} (1998) 369, hep-th/9711028.
\bibitem{1l2} A.E. Faraggi and M. Matone, Phys. Lett. {\bf B445} (1999)
357, hep-th/9809126.
\bibitem{Einstein} A. Einstein, Annalen der Phys. {\bf 49} (1916) 769.
\bibitem{Bohm} D. Bohm, Phys. Rev. {\bf 85} (1952) 166; ibidem 180.
\bibitem{Bornpq} M. Born, Rev. Mod. Phys. {\bf 21} (1949) 463.
\bibitem{M1} M. Matone, Phys. Lett. {\bf B357} (1995) 342, hep-th/9506102.
\bibitem{SW} N. Seiberg and E. Witten, Nucl. Phys. {\bf B426} (1994) 19,
hep-th/9407087; Nucl. Phys. {\bf B431} (1994) 484, hep-th/9408099.
\bibitem{FM} A.E. Faraggi and M. Matone, Phys. Rev. Lett. {\bf 78} (1997) 163,
hep-th/9606063.
\bibitem{Carroll} R. Carroll, hep-th/9607219; hep-th/9610216; hep-th/9702138;
hep-th/9705229; Nucl. Phys. {\bf B502} (1997) 561; Lect. Notes Phys. {\bf 502},
Springer (Berlin, 1998), pp. 33 -- 56.
\bibitem{Vancea} I.V. Vancea, gr-qc/9801072.
\bibitem{BOMA} G. Bonelli and M. Matone, Phys. Rev. Lett. {\bf 77} (1996) 4712,
hep-th/9605090.
\bibitem{3prima} A.E. Faraggi and M. Matone, Phys. Lett. {\bf 249A} (1998) 180,
hep-th/9801033.
\bibitem{3} A.E. Faraggi and M. Matone, Phys. Lett. {\bf B445} (1998) 77,
hep-th/9809125.
\bibitem{MugaSalaSnider} J.G. Muga, R. Sala and R.F. Snider, Phys. Scripta
{\bf 47} (1993) 732.
\bibitem{Floyd82b} E.R. Floyd, Phys. Rev. {\bf D26} (1982) 1339.
\bibitem{BFM} G. Bertoldi, A.E. Faraggi and M. Matone, paper in preparation.
\bibitem{geometricquantization} V.W. Guillemin and S. Sternberg, {\it Symplectic
Techniques in Physics}, Cambridge Univ. Press (Cambridge, 1990).\\ B. Kostant,
in ``Lectures in Modern Analysis and Applications III'', Lect. Notes Math.
{\bf 170} Springer (Berlin, 1970).\\ A.A. Kirillov, in ``Dynamical Systems
III'', Ed. by V.I. Arnold, Springer (Berlin, 1989) p. 137.\\ N.M.J. Woodhouse,
{\it Geometric Quantization}, Clarendon Press (Oxford, 1992).\\
J. Sniatycki, {\it Geometric Quantization and Quantum Mechanics},
Springer--Verlag (New York, 1980).\\ J.--L. Brylinski, {\it Loop Spaces,
Characteristic Classes and Geometric Quantization}, Birkh\"auser (Basel, 1993).
\bibitem{KlauderPerelomov} J.R. Klauder and B.-S. Skagerstam, {\it Coherent
States}, World Scientific (Singapore, 1985).\\ A.M. Perelomov, {\it Generalized
Coherent States and Their Applications}, Springer (Berlin, 1986).
\bibitem{NelsonGuerra} E. Nelson, {\it Quantum Fluctuations}, Princeton Univ.
Press (Princeton, 1985).\\ F. Guerra, Phys. Rept. {\bf 77} (1981) 263.
\bibitem{Gozzi} E. Gozzi, Phys. Lett. {\bf B158} (1985) 489; erratum, ibidem
{\bf B386} (1996) 495.
\bibitem{Periwal} V. Periwal, Phys. Rev. Lett. {\bf 80} (1998) 4366,
hep-th/9709200.
\bibitem{Anandan} J.S. Anandan, {\it Classical and Quantum Physical Geometry},
in ``Potential, Entanglement and Passion--at--a--Distance -- Quantum
Mechanics Studies for Abner Shimony'', Vol. 2. Ed. by R.S. Cohen, M. Horne and
J. Stachel, Kluwer (Dordrecht, Holland 1997), pp. 31 -- 52, gr-qc/9712015.
\bibitem{OnofrioViola} L. Viola and R. Onofrio, Phys. Rev. {\bf D55} (1997) 455,
quant-ph/9612039.\\ R. Onofrio and L. Viola, Mod. Phys. Lett. {\bf A12} (1997)
1411, quant-ph/9706004.
\bibitem{Cini} M. Cini, Annals Phys. {\bf 273} (1999) 99, quant-ph/9807001; {\it
Quantum Theory Without Waves: a Way of Eliminating Quantum Mechanical
Paradoxes?}, Roma 1 preprint n.1103.
\bibitem{GellMannHartle} M. Gell--Mann and J.B. Hartle, {\it Quantum Mechanics
in the Light of Quantum Cosmology}, in W.H. Zurek, ``Complexity, Entropy, and
the Physics of Information'', Reading, Addison--Wesley (MA, 1990), pp. 425 --
458.
\bibitem{GisinPercival} N. Gisin and I.C. Percival, J. Phys. {\bf A25} (1992)
5677.\\ I.C. Percival, Proc. Royal Soc. London {\bf A447} (1994) 189.
\bibitem{Adler} S.L. Adler, {\it Quaternionic Quantum Mechanics and Quantum
Fields}, Oxford Univ. Press (New York, 1995).
\bibitem{GRWP} G.C. Ghirardi, A. Rimini and T. Weber, Phys. Rev. {\bf D34}
(1986) 470.\\ P. Pearle, Phys. Rev. {\bf A39} (1989) 2277.
\bibitem{KochenSpecker} S. Kochen and E.P. Specker, Jour. Math. Mech. {\bf 17}
(1967) 59.
\bibitem{Bell1} J.S. Bell, {\it Speakable and Unspeakable in Quantum Mechanics},
Cambridge Univ. Press (Cambridge, 1987).
\bibitem{Brown} H.R. Brown, {\it Bell's Other Theorem and its Connection with
Non--Locality. Part 1.}, in ``Bell's Theorem and the Foundations of Modern
Physics'', A. van de Merwe, F. Selleri and G. Tarozzi, Editors, World Scientific
(Singapore, 1992).
\bibitem{IshamButterfield} C.J. Isham and J. Butterfield, quant-ph/9803055;
Int. J. Theor. Phys. {\bf 38} (1999) 827, quant-ph/9808067.
\bibitem{Arnold} V.I. Arnold, {\it Mathematical Methods of Classical Mechanics},
Springer--Verlag (New York, 1980).
\bibitem{Gunning} R.C. Gunning, {\it Lectures on Riemann Surfaces}, Princeton
Univ. Press (Princeton, 1966).
\bibitem{Cartan} E. Cartan, Ann. Ec. Normale, {\bf 21} (1904) 153.
\bibitem{GuillStern} V.W. Guillemin and S. Sternberg, Bull. Amer. Math. Soc.
{\bf 70} (1964) 16.
\bibitem{Fuks} D.B. Fuks, {\it Cohomologies of Infinite--Dimensional
Lie Algebras}, Consultants Bureau (New York, 1986).
\bibitem{Segal} G. Segal, Comm. Math. Phys. {\bf 80} (1981) 301.
\bibitem{Ovsienko1} V. Yu. Ovsienko, {\it Lagrange Schwarzian Derivative and
Symplectic Sturm Theory}, CPT -- 93/P.2890.
\bibitem{Messiah} A. Messiah, {\it Quantum Mechanics}, Vol. 1, North--Holland
(Amsterdam, 1961).
\bibitem{Holland} P.R. Holland, {\it The Quantum Theory of Motion}, Cambridge
Univ. Press (Cambridge, 1993).
\bibitem{Floyd8X} E.R. Floyd, Phys. Rev. {\bf D25} (1982) 1547; {\bf D29} (1984)
1842; Found. Phys. Lett. {\bf 9} (1996) 489, quant-ph/9707051; quant-ph/9708007;
Int. J. Mod. Phys. {\bf A14} (1999) 1111, quant-ph/9708026.
\bibitem{Floyd86} E.R. Floyd, Phys. Rev. {\bf D34} (1986) 3246; Phys. Lett.
{\bf 214A} (1996) 259.
\bibitem{RosenA} A. Rosen, Am. J. Phys. {\bf 32} (1964) 377.
\bibitem{BS} D.B. Berkowitz and P.D. Skiff, Am. J. Phys. {\bf 40} (1972) 1652.
\bibitem{HA} J.O. Hirschfelder, A.C. Christoph and W.E. Palke, J. Chem. Phys.
{\bf 61} (1974) 5435.
\bibitem{SchillerRosen} R. Schiller, Phys. Rev. {\bf 125} (1962) 1100; ibidem
1109.\\
A. Rosen, Am. J. Phys. {\bf 32} (1964) 597; Found. Phys. {\bf 16} (1986) 687.
\bibitem{Anderson} A. Anderson, Annals Phys. {\bf 232} (1994) 292,
hep-th/9305054.
\bibitem{Dirac} P.A.M. Dirac, {\it The Principle of Quantum Mechanics},
4th. ed., Oxford Univ. Press (Oxford, 1958).
\bibitem{Weyl} H. Weyl, {\it The theory of Groups and Quantum Mechanics},
2nd. ed., Dover (New York, 1950).
\bibitem{BHJ} M. Born, W. Heisenberg and P. Jordan, Ztschr. f. Phys. {\bf 35}
(1926) 557.\\ W. Heisenberg, Math. Ann. {\bf 95} (1926) 683.\\
P.A.M. Dirac, Proc. Roy. Soc. {\bf A110} (1926) 561.
\bibitem{FarkasKra} H.M. Farkas and I. Kra, {\it Riemann Surfaces}, 2nd. ed.,
Springer--Verlag (New York, 1992).
\bibitem{JonesSingerman} G.A. Jones and D. Singerman, {\it Complex Functions},
Cambridge Univ. Press (Cambridge, 1987).
\bibitem{Uniformization} M. Matone, Int. J. Mod. Phys. {\bf A10} (1995) 289.
\bibitem{Bers1} L. Bers, {\it On Moduli of Riemann Surfaces}, ETH, Z\"urich,
1964, (mimeographed); {\it Automorphic Forms and General Teichm\"uller Spaces},
Proceedings of the Conference on Complex Analysis, Minneapolis 1964,
Springer--Verlag (New York, 1965).
\bibitem{BeurlingAhlfors} A. Beurling and L.V. Ahlfors, Acta Math. {\bf 96}
(1956) 124.
\bibitem{Pekonen} O. Pekonen, Phys. Lett. {\bf B252} (1990) 555; J. Geom. Phys.
{\bf 15} (1995) 227, hep-th/9310045.
\bibitem{PRDM} M. Matone, Phys. Rev. {\bf D53} (1996) 7354, hep-th/9506181.
\bibitem{CallanWilczek} C.G. Callan and F. Wilczek, Nucl. Phys. {\bf B340}
(1990) 366.
\bibitem{tHooftSusskind} G. 't Hooft, {\it Dimensional Reduction in Quantum
Gravity}, Essay dedicated to Abdus Salam. Published in Salamfest 1993; 0284 --
296, gr-qc/93100026.\\ C.R. Stephens, G. 't Hooft and B.F. Whiting, Class.
Quant. Grav. {\bf 11} (1994) 621, gr-qc/9310006.\\
L. Susskind, J. Math. Phys. {\bf 36} (1995) 6377, hep-th/9409089.
\bibitem{Maldacena} J. Maldacena, Adv. Theor. Math. Phys. {\bf 2} (1998) 231,
hep-th/9711200.
\bibitem{SusskindWitten} L. Susskind and E. Witten, hep-th/9805114.
\bibitem{Bers2} L. Bers, {\it Universal Teichm\"uller Spaces}, in
``Analytic methods in mathematical physics'', R.P. Gilbert, R.G. Newton,
Editors, Gordon and Breach (New York, 1968).
\bibitem{Flanders} H. Flanders, J. Diff. Geom. {\bf 4} (1970) 515.
\bibitem{Ermakov} V.P. Ermakov, Univ. Izv. Kiev {\bf 20} (1880) 1.
\bibitem{Lewis} H.R. Lewis Jr., Phys. Rev. Lett. {\bf 18} (1967) 510; erratum,
ibidem 636; J. Math. Phys. {\bf 9} (1968) 1976.\\ H.R. Lewis Jr. and W.B.
Riesenfeld, J. Math. Phys. {\bf 10} (1969) 1458.
\bibitem{Milne} W.E. Milne, Phys. Rev. {\bf D35} (1930) 863.
\bibitem{svariati} M. Lutzky, Phys. Lett. {\bf 68A} (1978) 3; {\bf 75A} (1979)
8; J. Phys. {\bf A11} (1978) 249; {\bf A12} (1979) 973.\\ P.G.L. Leach, J. Math.
Phys. {\bf 18} (1977) 1902; {\bf 20} (1979) 96; {\bf 21} (1980) 300.\\ J.R. Ray
and J.L. Reid, Phys. Lett. {\bf 71A} (1979) 317; J. Math. Phys. {\bf 20} (1979)
2054; {\bf 21} (1980) 1583.\\ H.J. Korsch, Phys. Lett. {\bf 74A} (1979) 294;
{\bf 109A} (1985) 313.\\ H.J. Korsch and H. Laurent, J. Phys. {\bf B14} (1981)
4213.
\bibitem{Arnold2} V.I. Arnold, {\it Ordinary Differential Equations}, MIT Press
(1973).
\bibitem{Ovsienko2} V. Yu. Ovsienko, {\it Hook Law and Denogardus Great Number},
(Russian) Kvant, N.8, P.B -- 16 (1989).
\bibitem{Thirring} W. Thirring, {\it Quantum Mechanics of Atoms and Molecules},
Vol. 3, Springer--Verlag (New York, 1979).
\bibitem{MagnusOberhettinger} W. Magnus and F. Oberhettinger, {\it Formulas and
Theorems for the Functions of Mathematical Physics}, Chelsea (New York, 1954).
\bibitem{FloydKG} E.R. Floyd, Int. J. Theor. Phys. {\bf 27} (1988) 273.

\end{thebibliography}
\end{document}